\documentclass[twoside,12pt]{article}
\usepackage{epsfig}
\usepackage{float}

\usepackage{amsmath}
\usepackage{amssymb}
\usepackage{bbm}
\usepackage{booktabs}
\usepackage{graphicx}

\def\p{\vec p}
\def\P{\vec P}
\def\q{\vec q}

\def\e{\vec e}

\def\ps{\phi_S}
\def\pp{\phi_P}
\def\vpp{{\vec\phi}_P}

\def\x{\vec x}
\def\0{\vec 0}
\def\Zm{\mathbbm{Z}}

\def\nave{{\bar n}}

\def\K{{\mathbf K}}
\def\E{{\mathbf E}}
\def\F{{\mathbf F}}

\def\mk{\vec k}
\def\mL{\mathcal{L}}

\def\V{\mathcal{V}}
\def\pbp{\bar\psi\psi}
\newcommand{\eq}[1]{{Eq.~({\ref{#1}})}}
\newcommand{\Eq}[1]{{Eq.~({\ref{#1}})}}
\newcommand{\Fig}[1]{{Fig.~{\ref{#1}}}}
\newcommand{\one}{\mathbbm{1}}
\newcommand{\equalover}[1]{{\boldmath{\overset{#1}{=}}}}
\newcommand{\longrightarrowover}[1]{{\boldmath{\overset{#1}{\longrightarrow}}}}

\newcommand{\laplace}{\bigtriangleup}  

\usepackage{hyperref}
\hypersetup{
  colorlinks, breaklinks,
  linkcolor=blue,
  citecolor=blue,
  linktoc=all
}

\newcommand{\ave}[1]{{\langle{#1}\rangle}}

\newcommand{\beq}{\begin{equation}}
\newcommand{\eeq}{\end{equation}}
\newcommand{\bea}{\begin{eqnarray}}
\newcommand{\eea}{\end{eqnarray}}

\begin{document}

\title{ \vspace{1cm} Inhomogeneous chiral condensates} 
\author{Michael Buballa$^1$ and Stefano Carignano$^2$\\
\\
$^1$Theoriezentrum, Institut f\"ur Kernphysik, Technische Universit\"at Darmstadt, Germany\\
$^2$Department of Physics, The University of Texas at El Paso, USA}
\maketitle

\begin{abstract}
The chiral condensate, which is constant in vacuum, may become spatially modulated
at moderately high densities where in the traditional picture of the
QCD phase diagram a first-order chiral phase transition occurs.
We review the current status of this idea,
which originally dates back to Migdal's pion condensation,
but recently received new momentum through
studies on the nature of the chiral critical point
and by the conjecture of a quarkyonic-matter phase.
We discuss how these nonuniform phases emerge
in generalized Ginzburg-Landau analyses
as well as in specific calculations,
both within effective models and in
Dyson-Schwinger or large-$N_c$ approaches to QCD.
Questions about the most favored shape 
of the modulations and its dimension,
and about the effects of 
nonzero isospin chemical potential, 
strange quarks,
color superconductivity, 
and external magnetic fields
on these inhomogeneous phases
will be addressed as well.
\end{abstract}
\eject
\tableofcontents

\section{Introduction}

Quantum Chromodynamics (QCD)
is nowadays the widely accepted fundamental theory of strong interactions.
 Unlike the other theories composing the standard model,
 QCD exhibits some peculiar nonperturbative features at low energies
 which render its theoretical treatment extremely challenging.
 The two most prominent features in this context are confinement, which 
 binds quarks inside hadrons and makes it impossible to observe them 
 as asymptotic isolated particles, 
 and spontaneous chiral symmetry breaking, which 
 leads to the generation
 of a chiral condensate 
 $\ave{\bar\psi\psi} \neq 0$
 through which particles acquire 
 a dynamical mass.
 These features which characterize the QCD vacuum are naturally expected to be lost if the energy of the system increases, for example through
 the introduction of external factors
like a finite temperature or density.
 As such, strong-interaction matter is expected to experience 
 some sort of phase transition from a confined and chirally broken 
 to a deconfined and symmetric phase
 as these external parameters increase beyond given thresholds.

Thanks to its weak-coupling behaviour in the high-energy regime
\cite{Gross:1973,Politzer:1973}
the behaviour of QCD for very high temperatures and chemical potentials is nowadays well understood: 
the high T phase is a weakly-coupled quark-gluon plasma \cite{Rischke:2003mt}, 
while cold and dense matter is expected to form a color-superconductor 
\cite{Collins:1974ky,Barrois:1977,Bailin:1983},
for a review, see Ref.~\cite{Alford:2008}.

The intermediate region between vacuum and these asymptotic
phases 
is however still the subject of an intense debate and no definite agreement has yet been found.
 In particular, while great progress has been made in the past few years in characterizing
 strong-interaction matter at finite temperature thanks
to the development of ab-initio lattice calculations \cite{lattice2013}
 and to increasingly accurate data from heavy-ion collision experiments (see, e.g., \cite{qm2012}), 
 the nature of QCD at finite densities (or baryon chemical potentials) is still poorly understood
(see \cite{Fukushima:2010bq,Fukushima:2013rx} for reviews on recent theoretical developments).
On the experimental side, most heavy-ion data focuses on the high-temperature 
regime \cite{qm2012},
and only recently the attention has started shifting to the finite-density region,
 particularly with the new beam energy scan runs at RHIC in Brookhaven \cite{Kumar:2013cqa}
 and the upcoming facilities
 FAIR in Darmstadt~\cite{Friman:2011zz} and NICA in Dubna
\cite{NICA}.
From a theoretical point of view, since lattice simulations at
finite chemical potentials
are plagued by the sign problem \cite{deForcrand:2010ys},
in order to investigate the intermediate-density region
it is necessary to employ 
other
approaches.

The most widely employed tools in this context are effective models which, by sharing some relevant symmetries with QCD, are expected to reproduce 
some of its characteristic properties. 
Among them we find most notably the Nambu--Jona-Lasinio
and the Quark-Meson model (or linear sigma model with quarks), 
which are built by integrating out gluonic degrees of freedom 
and incorporating them in effective low-energy couplings.
While being rather crude simplifications compared to full QCD, 
these models are able to describe 
the phenomenon of spontaneous chiral symmetry breaking 
as well as its restoration by temperature and density, and provide
a numerically accessible framework which can be employed to get 
a first insight on the behaviour of strong-interacting matter. 

More recently, a whole new set of theoretical methods 
based on QCD allowed to obtain an unprecedented understanding on the 
fundamental properties of strong-interaction matter and the nature of its
phase transition. 
Functional methods like the Dyson-Schwinger equations and 
the functional renormalization group formalism provide a 
self-consistent non-perturbative framework which can be employed to 
systematically improve our knowledge of QCD at finite temperatures and densities (for recent reviews, see e.g. \cite{Fischer:2006ub,Pawlowski:2005xe,Gies:2006wv}).

Until relatively recently, the phase structure of QCD was believed to be 
quite simple. The widely accepted picture was that of a low-temperature and density confined and chirally broken phase, separated from a high-density and temperature deconfined and symmetric one by a phase transition \cite{Cabibbo:1975ig}. Lattice and experimental data support the idea that 
at low densities the transition is actually a smooth crossover (which would turn into a second-order transition if the low-energy degrees of freedom
were only two massless quark species), while 
at low temperatures a first-order transition was expected. The two different
transitions would then be separated by a (tri)critical point, whose location 
could in principle be determined experimentally \cite{Stephanov:1998dy,Gavai:2014ela}.
At high densities and low temperatures, the color-superconducting phase
would then set in, although estimates on its exact location and nature 
are model-dependent \cite{Alford:2008}.

In the past few years a growing number of indications has however 
been suggesting that
dense matter might actually exhibit a more complicated phase structure,
and in particular that the low temperature region might be
characterized by the formation of crystalline structures, an option that 
was mostly overlooked in the past.

The idea of inhomogeneous phases in dense systems
is certainly not a new one. 
Charge and spin density waves are commonly found in solid-state physics (for a review see, e.g., \cite{Gruner}), 
and crystalline 
phases
have also been discussed long time ago 
for superconductors
by Fulde and Ferrell as well as by Larkin and Ovchinikov \cite{Fulde:1964,LO}
and more recently for color superconductors \cite{Alford:2000ze}
(see also \cite{Casalbuoni:2003wh,Anglani:2013gfu} for reviews).
Inhomogeneous phases have also been shown to appear in imbalanced cold  atom Fermi gases \cite{Maeda:2012gw,Roscher:2013cma}.

In the context of the strong interaction, 
already in 1960 Overhauser discussed the idea of density waves in nuclear matter~\cite{Overhauser:1960},
while in the 1970s and early 1980s there was much activity related to inhomogeneous pion
condensation~\cite{BrownWeise,Migdal:1978az}, 
pioneered by Migdal~\cite{Migdal:1971cu,Migdal:1973zm}.

The Skyrme picture describing nucleons as chiral solitons
has also been successfully employed in describing dense 
strong-interaction matter, and seems to predict 
the formation of three-dimensional crystalline structures 
\cite{Klebanov:1985qi,Goldhaber:1987}. 

The present review focuses on the study of inhomogeneous chiral symmetry breaking, that is, on the formation of phases characterized by a 
nonzero spatially varying chiral condensate which breaks translational invariance. 
The idea that the QCD ground state at high densities might be characterized by such a structure 
was first proposed by Deryagin, Grigoriev and Rubakov in a pioneering study of quark matter in the limit of a large number of colors~\cite{DGR}.
Subsequent refinements on this study, focusing on the competition of this kind of inhomogeneous phase with color-superconductivity, have been discussed by Shuster and Son \cite{Shuster}, Park et al.\@ \cite{Park:1999bz} as well as by Rapp, Shuryak and Zahed \cite{Rapp:2000zd}. 
 Model analyses of 
inhomogeneous chiral symmetry breaking started in the early 1990s 
by Kutschera et al.\@ \cite{Kutschera:1989yz}
and have become increasingly systematic, particularly thanks to seminal papers by Nakano and Tatsumi \cite{NT:2004} and Nickel \cite{Nickel:2009wj,Nickel:2009ke}.

On a qualitative level the formation of inhomogeneous chiral condensates can be explained
by analogy to (color-) superconducting matter under stress.
The Cooper instability in dense Fermi systems is a consequence of the fact that directly
at the Fermi surface, particles can be created with no free-energy cost. Hence, if there
is an attractive force between the particles, the system can lower its free energy by 
creating a large number of
particle-particle pairs (Cooper pairs),  until eventually these pairs form a 
condensate. 
In BCS theory~\cite{BCS} the Cooper pairs consist of particles with opposite momenta, 
so that the total momentum of each pair vanishes and the condensate is constant in space. 
Pairs with nonvanishing total momentum are in general disfavored by their kinetic energy 
and by phase space.

The situation is more complicated, however, if the pairing takes place between two particle species 
with different Fermi momenta, like, e.g.,  electrons with opposite spin orientations in a magnetic field. 
For pairs with vanishing total momentum, the participants then cannot be both at their
respective Fermi surface. Hence, their creation now costs free energy, and
the pairing only occurs if this energy is outweighed by the condensation energy.
For weak coupling it was found by Chandrasekhar and Clogston that this is the case 
if the difference $\delta k_F$ of the Fermi momenta of the two species is less than 
$\Delta_\text{BCS}/\sqrt{2}$, where $\Delta_\text{BCS}$
is the pairing gap~\cite{Clogston:1962zz,Chandrasekhar}.
For higher values of $\delta k_F$, BCS pairing is disfavored against the normal-conducting phase,
which is reached in a first-order phase transition if only homogeneous phases are considered.

Allowing for inhomogeneous phases, on the other hand, corresponding to pairs with 
nonvanishing total momentum, both particles can stay near their Fermi surfaces, 
independent of $\delta k_F$. 
As a consequence, such solutions, which are disfavored against the BCS solution
at small $\delta k_F$, can become favored over both, the BCS and the normal conducting 
phase at higher $\delta k_F$, typically in some window around the Chandrasekhar-Clogston
limit. 
The emergence of crystalline (color-) superconducting phases is thus a consequence
of the stress exerted by $\delta k_F$ on the BCS pairs. 

In a similar way, inhomogeneous chiral condensates can become favored if there 
is stress on the usual homogeneous chiral condensate,
which in this case is exerted by a large chemical potential. 
This has been discussed nicely in Ref.~\cite{Kojo:2009}
and we briefly summarize their arguments:

In vacuum, chiral-symmetry breaking occurs through the pairing of left-handed quarks with
right-handed antiquarks and vice versa, resulting in a nonvanishing chiral condensate
$\ave{\bar\psi\psi} = \ave{\bar\psi_R\psi_L} + \ave{\bar\psi_L\psi_R}$.
Neglecting the small quark mass, quarks and antiquarks can be created without energy
cost, similar to the situation in the superconductor at $\delta k_F=0$.

\begin{figure}[htb]
\centering
\includegraphics[angle=0,width=.2\textwidth]{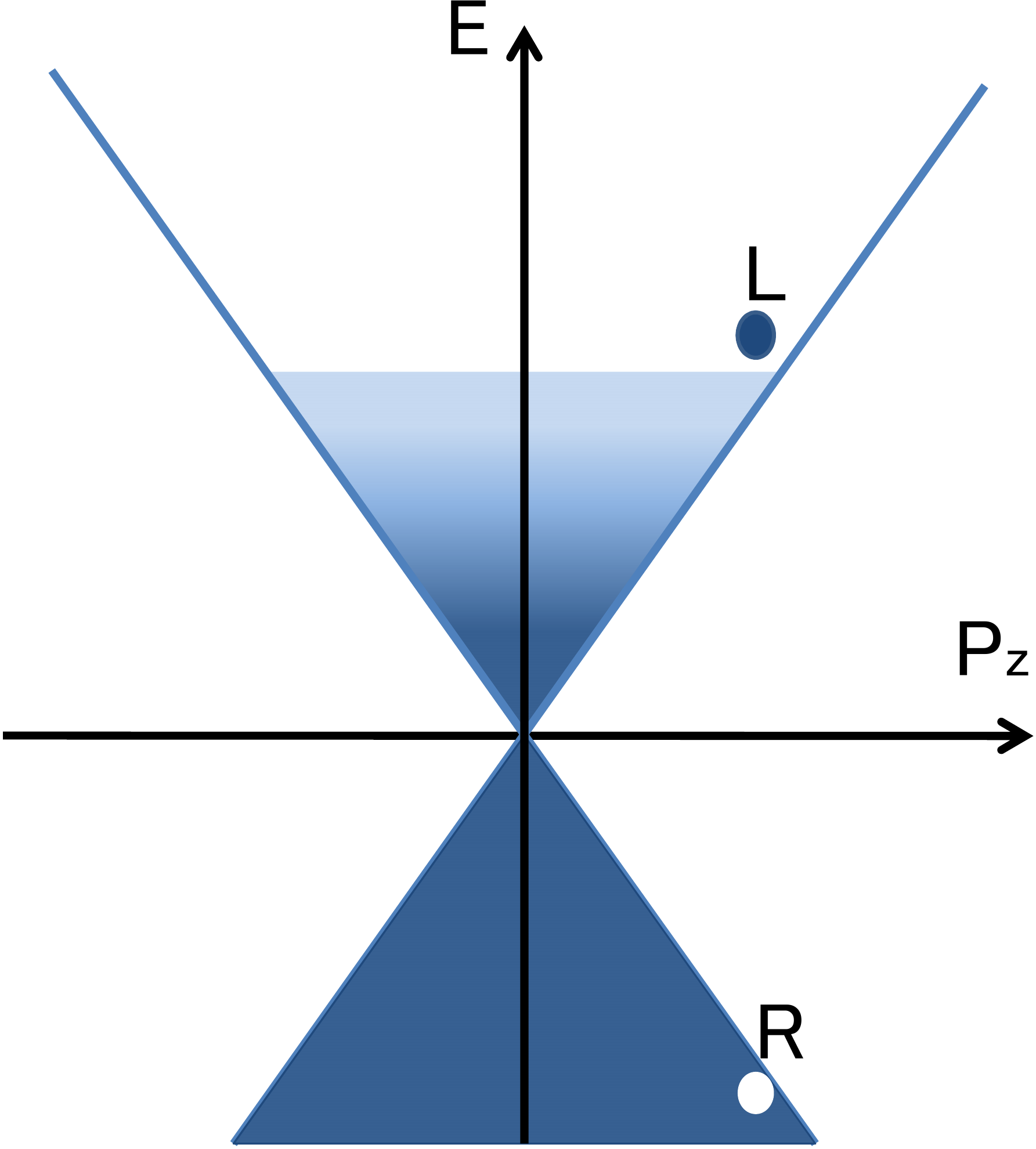}
\hspace{9mm}
\includegraphics[angle=0,width=.2\textwidth]{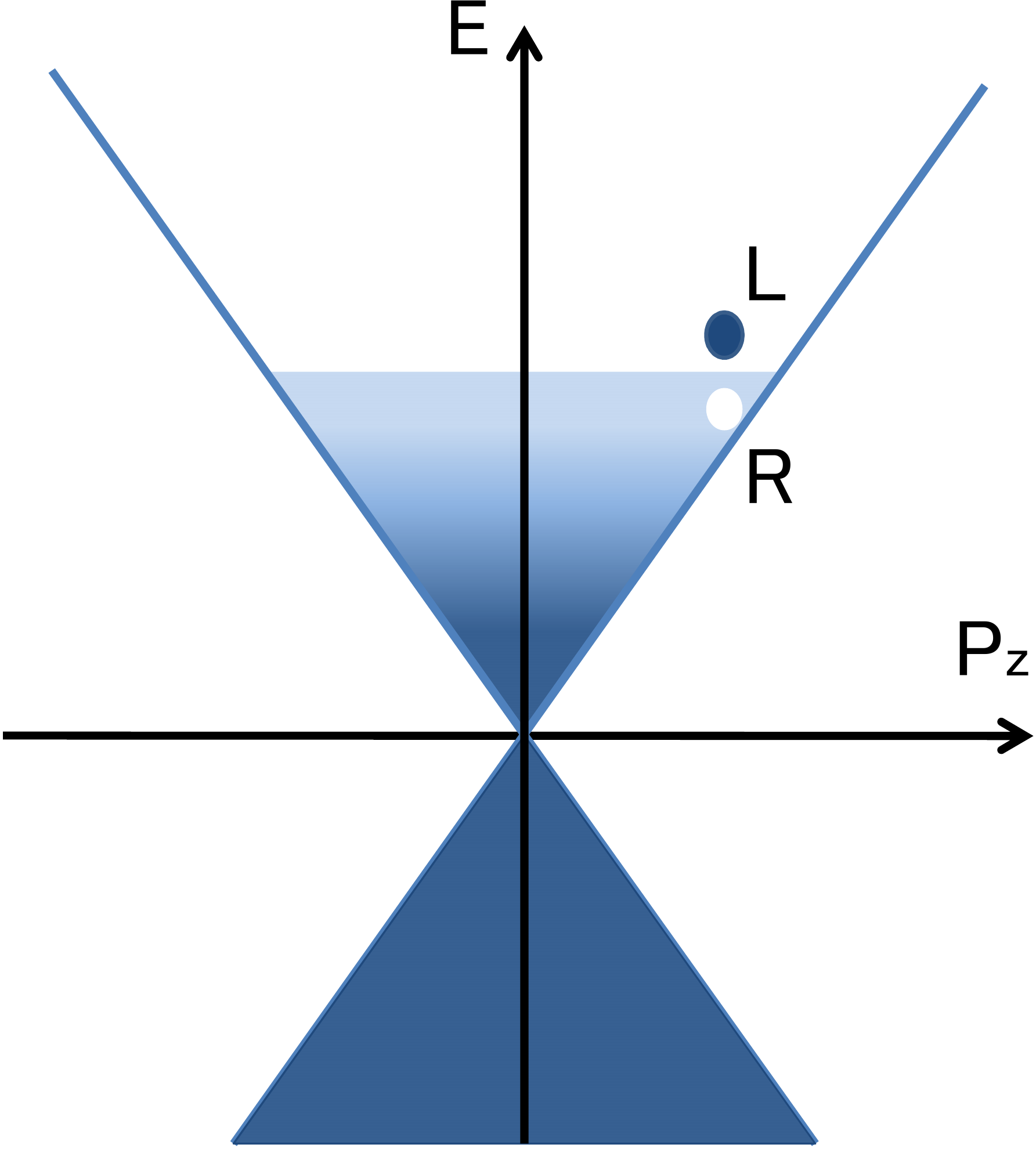}
\hspace{9mm}
\includegraphics[angle=0,width=.2\textwidth]{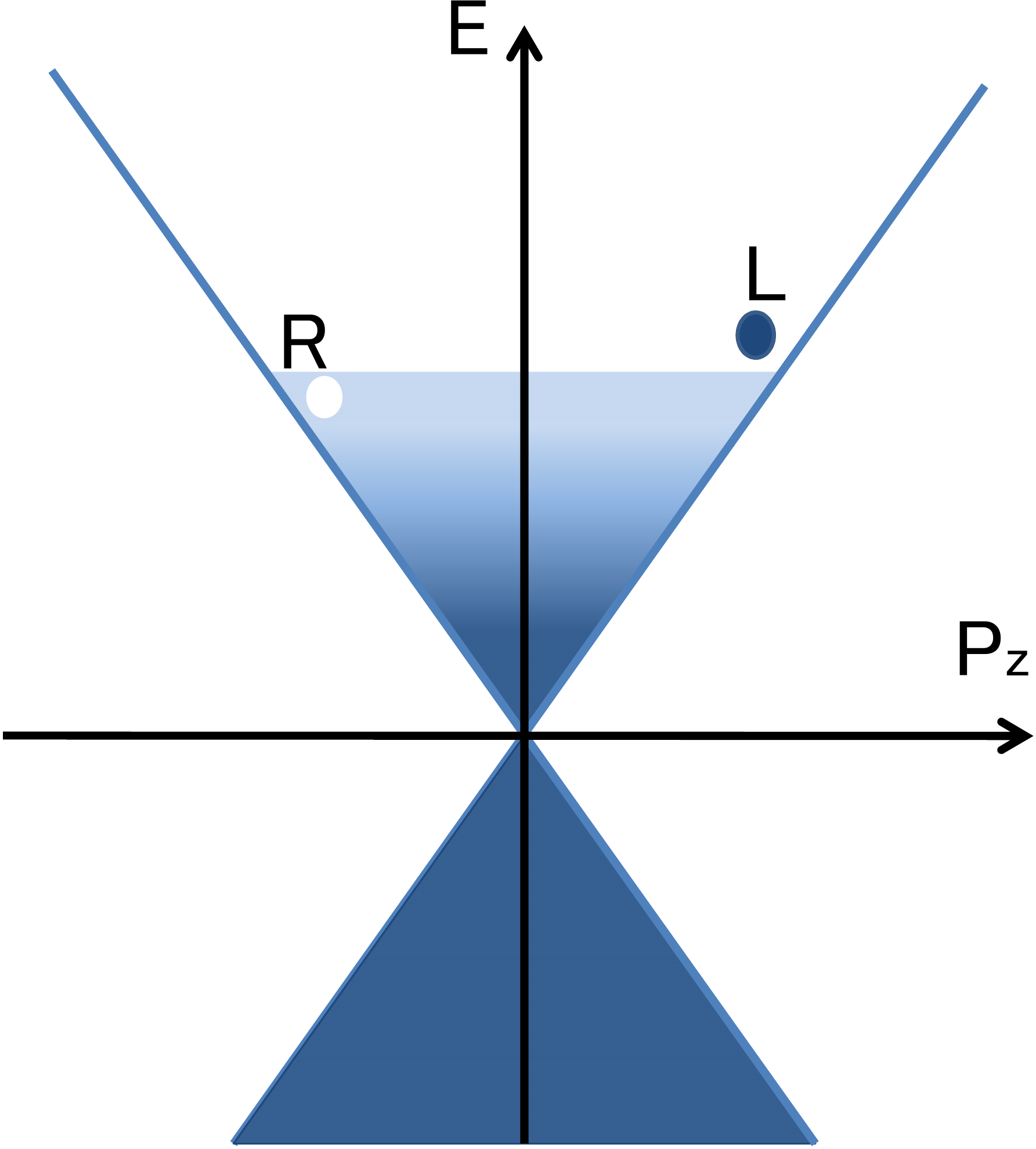}
\hspace{9mm}
\includegraphics[angle=0,width=.2\textwidth]{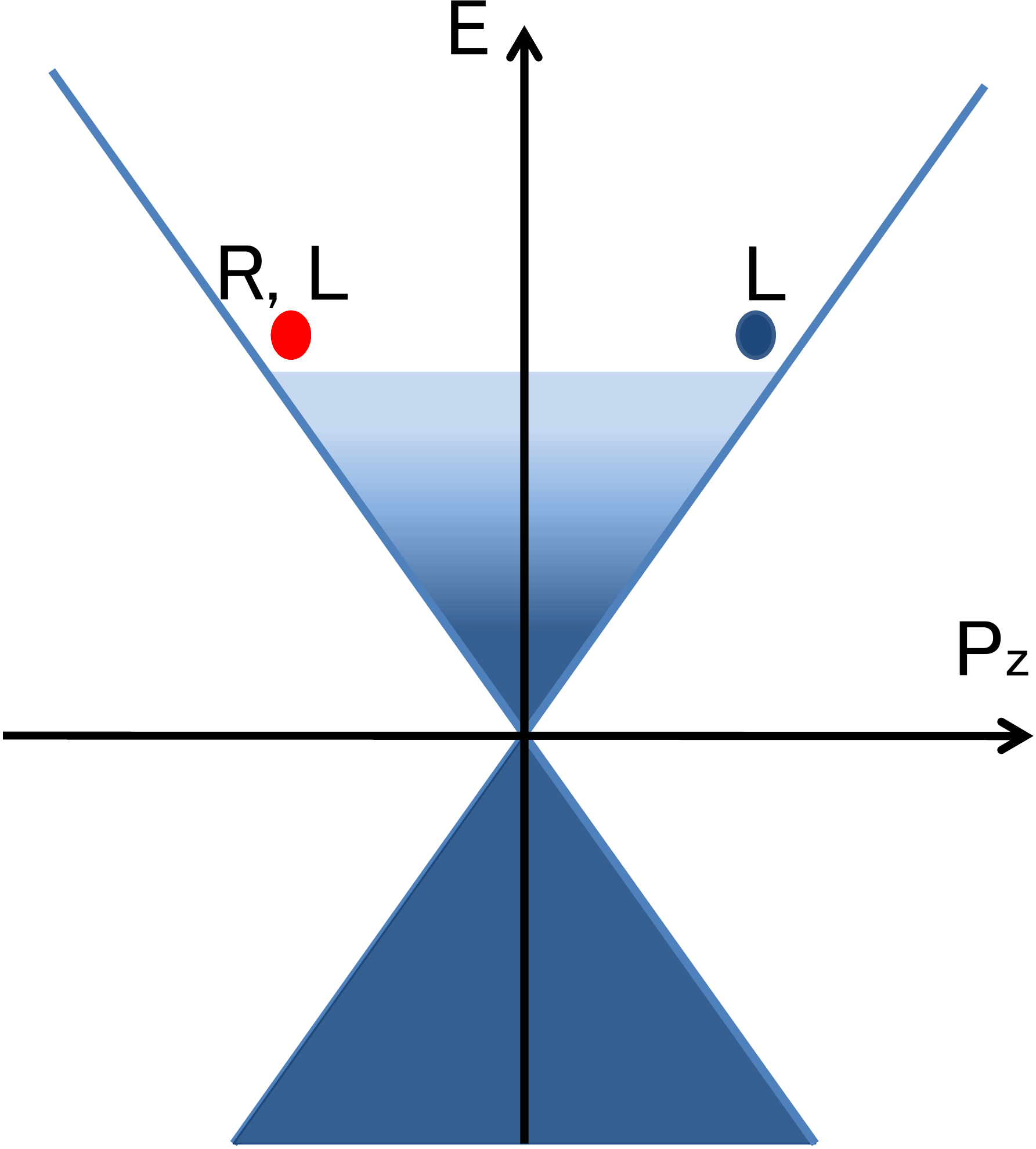}
\caption{Different pairing mechanisms in the presence of a Fermi sea. 
From left to right:
(a) quark-antiquark pairing,
(b) quark-hole pairing with vanishing total momentum (``exciton''),
(c) quark-hole pairing with nonzero total momentum,
(d) quark-quark pairing generating color superconductivity.
From Ref.~\cite{Kojo:2009}. 
}
\label{fig:kojo_pairing}
\end{figure}

If we now turn to finite density, corresponding to a quark chemical potential $\mu > 0$,
the energy needed to excite a quark-antiquark pair with vanishing total momentum 
(left panel of \Fig{fig:kojo_pairing}) is at least equal to $2\mu$.
At large values of $\mu$ this can no longer be outweighed by the condensation energy,
so that the mechanism which leads to chiral-symmetry breaking in vacuum eventually 
breaks down. 

Alternatively, chiral symmetry could be broken by pairing, e.g., a left-handed quark with a
right-handed quark hole, which costs almost no energy if the quark and the hole are close
to the Fermi surface. 
The pair can also have vanishing total momentum if we pair a quark with momentum $\p$
with a hole with momentum $-\p$ (corresponding to a removed quark with momentum $\p$
below the Fermi surface). 
This configuration, which is shown in the second panel of \Fig{fig:kojo_pairing}
 is called an exciton in condensed-matter physics and known to be an excitation 
 (hence the name), i.e., energetically not favored.
In QCD, this can be explained by  the fact that the relative momentum
between the particle and the hole is large at large density, $|\P_\text{rel}|=2|\p| \sim 2\mu$.
It is therefore difficult for them to form a collective state,
since the gluon propagator is suppressed at large momenta.

On the other hand, 
at least on the perturbative level,
the gluon propagator is singular in the infrared.
We may thus consider pairs where the momenta of the particle and the hole are equal,
e.g., by removing a right-handed quark with momentum $-\p$ below and
creating a left-handed quark with momentum $\p$ above the Fermi surface,
cf.~third panel of \Fig{fig:kojo_pairing}.
Since in this configuration the pairs have a nonvanishing total momentum 
$\P_\text{tot}=2\p$, the corresponding condensate is spatially non-uniform,
varying like $\exp(i\P_\text{tot}\cdot\x)$, with a wave number of the order 
$|\P_\text{tot}| \sim 2\mu$. 

Complementary insights into
the underlying mechanisms 
leading
to this kind of particle-hole pairing can be obtained from a typical chiral Lagrangian,
\beq
\label{eq:Lchi}
       \mathcal{L} = 
       \bar\psi\left( i\gamma^\mu\partial_\mu -g(\sigma + i\gamma^5\vec\tau\cdot \vec\pi)\right)\psi
       + \dots\,,       
\eeq
with a fermion (``quark'') field $\psi$ coupled to scalar and pseudoscalar boson (``meson'')
fields $\sigma$ and $\vec\pi$. 
From a QCD perspective, the latter should be interpreted as composite fields, e.g.,
$\sigma \sim \bar\psi\psi$ and $\vec\pi \sim \bar\psi i\gamma^5\vec\tau\psi$. 
The standard pattern for chiral-symmetry breaking in vacuum is then given by 
space-time independent mean fields $\ave{\sigma} \neq 0$ and $\ave{\vec\pi} = 0$. 
In this case the combination $M = g\sigma$ acts as an effective (``constituent'') quark mass,
so that the quarks get the dispersion relations $\pm E(\p) = \pm \sqrt{\p\,^2 + M^2}$.
In vacuum, where only the negative-energy states (``Dirac sea'') are filled, a larger 
quark mass then leads to a lowering of the Dirac-sea contribution to the energy,
thus favoring large values of $\ave{\sigma}$.\footnote{
In some models, the Dirac sea is not explicitly taken into account, but its effect is mimicked by 
a ``mexican-hat potential''. 
}
This effect is counterbalanced by other contributions, indicated by the ellipsis in \eq{eq:Lchi},
so that $\ave{\sigma}$ does not become arbitrarily large but takes an optimum value.
At nonzero density the Dirac-sea effect is further weakened by the occupation of 
positive-energy states (``Fermi sea''). 
Eventually, nonvanishing quark masses are no longer favored, and chiral symmetry gets restored.

\begin{figure}[htb]
\centering
\includegraphics[angle=0,width=.4\textwidth]{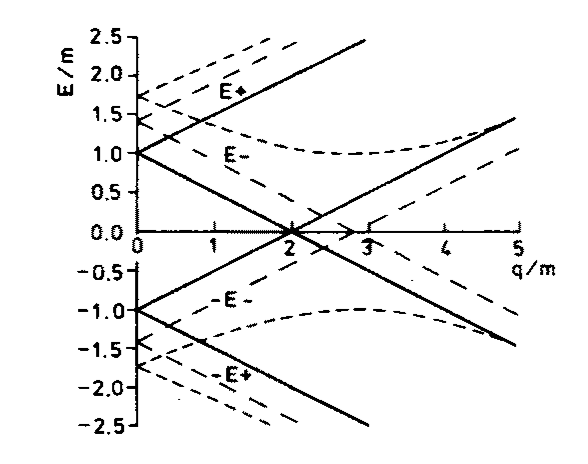}
\hspace{9mm}
\includegraphics[angle=0,width=.4\textwidth]{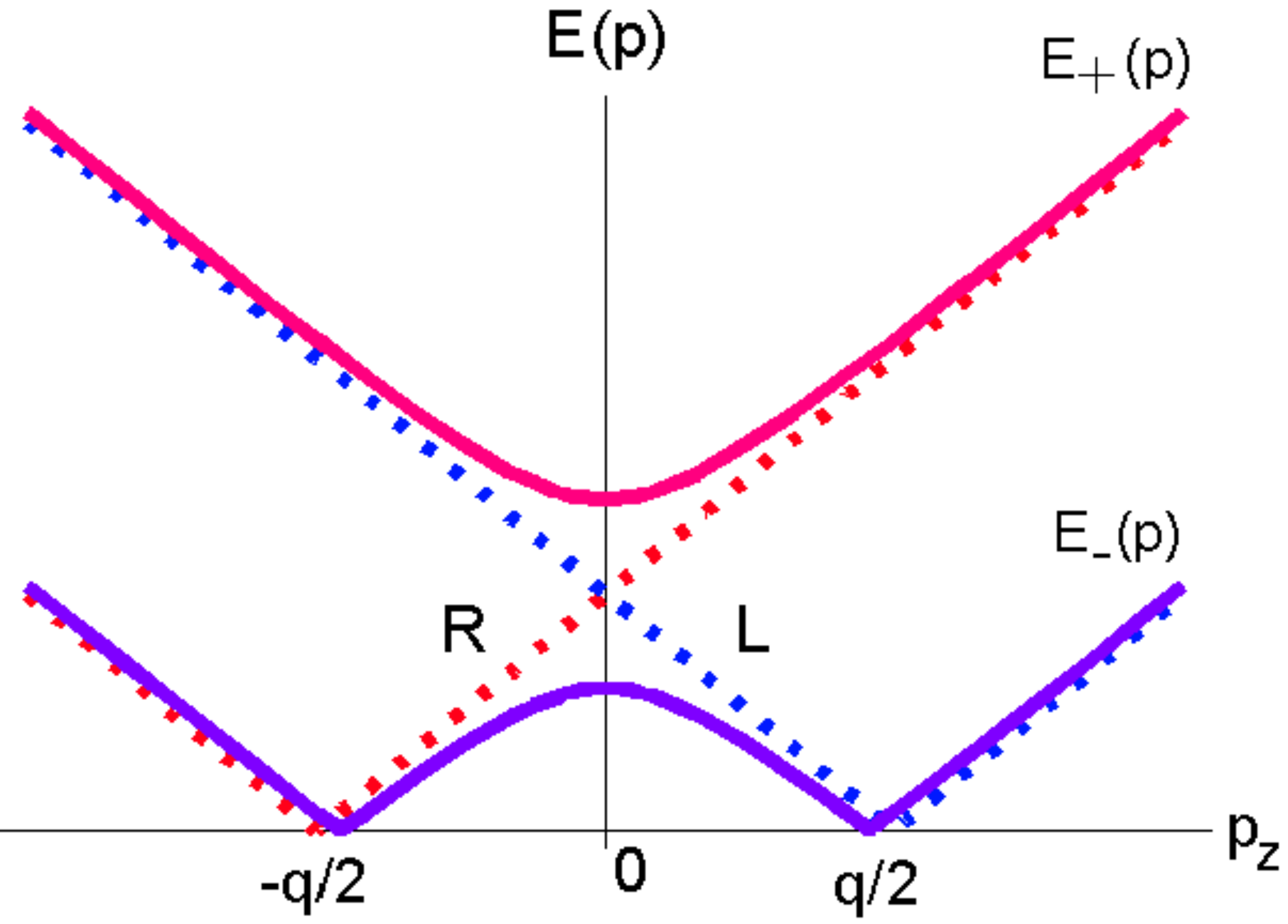}
\caption{
Left: 
Single-particle spectrum (\Eq{eq:CDWdispers}) for the ansatz of \Eq{eq:CDWansatz},
 plotted for
sample values of the quark momentum $\p$
as a function of the wave number $q$.
The positive-energy spectrum 
develops two branches:
by increasing $q$
the branches split, with
one increasing, and the other decreasing in energy.
The Fermi sea energy
can therefore be lowered when the lower branch is occupied.
The scale $m$ corresponds to the amplitude $\Delta$ in \Eq{eq:CDWdispers}.
From \cite{Broniowski:1990}.
Right: same spectrum as a function of $p_z$ for  $p_x=p_y=0$ 
and a fixed $q$.
Dashed (Solid) lines correspond to $\Delta=0$ ($\Delta\neq 0$). From \cite{NT:2004}.
\label{fig:CDWdispersion}
}
\end{figure}

As we will discuss in detail in this review, the analysis of arbitrary space dependent condensates 
is much more difficult, and a complete solution does not yet exist in $3+1$ dimensions, even in
mean-field models. However, important insights can be obtained from the ansatz, 
\beq
\label{eq:CDWansatz}
       \ave{\sigma(\x)} = \frac{\Delta}{g}\cos(\q\cdot\x)\,, \quad 
       \ave{\pi^3(\x)} = \frac{\Delta}{g}\sin(\q\cdot\x)\,, \quad      
       \ave{\pi^{1}(\x)} = \ave{\pi^{2}(\x)} = 0\,,
\eeq
which was studied already 35 years ago by Dautry and Nyman in the context of 
$p$-wave pion condensation in nuclear matter~\cite{Dautry:1979}.
Here $\q$ is a constant wave vector and $\Delta$ is an amplitude analogous to the 
constituent quark mass in the homogeneous case, which is recovered in the limit $\q = 0$.
For nonzero $\q$ the quark dispersion law $E(\p)$ splits into two branches~\cite{Dautry:1979},
\beq
\label{eq:CDWdispers}
        E_\pm(\p) = \sqrt{\p\,^2 + \Delta^2 + \frac{\q\,^2}{4} \pm\sqrt{\Delta^2\q\,^2 + (\q\cdot\p)^2}}\,,
\eeq
as illustrated in Fig.~\ref{fig:CDWdispersion}.

For small values of $|\q|$ it follows that $\partial E_-^2 / \partial |\q|^2 < 0$,
i.e., for a fixed value of $\Delta$, the energy of a particle with momentum $\p$
can be lowered by choosing a finite wave number.
At nonzero density, this offers the possibility to lower the energy of the system
by developing a nonvanishing value of $\q$ such that the quarks in the Fermi sea 
only populate the $E_-$ branch. 
The emergence of an inhomogeneous phase can thus be viewed as a cooperative effect of 
Dirac and Fermi sea, where the former is mainly responsible for the nonvanishing amplitude
of the chiral fields, while the latter favors a nonzero wave number.

In order to work out this argument more quantitatively, one has to take into account various competing
effects,
which depend in detail on the specific model.
The negative-energy states are of course modified in the same way as the positive-energy 
ones. Although the shifts of the $-E_+$ and $-E_-$ branches partially compensate each other,
this leads to an over-all enhancement of the Dirac-sea contribution to the energy.
Moreover, the ellipsis in \eq{eq:Lchi} may contain gradient terms, like kinetic terms of the meson 
fields, which also disfavor larger wave numbers. 

A detailed analysis has been performed for models in $1+1$ space-time dimensions, 
reviewed, e.g.,  in Refs.~\cite{Schon:2000qy,Thies:2006,Basar:2009fg}.
In $1+1$ dimensions, the mechanism described above is very effective and basically 
corresponds to the Peierls instability in one-dimensional electron chains~\cite{Peierls}.
As a consequence, inhomogeneous solutions are favored in large regions of 
the phase diagram at low temperatures. 

A very instructive example is the $1+1$ dimensional Nambu--Jona-Lasinio (NJL$_2$)
or chiral Gross Neveu model, given by the Lagrangian
\beq
\label{eq:NJL2}
       \mathcal{L}_\text{NJL$_2$}
       = 
       \bar\varphi \,i\tilde\gamma^\nu\partial_\nu \, \varphi
       + \frac{g^2}{2} \left[ (\bar\varphi \varphi)^2 + (\bar\varphi i\tilde\gamma^5 \varphi)^2 \right]\,.
\eeq
The fermion field $\varphi$ and its conjugate $\bar\varphi$ have two spinor components, 
and $\tilde\gamma^\nu$, $\nu=0,1$, 
are the corresponding $2\times 2$ gamma matrices.
A convenient representation is $\tilde\gamma^0 = \sigma^1$ and $\tilde\gamma^1 = -i\sigma^2$,
where $\sigma^i$ denote the Pauli matrices. 
In addition we define 
$\tilde\gamma^5 = \tilde\gamma^0\tilde\gamma^1 = \sigma^3$, which anticommutes with 
$\tilde\gamma^0$ and $\tilde\gamma^1$.

In order to describe the system at chemical potential $\mu$,
we must add a term $\mu \varphi^\dagger \varphi$ to the Lagrangian. 
The key observation is now that this term can be eliminated 
by performing a local chiral transformation \cite{Schon:2000qy}
\beq
\label{eq:chirot1p1}
       \varphi(x) = \exp(-i\mu z \tilde\gamma^5) \varphi'(x)\,,
\eeq
where $z\equiv x^1$ denotes the spatial coordinate.
One finds
\beq
       \mathcal{L}_\text{NJL$_2$} + \mu \varphi^\dagger \varphi
       =
        \bar\varphi'\, i\tilde\gamma^\nu\partial_\nu \, \varphi'
       + \frac{g^2}{2} \left[ (\bar\varphi' \varphi')^2 + (\bar\varphi' i\tilde\gamma^5 \varphi')^2 \right]\,,      
\eeq
i.e., the transformation (\ref{eq:chirot1p1})
provides an exact mapping of the system at chemical potential $\mu$ onto the system at $\mu=0$.
In particular, if at $\mu = 0$ the ground state is characterized by a nonvanishing homogeneous 
scalar condensate\footnote{This is assumption is not as innocent as it sounds, since in $1+1$ 
dimensions the spontaneous breaking of a continuous symmetry is forbidden by the
Mermin-Wagner theorem~\cite{Mermin:1966fe,Coleman:1973ci}. 
To avoid this problem, the NJL$_2$ model is therefore studied for an infinite number of
``flavors'', for which the mean-field results become exact.} 
and a vanishing pseudoscalar condensate, 
it follows that the same holds for the rotated fields at any $\mu$,

\beq
       \ave{\bar\varphi'\varphi'}_{\mu} = \ave{\bar\varphi\varphi}_{\mu=0} = \mathit{const.}\,,
       \quad 
       \ave{\bar\varphi' i\tilde\gamma^5\varphi'}_{\mu} = \ave{\bar\varphi i\tilde\gamma^5\varphi}_{\mu=0} = 0\,.
\eeq
Transforming this back to the original fields, one obtains
\beq
       \ave{\bar\varphi\varphi}_{\mu} =  \cos(2\mu z) \ave{\bar\varphi\varphi}_{\mu=0}\,, 
       \quad 
       \ave{\bar\varphi i\tilde\gamma^5\varphi}_{\mu} = \sin(2\mu z)\ave{\bar\varphi\varphi}_{\mu=0} \,.
\eeq
This solution, which has been termed ``chiral spiral''~\cite{Schon:2000he},
is obviously the $1+1$ dimensional analogue of the ansatz (\ref{eq:CDWansatz})
with $|\q| = 2\mu$.

\begin{figure}[htb]
\centering
\includegraphics[angle=0,width=.42\textwidth]{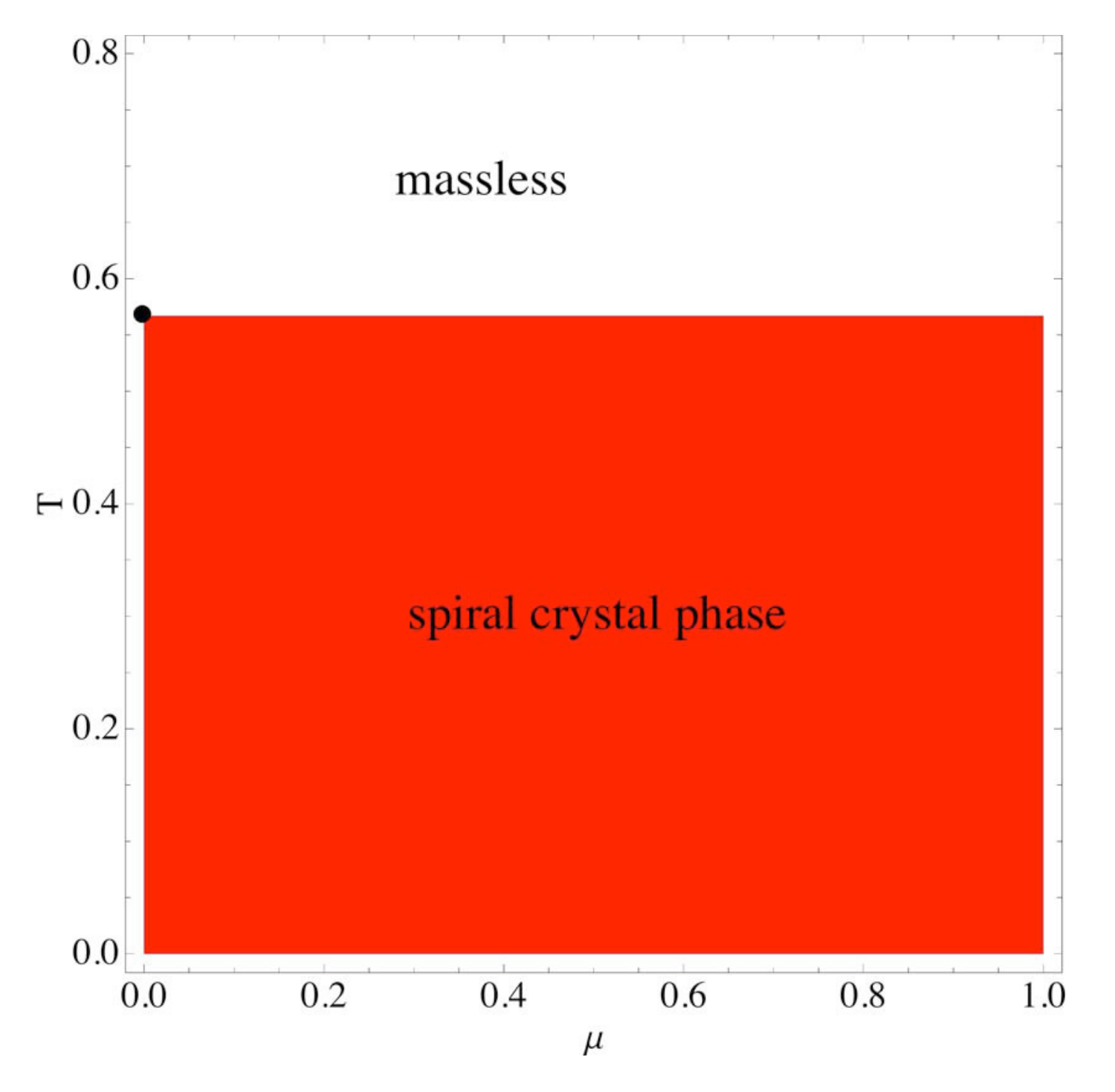}
\hspace{20mm}
\includegraphics[angle=0,width=.32\textwidth]{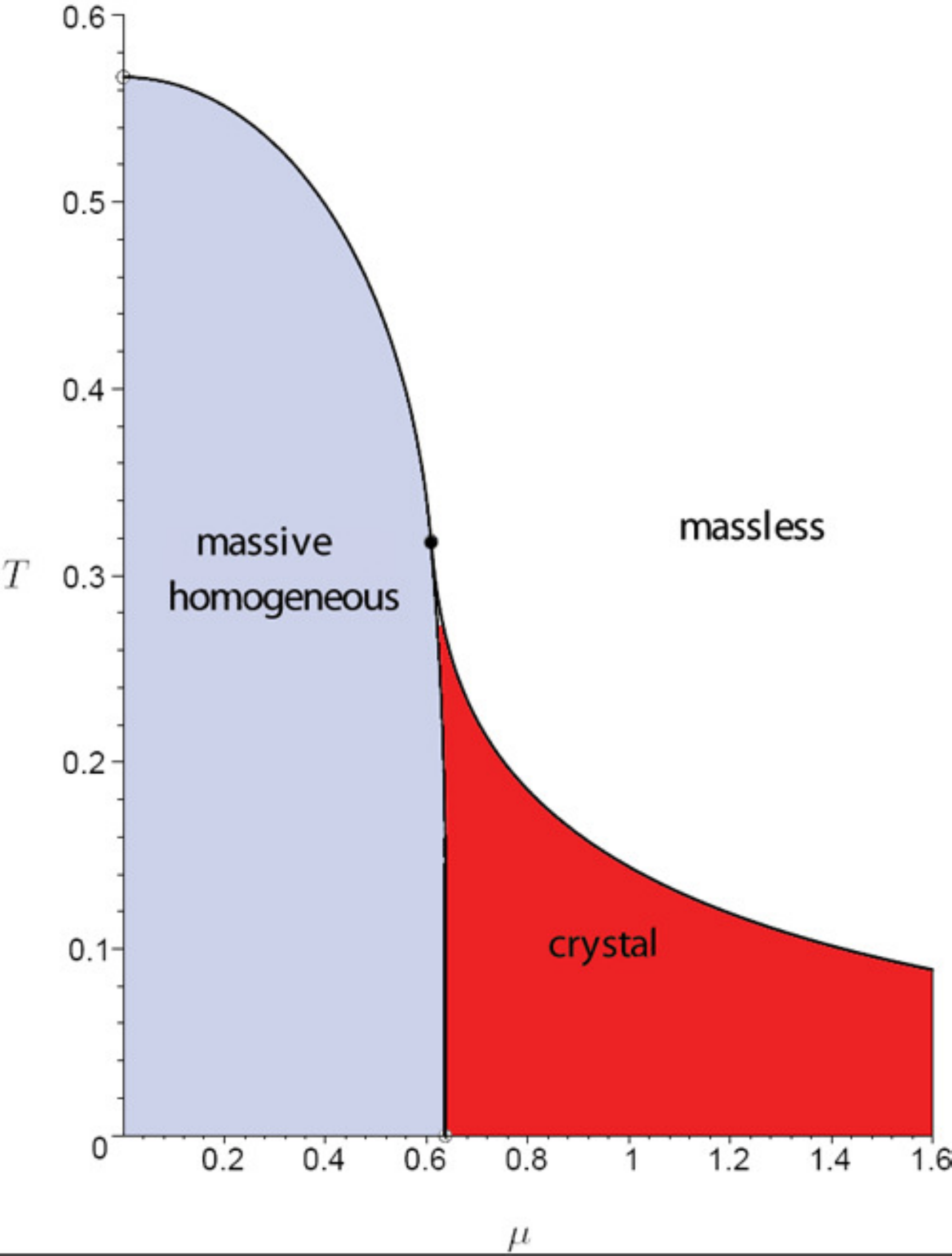}
\caption{Phase diagrams of two models in $1+1$ dimensions:
NJL$_2$ model (left) and GN model (right). 
Temperature and chemical potential are given in units of the 
vacuum constituent quark mass. From Ref.~\cite{Basar:2009fg}.
\label{fig:pd1d}
}
\end{figure}

This naive argument on the Lagrangian level is confirmed by a more thorough analysis, 
discussed in detail, e.g.,  in Refs.~\cite{Schon:2000qy,Basar:2009fg}. 
Employing the local chiral transformation, the chemical potential can be eliminated from the 
Hartree-Fock equation, showing that the chiral spiral is indeed a selfconsistent solution. 	 
Evaluating the free energies, it can furthermore be shown that this solution is favored over the 
homogeneous one.
In this context a crucial observation is that the chiral rotation does not completely remove the chemical potential 
from the thermodynamic potential but induces an anomalous term $-\mu^2/(2\pi)$ per flavor.\footnote{
Another important consequence of this term is a nonvanishing baryon density at $\mu > 0$,
$n = \mu/\pi $ per flavor, while a naive chiral rotation of $\ave{\varphi^\dagger \varphi}$ would yield $n=0$.
}

The phase diagram of the NJL$_2$ model has thus the following shape 
(see left panel of Fig.~\ref{fig:pd1d}). 
Below a critical temperature $T_c$, which is just the chiral restoration temperature at 
$\mu = 0$, the chiral spiral is the favored solution for arbitrary chemical potentials $\mu>0$. 
Above $T_c$, the system is in the restored phase, while a homogeneous phase with 
broken chiral symmetry only exists at $\mu=0$.

We note that the arguments which have led to these results are rather generic and apply to
any model of massless fermions in $1+1$ dimensions, as long as the interaction is 
invariant under the chiral transformation defined in \eq{eq:chirot1p1}. 
In particular this holds for $1+1$ dimensional QCD in the limit of infinitely many colors
('t Hooft model~\cite{tHooft:1974hx}), which therefore has the same phase structure~\cite{Schon:2000qy}.

On the other hand, the interaction term of the (non-chiral) Gross-Neveu model (GN model), 
defined by the Lagrangian \cite{GN}
\beq
\label{eq:LGN}
       \mathcal{L}_\text{GN}
       = 
       \bar\varphi \,i\tilde\gamma^\nu\partial_\nu \, \varphi
       + \frac{g^2}{2} (\bar\varphi \varphi)^2\,,
\eeq
is not invariant under \eq{eq:chirot1p1} but only under a discrete $Z_2$ symmetry.
As a consequence, the phase structure is different from the NJL$_2$ case
and has an extended region with homogeneous symmetry breaking
(see right panel of Fig.~\ref{fig:pd1d}). 
Above some (slightly $T$ dependent) critical chemical potential, there is, however, still 
an inhomogeneous phase,  which extends to infinite $\mu$ at $T=0$.  
In that phase, the scalar mean fields are given by Jacobi elliptic functions, which interpolate
smoothly between the two homogeneous phases, so that all phase boundaries 
correspond to second-order transitions.   

In $3+1$ dimensions, the $\mu$ dependence cannot be transformed away completely, 
even if the interaction is chirally symmetric.
The technical reason is that the relation $\tilde\gamma^1\tilde\gamma^5= \tilde\gamma^0$,
which was used above, has no exact correspondence in this case.
Physically, this is related to the presence of transverse momenta, so that the 
lowering of the single-particle energies by the mechanism
described above is partially washed out by the presence of transverse momenta.
This can be seen by writing \eq{eq:CDWdispers} as 
\beq
       E_{\pm}(\p) = \sqrt{\left(\sqrt{p_\parallel^2 + \Delta^2}  \pm \frac{|\q|}{2}
       \right)^2 + p^2_\perp}\,,
\eeq
where $\p\,_\parallel$  and $\p\,_\perp$ are the components of $\p$ parallel and perpendicular
to $\q$, respectively. 
For $p_\perp = 0$, which corresponds to the situation in $1+1$ dimensions,
and $\Delta \leq |\q|/2$, the dispersion $E_-$ is gapless, with nodes at 
$p_\parallel = \pm\sqrt{\q\,^2/4 - \Delta^2}$, cf.~\Fig{fig:CDWdispersion}.
A nonvanishing $\p_\perp$, on the other hand, acts as a gap, which weakens the influence of $\q$ 
on the spectrum. 

As a consequence, the exact properties found in $1+1$ dimensions are lost,
and the appearance of inhomogeneous phases in $3+1$ dimensions depends 
on the strength of the $q\bar q$ attraction and on the density of the system.
Nevertheless, such phases have been found to be present in many studies 
of strong-interaction matter.
As we will see, the typical result is more similar to the phase diagram of the GN model
(right panel of \Fig{fig:pd1d}), rather than to the NJL$_2$ model (left panel),
even if the underlying Lagrangian has a continuous chiral symmetry.

In this article, we review
the present understanding of inhomogeneous phases in strong interactions,
concentrating on the chiral condensate in $3+1$ dimensions.
The first part of this review focuses on results coming 
from model analyses of inhomogeneous phases.
In Section \ref{sec:general} 
we introduce 
the Nambu--Jona-Lasinio and the Quark-Meson model,
focusing on the procedure how to implement inhomogeneous phases.
Section \ref{sec:GL} is devoted to the discussion of 
the Ginzburg-Landau approach to study inhomogeneous phases close to
second-order phase transitions within these models.
In Section \ref{sec:CDW} we discuss explicitly the specific 
modulation of \Eq{eq:CDWansatz}, the so-called chiral density wave.
Section \ref{sec:dimred} describes a general approach to study one-dimensional
modulations of the chiral condensate, while higher-dimensional modulations 
are discussed in Section \ref{sec:highdim}. 
In Section \ref{sec:extensions} we discuss the effects of several
common model extensions on inhomogeneous phases.
Moving away from the effective model approaches, Section \ref{sec:largenc} is 
devoted to a discussion of quarkyonic chiral spirals in large $N_c$ QCD,
while in Section \ref{sec:DSE} the Dyson-Schwinger approach to 
inhomogeneous phases is presented.
We finally conclude with a brief outlook in Section \ref{sec:conclusions}.

\section{Model approaches to inhomogeneous matter}
\label{sec:general}

In recent years, the most widely used framework to explicitly study the properties
of inhomogeneous phases has been the Nambu-Jona Lasinio (NJL) model, 
which will as such be the main ``working horse'' in this review.
A second example is the quark-meson (QM) model, 
which differs from the NJL model by its renormalizability,
but is otherwise very similar.
In this section we briefly introduce both models and discuss how 
to apply them to study inhomogeneous matter.

\subsection{\it Nambu--Jona-Lasinio model}
\label{sec:NJL}

The original works of Nambu and Jona-Lasinio~\cite{NJL1,NJL2}
date back to 1961, when QCD and even the quark model did not yet exist.
The main idea was to provide a field-theoretical description of nucleons, 
featuring a mechanism for dynamical mass  generation
in analogy with the gap in the microscopic theory of superconductivity
\cite{BCS}.
With the advent of QCD, the fermionic fields in the Lagrangian have been reinterpreted 
as quarks~\cite{Kleinert:1976dsa,Volkov:1984kq,Hatsuda:1984jm}, incorporating color degrees of freedom
as additional quantum numbers. 
On the other hand, the NJL model contains no gluons and therefore no
running coupling and no confinement.
Instead, the quarks interact via local four-point vertices, 
which are chosen in such a way that the 
model shares the relevant global symmetries with QCD, 
in particular chiral symmetry. 

The NJL model has been employed to study low-energy properties of strong-interaction matter,
like the spectra and scattering cross sections of light hadrons, as well as the phase diagram 
at non-vanishing temperature or density, including the chiral phase transition and 
color superconductivity (for dedicated reviews, see, e.g., 
\cite{Vogl:1991qt,Klevansky:1992qe,Hatsuda:1994pi,Buballa:2005}).
Most of these investigations, however, were restricted to homogeneous phases.

In its standard form, the NJL-model Lagrangian reads \cite{NJL2}

\beq
\mL_\text{NJL} = \bar\psi (i\gamma^\mu\partial_\mu -  m )\psi 
+ G_S \left( (\pbp)^2 + (\bar\psi i \gamma^5 \vec\tau \psi)^2\right) \,,
\label{eq:LNJL}
\eeq
where $\psi$ represents a quark field with $N_f=2$ isospin and $N_c=3$ 
color degrees of freedom.
$\gamma^\mu$ and $\gamma^5$ are Dirac matrices, 
and $\tau^a$, $a = 1,2,3$, are Pauli matrices acting on the two-dimensional
isospin space.
The first part of the Lagrangian is nothing but the free Dirac Lagrangian
for fermions with bare mass $m$,
while the second part describes the four-fermion interactions.
The latter features a scalar iso-singlet $(\pbp)$ and a pseudo-scalar
iso-triplet $(\bar\psi i \gamma^5 \vec\tau \psi)$, 
both with the same coupling constant $G_S$,  so that the combination
is invariant under $SU(2)_L\times SU(2)_R$ chiral symmetry.
This symmetry is explicitly broken by the bare mass $m$, 
as well as spontaneously by nonvanishing condensates of the 
form $\langle\pbp\rangle$ or $\langle\bar\psi i \gamma^5 \tau^a \psi\rangle$.
Since we are mostly interested in spontaneous symmetry breaking,
we will often consider the so-called chiral limit, $m=0$.

As a consequence of the 4-point interactions, the NJL 
model is non-renorma\-li\-zable. This is reflected in the dimensionful
coupling constant $G_S$, which carries the units of an inverse energy squared.
The model therefore requires some regularization procedure to render 
its results finite, and the outcoming phenomenology will depend 
on the prescription used. In this sense, the model is to be interpreted as an effective
low-energy theory for strongly interacting quarks, and is as such 
not expected to be reliable above a given scale, identified by the regulator used.

\subsubsection{\it Mean-field approximation}
\label{sec:mfa}

The thermodynamic properties of the model 
at  temperature $T$ and the quark chemical
potential $\mu$ are encoded in the grand
potential per volume $V$,
\beq
\label{eq:Omega}
\Omega(T,\mu) 
=
-\frac{T}{V} \log\mathcal{Z}(T,\mu)\,,
\eeq
where $\mathcal{Z}(T,\mu)$ denotes the grand canonical partition
function.  
In path-integral formalism it is given by
\beq
\label{eq:Zpath}
\mathcal{Z}(T,\mu)\,
=  
\int \mathcal{D}\bar{\psi}\mathcal{D}\psi
\exp\left( \int_{V_4} d^4x_E \, (\mathcal{L}_\text{NJL}+\mu \bar{\psi}\gamma^0 \psi)\right)\,,
\eeq
where the integral in the exponent is performed in
Euclidean space-time, $x_E = (\tau, \x)$ with the imaginary time $\tau
= it$, and extends over the four-volume $V_4 = [0,\frac{1}{T}] \times
V$.

Because of the interaction terms, which are quartic in the quark fields, the path integral
cannot be performed in an exact way. 
In order to proceed, the most used approach is to
 employ the mean-field (or equivalently Hartree) 
approximation.
This basically amounts to linearize the Lagrangian \eq{eq:LNJL} 
in the field bilinears we are interested in, 
neglecting quadratic fluctuations \cite{Asakawa:1989}. 
Specifically, we introduce the expectation values for the scalar 
and pseudoscalar condensate,
\beq
\label{eq:phispdef}
       \ps \equiv \langle\pbp\rangle\,, \quad 
       \pp^a \equiv \langle\bar\psi i \gamma^5 \tau^a \psi\rangle\,,
\eeq
and write 
$\pbp = \ps + \delta\ps$, $\bar\psi i \gamma^5 \tau^a \psi = \pp^a + \delta\pp^a$.
After neglecting $\mathcal{O}(\delta\phi^2)$ terms, 
one obtains the mean-field Lagrangian 

\bea
\label{eq:mfL}
\mathcal{L_\text{MF}} 
&=&
\bar\psi S^{-1} \psi - \mathcal{V} \,, 
\eea
with the mean-field inverse quark propagator in coordinate space\footnote{
We use argument ``$x$''  as a short-hand notation to indicate the coordinate-space 
representation.  As clear from \eq{eq:mfL}, this also includes derivatives. 
A more careful definition of the quark propagator, depending on two space-time 
variables, will be introduced in Sec.~\ref{sec:DSE}.
}

\beq
\label{eq:Sm1}
S^{-1}(x) = i\gamma^\mu\partial_\mu - m + 2G_S (\ps + i\gamma^5\vec \tau\cdot \vpp) \,,
\eeq
and the condensate contribution

\beq
\mathcal{V} = G_S (\ps^2 + \vpp^{\,2}) \,. 
\eeq
Unlike the original Lagrangian, $\mathcal{L_\text{MF}}$ is bilinear in the quark fields, 
so that the path integral in \eq{eq:Omega} can be performed.
The resulting expression for the mean-field thermodynamic potential reads
\beq
\label{eq:OmegaS}
\Omega_\text{MF}(T,\mu;\ps, \vpp) 
=
-\frac{T}{V} \mathbf{Tr}\, \mathrm{Log} \left(\frac{S^{-1}}{T}\right)
+
\frac{T}{V}\int_{V_4} d^4x_E\, \V 
\equiv 
\Omega_\mathit{kin} + \Omega_\mathit{cond}
\,,
\eeq
where, for later convenience, we have identified  a kinetic and a condensate contribution.
The functional trace runs over $V_4$ and internal (color, flavor, and Dirac) degrees 
of freedom.

As indicated, besides temperature and chemical potential,
$\Omega_\text{MF}$ also depends on the scalar and pseudoscalar condensates, 
which have to be determined by minimization.
In this context we note that a completely equivalent procedure to the described 
linearization is to bosonize the model, e.g., via a Hubbard-Stratonovich transformation. 
The resulting bosonized Lagrangian takes exactly the same form as $\mathcal{L_\text{MF}}$, 
but with $\ps$ and $\pp$ being interpreted as auxiliary meson fields.
In fact, treating $\ps$ and $\pp$ as independent degrees of freedom,
$\mathcal{L_\text{MF}}$ is still equivalent to the original Lagrangian via the equations 
of motion. The mean-field approximation then consists in neglecting the fluctuations 
of the boson fields by performing a saddle point approximation.
This means, one does not integrate over the bosonic degrees of freedom
in the path integral, but determines their mean fields by minimizing $\Omega_\text{MF}$.

We thus have to solve
\beq
\label{eq:gapphi}
       \frac{\delta\Omega_\text{MF}}{\delta\phi_i} 
       = 0        
\eeq
for the condensates $\ps$ and $\pp^a$ ,where $\delta/\delta\phi_i$ denotes a
functional derivative. 
Since the derivative of the condensate term is again proportional to $\phi_i$,
this yields a coupled set of gap equations
which have to be solved selfconsistently.

At this point it is usually assumed that the mean fields are space and time independent, 
which simplifies the further evaluation considerably. 
In order to describe inhomogeneous phases, however, one must of course
retain the space dependence, 
\beq
       \ps = \ps(\x)\,, \quad \vpp = \vpp(\x)\,.
\eeq
On the other hand, we still consider the condensates to be static.
It is then useful to isolate the time derivative in the inverse 
quark propagator by writing 
\beq
\label{eq:SH}
       S^{-1}(x) = \gamma^0 (i\partial_0 - {H}(\x))
\eeq
with the effective Dirac Hamiltonian 
\begin{equation}
  \label{eq:mfH}
{H}(\x)  =  \gamma^0 \left[ -i\gamma^i\partial_i + m - 2G_S(\ps(\x) + i \gamma^5 \vec \tau \cdot \vpp(\x)) \right]\,.
\end{equation}
The latter is hermitean, so that, at least in principle, it can be diagonalized.
Moreover, since $H$ is time independent, the temporal 
part of the functional trace in \eq{eq:OmegaS}
can be evaluated in the usual way by summing fermionic 
Matsubara frequencies~\cite{Kap:89}.
One obtains
\beq
\label{eq:Omegakin}
\Omega_\mathit{kin} 
=
 -\frac{1}{V} \sum\limits_\nu \left[ \frac{E_\nu -\mu}{2} + T \log\left(1 + e^{-\frac{E_\nu -\mu}{T}}\right)\right]\,,
\eeq
where $\nu$ labels the eigenstates of $H$, and $E_\nu$ are the corresponding eigenvalues.
For the condensate term we can use the fact that the condensates are time independent, so that the integral over
imaginary time  is performed trivially:
\beq
\label{eq:Omegacond}
\Omega_\mathit{cond}
= 
\frac{1}{V}\int_{V} d^3x\, \V \,.
\eeq

If the eigenvalue spectrum of the model is known analytically,  \eq{eq:Omegakin} can be rewritten by introducing a ``spectral density'' function $\tilde\rho(E)$. Formally this is done by defining
\beq
\label{eq:rhotildedef}
\tilde\rho(E) =
\frac{1}{V}\sum_{\nu} \delta(E-E_\nu)
\eeq
leading to
\beq
\label{eq:OmegaeigenRHO}
\Omega_\mathit{kin} 
= 
-\int_{-\infty}^\infty dE \,\tilde\rho(E) \left[ \frac{E -\mu}{2} + T \log\left(1 + e^{-\frac{E-\mu}{T}}\right)\right]
\equiv 
-\int_{-\infty}^\infty dE \,\tilde\rho(E)  f(E)
\,.
\eeq
In most cases considered in the following, the eigenvalue spectrum 
of the model Hamiltonian is symmetric around $E=0$. In this case it is possible to rewrite \Eq{eq:OmegaeigenRHO} as 
\bea
\label{eq:OmegakinRhoSymm}
       \Omega_\mathit{kin}
       &=& 
       -\int_0^\infty dE \, \tilde\rho(E) \, \left[ f(E) + f(-E) \right]  \nonumber\\
      &=& 
         -\int_0^\infty dE \, \tilde\rho(E) 
       \left[ E 
       + T \log\left(1 + e^{-\frac{E -\mu}{T}}\right)  
       + T \log\left(1 + e^{-\frac{E +\mu}{T}}\right)\right]\,.
\eea
which allows to identify vacuum (i.e., $T$ and $\mu$ independent) and medium
contributions. Note that the integration runs over positive energies only.

\subsubsection{\it Regularization }
\label{app:regul}

Since the spectral density $\tilde\rho(E)$ grows asymptotically like ${\cal O}(E^2)$\footnote{This behaviour is independent from the shape of the chiral condensate.}, the vacuum contribution to the thermodynamic potential in NJL is quartically divergent. Due to the nature of the effective contact interaction, the model is non-renormalizable and some regularization prescription must be imposed. Since the model results reported in the following chapters have all been obtained within different regularization schemes, in this section we will briefly outline the most commonly employed procedures, while referring to the individual papers for more details.

When dealing with inhomogeneous condensates, the regularization procedure must be carried out with particular caution. In particular, the commonly employed three-momentum cutoff turns out to be inadequate, since the quark momenta are not conserved when scattering with the inhomogeneous condensate
and a sharp cut would inevitably lead to artifacts. 
In some special cases where the spatial dependence of the chiral condensate has a simple form, 
it is still possible to isolate a diverging part in $\Omega_{kin}$ which is independent 
of the inhomogeneity parameters and still employ a sharp momentum cutoff on it. This is the procedure 
followed in early works such as \cite{Sadzikowski:2000}.
An alternative and 
more generic approach is 
to employ covariant regularization schemes acting on the quark energies instead of their momenta. 
The starting point for most of these schemes is the 
 Schwinger proper-time representation of the logarithm in $\Omega_{kin}$ (\Eq{eq:OmegaS}),
\beq
{\rm Log}\,A \rightarrow -\int_0^\infty \frac{d\tau}{\tau} g(\tau) e^{-\tau A}~,
\label{eq:SchwingerPT}
\eeq
where $A$ is the inverse quark propagator and $g(\tau)$ is a blocking function acting as regulator. 

A common choice employed by several authors (for more details, see e.g. \cite{NT:2004})
is to act only on the vacuum contributions to $\Omega_{kin}$ and
 impose a sharp cutoff $g(\tau) = \theta(\tau - 1/\Lambda^2)$ on 
the lower bound of integration in \Eq{eq:SchwingerPT}.

In order to keep a structure which allows to
perform the Matsubara sum analytically, an alternative shape for the regulator function
is of the form 
\beq
g(\tau) = \sum_j c_j e^{-j\tau\Lambda^2} \,,
\eeq
where the $c_j$ must satisfy the conditions for Pauli-Villars coefficients \cite{Klevansky:1992qe}.

Limiting once again this procedure to the vacuum contributions, 
after performing the Matsubara sum one obtains the regularized expression 
first suggested in \cite{Nickel:2009wj}, which 
basically amounts to replacing 
in \Eq{eq:OmegakinRhoSymm} the Dirac sea term with 
\beq
E \,
\rightarrow \,
\sum_{j} c_j\sqrt{E^2+j\Lambda^2}
\,.
\eeq
In order to remove all divergencies in the thermodynamic potential three counterterms are required, 
while two are enough to render the gap equations finite.

In the following, even if not stated explicitly, a suitable regularization of the thermodynamic potential is always implied.

\subsubsection{\it Color, flavor and Dirac structure of the eigenvalue problem}
\label{sec:cfD}

As we have seen, the problem of determining the mean-field thermodynamic potential basically amounts
to calculating the eigenvalues of the effective Hamiltonian $H$.
Since the condensates do not depend on color, the color degrees of freedom simply yield a 
degeneracy factor of $N_c$.
A similar simplification arises in the flavor sector if we assume that the pseudoscalar condensate
$\vpp$ has a fixed direction in isospin space.
Exploiting the isospin invariance of the Lagrangian, we can choose the 3-direction,\footnote{Here 
we restrict ourselves to a common chemical potential
for up and down quarks. If isospin invariance is broken by a nonvanishing chemical potential
difference $\mu_I = \mu_u - \mu_d$, this assumption must be reconsidered, in particular in order 
to allow for charged pion condensates, cf.~Sec.~\ref{sec:isospin}.
}
i.e.,
\beq
\label{eq:ppa3}
        \pp^a(\x) = \pp(\x) \delta_{a3}\,.
\eeq
and the Hamiltonian becomes a direct product 
\beq
    H =  H_+ \otimes H_-\,,
\eeq
with
\beq
   H_\pm(\x)  =  \gamma^0 \left[ -i\gamma^i\partial_i + m - 2G_S(\phi_S \pm i \gamma^5 \phi_P) \right]\,, 
\eeq
acting on the up- and down quarks, respectively.
Again because of isospin symmetry,
$H_+$ and $H_-$ have the same eigenvalues, i.e., 
in the following we may restrict ourselves to 
$H_+$ and get another degeneracy factor of 2.

In chiral representation, the Dirac structure of $H_+$ is given by
\beq
\label{eq:Hchrep}
H_+(\x) = 
\left( 
\begin{array}{cc}
   i\sigma^i\partial_i &  m -2G_S(\ps + i\pp) \\
   m -2G_S(\ps -i\pp) & -i\sigma^i\partial_i \\
\end{array} 
\right)  \,,
\eeq
where $\sigma^i$ are $2\times 2$ Pauli matrices.
This motivates the introduction of a complex ``mass'' function
\beq
\label{eq:Mx}
M(\x) = m - 2G_S (\ps(\x) + i  \pp(\x)) \,,
\eeq
so that the Hamiltonian takes the form
\beq
\label{eq:Hplus}
H_+(\x)  
=
\left( 
\begin{array}{cc}
   i\sigma^i\partial_i &  M(\x) \\
   M^*(\x) & -i\sigma^i\partial_i \\
\end{array} 
\right)  \,.
\eeq
In this notation, the condensate part reads
\beq
\label{eq:VMx}
\mathcal{V} = \frac{|M(\x)-m|^2}{4G_S} \,.
\eeq

For homogeneous matter,  $M(\x) = M = \mathit{const.}$, the Hamiltonian is now
easily diagonalized with a plane-wave ansatz. 
The eigenstates correspond to free spinors of quarks and antiquarks with mass $M$,
thus being states of definite momentum $\p$ and energy eigenvalues 
$\pm E(\p) =\pm \sqrt{\p^{\,2} + M^2}$, each with a degeneracy factor of $2N_f N_c$.
The energy sum in \Eq{eq:Omegakin} can therefore be written as a momentum integration and one arrives at the standard expression 
\beq
\label{eq:omHomRho}
\Omega_\mathit{kin}^\mathit{hom} = -2 N_f N_c \int \frac{d^3p}{(2\pi)^3} \left[ f(E(\p)) + f(-E(\p)) \right] 
 = -\int_0^\infty dE \, \tilde\rho_\mathit{hom}(E) \left[ f(E) + f(-E) \right] \,,
\eeq
with $f(E)$ defined in \Eq{eq:OmegaeigenRHO} and the spectral density
\beq
\label{eq:rhohom}
\tilde\rho_\mathit{hom}(E,M) = \frac{N_f N_c}{\pi^2}  \, E \sqrt{E^2 - M^2} \, \theta(E^2 - M^2) \,.
\eeq

\subsubsection{\it Periodic condensates}
\label{sec:periodicconds}

For arbitrary space dependent mass functions, the problem is obviously much more involved.
In the following we assume that $M(x)$ is a periodic function, i.e., there exist three linear 
independent vectors $\vec a_i$, $i = 1,2,3$, so that
\beq
       M(\x + \vec a_i) = M(\x)\,.
\eeq
One can then perform a Fourier decomposition
\beq
\label{eq:Mxq}
       M(\x) = \sum_{\q_k} M_{\q _k} e^{i\q_k \cdot \x}\,,
\eeq
with momenta $\q_k$, which obey the condition $\frac{\q_k\cdot\vec a_i}{2\pi} \in \Zm$
and form a reciprocal lattice (RL) in momentum (wave vector) space.
Inserting this into \eq{eq:Hplus} and taking the Fourier transform
\beq
        H_{\p_m\p_n} = \frac{1}{V} \int d^3x\, e^{-i\p_m\cdot\x}\, H_+(\x) \,e^{i\p_n\cdot\x}\,,
\eeq        
we find that the momentum components of the Hamiltonian are given by
\begin{equation}
H_{\p_m,\p_n} =
 \left( 
\begin{array}{cc}
 -\vec\sigma\cdot\p_m\,\delta_{\p_m,\p_n} &  
 \sum\limits_{\q_k} M_{\q_k} \delta_{\p_m,\p_n+\q_k} 
 \\
 \sum\limits_{\q_k} M^*_{\q_k} \delta_{\p_m,\p_n-\q_k} &  
 \vec\sigma\cdot\p_m\,\delta_{\p_m,\p_n} 
\end{array} 
\right) \,.
\label{eq:Hmn}
\end{equation}
As one can see, the Fourier modes with $\q_k \neq 0$ 
couple unequal momenta and, hence, the Hamiltonian 
is a nondiagonal matrix in momentum space. 
It follows from \eq{eq:SH}
that also the mean-field quark propagator is nondiagonal 
in momentum space.
This is possible because the quarks can change their momenta by 
scattering off the condensates, which, being nonuniform in space,
carry nonvanishing momentum.
Indeed, for $M(\x) = \mathit{const.}$ only the $\q_k = 0$ mode of the mass 
function exists and, hence, the Hamiltonian is diagonal in momentum space.

Since, unlike the space of internal degrees of freedom,
momentum space is infinite, the diagonalization of $H$ in the presence of 
nonuniform condensates is obviously a highly nontrivial task.
However, since the reciprocal lattice is an in general infinite but discrete
set of momenta, not all in- and outgoing quark momenta are mutually 
connected but only those which differ by an element of the RL.
As a consequence, the Hamiltonian is still block diagonal.
Formally, this is related to Bloch's theorem, which states that in a crystal
all eigenstates which correspond to different vectors of the 
first Brillouin zone (BZ) are orthogonal. 
To exploit this fact, we write all momenta as
\beq
       \p_i = \mk_i + \q_i\,,
\eeq
where $\q_i$ belongs to the RL and $\mk_i$ is an element of the BZ.
Then two momenta $\p_m$ and $\p_n$ are only coupled
if $\mk_m = \mk_n$,
and the Hamiltonian can thus be 
written as a direct sum,
\beq
        H = \sum\limits_{\mk \in \text{BZ}} H(\mk)\,,
\eeq
where each block $H(\mk)$ corresponds to a vector $\mk$ in the BZ. 
Accordingly, the sum over eigenstates in \eq{eq:Omegakin} becomes 
\beq
       \frac{1}{V} \sum_\nu f(E_\nu) \rightarrow 
       \int_\text{BZ} \frac{d^3k}{(2\pi)^3} \sum_\lambda f(E_\lambda(\mk))\,,
\eeq
where $\lambda$ labels the (discrete) eigenstates of $H(\mk)$ and 
$E_\lambda(\mk)$ are the corresponding eigenvalues.
Taking care of the degeneracy factors for color and flavor,
as discussed above, we then get
\beq
\label{eq:OmegakinMq}
       \Omega_\mathit{kin}
       = 
       -2N_c  \int_\text{BZ} \frac{d^3k}{(2\pi)^3} \sum_\lambda
       \left[ \frac{E_\lambda(\mk) -\mu}{2} + T \log\left(1 + e^{-\frac{E_\lambda(\mk) -\mu}{T}}\right)\right]\,.
\eeq
As noted before, in most cases of interest, the eigenvalues come in pairs with with opposite sign,
$\pm |E_\lambda|$.
Hence, similar to \Eq{eq:OmegakinRhoSymm}, we can combine them to get 
\beq
\label{eq:Omegakinpm}
       \Omega_\mathit{kin}
       = 
       -2N_c  \int_\text{BZ} \frac{d^3k}{(2\pi)^3} \sum_{\lambda>0}
       \left[ E_\lambda(\mk) 
       + T \log\left(1 + e^{-\frac{E_\lambda(\mk) -\mu}{T}}\right)  
       + T \log\left(1 + e^{-\frac{E_\lambda(\mk) +\mu}{T}}\right)\right]\,,     
\eeq
where $\lambda > 0$ was introduced as a short-hand notation to indicate eigenstates with
positive eigenvalues. 

We can also evaluate the condensate part of the thermodynamic potential
in momentum space, which yields
\beq
\label{eq:OmegacondMq}
       \Omega_\mathit{cond} =
       \frac{1}{4G_S} \sum_{\q_k} \left| \tilde M_{\q_k}\right|^2\,,
\eeq
with $\tilde M_{\q_k}  =  M_{\q_k}  -m \delta_{\q_k,0}$.

\subsubsection{\it Gap equations}

In the formal set up developed above, the evaluation of the mean-field thermodynamic potential
is now a tedious, but numerically  tractable problem.
The main difficulty is however that the mass function is not a priori known, but must be found
by minimizing $\Omega_\text{MF}$. Until now, this problem has not been solved rigorously.
A necessary condition is that the condensates  fulfill the gap equations (\ref{eq:gapphi}),
which we now want to analyze further, basically following an analogous discussion for
inhomogeneous color superconductors~\cite{NB:2009}.

Expressing the condensates in terms of the mass function, the gap equations can be written 
as
\beq
       \frac{\partial \Omega_\text{MF}}{\partial M_{\q_k}} = 0\,,
       \qquad
       \frac{\partial \Omega_\text{MF}}{\partial M_{\q_k}^*} = 0\,,
\eeq
where $M_{\q_k}$ and $M_{\q_k}^*$, related to $\ps \pm i\pp$,
should be treated as independent variables. 
However, since $\Omega_\text{MF}$ is real, both sets of equations yield 
the same results, so that we may choose the second one,
without loosing any information.
We then obtain
\beq
       \tilde M_{\q_k} = 4G_S N_c  \int_\text{BZ} \frac{d^3k}{(2\pi)^3} \sum_\lambda
       \frac{\partial E_\lambda(\mk)}{\partial M_{\q_k}^*}
       \left( 1 - 2n_F(E_\lambda(\mk) -\mu) \right)\,,
\eeq
with the Fermi function $n_F(E) = [\exp(E/T) +1]^{-1}$.
Since the energy eigenvalues on the right-hand side in general depend on all
Fourier components,
this constitutes an infinite system of coupled selfconsistency equations.

Noting that $E_\lambda(\mk)$ is an eigenvalue of the 
Hamiltonian $H(\mk)$, we can write
\beq
       E_\lambda(\mk) =  w_\lambda^\dagger(\mk) H(\mk) w_\lambda(\mk)\,,
\eeq
where $w_\lambda(\mk)$ is the corresponding normalized eigenvector.
Taking the derivative, the terms related to $\partial w_\lambda/\partial M_{\q_k}^*$
and $\partial w_\lambda^\dagger/\partial M_{\q_k}^*$
cancel each other because of the
hermiticity of the Hamiltonian and the fixed normalization.
We are then left with
\beq
       \frac{\partial E_\lambda(\mk)}{\partial M_{\q_k}^*} = 
       w_\lambda^\dagger(\mk) P_{\q_k} w_\lambda(\mk)\,,
\eeq
with a matrix $P_{\q_k} =  \frac{\partial H(\mk)}{\partial M_{\q_k}^*}$,
which is easily obtained from~\eq{eq:Hmn}. One finds 
\beq
       (P_{\q_k})_{\p_m,\p_n} =
\left( 
\begin{array}{cc}
 0 & 0 \\ 1 & 0 
\end{array} 
\right) \delta_{\p_m,\p_n-\q_k}
= \frac{1}{2}\gamma^0 (1 - \gamma^5)\, \delta_{\p_m,\p_n-\q_k}
\,,
\eeq
i.e., $P_{\q_k}$ is independent of $\mk$ and the mass function,
and connects left- with right-handed eigenvector components,
corresponding to momenta which differ by $\q_k$.
Thus, writing chirality and momentum indices explicitly, 
the gap equations read 
\beq
       \tilde M_{\q_k} = 4G_S N_c  \int_\text{BZ} \frac{d^3k}{(2\pi)^3} \sum_{\lambda}
       \sum_{\p_n}
       \left({w_\lambda^R}(\mk)\right)^\dagger_{\p_n-\q_k}
       \left({w_\lambda^L}(\mk)\right)_{\p_n}       
       \left( 1 - 2n_F(E_\lambda(\mk) -\mu) \right)\,.
\eeq
Alternatively, we can Fourier transform this into coordinate space.
Using \eq{eq:Mxq}
and 
\beq
       w_\lambda(\mk;\x) = \frac{1}{\sqrt{V}} \sum_{\vec p_n} \left(w_\lambda(\mk)\right)_{\p_n}\,e^{i\p_n\cdot \x}\,,
\eeq
which keeps the eigenfunctions normalized in the volume $V$ of the unit cell,
we find
\beq
       M(\x) = m + 4G_S N_c  \,V \int_\text{BZ} \frac{d^3k}{(2\pi)^3} \sum_{\lambda}
       {w_\lambda^{R \, \dagger}}(\mk;\x) 
       {w_\lambda^L}(\mk;\x)   
       \left( 1 - 2n_F(E_\lambda(\mk) -\mu) \right)\,.
\eeq
This equation is the analogue to the selfconsistency condition for the 
Bogoliubov--de Gennes equation in superconductivity~\cite{DeGennes}.
In our case the ``pairing'' is between left handed quarks and right handed holes 
(and vice versa), thus breaking chiral symmetry.

\subsection{\it Quark-meson model}
\label{sec:qmmodel}

One of the major drawbacks of the NJL model is its non-renormalizability. 
This introduces ambiguities, since the results depend on the choice of the 
regularization prescription as well as on the value of the corresponding
cutoff parameters. 
A popular alternative approach is given by the quark-meson (QM) model,
see Ref.~\cite{Scadron:2013vba}
for a recent review. It
shares many features with the NJL model, but is renormalizable.

The QM model (Gell-Mann--L\'evy model or linear sigma model with quarks) 
is defined by the Lagrangian density 
\cite{GellMann:1960,Scavenius:2000qd,Schaefer:2006ds}
\beq
\mathcal{L}_\text{QM}
=
\bar{\psi}
\left(
i\gamma^\mu \partial_\mu
-
g(\sigma+i\gamma^5 \vec\tau \cdot \vec\pi)
\right)
\psi
+
\mathcal{L}_\mathit{M}^\mathit{kin}
-
U(\sigma,\vec\pi )
\ ,
\eeq
where $\psi$ denotes a again a quark field.
Unlike in the NJL model, the quarks do not interact with each other
directly but are coupled to a scalar-isoscalar field $\vec \sigma$
and a pseudoscalar-isovector field  $\vec \pi$, corresponding to
the sigma meson and the pion triplet, respectively.
The meson kinetic contribution reads 
\beq
\label{eq:LQM}
\mathcal{L}_\mathit{M}^\mathit{kin} = \frac{1}{2} \left(
\partial_\mu\sigma \partial^\mu\sigma 
+
\partial_\mu\vec\pi \partial^\mu\vec\pi
\right) \ ,
\eeq
and
\beq
\label{eq:Uqm}
U(\sigma,\vec\pi)
=
\frac{\lambda}{4}
\left(
\sigma^2
+
\vec\pi^2
-
v^2
\right)^{2}
- h\sigma
\,
\eeq
denotes the meson potential. 
The first term, proportional to the coupling constant $\lambda$, is 
invariant under $O(4)$ transformations, 
which is isomorphic to the chiral $SU(2)_L\times SU(2)_R$ symmetry. 
This symmetry is broken explicitly by the last term, which is linear 
in the $\sigma$ field.  
However, unless stated otherwise, we consider the chiral limit in
this review, i.e., $h=0$.
The three remaining model parameters, $g$, $\lambda$
and $v^2$, are fitted to vacuum properties as discussed  below.

\subsubsection{\it Mean-field thermodynamic potential}

The thermodynamic potential of the QM model is defined completely analogous to 
Eqs.~(\ref{eq:Omega}) and (\ref{eq:Zpath}),
with $\mathcal{L}_\text{NJL}$ replaced by $\mathcal{L}_\text{QM}$
and the path integral performed over the meson fields as well.
In this context we recall
that in the bosonized NJL model one also has to integrate over the meson fields,
as long as one does not resort to the mean-field approximation. 
The main difference between the two models is thus that the NJL model has no
kinetic terms and no quartic terms for the meson fields.

In mean-field approximation the meson fields are treated as
classical and replaced by their expectation
values,
neglecting both thermal and quantum fluctuations~\cite{Scavenius:2000qd, Schaefer:2006ds}.  
As in the NJL model, 
we assume these mean fields to be time independent but, in order to allow
for inhomogeneous phases, we retain their dependence on the spatial
coordinate $\x$.  
Since the action is bilinear in the quark fields, they can again be integrated out, 
and one obtains the mean-field thermodynamic potential 
\beq
\label{eq:OmegaQMsum}
        \Omega_\text{MF}(T,\mu;\sigma, \vec{\pi})
        = 
         \Omega_\text{q}(T,\mu;\sigma, \vec{\pi})
         + 
         \Omega_\text{mes} (\sigma, \vec{\pi})\,,
\eeq       
where the last term is given by
\beq
\label{eq:Omegamesx}
\Omega_\text{mes} 
=
\frac{1}{V}\int_V d^3x\,\mathcal{H}_\text{mes}\,,
\eeq
with the mesonic Hamiltonian density 
\beq
\label{eq:Hmes}
       \mathcal{H}_\text{mes}
       =
       \frac{1}{2}\left( (\nabla\sigma(\x))^2  + (\nabla\vec\pi (\x))^2 \right) 
       +  U(\sigma(\x), \vec\pi (\x))\,.
\eeq
The quark contribution $\Omega_\text{q}$  is exactly the same expression as 
the kinetic term $\Omega_\mathit{kin}$ in the NJL model if we identify
\beq
\label{eq:replace}
         m-2G_S \,\ps = g\sigma\,, \quad -2G_S \,\vpp  = g\vec\pi
       \,.
\eeq
This means that the difficult part of the calculation is identical in both models,
so that we can  proceed as before
and only need to replace the condensate term of the NJL model by $\Omega_\text{mes}$.

If we restrict ourselves to a single isospin component of the pion field,
$\pi_a = \pi \delta_{a3}$,  
we can again define a complex mass function for the dressed quarks.
Applying the prescription \eq{eq:replace} to \eq{eq:Mx} yields (in the chiral limit)
\beq
\label{eq:MxQM}
       M(\x) = g \left(\sigma(\x) + i\pi(\x)\right)\,,
\eeq
as one could have guessed already from the Lagrangian.
The meson Hamiltonian can then be written as
\beq
\label{eq:HmesM}
       \mathcal{H}_\text{mes}
       =
       \frac{1}{2g^2}\left| \nabla M \right|^2 
       + \frac{\lambda}{4}\left( \frac{|M|^2}{g^2} - v^2 \right)^2\,,
\eeq
where we have inserted the explicit form of the potential.

Turning to momentum space,  the quark part of  the thermodynamic potential is then given 
by \eq{eq:OmegakinMq},
while the meson part becomes
\beq
       \Omega_\text{mes} 
       =
       \sum_{\q_k} \left|M_{\q_k}\right|^2 \left( \frac{{\q}^{\;2}_k}{2g^2} - \frac{\lambda v^2}{2g^2} \right)
       +
       \frac{\lambda}{4g^4} \sum_{\q_k,\q_l,\q_m,\q_n}
       M_{\q_k}^*M_{\q_l}^*M_{\q_m}M_{\q_n} \, \delta_{\q_k+\q_l,\q_m+\q_n}
       \,,
\eeq
where we have dropped an irrelevant constant.

\subsubsection{\it Renormalization of the Dirac sea}

As we have seen already for the NJL model, the vacuum (or Dirac sea)
contribution to the kinetic part of the thermodynamic potential, \eq{eq:Omegakinpm} is divergent.
While in the NJL model, this term is essential for spontaneous chiral-symmetry breaking,
this is not the case in the QM model, where the symmetry breaking can be realized with the 
meson potential alone.  
Until quite recently, the divergent vacuum term was therefore simply omitted in QM-model studies,
expecting that a proper refit of the model parameters would compensate for it. 
This ``no-sea'' or ``standard mean-field'' approximation (sMFA) was also the basis of the first
investigations of inhomogeneous phases in the QM 
model~\cite{Kutschera:1989yz,Kutschera:1990xk,Broniowski:1990,Nickel:2009wj}. 

It was shown, however, that the sMFA leads to artifacts,
such as a first-order chiral phase transition in the chiral limit
for two flavors at vanishing chemical potential~\cite{Skokov:2010}.
In more recent calculations, the quark Dirac-sea contribution was therefore reinstalled
and effectively renormalized.
As a result the phase transition at $\mu=0$ becomes second order~\cite{Skokov:2010}. 
This framework was applied to inhomogeneous phases in Ref.~\cite{Carignano:2014jla}.

The basic idea is to determine the model parameters $g$, $\lambda$, and $v^2$,
by fitting them to three vacuum ``observables'', e.g., the pion decay constant $f_\pi$,
and the masses of the constituent quark and the sigma meson, $M_\text{vac}$ and
$m_\sigma$, respectively. 
In the sMFA one finds~\cite{Birse:1983gm} 
\beq
\label{eq:sMFAparameters}
        g^2 = \frac{M_\text{vac}^2}{f_\pi^2}\,, \quad
        \lambda = \frac{m_\sigma^2}{2f_\pi^2}\,, \quad
        v^2 = f_\pi^2\,,
\eeq
while including the Dirac sea, these relations get corrections.
Since the correction terms are divergent, one has to introduce a regularization
prescription in order to proceed.
As a consequence, the result of the parameter fit depends on the corresponding
cutoff parameter $\Lambda$.
However, keeping $f_\pi$,  $M_\text{vac}$ and $m_\sigma$ constant, while gradually 
increasing $\Lambda$, any observable should eventually converge against its renormalized 
value.
This also applies to the phase diagram, which
was found to be essentially cutoff independent for $\Lambda \gtrsim 2$~GeV
in the Pauli-Villars scheme~\cite{Carignano:2014jla}, see Sec.~\ref{sec:CDWmodels}.

\section{Ginzburg-Landau analysis}
\label{sec:GL}

Before discussing specific examples, we want to explore the phase structure 
within a Ginzburg-Landau (GL) analysis. 
In this approach, the thermodynamic potential is expanded around the symmetric ground state in terms 
of the order parameters and their gradients. Its validity is therefore restricted to regions where these 
quantities are small, i.e., close to second-order phase boundaries, and in particular to critical points.
In these regions the analysis can be kept rather general, e.g., 
without specifying the explicit form of the inhomogeneity.

Although often claimed to be model independent,
the results of the GL expansion ultimately depend on the underlying Lagrangian.
In this section we concentrate on the two models introduced above, i.e., the 
standard NJL model and the QM model.
More specific analyses, related to extensions or modifications of the model, 
will be discussed later. 

\subsection{\it Nambu--Jona-Lasinio model}
\label{sec:GLNJL}

We restrict ourselves to the chiral limit, $m=0$, and assume again that 
the pseudoscalar condensate has only one isospin component, which points to the 
3-direction, \eq{eq:ppa3}.
The expansion can then be done in terms of the mass function $M(\x)$, \eq{eq:Mx},
which contains the condensates $\ps$ and $\pp$ as real and imaginary parts, respectively. 
Treating the magnitude of $M$ and gradients to be of the same order,
the GL functional up to 4th order  takes the form
\beq
\label{eq:GLNJLgen}
\Omega_\text{MF}(M) = \Omega_\text{MF}(0)
+ \frac{1}{V}\int d^3x \, \Big\{  
    \frac{1}{2}\gamma_{2 } |M(\x)|^2
     + \frac{1}{4}\left(\gamma_{4,a }|M(\x)|^4 + \gamma_{4,b }|\nabla M(\x)|^2\right)  + \dots \Big\}
\,.     
\eeq
Note that odd terms vanish in the chiral limit because of chiral and rotational symmetry.
Higher-order terms, which are indicated by the ellipsis,  are important for the stability
of the system. We therefore assume that the corresponding coefficients are positive.

The GL coefficients are functions of temperature and chemical potential.
In this way they determine the $T$ and $\mu$ dependence of the favored mass functions 
and, thus, the phase structure.
If $\gamma_{4,b} > 0$, gradients of the mass function are disfavored and the analysis is 
reduced to the well-known case for homogeneous phases. 
If $\gamma_{4,a} > 0$ as well, 
the favored solution then has a nonzero mass 
$|M| \simeq \sqrt{-\gamma_2/\gamma_{4,a}}$  for $\gamma_2 <0$,
whereas $M=0$ for  $\gamma_2 >0$.
Hence, there is a second-order phase transition from the 
chirally broken to the restored phase at $\gamma_2 = 0$,
which defines a line in the $T-\mu$ phase diagram.
For $\gamma_{4,a} < 0$ on the other hand, but still restricting the analysis
to homogeneous mass functions, the phase transition is first order.
The critical point (CP) where the first-order phase transition
turns into second order is thus given by the condition
\beq
\label{eq:NJLGLCP}
        \gamma_2|_\text{CP} = \gamma_{4,a}|_\text{CP} = 0\,.
\eeq

However, for $\gamma_{4,b} < 0$ inhomogeneous solutions, i.e.,
solutions with nonzero gradients can become favored. 
A particularly interesting point is then the point where the two homogeneous
phases, i.e., the chirally broken and the restored one, meet with the 
inhomogeneous phase.  As we will see later, at least for the most favored
inhomogeneous solutions known so far (cf.~Sec.~\ref{sec:solitons}), the three 
phase boundaries which meet at this point correspond to second-order
phase transitions.  In the literature, such a point is called a ``Lifshitz point''.
In the present context it has become customary, to use this term more generally
for the meeting point of the three phase boundaries, even if one of them does
not correspond to a second-order transition if one slightly goes away from the point. 
We will follow this custom throughout this review.
The Lifshitz point (LP) is then given by the condition
\beq
\label{eq:NJLGLLP}
        \gamma_2|_\text{LP} = \gamma_{4,b}|_\text{LP} = 0\,.
\eeq

For inhomogeneous phases in the NJL model the GL coefficients have been 
calculated first by Nickel~\cite{Nickel:2009ke},
generalizing similar analyses from the $1+1$ dimensional Gross-Neveu 
model~\cite{Schnetz:2004} to $3+1$ dimensions. 
Starting point is the general expression for the mean-field thermodynamic potential,
\eq{eq:OmegaS}.
Accordingly, the GL coefficients can be written as a sum
\beq
       \gamma_i = \alpha_i + \beta_i\,,
\eeq
where $\alpha_i$ and $\beta_i$  correspond to the 
condensate and kinetic contributions,
$\Omega_\mathit{cond}$ and $\Omega_\mathit{kin}$, respectively.
From Eqs.~(\ref{eq:Omegacond}) and (\ref{eq:VMx}) it is 
then easy to see that in the chiral limit
the condensate part only contributes to $\alpha_2$,
\beq 
\label{eq:alphaNJL}
       \alpha_i = \frac{1}{2G_S} \delta_{i2}\,.
\eeq       
In order to derive the kinetic contributions to the coefficients, 
the inverse dressed propagator is decomposed  
as $S^{-1} = S_0^{-1} - \Sigma$,
with the inverse bare propagator $S_0^{-1}$ and the selfenergy 
\beq
\label{eq:SigmaGL}
       \Sigma(\x) = -2G_S (\ps(\x) +i\gamma^5 \tau^3\pp(\x))
                    = \mathrm{Re}\,M(\x) + \mathrm{Im}\,M(\x)\, i\gamma^5 \tau^3      
       \,,
\eeq
which is treated as a small perturbation. 
Factorizing out the bare part, the logarithm in \eq{eq:OmegaS} is then expanded
in a Taylor series,
\beq
      \mathrm{Log}\,S^{-1}
       = 
       \mathrm{Log}\left(S_0^{-1}(1 - S_0\Sigma)\right)
      = 
      \mathrm{Log}\,S_0^{-1} - \sum_{n=1}^\infty \frac{1}{n} (S_0\Sigma)^n
      \,,
\eeq
with the bare propagator $S_0$. One obtains
\beq
\label{eq:Omegakinsum}
       \Omega_\mathit{kin}(\Sigma) = \Omega_0
       + \sum_{n=1}^\infty \Delta\Omega_\mathit{kin}^{(n)}(\Sigma)\,,
\eeq
where $\Omega_0 \equiv \Omega_\mathit{kin}(\Sigma = 0)$ corresponds to the thermodynamic
potential in the restored phase and
\beq
       \Delta\Omega_\mathit{kin}^{(n)}(\Sigma)
       =
       \frac{1}{n }\frac{T}{V}\mathbf{Tr}\, (S_0\Sigma)^n
\eeq
describes the corrections due to the condensates. 
More explicitly, the functional trace reads 
\beq
        \mathbf{Tr}\, (S_0\Sigma)^n
        =
        \int_{V_4} d^4 x 
        \int_{V_4} d^4 x_2 \dots \int_{V_4} d^4 x_n \, \mathrm{tr}
        \left[S_0(x,x_2) \Sigma(\x_2) S_0(x_2,x_3) \Sigma(\x_3) \dots
               S_0(x_n,x) \Sigma(\x)  
        \right]\,,
\eeq
where $\mathrm{tr}$ denotes the trace over internal degrees of freedom.
Since the bare propagator $S_0(x,x')$ is defined in the homogeneous phase, 
it only depends on the distance $x-x'$ and takes the standard textbook form. 
Performing a gradient expansion of the selfenergies,
\beq
       \Sigma(\x_j) =
      \Sigma(\x) + \nabla\Sigma(\x) \cdot (\x_j - \x) + \dots \,,  
\eeq
and turning to momentum-space representation of the propagators, 
the internal traces and the integrals over $x_j$ are readily evaluated. 
The terms can then be organized in the same way as in \eq{eq:GLNJLgen},
so that the coefficients $\beta_i$ can be read off.

For the explicit expressions we refer to Ref.~\cite{Nickel:2009ke}.
The most important result is that the two 4th-order coefficients are equal,
\beq
\label{eq:GLNJLCPeqLP}
       \beta_{4,a} = \beta_{4,b} \quad \Rightarrow \quad
       \gamma_{4,a} = \gamma_{4,b} \equiv \gamma_4\,,
\eeq
and, hence, the Lifshitz point coincides with the critical point 
(cf.~Eqs.~(\ref{eq:NJLGLCP}) and (\ref{eq:NJLGLLP})).
Moreover, since $ \gamma_{4,a} < 0$ is a condition for a first-order phase transition
between homogeneous phases, while  for $\gamma_{4,b} < 0$ inhomogeneous solutions 
are favored, this suggests that the first-order phase boundary which would be present
in the homogeneous case is covered by an inhomogeneous phase~\cite{Nickel:2009ke}.
This prediction has been confirmed by numerical calculations~\cite{Nickel:2009wj},
and, as we will see later, seems to hold even beyond the range of validity of the GL expansion.  

A relation analogous to \eq{eq:GLNJLCPeqLP} has been found earlier for the 
$1+1$ dimensional Gross-Neveu model~\cite{Schnetz:2004}.
In both cases, the equality of $\beta_{4,a}$ and $\beta_{4,b}$
can be shown by an integration by parts.
However, while this is straightforward in $1+1$ dimensions, in $3+1$ dimensions,
the integrals are divergent and must be regularized.
In order to arrive at \eq{eq:GLNJLCPeqLP},  Nickel had to make the additional 
assumption that for the regularized integrals there are no surface contributions
when the integration by parts is performed.
This excludes for instance a simple cutoff regularization,
whereas Pauli-Villars and proper-time regularization turned out to be 
suitable schemes.
We also stress again that the above results rely on the assumption that the
the 6th- and higher-order terms of the GL expansion are positive. 
In Ref.~\cite{Abuki:2011} Abuki, Ishibashi and Suzuki have extended Nickel's analysis
to include terms up to 8th order in the expansion.
Assuming that the 8th-order terms are positive, they have then studied the influence
of the sign of the 6th-order coefficient $\gamma_6$ on the phase diagram.\footnote{
As in 4th order, all 6th-order terms are proportional to a single coefficient $\gamma_6$.} 
While for $\gamma_6>0$ Nickel's results remain qualitatively unchanged, 
a much richer phase structure showed up for $\gamma_6<0$. 
In particular there is a second homogeneous chirally broken region and, related to this,
a new critical endpoint, a triple point, and a Lifshitz point, which are all distinct. 
Abuki et al.\@ estimated, however, that this only occurs for unrealistically large
values of the coupling constant. Indeed, the explicit model calculations presented later 
in this review all favor Nickel's scenario if a proper regularization scheme is used.

Another interesting observation can be made in the vicinity of the second-order phase boundary
between the inhomogeneous and the chirally restored phase. 
At this boundary the amplitude of the mass function goes to zero, so that it is sufficient
to take into account two self-energy insertions, i.e., stop the expansion in \eq{eq:Omegakinsum}
at $n=2$.
However, except for the LP, the gradients of $M$ do not vanish at the phase boundary, and we
must keep higher orders. The corresponding expansion thus reads
\beq
\label{eq:GLNJLbound}
\Omega = 
\Omega_0 + \Delta\Omega^{(2)}
=
\Omega_0
+ \frac{1}{V}\int d^3x \, \Big\{  
    \frac{1}{2}\gamma_{2 } |M(\x)|^2
     + \frac{1}{4}\gamma_{4}|\nabla M(\x)|^2   
     + \frac{1}{12}\gamma_{6}|\nabla^2 M(\x)|^2  +      
     \dots \Big\}
\,,    
\eeq
where we assume again that $\gamma_6$ and all
higher-order coefficients are positive.
Inserting the Fourier decomposition \eq{eq:Mxq}
and performing the integration,
one finds that the different modes decouple at this order: 
\beq
\Delta\Omega^{(2)}
= \sum_{\q_k} |M_{\q_k}|^2 \, \Big\{  
    \frac{1}{2}\gamma_{2 } 
     + \frac{1}{4}\gamma_{4}\q_k^{\;2}   
     + \frac{1}{12}\gamma_{6}\q_k^{\;4}  +      
     \dots \Big\}
\equiv      
\sum_{\q_k} |M_{\q_k}|^2 \, \mathcal{F}(\q_k^{\;2}; \{\gamma_i\})
     \,.
\eeq
Minimization with respect to $|\q_k|$ yields an optimized length $q_\text{opt}$ 
of the wave vector, given by
$\frac{\partial\mathcal{F}}{\partial\q_k^{\;2}}|_{\q_k^{\;2}  =q_\text{opt}^2} = 0$. 
This restricts the sum to wave vectors which all have the same modulus,\footnote{
In principle, the equation $\frac{\partial\mathcal{F}}{\partial\q_k^{\;2}}|_{\q_k^{\;2}  =q_\text{opt}^2} = 0$
can have more than one solution. However, it is very unlikely that two different solutions coincide
simultaneously with one or more points of the (discrete) reciprocal lattice.}
\beq
       \Delta\Omega^{(2)}_\text{opt} = 
       \mathcal{F}(q_\text{opt}^{\;2}; \{\gamma_i\}) \sum_{|\q_k|=q_\text{opt}} |M_{\q_k}|^2 
     \,,
\eeq
i.e., the amplitudes of higher harmonics vanish. 
Still, the wave vectors can have different directions, 
so that various configurations are possible,
e.g., a single plane wave, a one-dimensional sine, or higher-dimensional superpositions of 
plane waves with a fixed wave number. 
The remarkable result is, however, that for all these configurations the phase boundaries 
to the restored phase coincide, as they are given by the condition that 
$\mathcal{F}(q_\text{opt}^{\;2}; \{\gamma_i\})$ changes its sign.

This can be worked out more explicitly in the vicinity of the LP, where gradients become small as 
well and terms of the order $\q_k^{\;6}$ can be neglected.
We then find $q_\text{opt}^2 = -\frac{3}{2}\frac{\gamma_4}{\gamma_6}$ and 
$\mathcal{F}(q_\text{opt}^{\;2}; \{\gamma_i\}) =     
\frac{1}{2}\gamma_{2 }  - \frac{3}{16}\frac{\gamma_{4}^2}{\gamma_6}$.
Since $\gamma_4$ must be negative to allow for an inhomogeneous
phase on one side of the phase boundary, while $\gamma_2$ must be positive to allow for a
restored phase on the other side, this yields that the phase boundary is given by
$\gamma_{4} = - \sqrt{\frac{8}{3}\gamma_2\gamma_6}$~\cite{Nickel:2009ke}. 
In particular the phase transition is of second order, if the GL expansion is not
invalidated by higher-order terms.

The GL analysis of the phase transition between the homogeneous chirally broken to the
inhomogeneous phase is more difficult but has been done as well~\cite{Nickel:2009ke,Abuki:2011}.
We will come back to this in Sec.~\ref{sec:dimred}.

\subsection{\it Quark-meson model}
\label{sec:GLQM}

The GL expansion for the QM model can be done in exactly the same manner. 
As we have seen in Sec.~\ref{sec:qmmodel},
the thermodynamic potential, \eq{eq:OmegaQMsum},
is now a sum of a mesonic part and a quark part,
where the latter is identical to the kinetic contribution to the NJL thermodynamic 
potential if we make the identifications given in \eq{eq:replace}.
Hence, in terms of the mass function $M(\x)$, defined in \eq{eq:MxQM},
the GL expansion takes again the form of \eq{eq:GLNJLgen}, 
with $\gamma_i = \alpha_i + \beta_i$ being a sum of the contributions from 
$\Omega_\text{mes}$ and $\Omega_\text{q}=\Omega_\mathit{kin}$, respectively.
For the $\beta$-coefficients we can then directly take over the NJL-model results,
so that we only need to compute the $\alpha_i$.

To that end we insert \eq{eq:HmesM} into \eq{eq:Omegamesx} and expand the integral
analogously to \eq{eq:GLNJLgen}. 
Comparing the integrands, one immediately sees that the nonvanishing coefficients 
are~\cite{Nickel:2009wj} 
\beq
       \alpha_2 = -\frac{\lambda v^2}{g^2}\,, \quad
       \alpha_{4,a} = \frac{\lambda}{g^4}\,, \quad
       \alpha_{4,b} = \frac{2}{g^2}\,. 
\eeq
Hence, in contrast to the condensate part of the NJL model, \eq{eq:alphaNJL},
the meson part of the QM model also contributes to the 4th-order GL coefficients.
As a consequence, although $\beta_{4,a}$ and $\beta_{4,b}$ are still equal,
$\gamma_{4,a}$ and $\gamma_{4,b}$ 
are in general not: 
\beq
       \gamma_{4,a} - \gamma_{4,b} = \alpha_{4,a}-\alpha_{4,b} = \frac{1}{g^4} (\lambda-2g^2)\,.
\eeq
This means, the LP only coincides with the CP if $\lambda = 2g^2$, which is not 
necessarily the case.
A closer inspection reveals that the relative location of the two points strongly depends on the
ratio between the vacuum masses of the constituent quark and the sigma meson~\cite{Carignano:2014jla}.
Indeed, for the sMFA relations, given in \eq{eq:sMFAparameters}, one finds
\beq
       \gamma_{4,a} - \gamma_{4,b}
=    
       2\frac{f_\pi^2}{M_\text{vac}^2}  \left( \frac{m_{\sigma}^2}{4M_\text{vac}^2}-1 \right)\,,
\eeq
while including the Dirac sea only leads to a modification of the prefactor~\cite{Carignano:2014jla}.
Thus, in both cases, LP and CP coincide if $m_\sigma = 2M_\text{vac}$.
Remarkably, this is just the NJL-model relation, where $m_\sigma$ is always equal to
$2M_\text{vac}$ in the chiral limit~\cite{NJL1,Klevansky:1992qe}.
In contrast to the NJL model, however, the ratio of $m_\sigma$ and $M$ can be chosen freely
in the QM model, and we have therefore the possibility to 
separate the two points. 
This is confirmed by numerical calculations, which will be presented in Sec.~\ref{sec:CDWmodels}.

The GL analysis has also given insights in the importance of the Dirac-sea 
contributions for the phase structure. 
According to the vacuum and medium parts of $\Omega_\mathit{kin}$, \eq{eq:Omegakinpm},
the $\beta$-coefficients can be decomposed as
\beq
       \beta_i(T,\mu) = \beta_i^{\text{vac}} + \beta_i^{\text{med}}(T,\mu)\,. 
\eeq
In the chiral limit, the coefficients $\beta_4^\text{vac}$ and $\beta_4^\text{med}$ both
exhibit logarithmic divergencies in the infrared, which cancel each other if they are summed.  
In the sMFA, however, the vacuum term is dropped, and therefore this cancellation does not 
occur~\cite{Skokov:2010}. 
As a consequence, the phase transition at $\mu=0$, which is expected to be of second order 
for two massless flavors, becomes first order, if the analysis is restricted to homogeneous phases.
Allowing for inhomogeneous solutions one could thus have an inhomogeneous phase reaching
down to vanishing chemical potential.
Although the GL analysis is, strictly speaking, invalidated by the diverging coefficients, this is
indeed what one finds numerically in the sMFA, while this artifact disappears 
when the Dirac sea is included (cf.~Sec.~\ref{sec:CDWmodels}).

\section{Chiral density wave}
\label{sec:CDW}

After having discussed some generic features of inhomogeneous phases in the NJL and QM model, we are now ready to discuss specific examples for spatially modulated condensates and investigate the corresponding 
structure. 
As pointed out in Sec.~\ref{sec:general}, the general solution of the problem 
involves the diagonalization of infinite matrices in momentum space
for arbitrary condensate functions
and the subsequent minimization of the free energy with respect to these functions.
Equivalently, an infinite set of coupled gap equations in the various Fourier components
would have to be solved.
Until now, this problem has not been solved rigorously in $3+1$ dimensions.
Inspired by the GL analysis however, one can expect that many qualitative features of inhomogeneous phases will be independent
of the explicit shape chosen for the chiral condensate. This suggests us that in order to get a first
glimpse of the model phase structure in presence of space-dependent condensates, it is possible to consider simpler variational ans{\"a}tze 
which will enforce a less involved structure in momentum space, thus dramatically reducing the complexity of the problem. 

With this spirit in mind, the simplest possible ansatz is given by a single plane wave, 
\beq
\label{eq:Mcdw}
M(\x) = \Delta e^{i\q \cdot \x} \,,
\eeq
with two variational parameters $\Delta$ and $\q$, characterizing the amplitude and the wave vector
of the modulation. 
This ansatz is  formally analogous to the Fulde-Ferrell solutions 
introduced in the context of superconductivity~\cite{Fulde:1964zz} 
and color superconductivity~\cite{Alford:2000ze}.
According to Eqs.~(\ref{eq:Mx}) and (\ref{eq:MxQM}) it corresponds to  a sinusoidal behavior
of the scalar and pseudoscalar condensates in the NJL model,
\beq
\label{eq:condCDW}
       \ps(\x) = -\frac{\Delta}{2G_S} \cos(\q\cdot\x)\,, \qquad 
       \pp(\x) = -\frac{\Delta}{2G_S} \sin(\q\cdot\x)\,,       
\eeq
or of the scalar and pseudoscalar meson fields in the QM model,
\beq
       \sigma(\x) = \frac{\Delta}{g} \cos(\q\cdot\x)\,, \qquad 
       \pi(\x) = \frac{\Delta}{g} \sin(\q\cdot\x)\,,      
\eeq
which is just the example we have already addressed in the Introduction,
cf.~\eq{eq:CDWansatz}.
In analogy to the spin-density waves in condensed matter systems~\cite{Overhauser:1962},
this ansatz is called ``chiral density wave'' (CDW) or
``dual chiral density wave" (DCDW), where ``dual'' refers to the presence of two 
(scalar and pseudoscalar) standing waves~\cite{NT:2004b,NT:2004}.\footnote{
From a technical point of view an ansatz with a single cosine, e.g., in the scalar channel,
is actually much more involved, as it corresponds to a superposition of two plane waves,
$M \sim (e^{i\q\cdot\x} + e^{-i\q\cdot\x})$. 
}
The ansatz is also sometimes called ``chiral spiral'' because $M(\x)$ describes a spiral in the
complex plane, when going along the $\q$ direction. 
In the following, however, we reserve this term for the $1+1$ dimensional analogue, for which 
it was originally introduced~\cite{Schon:2000he}, and stick to the name CDW in the $3+1$ dimensional
case.

\subsection{\it Eigenvalue spectrum  and expectation values}
\label{sec:CDWev}

Plugging the  CDW ansatz into \eq{eq:mfH}, the effective Dirac Hamiltonian takes the form 
\beq
\label{eq:HCDW}
       H(\x) = -i\gamma^0\gamma^i\partial_i + \gamma^0 \Delta e^{i\gamma^5\tau^3 \q\cdot\x}\,.
\eeq
The remaining task is then to find the eigenvalues of this operator and to minimize the 
corresponding thermodynamic potential with respect to the two variational parameters 
$\Delta$ and $q = \vert \q \vert$.

The (almost unique) property of the CDW ansatz which makes it historically so important is
that the eigenvalues of $H$ can be computed analytically~\cite{Dautry:1979}.
The most elegant way is to solve the corresponding Dirac equation by performing
a local chiral rotation of the quark fields~\cite{Dautry:1979,Kutschera:1989yz,Kutschera:1990xk}
\beq
\label{eq:localchiralrotation}
\psi(\x) = U(\x) \psi'(\x) 
=
\exp(-\frac{1}{2} i\gamma^5 \tau^3 \q\cdot\x) \psi'(\x) \,.
\eeq
The Dirac equation $H\psi = E\psi$ is then equivalent to the equation 
$H'\psi' = E\psi'$ for the rotated fields with the rotated Hamiltonian 
\beq
        H' 
        = U^\dagger(\x) H(\x) U(\x)
        =-i\gamma^0\gamma^i\partial_i  -\frac{1}{2}\gamma^0\gamma^i \gamma^5 \tau^3 q^i
        + \gamma^0 \Delta\,.
\eeq
Since this Hamiltonian does not depend on $\x$ 
the eigenvalue problem can
be solved with a standard plane-wave ansatz 
(for more details, see, e.g., Ref.~\cite{Kutschera:1990xk}).
The resulting eigenvalues are the positive and negative square roots of
\beq
\label{eq:CDWdispers2}
        E^2_\pm(\p) = \p\,^2 + \Delta^2 + \frac{\q\,^2}{4} \pm\sqrt{\Delta^2\q\,^2 + (\q\cdot\p)^2}\,,
\eeq
as already quoted in \eq{eq:CDWdispers}.

Alternatively, we could follow the framework outlined in Sec.~\ref{sec:general} and diagonalize 
$H$ in momentum space.
Compared with the general case,
this is greatly simplified by the fact that the mass function, \eq{eq:Mcdw}, has only a single
Fourier component,
so that only the diagonal and the first off-diagonal momentum blocks of the matrix  \eq{eq:Hmn} 
are filled.
Moreover, a closer inspection of the Dirac structure reveals that the Hamiltonian can be decomposed
into finite-dimensional blocks, whose eigenvalues can be computed analytically. 
The result is, of course, the same as with the other method.\footnote{From the method of 
Sec.~\ref{sec:general}, we obtain an eigenvalue spectrum $E_\lambda(\mk)$, where
$\lambda$ labels an infinite discrete set and
$\mk$ is restricted to the Brillouin zone. For the CDW the latter is infinite in the transverse directions,
but $k_\parallel$ is limited to the interval $[0,q)$. 
It is possible, however, to join the infinite number of discrete pieces in such a way that one finally ends up with
the spectrum given by \eq{eq:CDWdispers2}.
}

Besides the eigenvalues, it is also interesting to study the properties of the eigenstates, i.e.,
of the quasiparticles in the CDW background.
This has already been done by Dautry and Nyman \cite{Dautry:1979}
for the non-relativistic limit, and later by 
Kutschera et al.~\cite {Kutschera:1989yz,Kutschera:1990xk}
and by Nakano and Tatsumi~\cite{NT:2004}
for the relativistic case. 
In this context we remind that in the CDW ground state only the $E_-$ branch is populated.
In the non-relativistic limit, one finds that the corresponding quasiparticles are up quarks
with spin parallel and down quarks with spin antiparallel to the wave vector $\q$.
It was pointed out in Refs.~\cite{Kutschera:1989yz,Kutschera:1990xk} that the unequal magnetic 
moments of the up and down quarks should give rise to a nonvanishing magnetization of the matter, 
even though the total spin is zero in the isospin-symmetric case.
This effect is somewhat weaker in the relativistic case, but still present.

Using the chirally rotated states, \eq{eq:localchiralrotation}, general expectation values of an
operator $\Gamma$ are given by
\beq
       \ave{\bar\psi|\Gamma|\psi} = 
       \ave{\bar\psi' | e^{-\frac{1}{2} i\gamma^5 \tau^3 \q\cdot\x}
       \,\Gamma\,
        e^{-\frac{1}{2} i\gamma^5 \tau^3 \q\cdot\x} | \psi'}\,.
\eeq
This becomes particularly simple if $\Gamma$ does not contain derivatives and anticommutes
with $\gamma^5\tau^3$. In this case the exponentials drop out, i.e., 
$\ave{\bar\psi|\Gamma|\psi}  =  \ave{\bar\psi'|\Gamma|\psi'}$.
Unless $\Gamma$ has an explicit space dependence, the result is then constant in space.
An important example is the quark number density, corresponding to 
$\Gamma = \gamma^0$, which is therefore constant, despite the oscillating chiral condensates.

On the other hand, since $\Gamma = 1$ and $\Gamma = i\gamma^5\tau^3$ commute with 
$\gamma^5\tau^3$, the scalar and pseudoscalar condensates are not constant in space.
In fact, selfconsistency of the problem requires that they are just given by $\ps$ and 
$\pp$, respectively, and it can be shown that this is the case for the stationary points of
the thermodynamic potential with respect to $\Delta$.

In this context we note that the CDW ansatz only has selfconsistent solutions in the chiral limit.
Attempts to accommodate a nonzero bare quark mass at least in an approximate way have 
been performed in Ref.~\cite{Maedan:2009yi}
and recently in Ref.~\cite{Karasawa:2013zsa}
where a generalized CDW ansatz was used in a variational approach.
However, as we will see in Sec.~\ref{sec:dimred},
there exists a different class of modulations, which can straightforwardly be extended to
nonvanishing bare quark mass in a selfconsistent way.

\subsection{\it Model results}
\label{sec:CDWmodels}

\begin{figure}[H]
\centering
\includegraphics[angle=0,width=.4\textwidth]{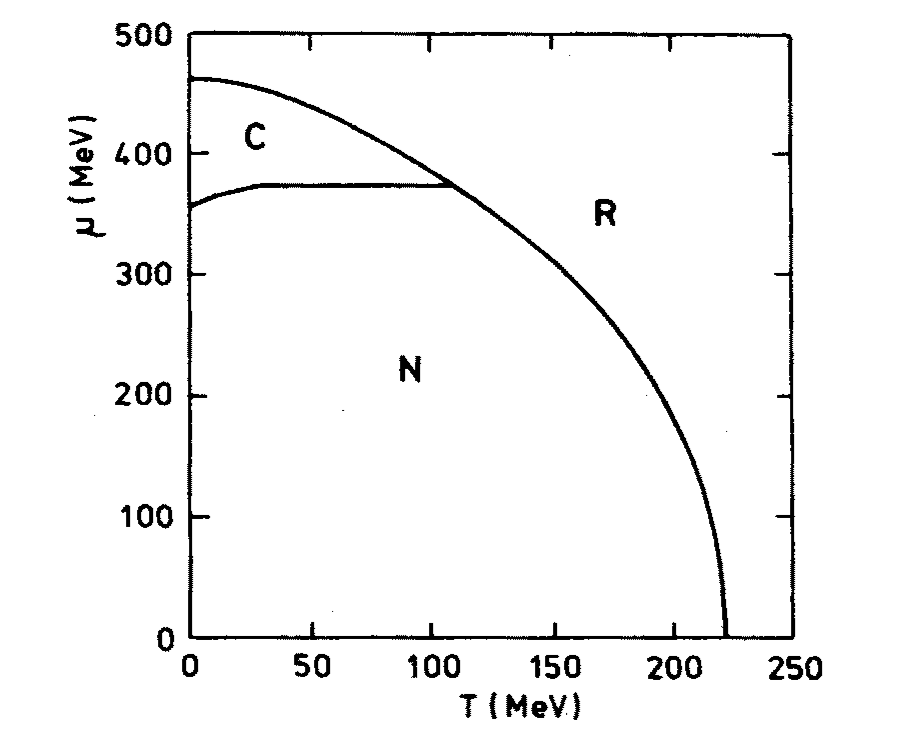}
\caption{Phase diagram presented in Ref.~\cite{Broniowski:1990}  for the QM model,
allowing for CDW-type modulations. 
The homogeneous phases with broken and restored chiral symmetry are labelled with
``N'' and ``R'', respectively, whereas ``C'' indicates the region where the CDW solution is
favored. Notice that the temperature and chemical-potential axes are interchanged 
compared with today's conventions.
}
\label{fig:pdbron1}
\end{figure}

Being an analytically treatable case, the plane wave-ansatz has been
the object of intense investigations during the course of the last 25 years. 
Aside from its historical relevance, the CDW provides us with an excellent prototype 
for many generic features of inhomogeneous condensation in quark matter.
As will become apparent in
the following chapters, even when considering more involved types of modulations, 
many qualitative features of the phase structure will remain the same. Indeed, the reader 
will notice that most of the phase diagrams shown throughout this review bear a remarkable
resemblance to \Fig{fig:pdbron1},
which was obtained by Broniowski et al.~\cite{Broniowski:1990} already in 1990 
(published 1991) in the QM model.
The most striking feature here is the existence of an inhomogeneous ``island'' in the ($T,\mu$) phase diagram 
at low temperatures and intermediate densities, where a CDW solution with nonvanishing amplitude
and wave number is favored over the homogeneous chirally broken and restored phases.

\begin{figure}[htp]
\centering
\includegraphics[angle=0,width=.5\textwidth]{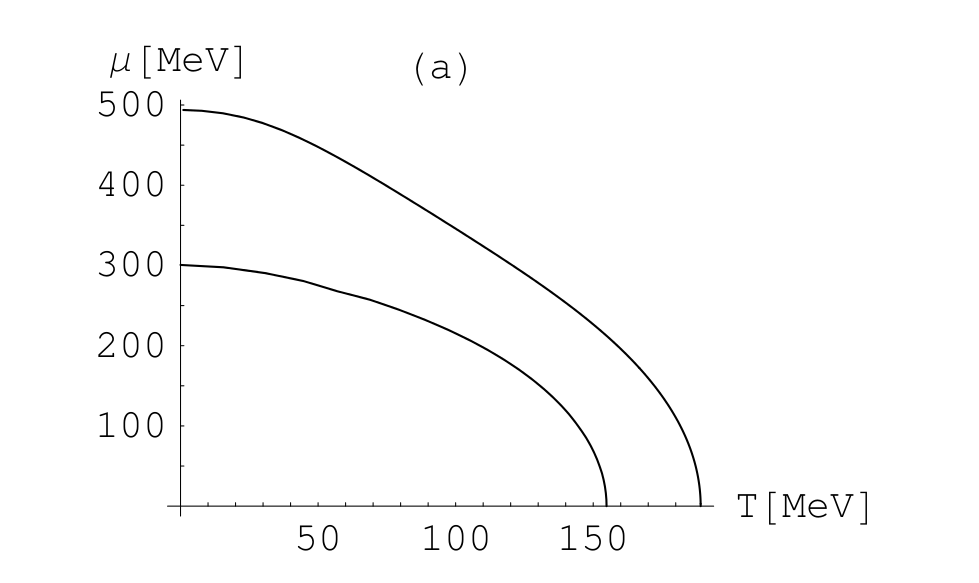}
\includegraphics[angle=0,width=.4\textwidth]{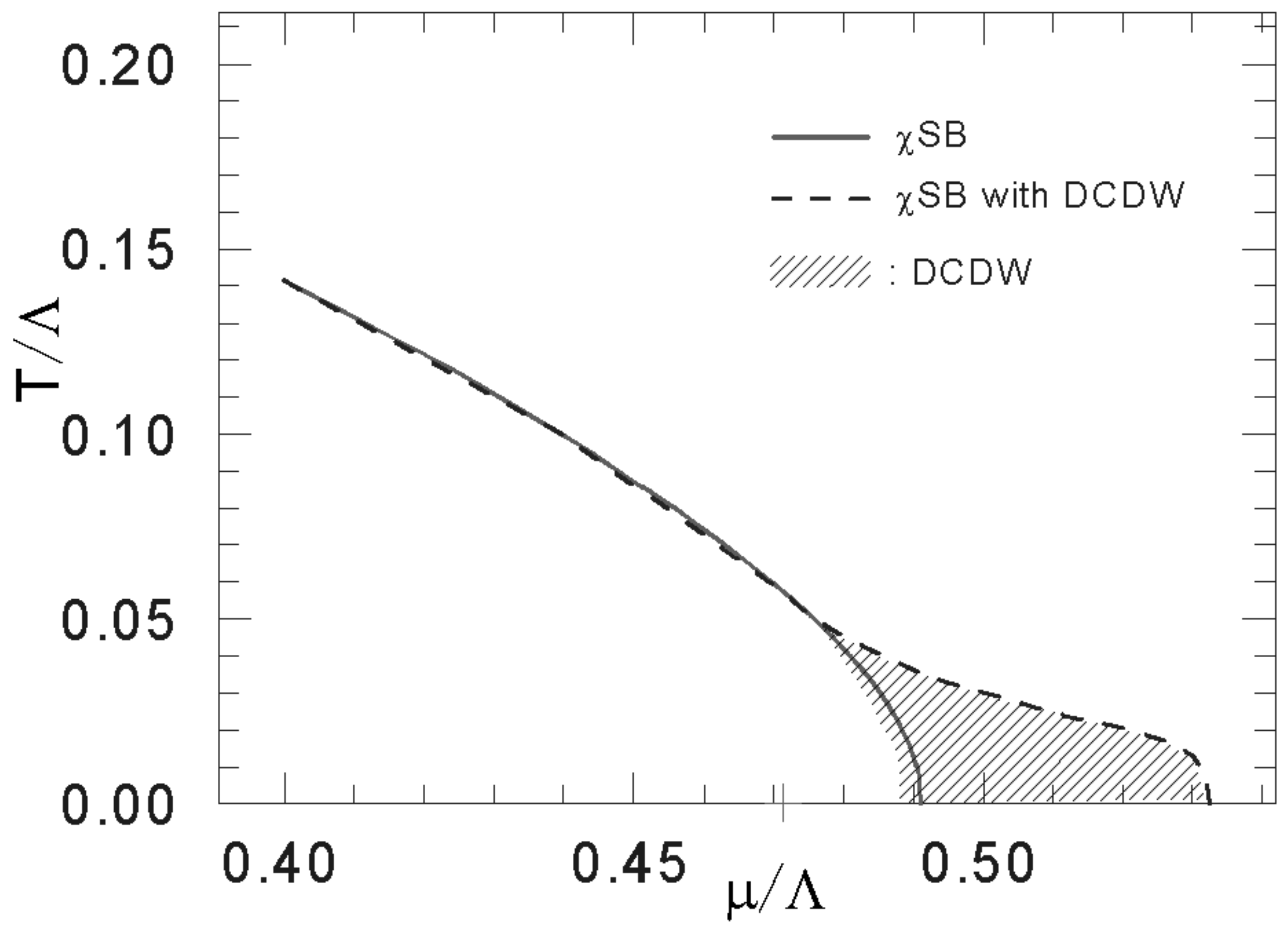}
\caption{Phase diagrams presented in Refs.~\cite{Sadzikowski:2000} (left) and 
\cite{NT:2004} (right) for the NJL model, allowing for CDW-type modulations. 
In both figures, there is a homogeneous phase with broken chiral symmetry 
in the lower left region and the restored phase in the upper right region,
while the phase in between (shaded in the right figure) corresponds to a
favored CDW solution. 
In the left figure the axes are again interchanged compared with today's 
conventions. The scale in the right figure is $\Lambda = 850$~MeV.
}
\label{fig:pdSadBroNakTat}
\end{figure}

Most of the work performed in the field since then
has been devoted to determining the size of this island, the properties of quark matter inside it,
and the nature of the phase transitions.
Two further examples of phase diagrams are shown in \Fig{fig:pdSadBroNakTat},
summarizing the NJL-model results of Sadzikowski and Broniowski (left) 
and of Nakano and Tatsumi (right).
Again, both groups find that there is an inhomogeneous phase located in the 
intermediate-density region, but obviously there are also quantitative differences.

\begin{figure}[H]
\centering
\includegraphics[width=.4\textwidth]{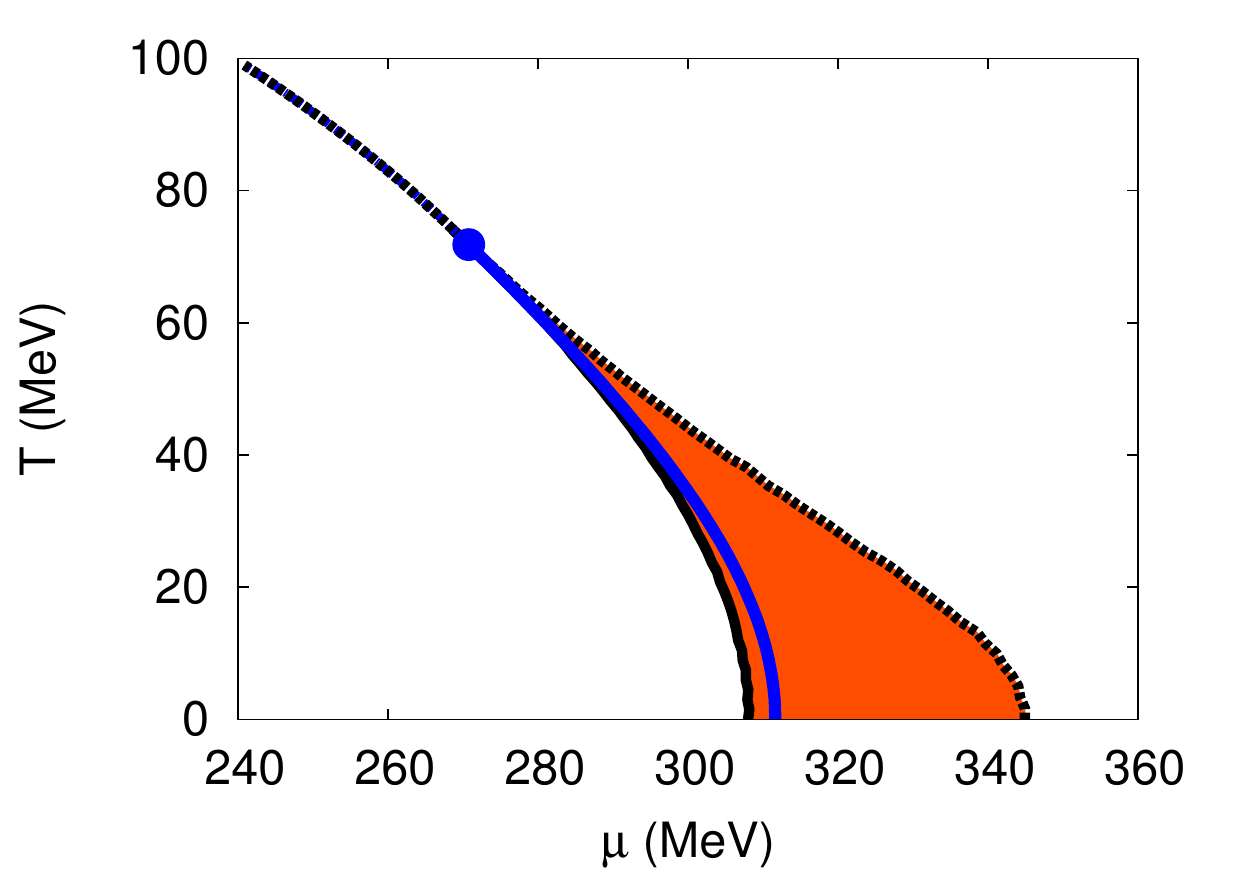}
\includegraphics[width=.37\textwidth]{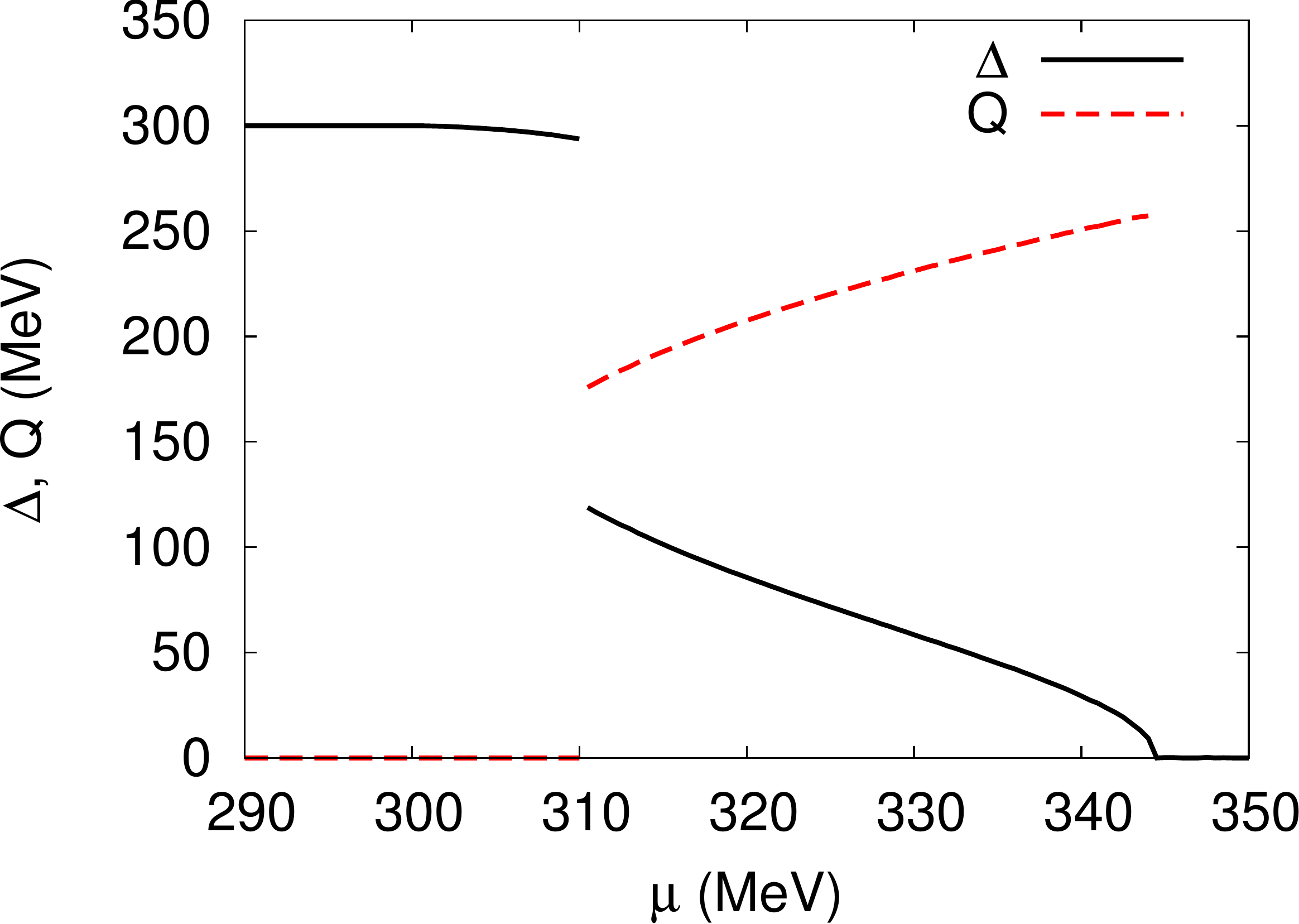}
\caption{Left: Phase diagram for the NJL model, allowing for CDW-type modulations. 
The inhomogeneous phase corresponds to the the shaded region. 
Solid (dashed) lines indicate first (second) order phase boundaries. 
The blue solid line inside the inhomogeneous region indicates the first-order phase
boundary which is obtained if the analysis is restricted to homogeneous phases.
The corresponding CP, which coincides with the LP, is marked by a dot.
Right: Favored values of the amplitude $\Delta$ (black solid line) and 
the wave number (divided by 2)
$Q = q/2$
(red dashed line)
for $T=0$ as functions of $\mu$.
}
\label{fig:njlcdw}
\end{figure}

As a reference point for a more detailed discussion, we show in \Fig{fig:njlcdw} our own NJL-model results,
which have been obtained employing  the formalism described above, together with the
Pauli-Villars-type regularization scheme developed in Ref.~\cite{Nickel:2009wj}.
In the phase diagram, displayed in the left panel, we have also indicated the first-order 
chiral phase boundary one finds if the possibility of inhomogeneous phases is not
taken into account (blue solid line). 
As one can see, this phase boundary is completely covered by the inhomogeneous phase
(shaded region). 
Moreover, in agreement with the GL prediction discussed in Sec.~\ref{sec:GLNJL},
the LP where the two homogeneous and the inhomogeneous phase meet, 
sits on top of the CP, where in the homogeneous case the first-order phase transition 
turns into second order.

Although at that time they were unaware of this property, CP and LP should also coincide 
in the phase diagram of Nakano and Tatsumi (right panel of \Fig{fig:pdSadBroNakTat})
who employed a proper-time regularization scheme.
In contrast, there is no CP in Sadzikowski and Broniowski's phase diagram (left panel of \Fig{fig:pdSadBroNakTat}), where the CDW region extends all the way down to the 
temperature axis, even though there is a CP in the corresponding homogeneous 
analysis~\cite{Schwarz:1999dj}.
This is most likely an artifact of the cutoff regularization used in these references, which
spoils the GL property.\footnote{In addition, Sadzikowski and Broniowski expanded the thermodynamic 
potential in powers of $\q\,^2$ and neglected terms of $\mathcal{O}(\q\,^4)$. 
In this way the model becomes effectively a QM model where the Dirac sea is regularized by a finite
cutoff.}

The various model studies all agree that the transition from the homogeneous chirally broken
phase to the CDW phase is of first order.
For $T=0$ this can be seen from the right figure in \Fig{fig:njlcdw}, where the amplitude and
the half-wave number
$Q = q/2$\footnote{
In the literature there are two standard definitions of the CDW modulations, which differ
by a factor of 2 in the periodicity. 
In this review, we reserve the letter $q$ for the true wave number, cf.~\Eq{eq:Mcdw},
while the introduction of $Q=q/2$ is motivated by the fact that in $1+1$ dimensions
one finds $q=2\mu$ and hence $Q = \mu$. In $3+1$ dimensions $Q$ is lower but expected 
to approach $\mu$ at high chemical potentials.
} 
of the favored solution are plotted as functions of the chemical potential.
At the onset of the inhomogeneous phase both quantities change discontinuously, clearly
signalling a first-order phase transition. 
In particular the wave number jumps from zero to a finite value and grows further when the 
chemical potential is increased.
The amplitude, on the other hand, decreases and eventually goes to zero, marking the transition
to the chirally restored phase.
Since this happens in a smooth way, we conclude that the phase transition is of second order,
which is in agreement with the findings of Sadzikowski and Broniowski~\cite{Sadzikowski:2000}.
In contrast, Nakano and Tatsumi report a weak first-order transition to the restored 
phase~\cite{NT:2004}.
Although the model details are not completely identical, we believe that this discrepancy is
more a numerical issue, since it is always a delicate problem to distinguish a weak first-order
from a second-order phase transition numerically. 
In this context we remind that the GL analysis to 6th order predicts a second-order transition 
to the restored phase~\cite{Nickel:2009ke,Abuki:2011}, 
although the relevance of higher-order terms cannot be excluded.\footnote{
Indeed, in Ref.~\cite{Abuki:2011} the 8th-order coefficient was found to be negative in the 
relevant region. 
A similar effect was found in Ref.~\cite{NB:2009} in the context of color superconductivity.}

We also
note that in the ``historic'' phase diagram of Broniowski et al., \Fig{fig:pdbron1},
all phase transitions are first order. Concerning the transition between CDW and restored phase
this is simply due to the fact that, for numerical reasons, the amplitude of the modulation was 
kept constant in the CDW, and the thermodynamic potential was only minimized with respect
to $q$. 
A second-order phase transition to the restored phase is therefore obviously excluded by
construction.
The first-order phase transition between the two homogeneous phases in \Fig{fig:pdbron1},
on the other hand, can be traced back to the omission of the Dirac sea, as discussed in
Sec.~\ref{sec:GLQM}.

\begin{figure}[htp]
\centering
\includegraphics[angle=270,width=.32\textwidth]{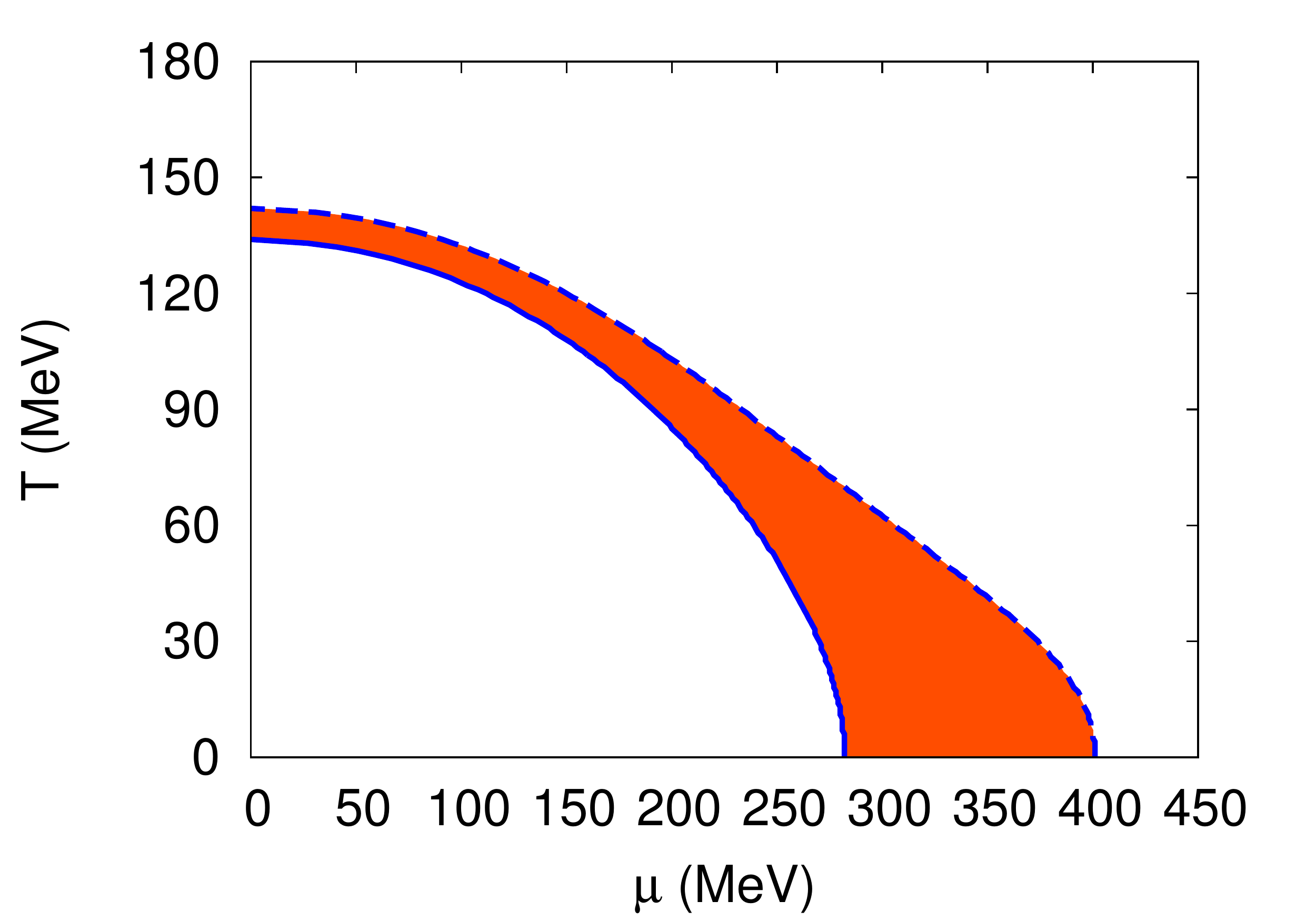}
\includegraphics[angle=270,width=.32\textwidth]{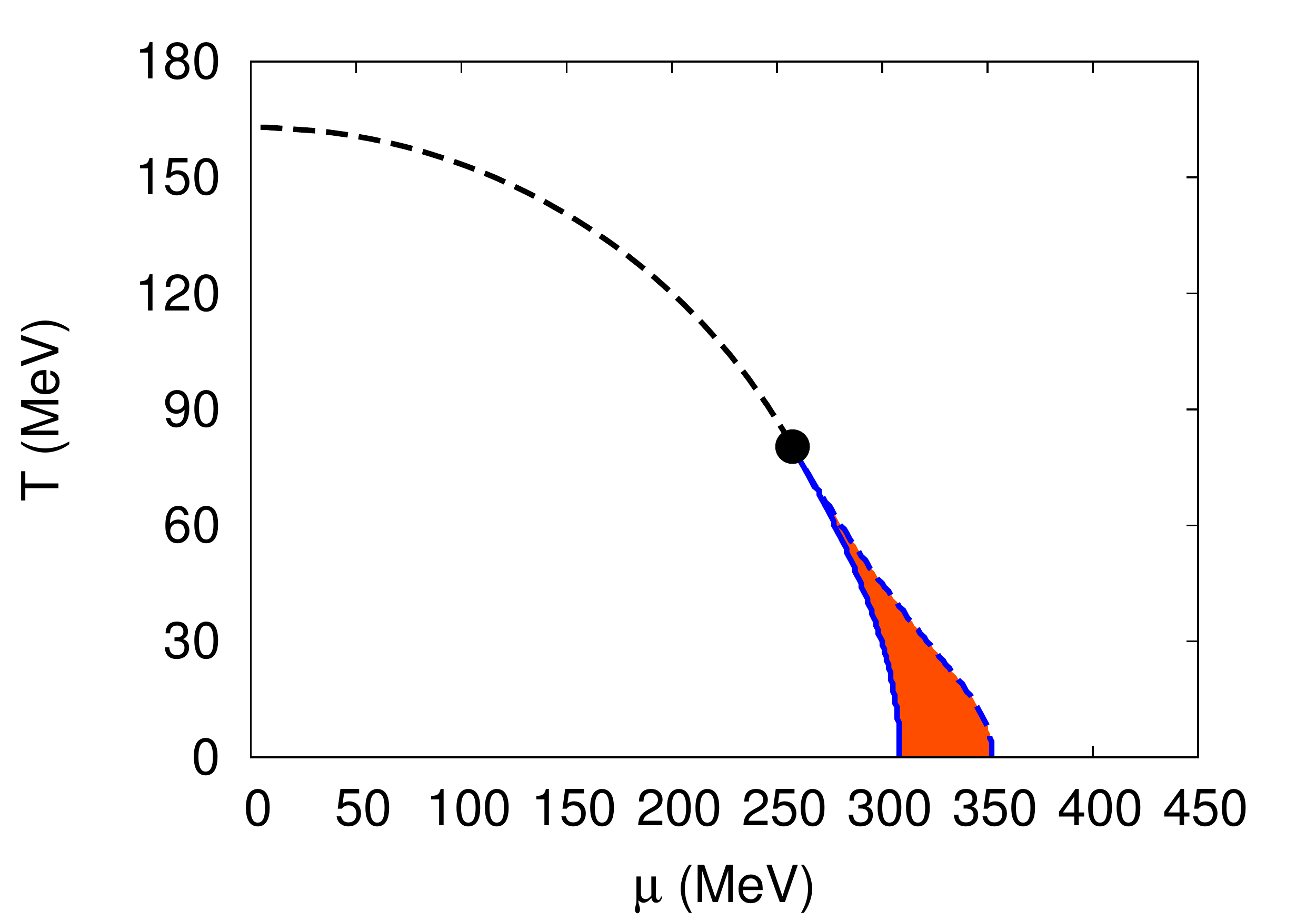}
\includegraphics[angle=270,width=.32\textwidth]{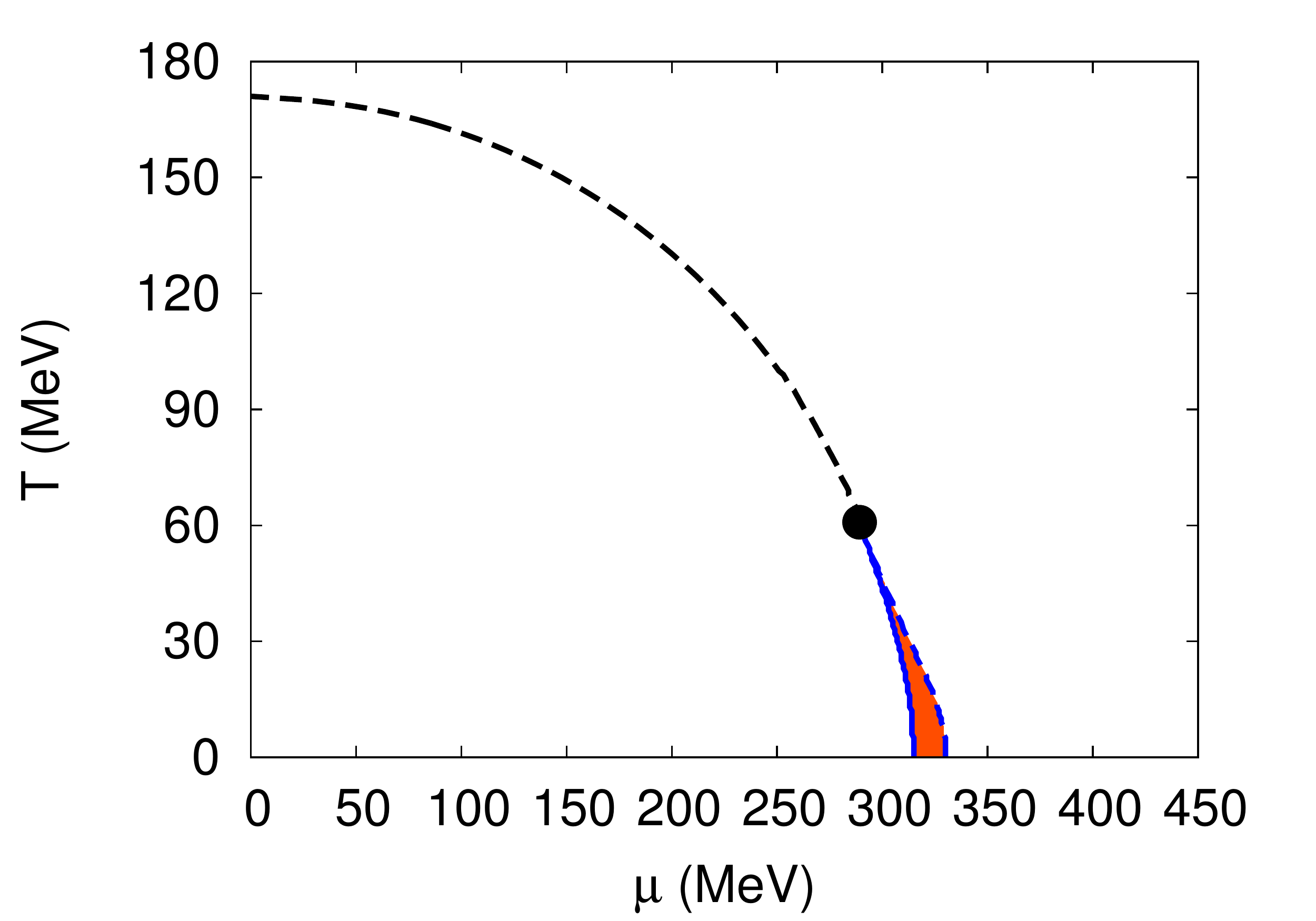}
\caption{Phase diagram of the QM model, allowing for CDW-type modulations.
The different panels correspond to different values of the Pauli-Villars regulator $\Lambda$,
which controls the extent of the Dirac-sea contribution to the thermodynamic potential.
 Left: sMFA ($\Lambda=0$).  Middle: $\Lambda=600$~MeV.  
 Right: ``renormalized''
  results, obtained with $\Lambda=5$~GeV.  First-order phase
  transitions are indicated by solid lines, second-order transitions
  by dashed lines.  The shaded areas indicate the inhomogeneous phase,
  the black dot denotes the Lifshitz point, coinciding with the
  location of the critical point for homogeneous phases.  
  From Ref.~\cite{Carignano:2014jla}.}
\label{fig:pdqmrenorm}
\end{figure}

Only very recently the study of inhomogeneous phases has been performed in a properly 
``renormalized'' quark-meson model \cite{Carignano:2014jla}. 
The main result is shown in \Fig{fig:pdqmrenorm}, where the influence of the Dirac sea on 
the phase diagram is discussed in a systematic way by increasing the corresponding Pauli-Villars
regulator from zero (no Dirac sea, left panel) via an intermediate value (middle) to essentially infinity
(``renormalized result'', right). 
Since the calculations have been performed for $m_\sigma = 2M_\text{vac}$, the LP
again coincides with the CP (see Sec.~\ref{sec:GLQM}).
The latter is known from homogeneous studies to be absent in the sMFA , 
while it exists when the Dirac sea is included~\cite{Skokov:2010},
and hence the LP behaves in the same way.

\begin{figure}[hbt]
\centering
\includegraphics[angle=270,width=.48\textwidth]{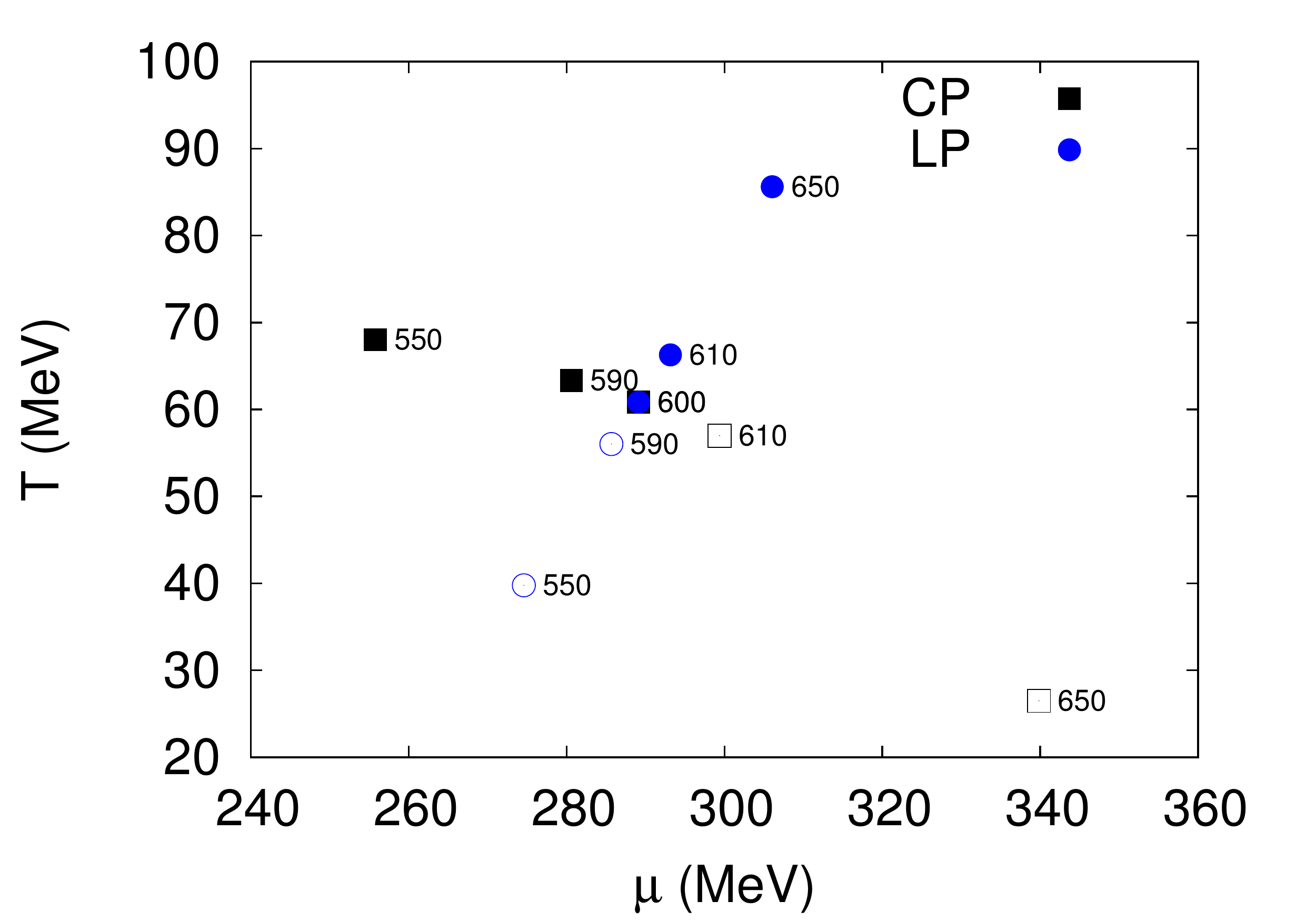}
\includegraphics[angle=270,width=.48\textwidth]{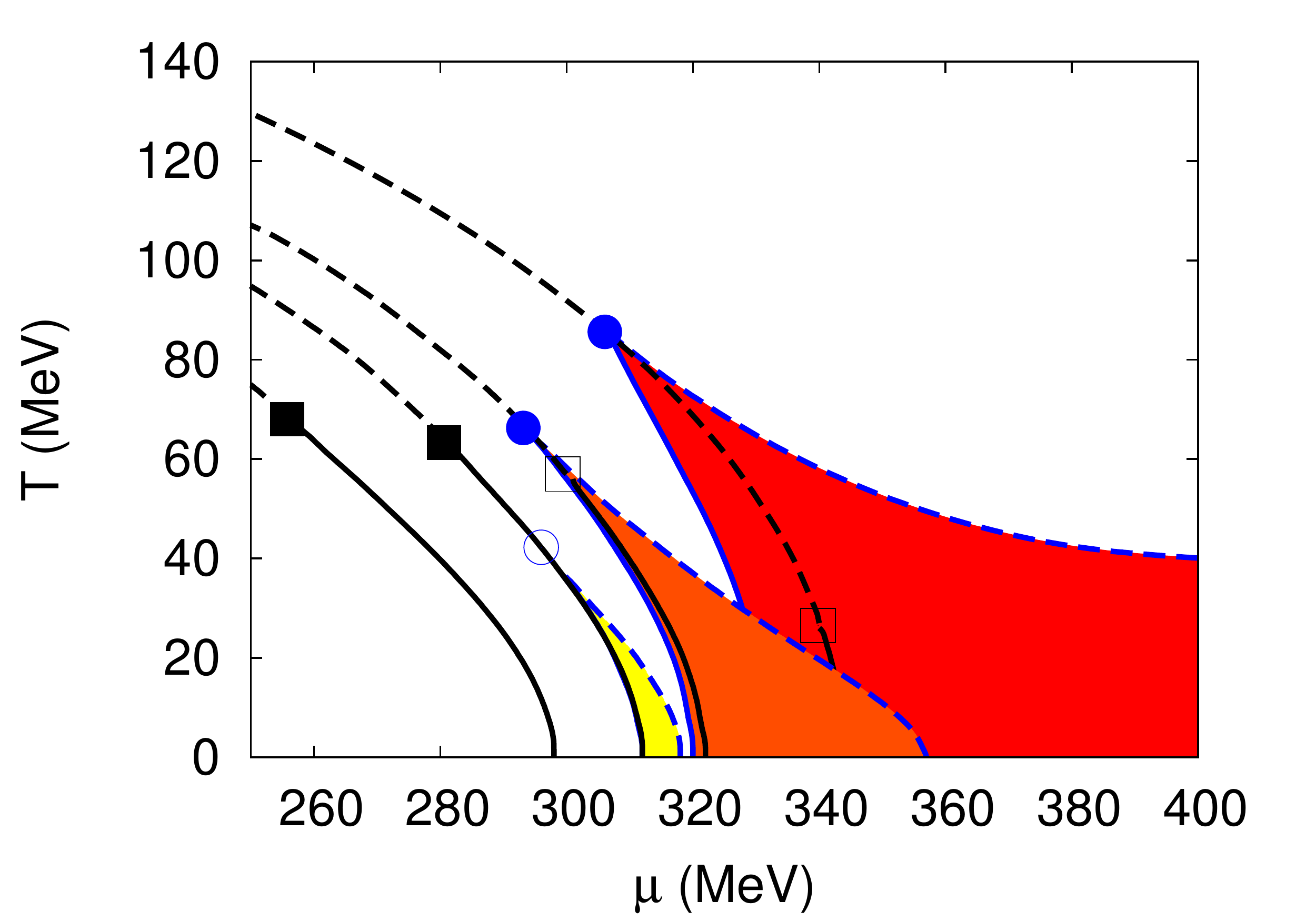}
\caption{Influence of the sigma-meson mass on the phase diagrams of the QM model 
with ``renormalized'' Dirac sea and $M_\text{vac} = 300$~MeV. 
Left:  
Location of the critical (black squares) and Lifshitz (blue circles) points 
according to the Ginzburg-Landau analysis.  
The numbers indicate the corresponding values of
the sigma mass in MeV.  The full circles are reliable, the empty
ones are not because the analysis is invalidated by the homogeneous
first-order phase transition.  The critical points indicated by open
squares are only relevant if the analysis is restricted to homogeneous phases.
Right:  
Full numerical results for the phase diagrams.  
  From left to right: $m_\sigma~=~550, 590, 610$
  and 650 MeV.  The shaded areas indicate the regions where the CDW is
  favored.  The black lines denote the phase boundaries between
  homogeneous phases (solid: first-order; dashed: second-order).
  Filled squares are the corresponding true CPs, while open squares
  indicate the locations of the CPs which would be present in a
  homogeneous analysis but are covered by the inhomogeneous phase.
  Filled circles are the LPs which agree with the predictions of the
  GL analysis, while the open circle indicates the point where the
  inhomogeneous phase ends on the first-order phase boundary between
  homogeneous phases, and which does not agree with the LP of the GL
  analysis.  
  From Ref.~\cite{Carignano:2014jla}.
  }
\label{fig:pdmsigma}
\end{figure}

However, as we have seen in Sec.~\ref{sec:GLQM}, CP and LP do not need to coincide in the 
QM model and can be separated by choosing $m_\sigma$ different from $2M_\text{vac}$.
This is illustrated in \Fig{fig:pdmsigma}, where the effect of a varying sigma meson mass is
demonstrated.
In the left panel we show the GL results for CPs and LPs, while the full phase diagrams are shown
on the right. 
Recall that the GL analysis is only reliable for small amplitudes and gradients, so that the prediction
for the LP can be invalidated by the presence of a near-by first-order transition.
As seen in the right figure,  the size and even the existence of the inhomogeneous phase is 
very sensitive to the value of $m_\sigma$.

\subsection{\it Chiral density waves in nuclear matter}
\label{sec:CDWNM}

While the main focus of this review is on quark matter,
inhomogeneous chiral condensates have also been investigated in nuclear matter.
As mentioned earlier, the idea of (spin-) density waves in nuclear matter dates back 
to the seminal paper by Overhauser~\cite{Overhauser:1960} 
and has been studied intensively in the 1970s and early 1980s in the context of $p$-wave pion 
condensation pioneered by Migdal~\cite{Migdal:1971cu,Migdal:1973zm}.
(For dedicated reviews, see e.g.\@ Refs.~\cite{Migdal:1978az,BrownWeise,Migdal:1990vm,Kunihiro:1993pz}.)

Similar to the other condensation phenomena discussed in the Introduction,
pion condensation is a consequence of an instability which arises
when attractive interactions make it energetically favorable to add pions to the system. 
Formally, this can be discussed in terms of the dispersion relations, which are given by the 
poles of the dressed pion propagator, 
\beq
       \omega^2 = m_\pi^2 + \mk\,^2 +\Pi(\omega, \mk)\,.
\label{eq:pidisp}
\eeq
Here $\omega$ and $\mk$ are energy and momentum of the pion, and 
$\Pi(\omega, \mk)$ denotes the polarization function due to interactions with the surrounding 
medium.
For attractive interactions $\Pi$ is negative, and its magnitude typically grows with increasing density. 
At sufficiently high densities the right-hand-side of \Eq{eq:pidisp} can thus become negative, 
signalling an instability. 
More precisely, the onset of the instability is given by the condition 
\beq
       m_\pi^2 + \mk_0^{\,2} +\Pi(0, \mk_0) = 0\,,
\eeq
where $\mk_0$ is the momentum which minimizes the combination $\mk_0^{\,2} +\Pi(0, \mk_0)$
and corresponds to the wave vector of the unstable mode.
Since the $\pi$-$N$ interaction is predominantly $p$-wave, one finds $\mk_0 \neq 0$,
i.e., the system develops an instability towards a spin-density wave with a nonvanishing wave vector.

The question whether and at which density pion condensation takes place in nuclear matter
obviously depends on the polarization function $\Pi$. 
Considering particle-hole and $\Delta$-hole excitations, which are the most important attractive
contributions, Migdal originally concluded that the critical density is close to or even below
nuclear saturation density~\cite{Migdal:1973zm,Migdal:1978az}.  
It turned out however
that the instability is pushed to considerably higher densities if short-range correlations
in the spin-isospin channel are included~\cite{Barshay:1974zw,Weise_Brown}. 
Although the exact results sensitively depend on further details of the $NN$ interaction~\cite{Kunihiro:1993pz}, 
it is nowadays generally accepted that 
there is no pion condensation at nuclear saturation density, both in isospin symmetric and in 
pure neutron matter~\cite{Migdal:1990vm,Akmal:1997ft}.

In the arguments outlined above, chiral symmetry does not enter explicitly but is hidden in the
$p$-wave nature of the $\pi$-$N$ interaction.
Quite early on, pion condensation in nuclear matter was therefore also studied 
within the non-linear~\cite{Dashen:1974ff} 
and the linear sigma model~\cite{Dautry:1979,Campbell:1974qt,Baym:1975tm} , 
where the role of chiral symmetry is more obvious.
Having seen in the phenomenological approaches that a correct description of the $NN$ interaction
is crucial to produce reliable results, a major concern at that time was the difficulty to incorporate
the relevant details in this framework. 
In this context the analysis was restricted to the vicinity of the chiral circle, thus excluding the 
possibility of a chiral phase transition.
This is quite opposite to most quark-model studies in the past 25 years, which focus on the chiral phase 
transition but usually neglect the existence and phenomenology of nuclear matter.

Indeed, the description of both, bound nuclear matter and a chiral phase transition at higher
density within a single framework is known to be a nontrivial task, even if the analysis is restricted 
to homogeneous matter. 
The problem is closely related to the fact that a straightforward chiral extension of 
the Walecka model typically leads to saturation in the restored phase 
(``Lee-Wick phase''~\cite{Lee:1974ma}), rather than a phase with broken chiral symmetry. 
The same happens in the standard NJL model with nucleon degrees of 
freedom~\cite{Koch:1987py,Buballa:1996tm}. 
A naive replacement of the quark fields in the models discussed in the previous section
by nucleon fields would therefore lead to the appearance of inhomogeneous condensates 
already at saturation density,  contradicting the empirical evidence and the lessons learned from 
the ``old'' studies of pion condensates.

Recently Heinz, Giacosa and Rischke (HGR) have therefore revisited the question 
of inhomogeneous chiral condensates in nuclear matter~\cite{Heinz:2013hza} 
using an extended version of the linear sigma model (eLSM), developed in Refs.~\cite{Gallas:2009qp}
(see also \cite{Zschiesche:2006zj} for an earlier version). 
This model successfully describes the low-energy meson and baryon spectra in vacuum
as well as the basic ground-state properties  of nuclear matter
(density $\rho_0$, binding energy, and compressibility) 
and has a first-order phase transition to the (approximately) chirally restored phase at about 
2.5 $\rho_0$ if the condensates are assumed to be homogeneous~\cite{Gallas:2011qp}.

The relevant piece of the model Lagrangian can be written as the sum
\beq
       \mathcal{L}_{\text{eLSM}} =  \mathcal{L}_{\text{mes}}   + \mathcal{L}_{\text{bar}}\,,   
\eeq
with a mesonic part 
\bea
\label{eq:Lmes}
\mathcal{L}_{\text{mes}}   
&=& \frac{1}{2}\partial_{\mu}\sigma\partial^{\mu
}\sigma +\frac{1}{2}\partial_{\mu}\pi\partial^{\mu}\pi
+\frac{1}{2}m^{2}(\sigma^{2}+\pi^{2})
-\frac{\lambda}{4}(\sigma^{2}+\pi^{2})^{2}
+\varepsilon\sigma
\nonumber\\
&&
 -\frac{1}{4}(\partial_{\mu}\omega_{\nu}-\partial_{\nu}\omega_{\mu})^{2}
+\frac{1}{2}m_{\omega}^{2}\omega_\mu\omega^\mu 
 +\frac{1}{2}\partial_{\mu}\chi\partial^{\mu}\chi
-\frac{1}{2}m_{\chi}^{2}\chi^{2}+g\chi(\sigma^{2}+\pi^{2})
\eea
and a baryonic part
\bea
  \mathcal{L}_{\text{bar}}  
  &=& 
  \overline{\Psi}
  \left( \begin{array}{cc}
   i\gamma^{\mu}\partial_{\mu}
  -\frac{\widehat{g}_{1}}{2}(\sigma+ i \gamma^5 \tau^3 \pi)
  -g_{\omega}\gamma^{\mu}\omega_{\mu}
  & a \chi \gamma^5
  \\
  - a \chi \gamma^5
  &
  i\gamma^{\mu}\partial_{\mu}
 -\frac{\widehat{g}_{2}}{2}(\sigma- i \gamma^5\tau^3 \pi)
-g_{\omega}\gamma^{\mu}\omega_{\mu}
 \end{array} \right) \Psi\,.
\eea
The first line of the mesonic part corresponds to the usual linear-sigma model Lagrangian,
with a sigma and a pion field in a mexican-hat potential (parametrized by the constants
$m^2$ and $\lambda$) and a term linear in $\sigma$, introduced to break chiral symmetry 
explicitly.  
As before, $\pi$ denotes the isospin-3 component of the pion triplet.
We have already dropped the two other components, which do not 
contribute in mean-field approximation.
In addition, $\mathcal{L}_{\text{mes}}$ contains a vector-isoscalar field $\omega$
(the $\omega(782)$ meson) and a dilaton field $\chi$.
The latter is a chiral singlet and is identified by HGR with lowest  scalar-isoscalar resonance 
$f_0(500)$, while the $\sigma$ is interpreted as the heavier $f_0(1370)$.

The baryonic Lagrangian describes the dynamics of the chiral doublet $\Psi = (\Psi_1,\Psi_2)^T$,
related to the nucleon and its chiral partner.
These fields are treated in the ``mirror assignment'' of chiral symmetry, which means  that
the left handed part of $\Psi_1$ transforms like the right handed part of $\Psi_2$ and vice versa.
This is the reason why the off-diagonal terms in $\mathcal{L}_{\text{bar}}$,
\beq
         a\chi \left(\overline{\Psi}_1 \gamma^5 \Psi_2 - \overline{\Psi}_2 \gamma^5 \Psi_1\right)
        = 
        a\chi \left(
         \overline{\Psi}_{1,L} \Psi_{2,R}
        -\overline{\Psi}_{1,R} \Psi_{2,L}
        -\overline{\Psi}_{2,L} \Psi_{1,R}
        +\overline{\Psi}_{2,R} \Psi_{1,L} \right)\,,
\eeq
which are controlled by a parameter $a$, are allowed by symmetry. These terms
have the important consequence to give the fermions a nonvanishing mass even in the 
chirally restored phase, once the dilaton field condenses.

In order to study inhomogeneous phases in this model, HGR 
make an CDW-like ansatz for the sigma and pion mean fields,
\beq
\label{eq:CDWNM}
        \ave{\sigma(x)} = \phi\cos(2fx), 
        \qquad
        \ave{\pi(x)} = \phi\sin(2fx)\,,
\eeq
depending on an amplitude $\phi$ and a wave number $q = 2f$.
In addition they allow the dilaton field and the zero-component of the omega field to
develop non-zero condensates $\ave{\chi} = \bar\chi$ and 
$\ave{\omega_\mu} = \bar\omega_0 \delta_{\mu 0}$.
Plugging this into \eq{eq:Lmes},
the mesonic contribution to the thermodynamic potential becomes
\beq
       \Omega_\text{mes} = 
       2f^2\phi^2 - \frac{1}{2}m^2 \phi^2 + \frac{\lambda}{4}\phi^4        
       -\frac{1}{2} m_\omega^2 \bar\omega_0^2 
       +\frac{1}{2} m_\chi^2 \bar\chi^2 
       -g\bar\chi\phi^2      
       - \varepsilon \phi\, \frac{1}{V} \int\limits_V d^3x\, \cos(2fx)
\eeq
where the last term averages to zero in the inhomogeneous phase ($f\neq 0$), while 
it yields a contribution $- \varepsilon \phi$ in the homogeneous phase ($f=0$).\footnote{
As noted earlier, the CDW ansatz only allows for a selfconsistent solution in the chiral limit,
i.e., $\varepsilon = 0$. Still, \eq{eq:CDWNM} can be viewed as a reasonable  variational ansatz.
}

The baryonic contribution is again obtained from the eigenvalue spectrum of the effective
Dirac Hamiltonian.
As before, it is possible to remove the space dependence from the 
Hamiltonian by performing chiral rotations,
\beq
       \Psi = 
  \left( \begin{array}{cc}
   \exp(- i \gamma^5 \tau^3 fx) & 0
  \\
  0 & \exp( i \gamma^5 \tau^3 fx)
 \end{array} \right) \Psi'\,.            
\eeq
where, because of the mirror assignment,  $\Psi_1$ and $\Psi_2$ are
transformed in opposite directions.
Treating the baryonic contribution in no-sea approximation, the total thermodynamic potential
at $T=0$ and baryon chemical potential $\mu$ is then given by~\cite{Heinz:2013hza} 
\beq
\Omega(T=0,\mu;\phi,\bar\chi,\bar\omega_0,f)
= 
2\sum_{k=1}^4 \int\frac{d^3p}{(2\pi)^3}[E_k(\p) - \mu^*]\theta(\mu^*- E_k(\p)) 
+
\Omega_\text{mes}\,, 
\eeq
with $\mu^* = \mu - g_\omega\bar\omega_0$.
$E_k(\p)$ are the dispersion relations of the quasiparticles. 
Due to the doublet structure of $\Psi$, there are now four modes
instead of two as in \Eq{eq:CDWdispers2}.

\begin{figure}[hbt]
\centering
\includegraphics[angle=0,width=.45\textwidth]{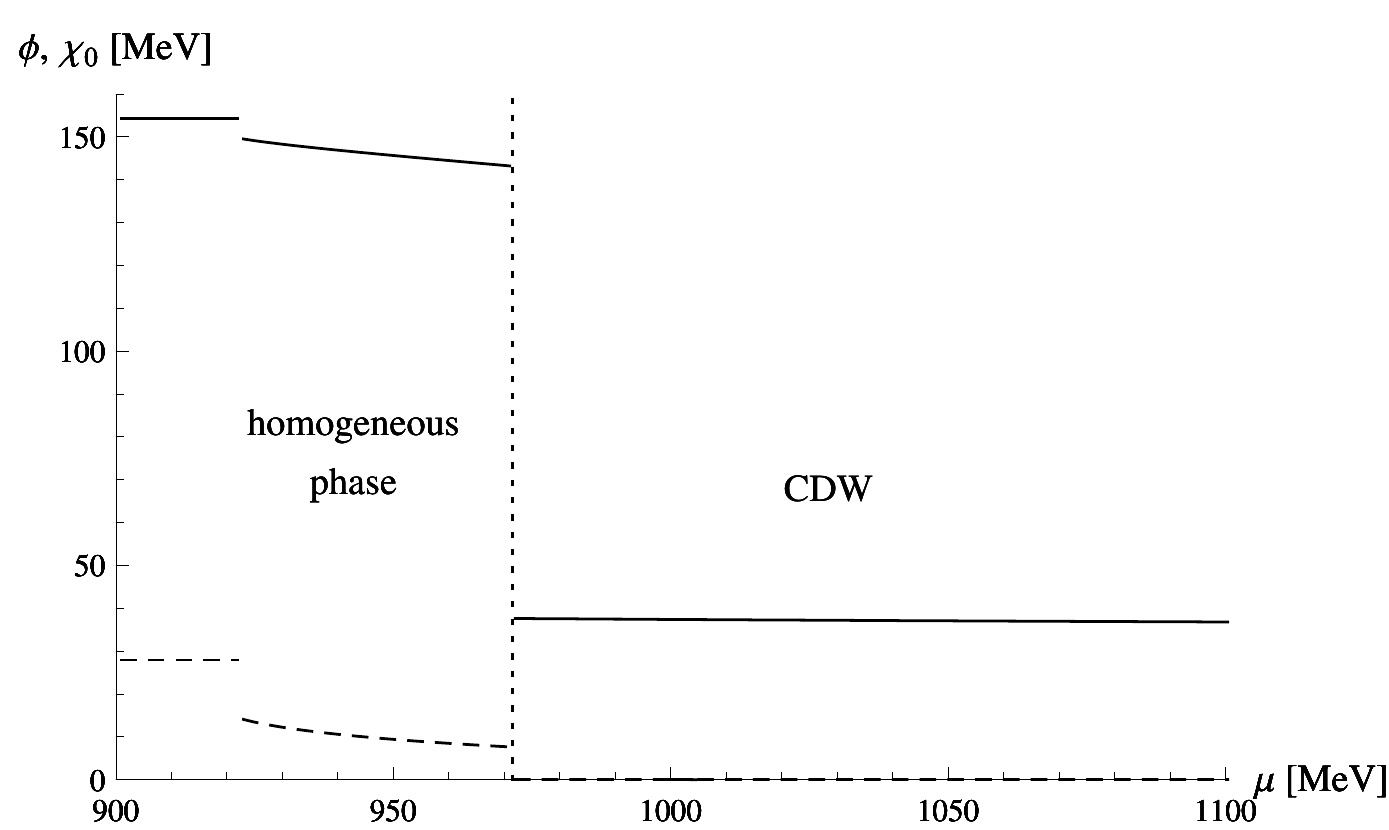}
\qquad
\includegraphics[angle=0,width=.45\textwidth]{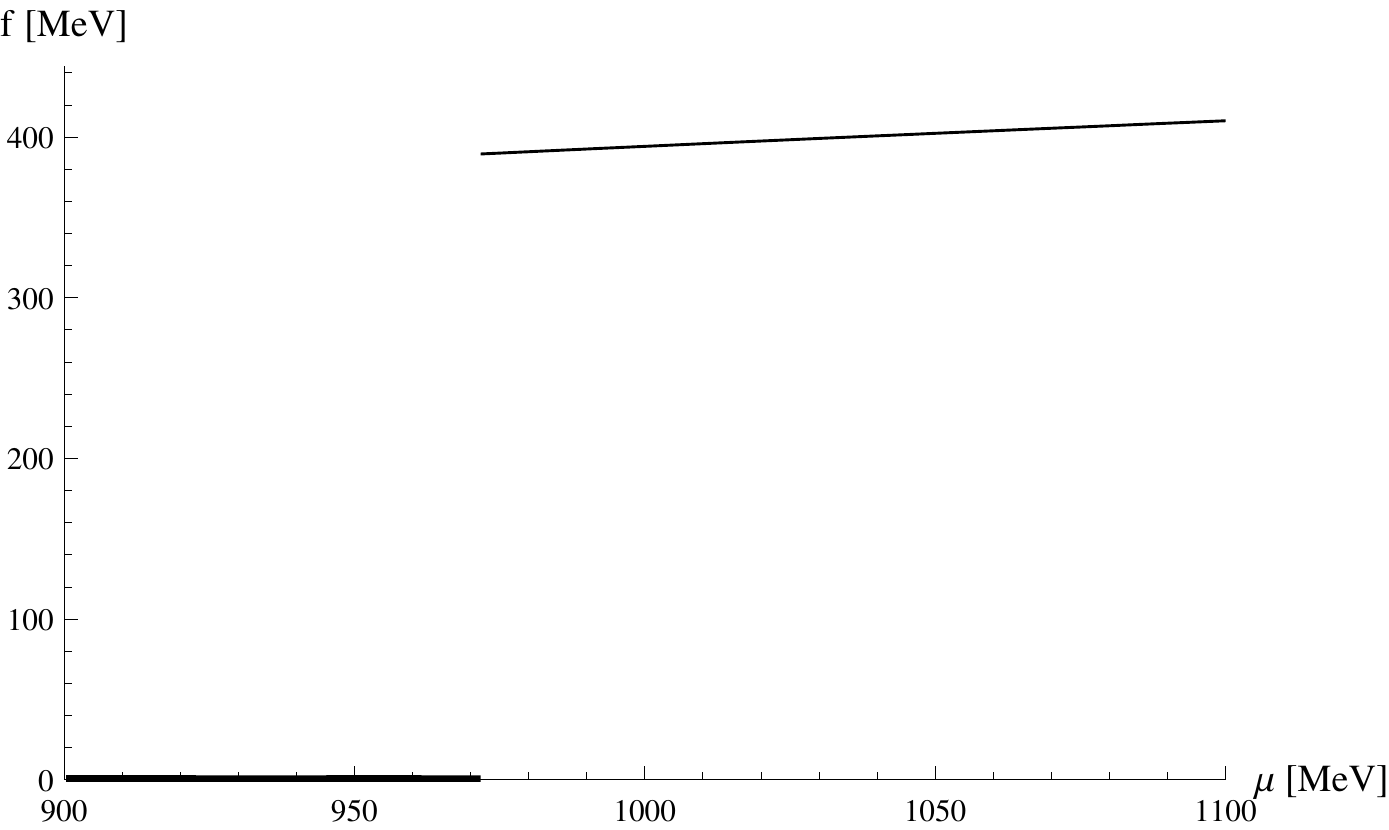}
\caption{Mean-field quantities characterizing the properties of nuclear matter
at $T=0$ as functions of the baryon chemical potential $\mu$.
Left: $\phi$ (solid line) and $\bar\chi$ (dashed line). Right: $f$.  
From Ref.~\cite{Heinz:2013hza}. 
}
\label{fig:cdwnm}
\end{figure}

Minimizing the thermodynamic potential with respect to $\phi$, $\bar\chi$ and $f$, and 
maximizing it with respect to $\bar\omega_0$,
HGR have obtained the results shown in \Fig{fig:cdwnm} as functions of the baryon chemical
potential $\mu$. 
As one can see in the left panel, 
there are two first-order phase transitions, where the condensates
$\phi$ and $\bar\chi$ discontinuously drop to lower values. 
The first one corresponds to the liquid-gas phase transition 
from the vacuum phase to nuclear matter.
As seen in the right panel, the wave number is zero in both phases, i.e., 
they are homogeneous, in agreement with phenomenology. 
Since in Ref.~\cite{Gallas:2011qp} 
the model parameters have been fixed in such a way that the main properties of 
nuclear matter at saturation density (corresponding to the onset point)
are correctly reproduced when the analysis is restricted to homogeneous matter, 
this feature remains thus unchanged when allowing for CDWs.

At higher chemical potential, on the other hand, where in Ref.~\cite{Gallas:2011qp}
a transition to an almost restored phase ($\phi\approx 0$) was reported,
HGR find that a CDW, characterized by nonvanishing values of $\phi$ and $f$, is 
favored. At the onset the baryon density jumps from $2.4\,\rho_0$ to $10.4\,\rho_0$,
which seems to be too high to trust the hadronic description.  
Nevertheless, it is an important qualitative
result that homogeneous bound nuclear matter at lower
densities and an inhomogeneous solution at higher densities can be realized within 
the same framework. 
Of course, one should not forget the lessons from the earlier investigations, and
effects from $\Delta$ degrees of freedom and short-range correlations must be included 
in order for these results to be quantitatively reliable.

\section{One-dimensional modulations}
\label{sec:dimred}

While being an excellent prototype for most properties of 
chiral crystalline phases, the chiral density wave remains a
quite simplistic ansatz and it is natural to expect that a more general solution
would provide a better description of the true ground state of quark matter.
The study of general periodic structures remains however an extremely challenging task: although the 
numerical diagonalization procedure outlined in chapter \ref{sec:periodicconds} is in principle 
straightforward,
its practical implementation is extremely demanding from a computational point of view.
However, as pointed out by Nickel~\cite{Nickel:2009wj}, if the generality of the ansatz 
\eq{eq:Mxq} is limited to lower-dimensional modulations, the problem can be mapped onto
a lower-dimensional one.
In particular, the analytically known solutions of the $1+1$ dimensional GN and NJL$_2$ 
models~\cite{Schon:2000qy,Thies:2006,Basar:2009fg}
can be utilized to construct solutions with one-dimensional modulations in $3+1$ dimensions.
The details of this very powerful method are discussed in the following.

\subsection{\it Boosting and dimensional reduction}
\label{sec:boost}

If the mass function is 
modulated only in one or two spatial directions, due to the remaining translational invariance in the ``transverse'' directions along which the condensate is spatially constant,
 the problem can be factorized and 
the diagonalization can be performed on a dimensionally-reduced Hamiltonian~\cite{Nickel:2009wj}. 
To see how this works, let us assume that the condensates which enter the original Hamiltonian $H$
vary only in $d_\parallel$ dimensions and
do not depend on the $d_\perp = 3- d_\parallel$ transverse coordinates.
A quark moving in this system can have arbitrary momenta $\p$, containing longitudinal
and transverse components. However, since the condensates vary only in longitudinal direction,
the transverse momentum $\p_\perp$ of the quark is conserved.
Formally this means that the Hamiltonian commutes with the transverse momentum operator 
$\vec{P}_\perp$, i.e., there are common eigenfunctions $\psi_{E,\p_\perp}$ of $H$ and 
$\vec{P}_\perp$ with energy $E$ and transverse momentum $\p_\perp$.
We may thus choose a frame where $\p_\perp = 0$ and solve the simpler eigenvalue problem 
\beq
       H \psi_{\varepsilon,\p_\perp=0} = \varepsilon\, \psi_{\varepsilon,\p_\perp=0}\,,
\eeq
where the transverse coordinates do not enter, so that it effectively
corresponds to a dimensionally reduced problem. 
The full set of eigenvalues and eigenfunctions of the $3+1$ dimensional system,
\beq
       H \psi_{E_\varepsilon(\p_\perp),\p_\perp} = 
       E_\varepsilon(\p_\perp) \, \psi_{E_\varepsilon(\p_\perp),\p_\perp}\,,   
\eeq
is then obtained by a subsequent boost to arbitrary values of $\p_\perp$.
In particular the energy eigenvalues are simply given by boosting the energies $\varepsilon$
of the dimensionally reduced problem~\cite{Nickel:2009wj},
\beq
\label{eq:boost}
       E_\varepsilon(\p_\perp) = \mathrm{sign}(\varepsilon)\sqrt{\varepsilon^2 + \p_\perp^{\,2}}\,.
\eeq
 The kinetic part of the thermodynamic potential, \Eq{eq:Omegakin}, can then be calculated as 
\beq
\label{eq:OmegaLowD}
\Omega_\mathit{kin}
=
- 2N_c
\frac{1}{V_\parallel}
\sum_{\varepsilon}
\int\frac{d^{d_\perp}p_\perp}{(2\pi)^{d_\perp}}
 \left[ \frac{E_\varepsilon(\p_\perp) -\mu}{2} 
 + T \log\left(1 + e^{-\frac{E_\varepsilon(\p_\perp) -\mu}{T}}\right)\right]  \,,
\eeq
where $2N_c$ is the color-flavor factor discussed in Sec.~\ref{sec:cfD}
and
$V_\parallel$ is the ``volume'' of the unit cell in longitudinal direction.

The simplest case one can consider at this point is obviously a one-dimensional modulation
of the chiral order parameter,
\beq
M(\x) = M(z) \,.
\eeq
In the frame where the transverse quark momentum vanishes,
the Hamiltonian of \Eq{eq:Hplus}  is effectively reduced to
\bea
\label{eq:mfH1d}
H_+
\rightarrow
H_{1D}
&=&
\left(
\begin{array}{cccc}
i\partial_z & & M(z)\\
&-i\partial_z && M(z)\\
M(z)^* & &-i\partial_z&\\
&M(z)^* & &i\partial_z
\end{array}
\right)
\,.
\eea
Performing a unitary transformation,  this can be turned into a block-diagonal form, 
\beq
\label{eq:mfH1dBLOCK}
H_{1D}
\rightarrow
H_{1D}'
=
\left(
\begin{array}{cc}
H_\text{BdG}(M) & \\
 & H_\text{BdG} (M^*) \\
\end{array}
\right)
\,, 
\eeq
where 
\beq
\qquad H_\text{BdG}(M) = 
\left(
\begin{array}{cc}
-i\partial_z & M(z) \\
 M^*(z) & i\partial_z \\
\end{array}
\right)
\eeq
is the Bogoliubov--de Gennes Hamiltonian \cite{DeGennes},
which corresponds to the effective mean-field Hamiltonian of the 1+1-dimensional NJL$_2$ model
or, if the mass function $M(z)$ is real, 
of the Gross-Neveu model defined in \Eq{eq:LGN}. 

After the boosting procedure, the dimensionally-reduced Hamiltonian for the 3+1-dimensional NJL model can therefore be reduced to the direct product of two Hamiltonians of its 1+1-dimensional counterpart.
 This in turn implies that from the knowledge of the eigenvalues of the NJL$_2$ model one can obtain the spectrum 
 for the full 3+1-dimensional model via \eq{eq:boost}. 
 It is worth noting however that,  since the 
 total Hamiltonian in \eq{eq:mfH1dBLOCK} is the direct sum of
 two one-dimensional Hamiltonians, one evaluated
 at $M$ and the other at $M^*$, the thermodynamically favored solution
 for the full 3+1-dimensional system will not necessarily 
 be the same as for the NJL$_2$ model,
 even if we restrict ourselves to one-dimensional modulations.
 In fact, as we will see later, it turns out that this is not the case. 

 For most cases considered
 in the following, an analytical expression for the spectral density  
 $\rho_M(\varepsilon) = \frac{1}{V_\parallel} \sum_{\varepsilon'} \delta(\varepsilon -\varepsilon')$  
  of $H_\text{BdG}(M)$ is known (for a classification of lower-dimensional solutions see e.g. \cite{Correa:2009xa,Basar:2009fg})
so that \Eq{eq:OmegaLowD} can be written as
\beq
\label{eq:OmegaAnalytics}
\Omega_{kin}
= - 2N_c\int_{-\infty}^\infty d\varepsilon \int \frac{d^2p_\perp}{(2\pi)^2} 
\left( \rho_M(\varepsilon) + \rho_{M^*}(\varepsilon) \right)  
\left[ \frac{E_\varepsilon(\p_\perp) -\mu}{2}  
+ T \log\left(1 + e^{-\frac{E_\varepsilon(\p_\perp) -\mu}{T}}\right)\right]  \,.
\eeq
Switching to polar coordinates in the $\varepsilon-p_\perp$-plane
and performing the angular integration, 
the thermodynamic potential can then be rewritten as a single integral over the boosted energy 
$E_\varepsilon(\p_\perp) \equiv E$.
Comparing the result with \eq{eq:OmegaeigenRHO} in Sec. \ref{sec:mfa} ,
this allows to identify the ``three-dimensional'' density of states 
as~\cite{Nickel:2009wj}\footnote{To be consistent with \eq{eq:rhotildedef} 
we have included the color-flavor degeneracy factors $2N_c$ in the definition of the effective 
density of states. In some literature, e.g., in \cite{Nickel:2009wj}, these
factors are kept out of the definition of $\tilde\rho$.}

\beq
\label{eq:rhotilde}
\tilde\rho(E) = \frac{2N_c}{4\pi} E^2 \int_{-1}^1 du\, \left[ \rho_M(E u) + \rho_{M^*}(E u) \right] \,.
\eeq
If the spectrum is symmetric, the kinetic part of the thermodynamic potential is then given 
by~\eq{eq:OmegakinRhoSymm}.
As an example let us once again consider a simple CDW-type plane wave ansatz, \Eq{eq:Mcdw}.
The NJL$_2$ spectral density for this kind of modulation (``chiral spiral'')
has been calculated in \cite{Basar:2009fg} and reads
\beq
\label{eq:rhoCDW1d}
\rho_M(\varepsilon) = \frac{1}{\pi} \frac{ \vert \varepsilon - Q \vert}{\sqrt{(\varepsilon-Q)^2 - \Delta^2}} \, \theta\left((\varepsilon-Q)^2 - \Delta^2\right) \,,
\eeq
where $Q \equiv q/2$, and the step function reflects the fact that the eigenvalue spectrum contains two bands. 
As one can see, it
is not symmetric around $\varepsilon = 0$, but is centered around $Q$ instead \cite{Basar:2009fg}. The expression $\rho_M + \rho_{M^*}$ appearing in \Eq{eq:OmegaAnalytics} is however symmetric around $\varepsilon = 0$.
As a consequence, the three-dimensional effective density of states resulting from 
\Eq{eq:rhotilde}~\cite{Nickel:2009wj}\footnote{Note that the derivation
of the density of states for the CDW modulation is absent in the arXiv version of Ref.~\cite{Nickel:2009wj} and can only be found in the published version.}
\bea
\label{eq:rhoCDW}
{\rho}_{CDW}(E)
=
2 N_c \, \frac{E}{2\pi^2} \Big\{&&
\theta(E-{Q}-\Delta) \sqrt{(E-{Q})^2-\Delta ^2}
\nonumber\\
&+&
\theta(E-{Q}+\Delta)\theta(E+{Q}-\Delta)\sqrt{(E+{Q})^2-\Delta ^2}
\nonumber\\
&+&
\theta({Q}-\Delta-E)
\left(\sqrt{(E+{Q})^2-\Delta^2}-\sqrt{(E-{Q})^2-\Delta^2}\right)
\Big\}
\,,
\eea
is also symmetric, so that \eq{eq:OmegakinRhoSymm} can be used to calculate the 
thermodynamic potential. 
The result is, of course, identical to one presented in Chapter \ref{sec:CDW}.

\subsection{\it Solitonic solutions}
\label{sec:solitons}

In this section we introduce an important class of one-dimensional real 
modulations
which arise as solution of the Dirac-Hartree-Fock eigenvalue equation
in the 1+1-dimensional Gross-Neveu model.
The mass function is given by~\cite{Schnetz:2004}
\beq 
\label{eq:Mzsolitons}
M(z) = \Delta\nu \frac{\text{sn}(\Delta z \vert \nu) \text{cn}(\Delta z \vert \nu)}{\text{dn}(\Delta z\vert \nu)} \,,
\eeq
where sn, cn and dn are Jacobi elliptic functions. 
This kind of solution is characterized by two parameters, $\Delta$ and $\nu$.
The latter, the so-called elliptic modulus (ranging from 0 to 1)
determines the shape of these  functions. The interpretation of the two parameters as amplitude and wave vector is not as straightforward as for the CDW case.
For this it is useful to rewrite \Eq{eq:Mzsolitons} in a simpler way by performing a Landen transformation \cite{Abramowitz}, 
\beq
\label{eq:Mzsolitonsalt}
M(z) = \Delta' \sqrt{\nu'}\, \text{sn} (\Delta' z \vert \nu' ) \,,
\eeq
with $\Delta' = \Delta(1+\sqrt{1-\nu})$ and $\nu' = \left(\frac{1-\sqrt{1-\nu}}{1 + \sqrt {1-\nu}}\right)^2$. 
As one can see from this expression,
the amplitude is given by $\Delta'\sqrt{\nu'} = \Delta(1-\sqrt{1-\nu})$,
while the period of $M(z)$ is given by the combination 
$L = 4 \K (\nu') / \Delta' =  2 \K (\nu) / \Delta$, where
$\K(\nu)$ is a complete elliptic integral of the first kind.

\begin{figure}[hbt] 
\includegraphics[width=.33\textwidth]{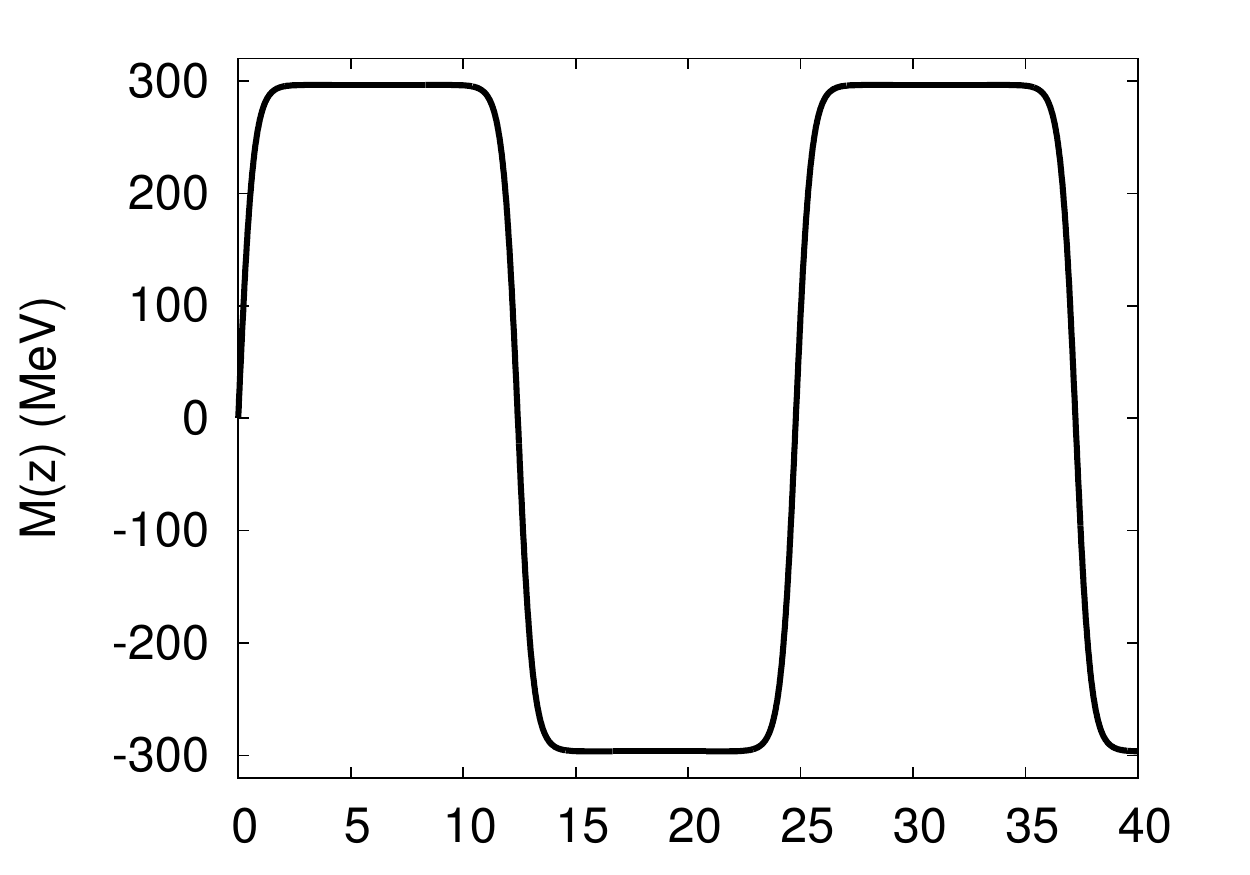}
\includegraphics[width=.33\textwidth]{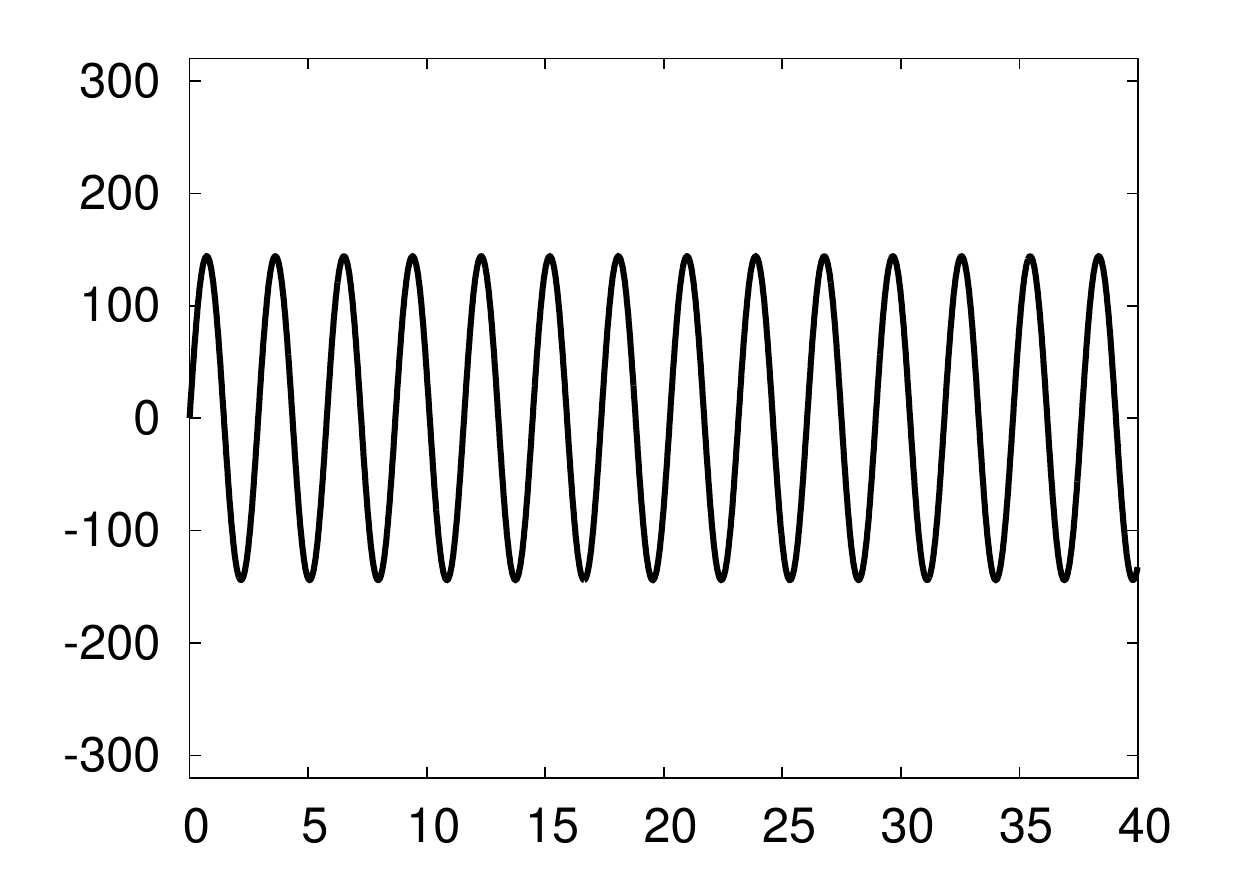}
\includegraphics[width=.33\textwidth]{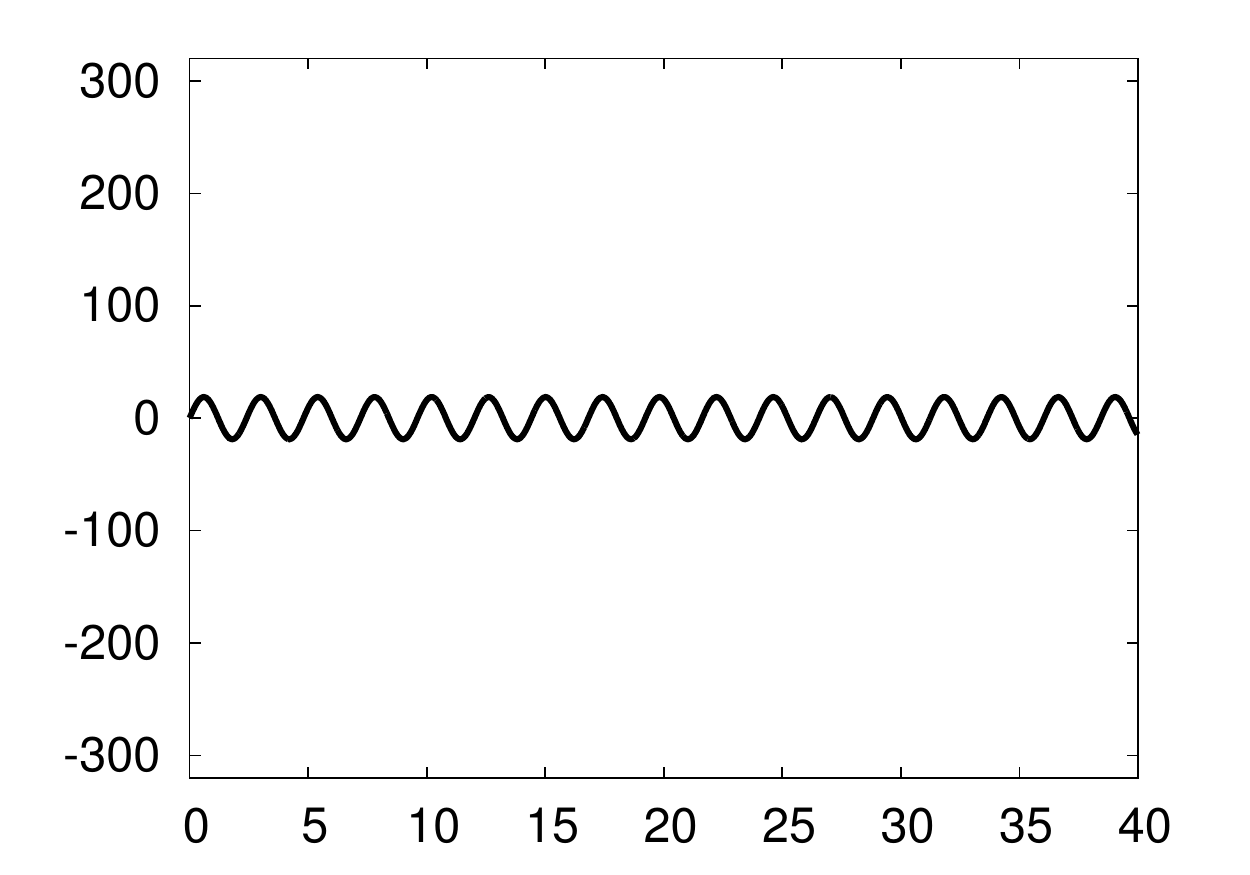}

\includegraphics[width=.33\textwidth]{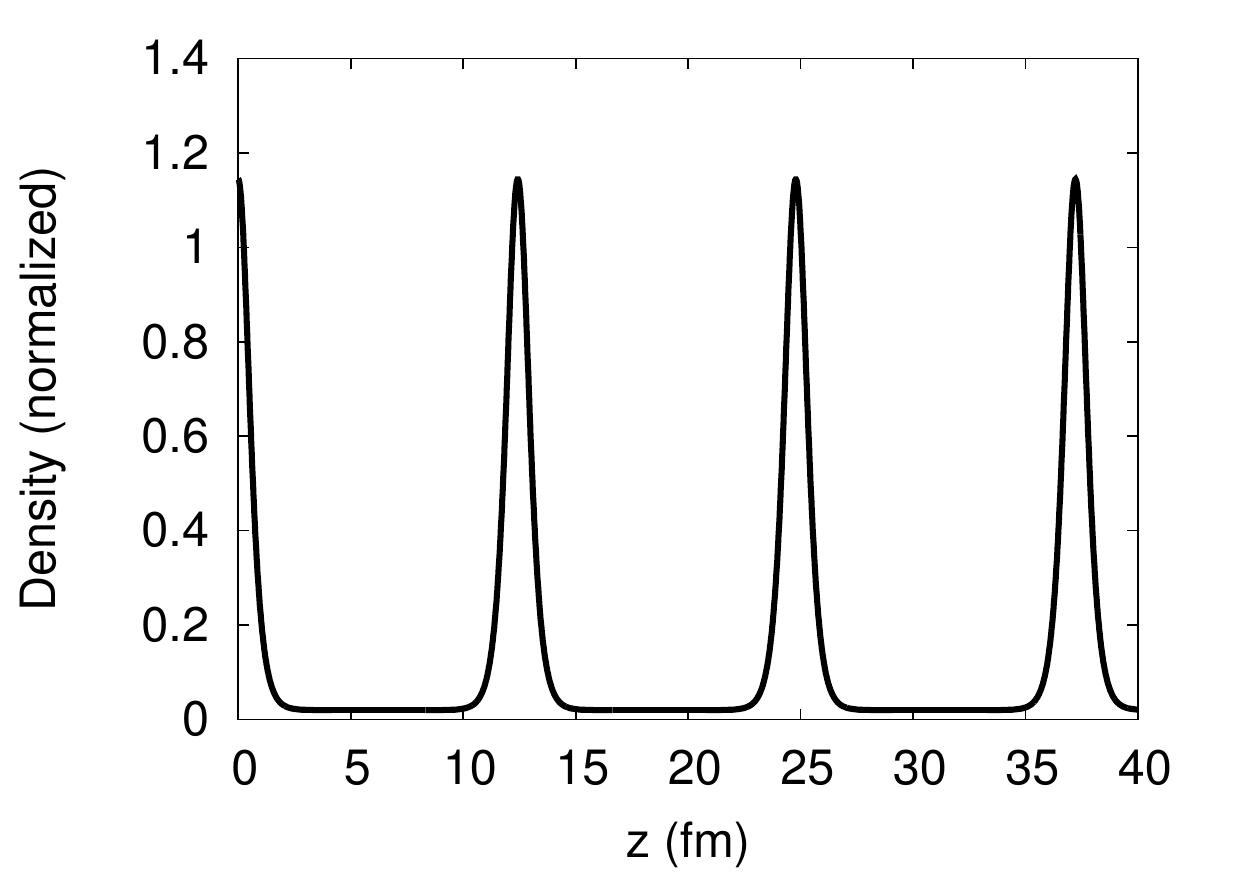}
\includegraphics[width=.33\textwidth]{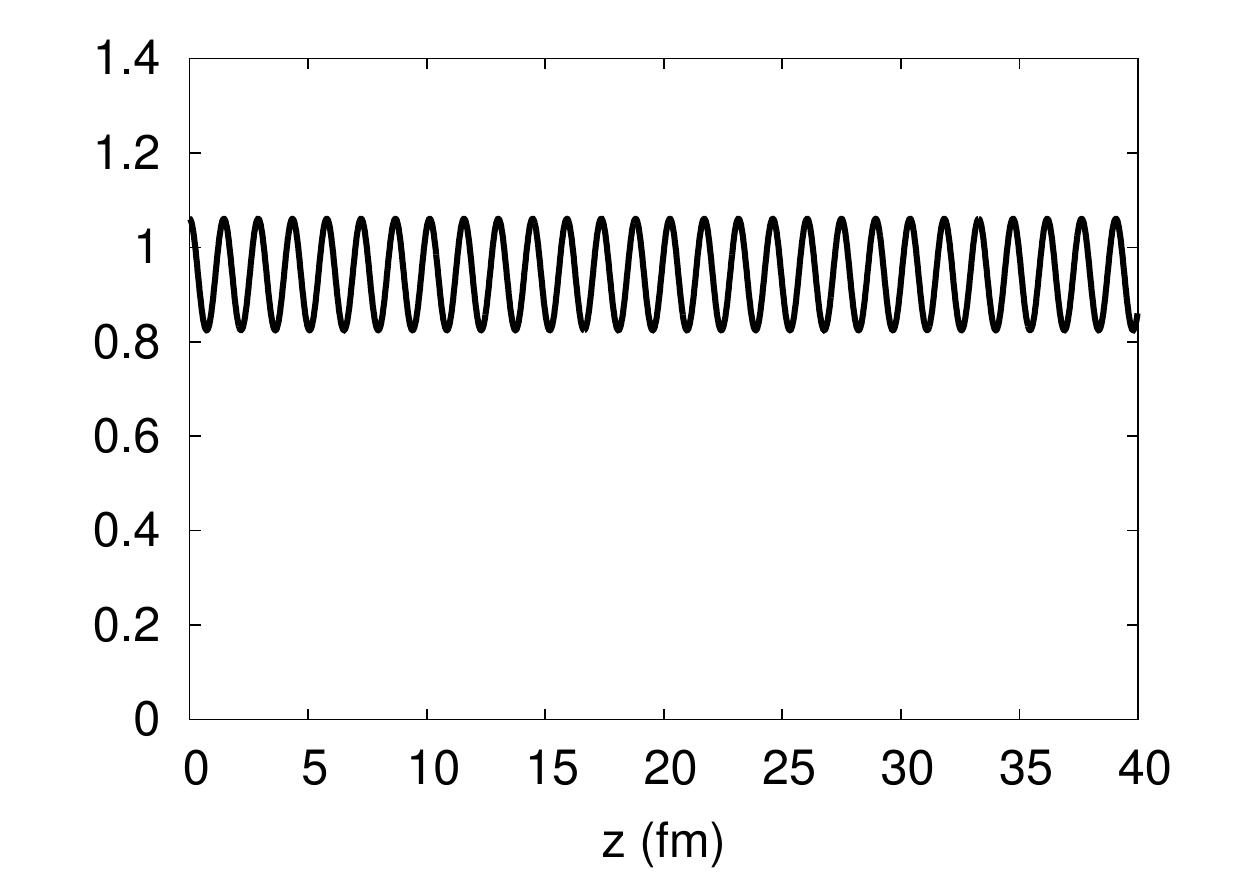}
\includegraphics[width=.33\textwidth]{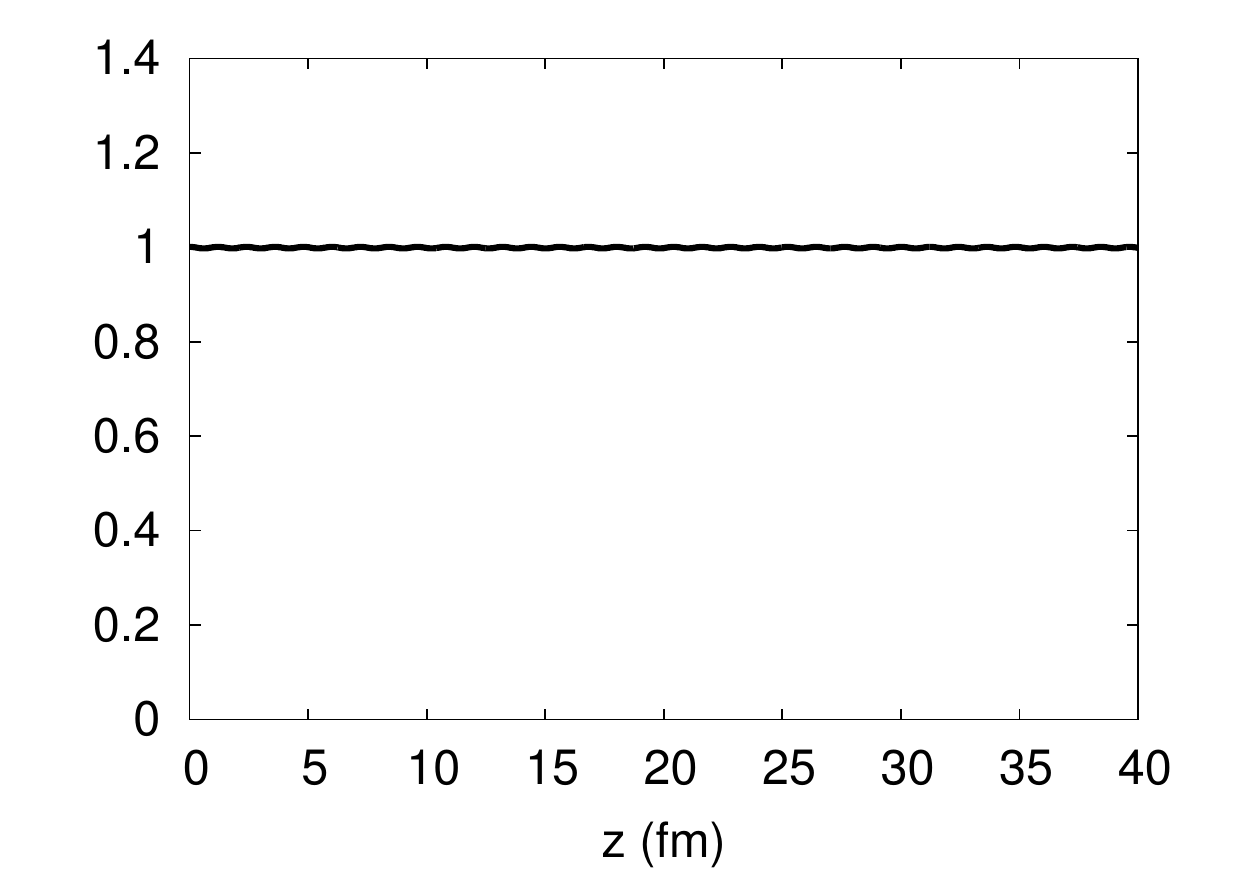}
\caption{Shape of the mass modulation (upper row) and 
corresponding quark number density (lower row) for the solitonic solutions 
at different values of the elliptic modulus.
From left to right: $\nu = 0.999$, 0.8  and 0.35.
The figure was taken from Ref.~\cite{CNB:2010}, where
these values of $\nu$ correspond to the zero-temperature minima of thermodynamic 
potential at $\mu =308$, 325, and 345~MeV, respectively, while the value of $\Delta$ is almost constant
(cf.~\Fig{fig:solDQ}).
The densities in the lower row have been normalized to the densities in the restored phase
at the given values of the chemical potential.
}
\label{fig:densvsmass}
\end{figure}

In order to get further insights into the shape of the mass modulations, let us briefly discuss
how it evolves when the elliptic modulus $\nu$ is gradually decreased from 1 to 0.
For $\nu=1 $, the function sn  assumes the shape of a hyperbolic tangent, 
sn$(\alpha,\nu=1) = \tanh(\alpha)$, i.e., the mass function is given by 
\beq
      M(z) \,\equalover{\nu = 1}\, \Delta\tanh(\Delta z)\,.
\eeq            
This can be interpreted as a single kink (or domain-wall soliton in 3+1 dimensions, which is why in the 
following we will refer to this kind of solution as ``solitonic''), interpolating between the two degenerate chirally broken solutions for a homogeneous system. As $\nu$ decreases from 1, these domain-wall solitons begin to overlap, forming an array of kinks and antikinks, as shown on the left-hand side of \Fig{fig:densvsmass}. Further decreasing $\nu$, the shape of the mass modulation rapidly assumes a sinusoidal shape, as can be seen in 
the center panel. At the same time the amplitude melts, which becomes the main effect when $\nu$ is
further decreased (right). 
At $\nu=0$, the function sn becomes exactly a sine,
sn$(\alpha,\nu=0) = \sin(\alpha)$, 
but the amplitude of the mass function is zero in that limit.

The lower part of \Fig{fig:densvsmass} shows the quark number density for 
the solitonic solutions, which is proportional to \cite{Schnetz:2004,Schnetz:2005ih,Thies:2006,Basar:2009fg}
\beq
\psi^\dagger_{\varepsilon}(z)\psi_{\varepsilon}(z)
=
\frac{(\varepsilon/\Delta)^2+\frac{1}{2}((M(z)/\Delta)^2 + \nu -2)}
{(\varepsilon/\Delta)^2 -\E(\nu)/\K(\nu)}
\,,
\eeq
$\psi_{\varepsilon}(z)$ being the one-dimensional quark wave-function in the Gross-Neveu model.
Here $\E(\nu)$ denotes a complete elliptic integral of the second kind.
A particularly interesting behaviour occurs close to the $\nu=1$ limit,
where the solitons are well separated
and the corresponding density is given by an array of
strongly localized peaks, located at the position of the 
zero-crossings of the mass functions. 
At the maxima the density overshoots the value it would have in the restored phase
for the same chemical potential.
In a bag-model like picture, this can be interpreted as the quarks being 
squeezed by the bag pressure of the domains with broken chiral symmetry 
into the regions of space where chiral symmetry is almost restored. 
A detailed analysis of the properties of the domain-wall solitons in this
regime can be found in Ref.~\cite{Buballa:2012vm}.
At lower values of $\nu$ where the solitons overlap, the density profiles become
more and more washed out, gradually approaching the homogeneous limit.

From \Fig{fig:densvsmass} we can also see a fundamental difference between
the solitonic inhomogeneous solution and a mixed phase constructed from two 
homogeneous solutions.
Although in both cases, ``lasagne-type'' density profiles, i.e., 
alternating parallel plates with high and low density, exist,
the mass functions are rather different:
In the mixed phase it oscillates between 
high and low (or vanishing) values, whereas the solitonic mass function
oscillates between positive and negative values and vanishes only 
``in passing''.
In fact, in contrast to the solitonic phase,
mixed phases have a higher free energy than their homogeneous
constituents because of the surface tension, which has been estimated to be 
positive for the NJL model~\cite{Molodtsov:2011ii,Lugones:2011xv,Pinto:2012aq}.

The quark spectral density in the solitonic
background has been determined by Schnetz et al.\@ in \cite{Schnetz:2004} for the Gross-Neveu model, and its three-dimensional counterpart, obtained by Nickel through the boosting procedure, \Eq{eq:rhotilde},
 is given by \cite{Nickel:2009wj}
\bea
\label{eq:rhosolitons}
\tilde{\rho}_{sn}(E)
=
2 N_c \frac{E \Delta}{\pi^2} \, \Big\{\hspace{-2mm}&&\hspace{-2mm}
\theta(\sqrt{\tilde{\nu}}\Delta-E)
\left[
 \E(\tilde{\theta} |\tilde{\nu})+\left(\frac{\E(\nu)}{\K(\nu)}-1\right) \F(\tilde{\theta} |\tilde{\nu})\right] 
\nonumber\\
\hspace{-2mm}&+&\hspace{-2mm}
\theta(E-\sqrt{\tilde{\nu}}\Delta)
\theta(\Delta-E)
\left[ \E(\tilde{\nu})+\left(\frac{\E(\nu)}{\K(\nu)}-1\right) \K(\tilde{\nu})\right]
\nonumber\\
\hspace{-2mm}&+&\hspace{-2mm}
\theta(E-\Delta)
\left[\E(\theta |\tilde{\nu})
+ \left(\frac{\E(\nu)}{\K(\nu)} -1\right) \F(\theta |\tilde{\nu})
+\frac{\sqrt{(E^2 - \Delta^2) (E^2 - \tilde{\nu}\Delta^2)}}{E\Delta} \right]
\Big\}
\,.
\nonumber\\
\eea
Here
$\tilde{\nu}=1-\nu$, $\tilde{\theta}=\arcsin(E/(\sqrt{\tilde{\nu}}\Delta))$, $\theta=\arcsin(\Delta/E)$, and 
$\F(\theta|\nu)$  and  $\E(\theta|\nu)$ are incomplete elliptic integrals of the first and of the second kind, respectively \cite{Abramowitz}. 
 In the $\nu=1$ limit, \Eq{eq:rhosolitons} reduces to the density of states for homogeneous order parameters, \Eq{eq:rhohom}. This indicates that in the limit of a single domain wall, \Eq{eq:Mzsolitons} is 
 thermodynamically degenerate with a homogeneous solution with mass gap~$\Delta$.

Before moving on to the numerical results, we would like to mention that the  
solitonic mass function can also be obtained in a completely different way from  minimizing 
the GL free energy. This has been shown in Ref.~\cite{Abuki:2011} by Abuki et al.\@, 
following a similar analysis in the context of superconductivity~\cite{1997PhLA..225..341B}.
Writing the GL functional \eq{eq:GLNJLgen} as
\beq
       \Omega_\text{MF}(M) = \frac{1}{V} \int d^3x \, \omega(M(\x))\,,
\eeq
the minimization of the free energy $\delta \Omega_\text{MF}(M) = 0$, yields the
Euler-Lagrange equation
\beq
       \frac{\delta\omega}{\delta M}
       - \nabla_k  \frac{\delta\omega}{\delta (\nabla_k M)} 
       + \laplace  \frac{\delta\omega}{\delta(\laplace M)} 
       - \dots
       = 0~,
\eeq
where the ellipsis indicates higher-order derivative terms.
Evaluating this at 6th order of the GL expansion 
for mass functions with a one-dimensional modulation, 
this becomes a 4th-order nonlinear differential equation,
and it was shown in Ref.~\cite{Abuki:2011} that \eq{eq:Mzsolitonsalt} is a particular
solution of it.

\subsubsection{\it Phase diagram}

\begin{figure}[htb] 
\begin{center} 
\includegraphics[width=.3\textwidth,angle=270]{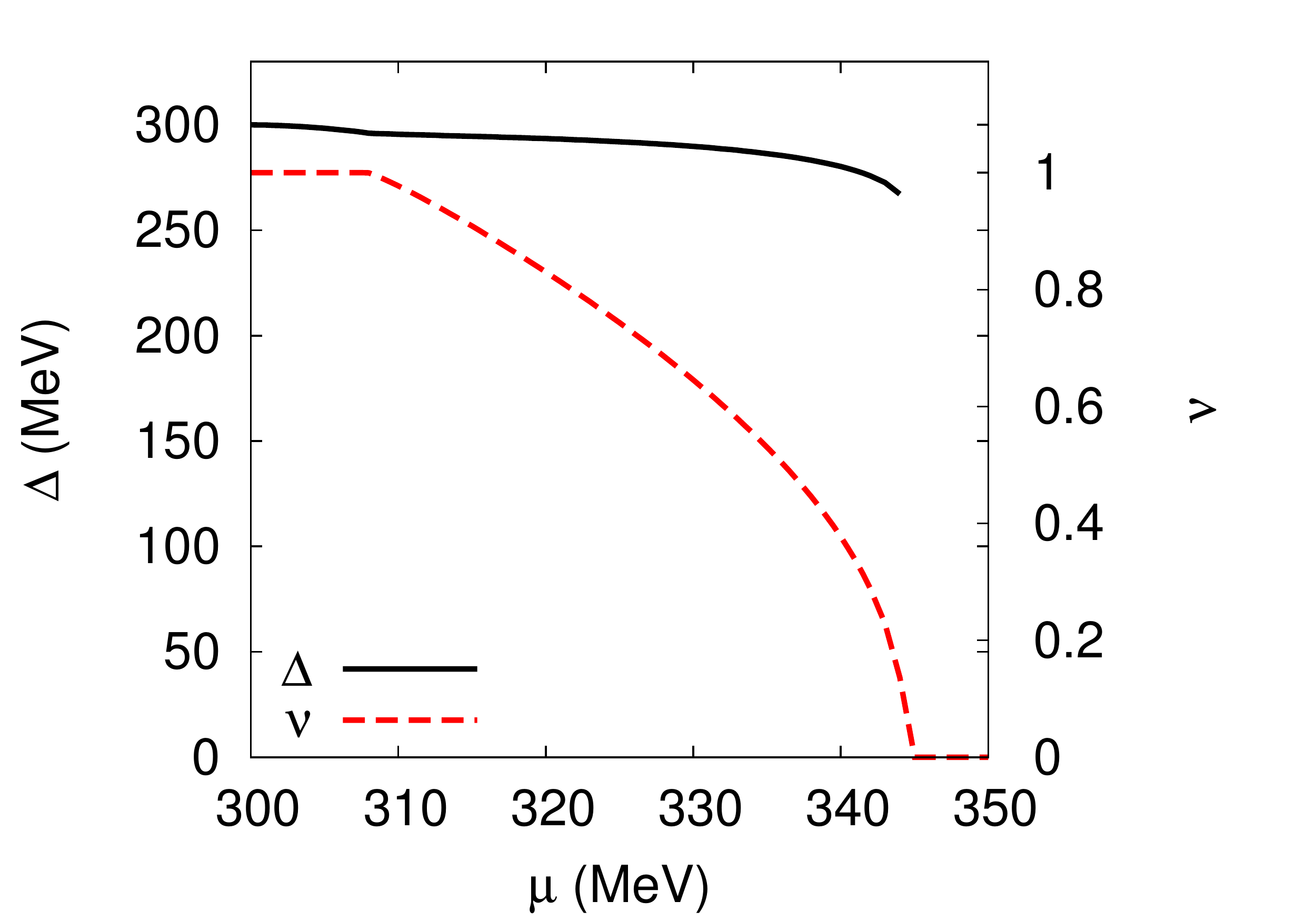}
\includegraphics[width=.3\textwidth,angle=270]{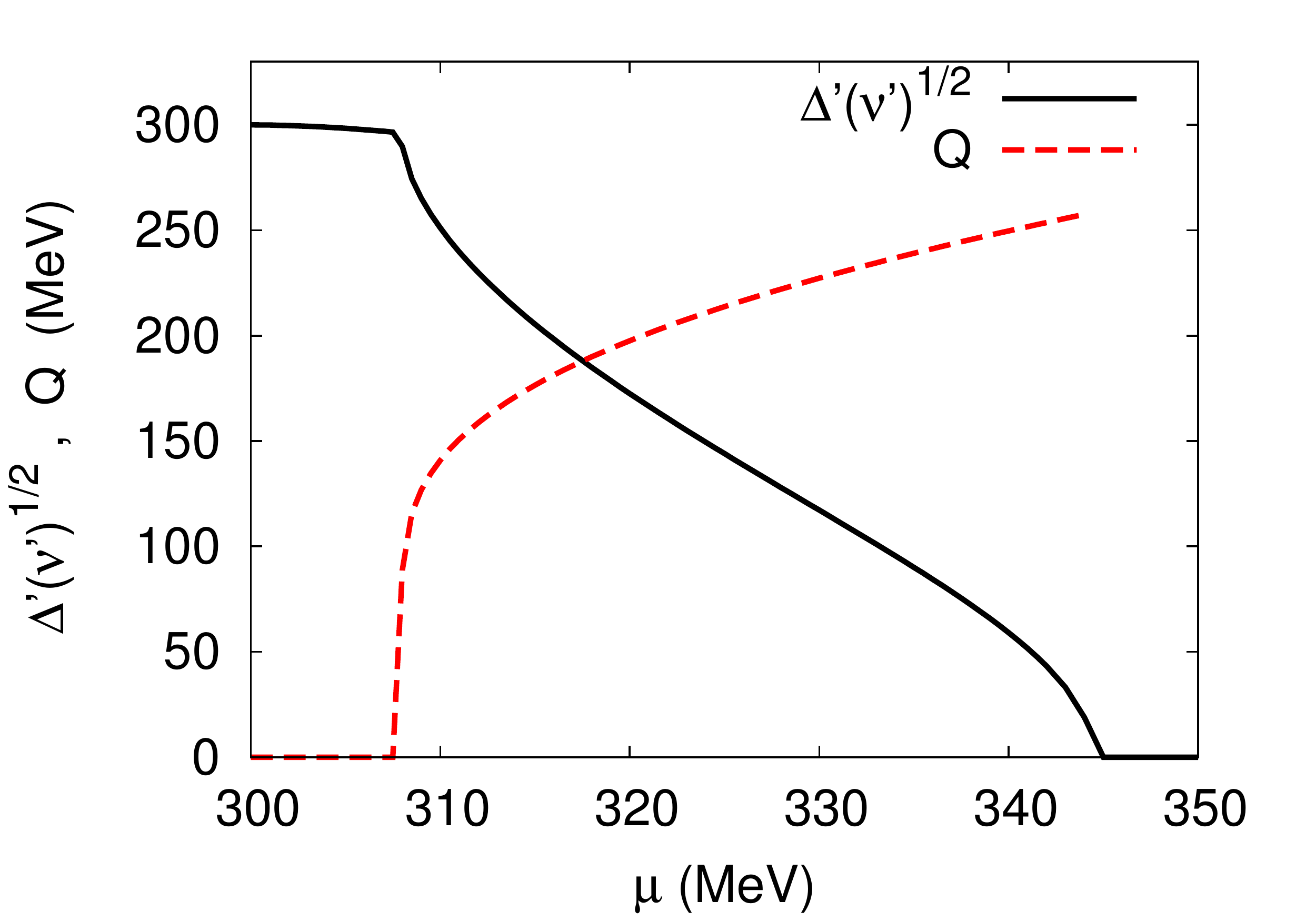}
\caption{Left: chemical-potential dependence of the variational parameters $\Delta$ and $\nu$ at $T=0$. 
Right: corresponding amplitude 
$\Delta'\sqrt{\nu'} = \Delta(1-\sqrt{1-\nu})$ and $Q=\pi/L$, where $L = 2 \K (\nu) / \Delta$
is the spatial period of the modulation.  
\label{fig:solDQ}}  
\end{center} 
\end{figure}

Results for the chemical-potential dependence 
of the two variational parameters $\Delta$ and $\nu$ 
at $T=0$
are shown on the left-hand side of \Fig{fig:solDQ}. The corresponding effective amplitude 
$\Delta'\sqrt{\nu'} = \Delta(1-\sqrt{1-\nu})$ as well as $Q = \pi/L$ are shown on the right.
The latter has been defined in analogy to $Q = q/2$ we have introduced for the CDW, and 
is thus expected to be slightly lower than the chemical potential at high values of $\mu$.

The most important result is that both phase transitions, from the homogeneous broken to the 
inhomogeneous phase and from the inhomogeneous to the chirally restored phase, are continuous,
i.e., second-order transitions, as in the 1+1-dimensional Gross-Neveu model~\cite{Schnetz:2004}.
While for the transition to the restored phase, this was also the case
for the CDW (according to most authors), the continuous transition from the chirally broken to the
inhomogeneous phase is a rather unique feature of the solitonic mass functions.
This is possible because the quarks concentrated in the single domain-wall at $\nu = 1$ are 
thermodynamically irrelevant, both, for the free energy and for the total density, when
averaged over the whole space.

 At the onset of the inhomogeneous phase, the elliptic modulus decreases smoothly from 1, while $\Delta$ remains roughly constant. With increasing chemical potential, $\nu$ keeps decreasing and the 
 amplitude gradually melts from the homogeneous value to zero, so that 
chiral restoration is smoothly reached. 
 The evolution of the mass modulation within the inhomogeneous phase is therefore exactly the one described in \Fig{fig:densvsmass}: 
 starting from a single domain-wall, the solitons quickly begin to overlap 
 as $\mu$ increases, and the mass rapidly assumes a sinusoidal shape.
Close to the second-order transition to the restored phase, the amplitude
melts and the density profile smoothly approaches the uniform 
density of the restored phase. The density of the system therefore
approaches a constant value as we move towards the chiral restoration transition.
This remains true at finite temperature, and in particular in the proximity of the Lifshitz point,
an aspect which will become important in Sec.~\ref{sec:vector} for our discussion
on the effect of vector interactions on this kind of solution.

\begin{figure}[hbt]
\centering
\includegraphics[angle=0,width=.5\textwidth]{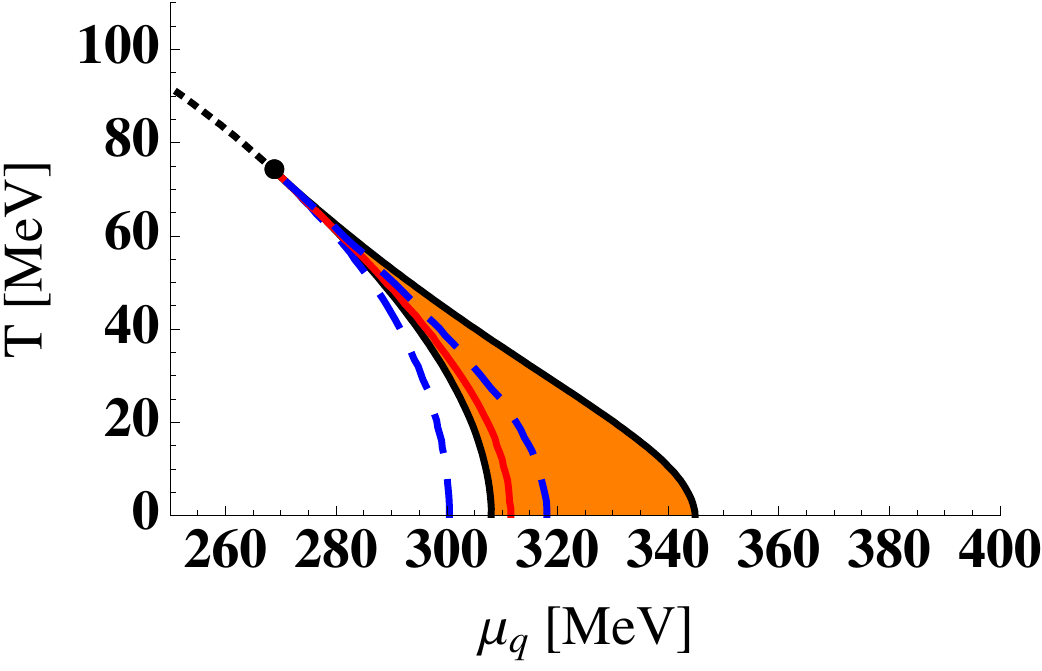}
\caption{Phase diagram when allowing for the solitonic solutions. The shaded region denotes the 
inhomogeneous phase, the red solid line is the first-order phase transition
when limiting to homogeneous phases, while the two dashed lines denote the boundaries of the spinodal region.
The dot at the tip of the inhomogeneous region indicates the LP, which coincides with the CP.
From \cite{Nickel:2009wj}.
\label{fig:pdsolitonsDN}
}
\end{figure}

The full NJL phase diagram for the solitonic modulations
has been calculated by Nickel in \cite{Nickel:2009wj} and is reported in \Fig{fig:pdsolitonsDN}. As already anticipated in the previous chapter, it looks qualitatively very similar to the one obtained for a plane wave modulation, the main difference being the second-order transition 
from the homogeneous broken to the inhomogeneous phase, instead of a first-order one. 
In agreement with the GL analysis, the LP is again found to coincide with the CP.
From \Fig{fig:pdsolitonsDN} it is also possible to see that 
the location of the inhomogeneous phase does not coincide with the 
spinodal region\footnote{That is, the region where the 
model thermodynamic potential has different local minima, 
associated with chirally broken and restored solutions.}
appearing around the first order transition line for homogeneous phases.

\subsubsection{\it Finite quark masses}
\label{sec:massive}

While all results presented so far have been obtained in the chiral limit,
for the case of solitonic solutions a self-consistent extension of the order parameter
to include nonzero current quark masses is known from studies of the GN model.
For this, \Eq{eq:Mzsolitons} is modified into 
\cite{Schnetz:2005ih,Nickel:2009wj}
\bea
\label{eq:MzReal}
M_{\rm sn, m}(z)
&=&
\Delta
\left(
\nu\,
\mathrm{sn}(b\vert \nu)
\mathrm{sn}(\Delta z\vert \nu)
\mathrm{sn}(\Delta z+b\vert \nu)
+
\frac{
\mathrm{cn}(b\vert \nu)\mathrm{dn}(b\vert \nu)
}{
\mathrm{sn}(b\vert \nu)
}
\right)
\,.
\eea
Here $b$ is an additional parameter to be varied together with $\Delta$ and
$\nu$ when minimizing the thermodynamic potential.
For the solitonic solutions
the inclusion of finite current quark masses simply 
amounts to modifying the density of states \Eq{eq:rhosolitons}
 into \cite{Nickel:2009wj}

\beq
{\tilde\rho}_{\rm sn,m}(E)
=
\frac{E}{\sqrt{E^2-\delta \Delta^2}}{\tilde\rho}_{\rm sn}
(\sqrt{E^2-\delta \Delta^2})\theta(E-\sqrt{\delta}\Delta)\,,
\eeq
where $\delta\in [0,\infty]$ is defined through the relation 
$\mathrm{sn}(b\vert \nu) = \frac{1}{\sqrt{1+\delta}}$. The chiral limit, i.e. $m=0$, corresponds to $\delta=0$ or equivalently $b=\K(\nu)$.

\begin{figure}[htb]
\centering
\includegraphics[angle=0,width=.5\textwidth]{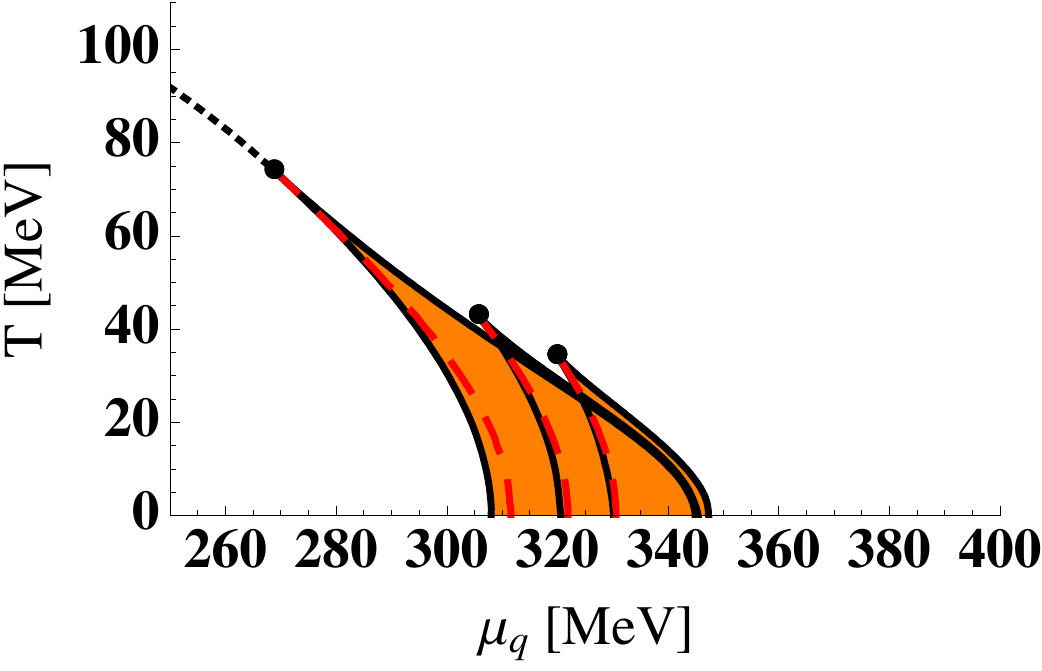}
\caption{Phase diagram for the solitonic solutions 
calculated for three different values of the bare quark mass. From
left to right: $m = 0$ (chiral limit), $m = 5$ and $ m = 10$ MeV.
From Ref.~\cite{Nickel:2009wj}.
\label{fig:pdinhmassive}
}
\end{figure}

The phase diagram for nonzero bare quark masses has been calculated in \cite{Nickel:2009wj}
and is shown in \Fig{fig:pdinhmassive}. There  
it can be seen that the inclusion of a finite current quark mass reduces the 
size of the inhomogeneous window, which however is not destroyed
for reasonable values of $m$. The Lifshitz 
point\footnote{
Since there is no exact order parameter to distinguish the homogeneous broken 
from the restored phase in this case, there are strictly speaking only
two phases, a homogeneous and an inhomogeneous one, and, hence,
no Lifshitz point. Nevertheless we can easily identify the remnant of the
LP as a cusp in the phase boundary. 
For simplicity, we 
call this point a Lifshitz point as well. 
} is shifted towards lower temperatures and higher chemical potentials. 

As in the chiral limit, all phase transitions are second order. 
In particular, the transition to the almost restored phase is still characterized by
$\nu$ going to zero, so that the amplitude of the oscillating part of the mass function
melts away. The $z$-independent term, on the other hand, survives and
approaches $\Delta\cot(b)$, thus allowing for a smooth phase transition.

\subsection{\it Comparison with other one-dimensional modulations}

\label{sec:comparison1D}

\begin{figure}[hbt] 
\centering
\includegraphics[angle=0,width=.45\textwidth]{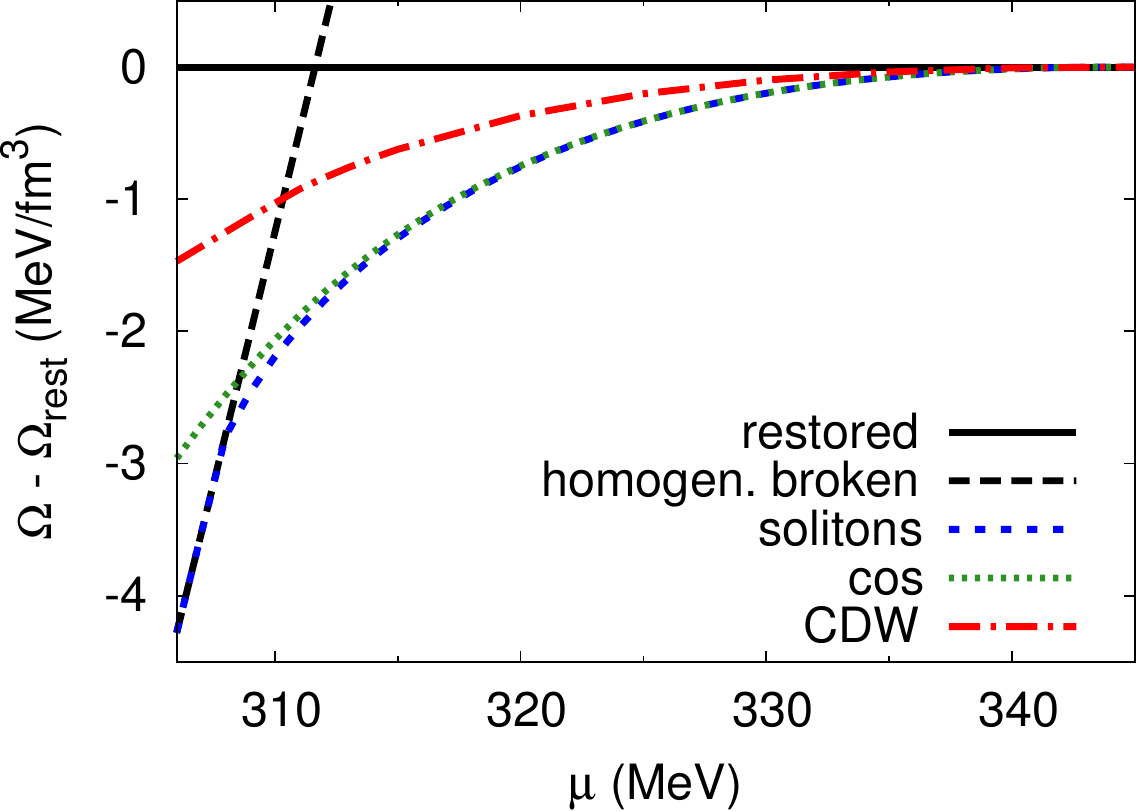}
\caption{Comparison of free energies for different one-dimensional modulations. The solitonic 
solution is most favored over the whole inhomogeneous window, and is
almost degenerate with the single cosine ansatz. The complex CDW solution is always disfavored against these real solutions. 
\label{fig:omegas1d}
} 
\end{figure}

Since both, the CDW and the solitonic solutions, are favored over the
homogeneous ones in a very similar chemical-potential window, it is 
natural to ask which of the two leads to the biggest gain in 
free energy and is therefore thermodynamically favored. As found by Nickel, the real solitonic solutions turn out to be energetically favored in the whole inhomogeneous window \cite{Nickel:2009wj}, contrary to what happens in 1+1 dimensions \cite{Schon:2000qy}. 
This can be seen from \Fig{fig:omegas1d} (a similar figure is also shown in Ref.~\cite{Nickel:2009wj}), 
where a comparison
of free energies associated with different kinds of one-dimensional solutions is presented.

As discussed in \cite{Nickel:2009wj}, the explanation of this difference lies in the particular structure of the the 3+1-dimensional NJL Hamiltonian, which, after boosting to the frame of vanishing transverse momentum,
is given by the direct sum of two BdG Hamiltonians, $H_\text{BdG}(M)$ and $H_\text{BdG}(M^*)$,
see \Eq{eq:mfH1dBLOCK}.
This particular symmetry leads to an eigenvalue spectrum which resembles more that of the (non-chiral)
Gross-Neveu model than the NJL$_2$ one.\footnote{We recall that the Hamiltonians 
for NJL$_2$ and GN both look like
$H_\text{BdG}(M)$, the difference being that 
in the former case the mass function is given by the sum of scalar and 
pseudoscalar condensates, $M \propto (\phi_S + i\phi_P)$, and is therefore complex,
whereas in the latter it is real since there is no pseudoscalar part.}
The difference can be seen most clearly by looking at the respective GL expansions.
For the NJL$_2$ model it is given by~\cite{Boehmer:2007ea}
\beq
\label{eq:GLNJL2gen}
\Omega_{\text{NJL}_2}(M) = \Omega_{\text{NJL}_2}(0)
+ \frac{1}{L}\int dx \, \Big\{  
    \frac{1}{2}\gamma_{2} |M|^2
    + \frac{1}{3}\gamma_{3} \mathrm{Im}\left(M {M'}^* \right)
     + \frac{1}{4}(\gamma_{4}\left(|M|^4 +|M'|^2\right)  + \dots \Big\}
\,,     
\eeq
where $M' = \frac{dM}{dx}$.
The striking difference to the NJL model in $3+1$ dimensions, cf.~\Eq{eq:GLNJLgen},
is the existence of odd-order terms.
As explicitly seen for the $\gamma_3$ term, these terms are only present for complex mass functions,
so that
for the GN model, where $M(x)$ is real, they vanish.
The GL expansion of the NJL model in $3+1$ dimensions, on the other hand, essentially corresponds
to the sum of $\Omega_{\text{NJL}_2}(M) +  \Omega_{\text{NJL}_2}(M^*)$.
As a consequence, the odd terms drop out, and the expansion takes the same form as for the
GN model, even for complex mass functions. 

Still, this does not a priori exclude the possibility that in the $3+1$ dimensional NJL model
complex mass functions lead to a lower free energy than real ones. 
By explicitly evaluating the GL functional for different shapes of the modulation, 
Nickel showed however that real sinusoidal mass functions are favored over the plane-wave ansatz.
The same behavior is also known from (color) superconductivity,
where it has been found as well that the favored shape for the order parameter in higher 
dimensional systems is a real one \cite{LO,Bowers:2002,NB:2009}.

A thorough comparison of different condensate shapes with real one-dimensional modulations
has been performed by Abuki et al.~\cite{Abuki:2011}
who considered a harmonic expansion  
of the form
\beq
       M(z;\{L,M_n\}) = \sum\limits_{n} M_n\,\sin\left(\frac{2\pi nz}{L}\right)
\eeq
for the mass functions
and minimized the GL free-energy functional at 6th order with respect to the variational 
parameters $L$ and $M_n$ for $n\leq 5$. 
It turned out that components with even $n$ vanish, i.e., the favored solution has the symmetry
$M(z + L/2) = -M(z)$, just as the solitonic mass function.
Moreover, except for very close to the onset of the inhomogeneous phase, the favored
values of the odd coefficients, $M_1$, $M_3$, and $M_5$ and of $L$, are very similar 
to the Fourier components and the period of the favored solitonic solution,
and the latter was always lower in free energy.
The authors conclude from this that Jacobi elliptic functions are the most favored mass shapes
with one-dimensional modulations~\cite{Abuki:2011}.

In \Fig{fig:omegas1d} we also show the free energy obtained by minimization of
the thermodynamic potential for a purely sinusoidal mass function, $M(z) = \Delta \cos(qz)$,
neglecting higher harmonics completely.
The free energy was calculated without GL expansion by numerical diagonalization of the 
Hamiltonian and minimization of the thermodynamic potential.

As seen in \Fig{fig:densvsmass}, 
with the exception of a narrow region close to the onset of the inhomogeneous phase, 
the Jacobi elliptic functions have an almost perfect sinusoidal shape, with higher harmonics
playing a minor role.  
It is therefore not very surprising that the free energies of the solitonic mass  function and 
the cosine lie almost on top of each other in the major part of the inhomogeneous region. 
In fact, even at the onset of the inhomogeneous phase, the difference remains 
small, in particular when we compare it with the difference to the CDW free energy.
It is nevertheless worth mentioning that, in spite of their marginal role 
for the free-energy gain, higher harmonics are crucial to allow for a smooth interpolation
between the homogeneous chirally broken and the inhomogeneous phase. In that sense, 
their presence affects the order of the phase transition, which can be second order only 
for the solitonic solutions. 

\begin{figure}[htp]
\centering
\includegraphics[angle=270,width=.4\textwidth]{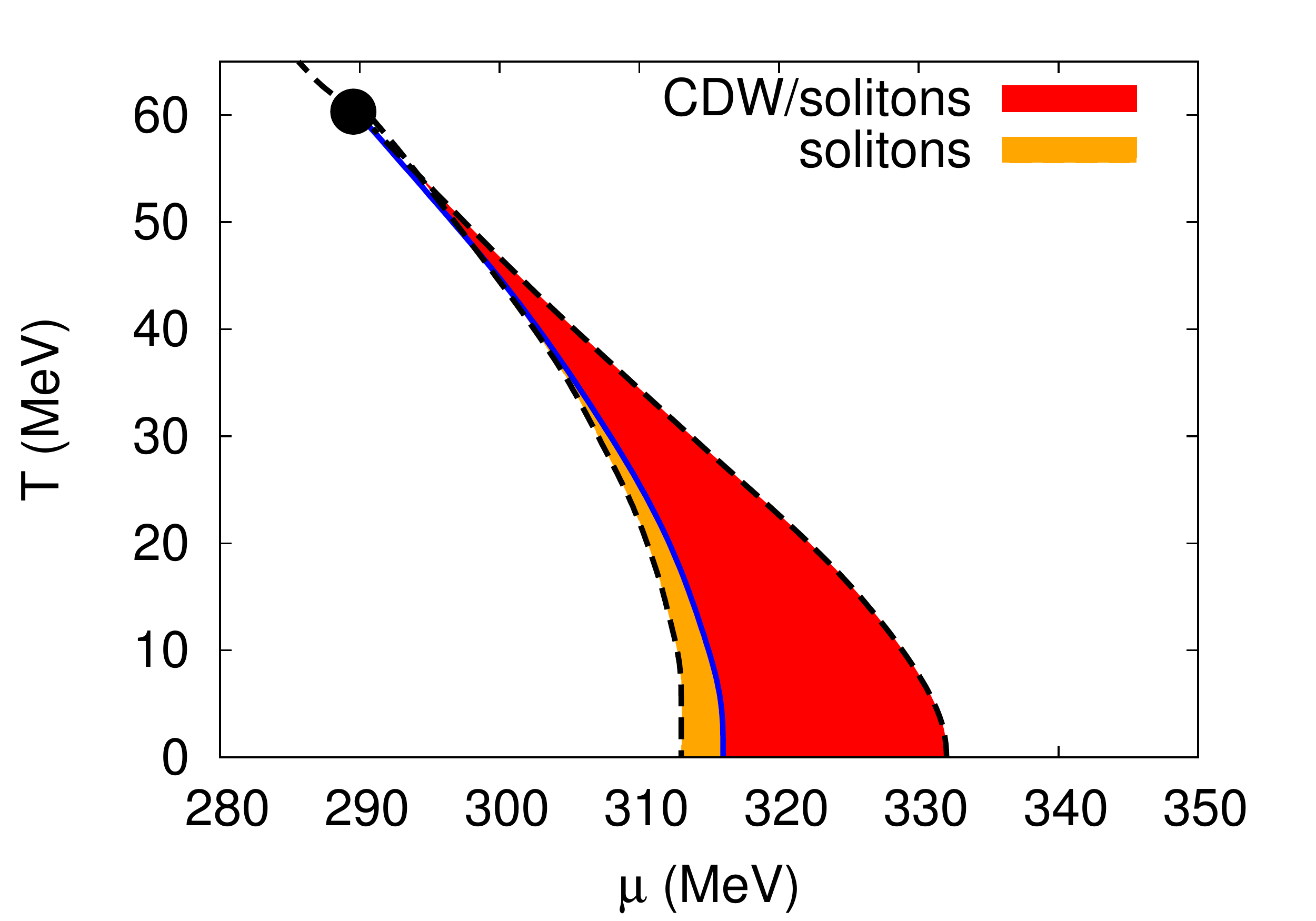}
\caption{Comparison of the phase diagrams for the solitons and the CDW
  modulation in the renormalized QM model.  The light-shaded area denotes the region where only
  solitonic solutions are favored over the homogeneous
   phases. In the dark-shaded area
    the CDW is favored over the homogeneous phases as well, but still
    disfavored against the solitonic modulation. From Ref.~\cite{Carignano:2014jla}. 
  }
\label{fig:QMsoliCDW}
\end{figure}

Being energetically more favored,
allowing for solitonic solutions leads to a larger inhomogeneous region in the phase diagram
than one finds if the possible modulations are restricted to the CDW type.
Such a comparison has been performed in \cite{Carignano:2014jla} within the
renormalized QM model and is shown in \Fig{fig:QMsoliCDW}. 
There it can be seen that the onset of the inhomogeneous phase 
for the solitonic solutions occurs at slightly lower chemical potentials compared to the CDW,
while the transition to the chirally restored phase and the position of the LP are
the same, as predicted by the GL analysis.

 Finally, we note that the favored shape of the condensate modulation may change if the 
 the Lagrangian is modified, e.g., by adding vector interactions~\cite{Marco:MSc}
 or in an external magnetic field~\cite{Tatsumi:2014wka}.
 We will come back to this in Sec.~\ref{sec:extensions}.

\section{Higher-dimensional modulations}
\label{sec:highdim}

Having determined that the most favored one-dimensional shape for the chiral order parameter is given 
by the solitonic ansatz \Eq{eq:Mzsolitons}, we now move on and consider higher-dimensional modulations. 
Unfortunately, there are no analytically known solutions, not even from $2+1$-dimensional models.
All studies therefore start from some ansatz for the condensate function in order
to find the most favorable crystalline structure.
This has been done within a GL analysis~\cite{Abuki:2011}
as well as by numerical diagonalization of the Hamiltonian, following the prescription outlined in 
Sec.~\ref{sec:periodicconds}~\cite{Carignano:2012sx}.
Below we will review the results of both approaches.

\subsection{\it Ginzburg-Landau analysis}

In Ref.~\cite{Abuki:2011} Abuki et al.\@ 
investigated the effect of multidimensional modulations of the condensate functions
within a GL expansion of the thermodynamic potential to 6th order.
Thereby they considered the following real mass functions,
\bea
\label{eq:ndmodGL}
  M_{\mathrm{LO;1D}}({\x})&=&\sqrt{2}M_0\sin(kz)\,,
  \nonumber\\[2ex]
  M_{\mathrm{LO;2D}}({\x})&=&M_0(\sin(kx)+\sin(ky))\,,
  \nonumber\\
  M_{\mathrm{LO;3D}}({\x})&=&\sqrt{\frac{{2}}{{3}}}M_0
  \left(\sin(kx)+\sin(ky)+\sin(kz)\right)\,,
\eea
corresponding to a one-dimensional modulation, a quadratic, and a cubic lattice, respectively.
Minimizing the respective free energies with respect to the wave number $k$, the results for 
the $d$-dimensional modulation were found to be
\beq
  \Omega_{\mathrm{LO;dD}}(M_0,k_\text{opt})
  =
 \left(\frac{1}{2}\gamma_2 - \frac{3}{16}\frac{\gamma_4^2}{\gamma_6}\right) M_0^2
 + \frac{2d-1}{4d} \,|\gamma_4|\, M_0^4 + \mathcal{O}(M_0^6)\,, 
\eeq
where $\gamma_i$ are the GL coefficients introduced in Sec.~\ref{sec:GL}.
As already discussed there, the quadratic term is universal, so that the 
location of the second-order phase transition is independent of the shape of the modulation.
From the coefficient of the quartic term it is then possible to see that the free energy
rises with the dimension, 
$ \Omega_{\mathrm{LO;1D}} <  \Omega_{\mathrm{LO;2D}} <  \Omega_{\mathrm{LO;3D}}$, 
i.e., the one-dimensional modulations are most favored.
Neglecting the $\mathcal{O}(M_0^6)$ terms, minimization with respect to the amplitude $M_0$ 
yields
\beq
       \Omega_{\mathrm{LO;dD}}(M_{0, \text{opt}},k_\text{opt})
       =
       -\frac{d}{2d-1} \frac{1}{|\gamma_4|} \left(\frac{1}{2}\gamma_2 - \frac{3}{16}\frac{\gamma_4^2}{\gamma_6}\right) ^2\,,
\eeq
i.e., the free-energy gain for $d =$ 1, 2 and 3 behaves like $1 : \frac{2}{3} :\frac{3}{5}$.

\begin{figure}[htp]
\centering
\includegraphics[angle=0,width=.4\textwidth]{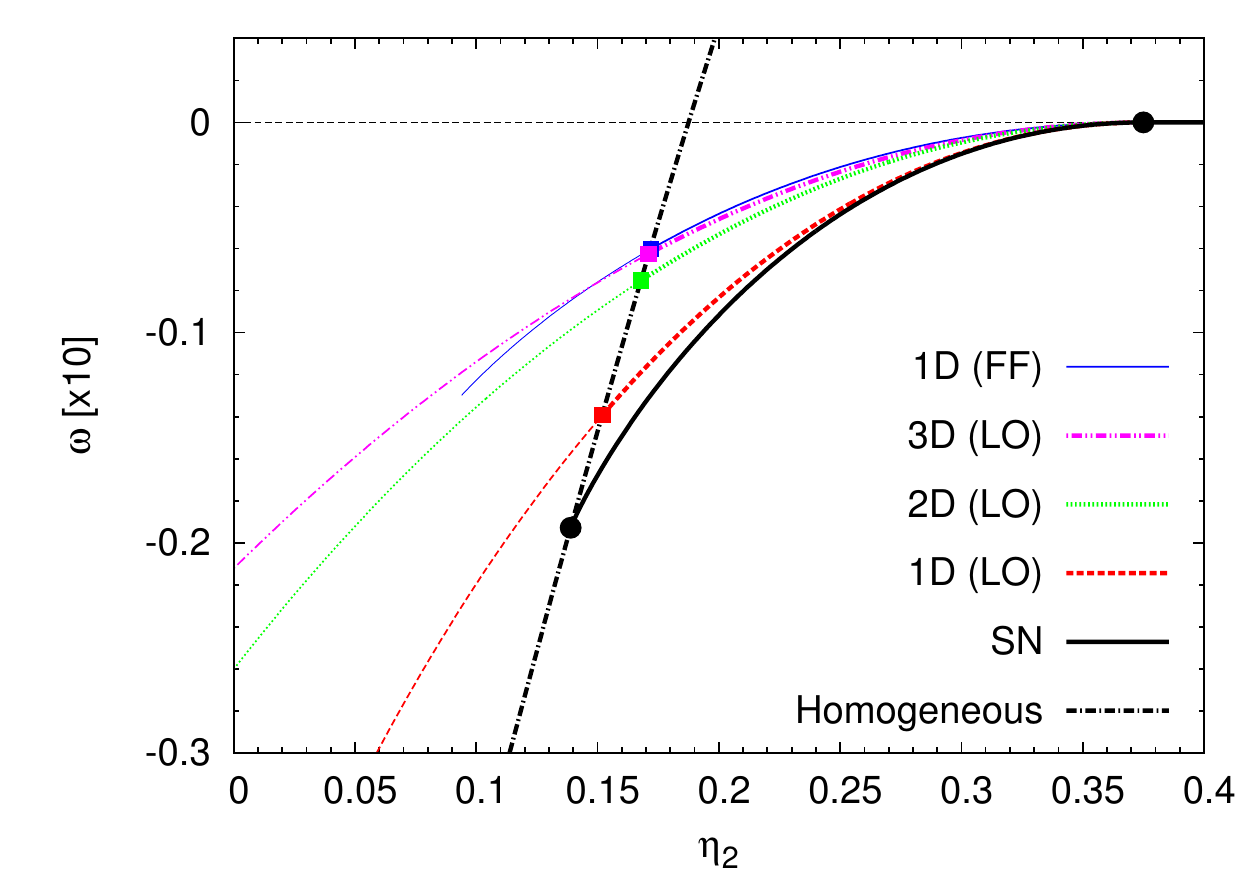}
\caption{Comparison of the free energies $\omega \equiv \Omega / [\frac{\gamma_4|^3}{\gamma_6^2}]$
for the various mass functions defined in Eqs.~(\ref{eq:ndmodcGL}) and (\ref{eq:ndmodGL}) 
as well as for the the one-dimensional solitonic solution (SN), \Eq{eq:Mzsolitons}
as functions of $\eta_2 \equiv \frac{\gamma_2\gamma_6}{\gamma_4^2}$. 
From Ref.~\cite{Abuki:2011}. 
 }
\label{fig:GLmultiD}
\end{figure}

Abuki et al.\@ also considered complex mass functions of the form
\beq
\label{eq:ndmodcGL}
       M_{FF;dD}(\x) = \sqrt{\frac{1}{d}} M_0 \left( e^{ikx_1} + \dots + e^{ikx_d}\right)\,.
\eeq
The corresponding free energies for the optimized values of $k$ turn out to be degenerate
up to quartic order in $M_0$, where the prefactor is given by $\frac{1}{2}|\gamma_4|$.
The complex mass functions are thus disfavored against all real functions considered before,
as we have seen already for one-dimensional modulations.  
This is also shown in \Fig{fig:GLmultiD}, where the free energies associated with the various 
mass functions are compared.

\subsection{\it Numerical analysis of  two-dimensional solutions}

The GL study presented above is expected to be most reliable near the Lifshitz point.
Complementary to that, a numerical study at zero temperature has been performed
in Ref.~\cite{Carignano:2012sx}. 
The analysis was restricted to two-dimensional modulations, which allows to 
reduce the numerical effort by using the boosting method outlined in Sec.~\ref{sec:boost}.

\begin{figure}[h]
\begin{center}
\includegraphics[height=.4\textwidth,angle=270]{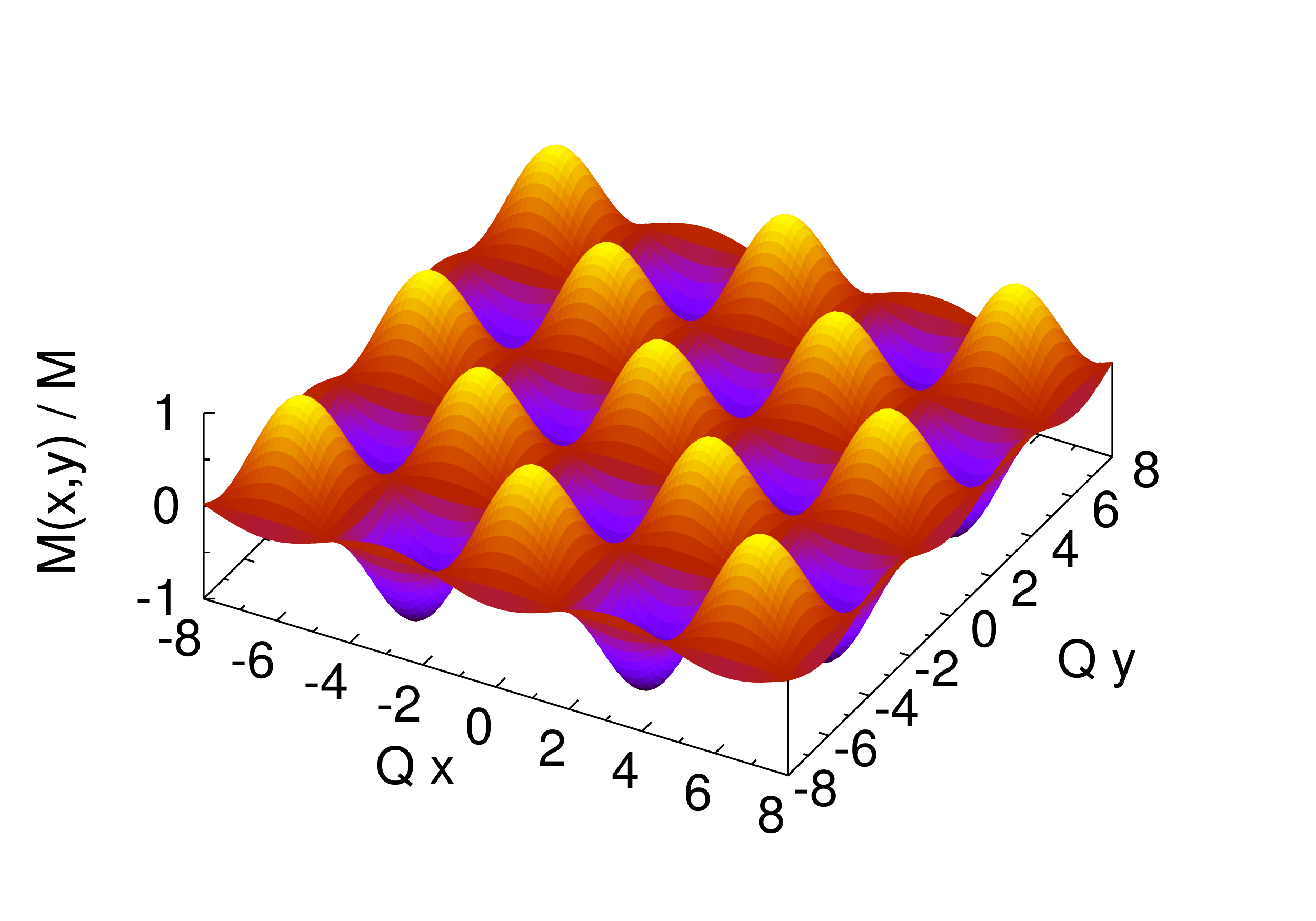}
 \includegraphics[height=.4\textwidth,angle=270]{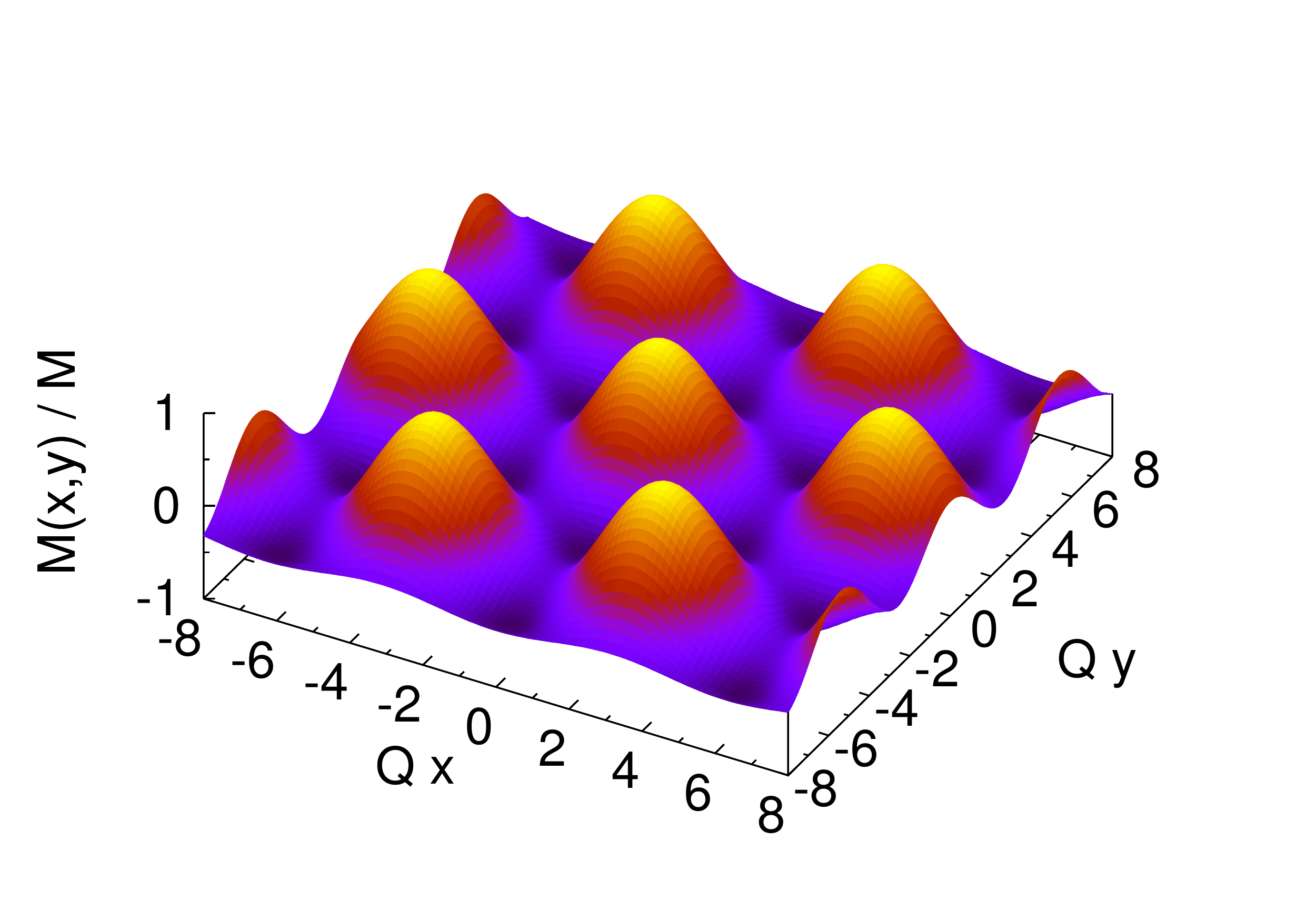}
\end{center}
\caption{Shapes of the considered mass functions $M(x,y)$ in coordinate space.
Left: ``egg-carton'' modulation, \eq{eq:Mcos2d}. 
Right: ``honeycomb'' ansatz, \eq{eq:Mhex}.
From \cite{Carignano:2012sx}. 
}
\label{fig:2dshape}
\end{figure}

Thereby two kinds of two-dimensional crystal structures have been considered. 
The first is a square lattice with
\begin{equation}
  \label{eq:Mcos2d}
 M(x,y) = M \cos(Q x)\cos(Q y) \,,
\end{equation}
which has an egg-carton-like shape (see \Fig{fig:2dshape}, left)
and is symmetric under discrete rotations by $\pi/2$.\footnote{
This ansatz is equivalent to the 2D modulation in \eq{eq:ndmodGL}
with $M_0 = \Delta/2$ and $k = \sqrt{2} Q$, 
after performing a rotation of $\pi/4$ about the $z$-axis followed by
a shift of the origin.}
The second is a ``honeycomb-shaped'' mass function with hexagonal symmetry.
\beq
\label{eq:Mhex}
M(x,y) = 
 \frac{M}{3} \left[ 
2 \cos\left(Q x\right) \cos\left( \frac{1}{\sqrt{3}} Qy\right) 
+
\cos(\frac{2}{\sqrt{3}} Q y) 
\right] \,,
\eeq
which is symmetric under discrete rotations by $\pi/3$ 
(see \Fig{fig:2dshape}, right).

\begin{figure}[h]
\begin{center}
  \includegraphics[height=.4\textwidth,angle=270]{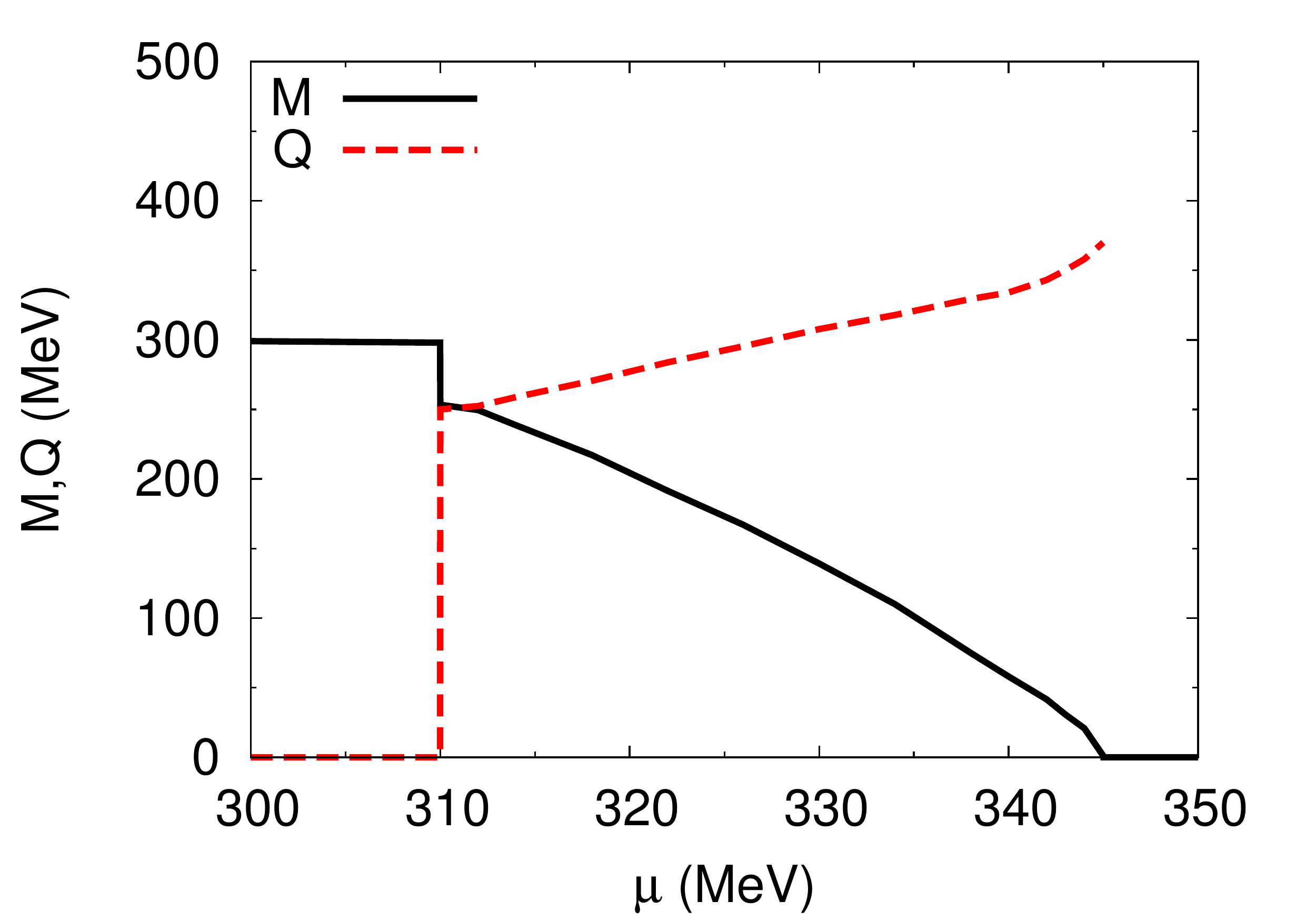}
  \includegraphics[height=.4\textwidth,angle=270]{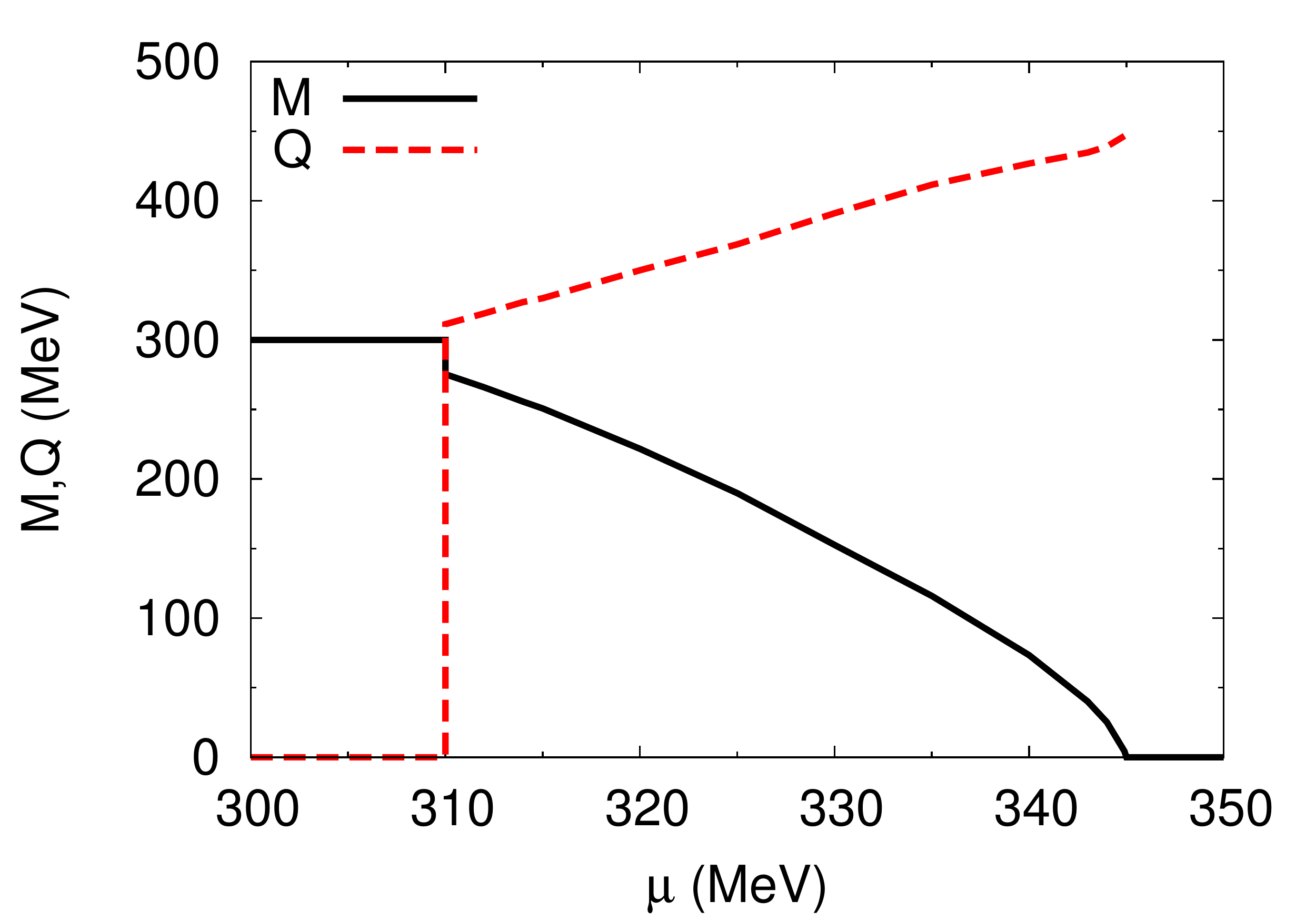}
\end{center}
\caption{Results of the numerical minimization of the thermodynamic potential
for the variational parameters $Q$ and $M$.
Left: ``egg-carton'' modulation, \eq{eq:Mcos2d}. 
Right: ``honeycomb'' ansatz, \eq{eq:Mhex}.
From \cite{Carignano:2012sx}. 
\label{fig:2dMQ}
}
\end{figure}

Results for the numerical minimization in the parameters $Q$ and $M$ 
are shown in \Fig{fig:2dMQ}. It is possible to see
that the variational parameters for both kinds of two-dimensional
modulations are rather similar to each other and
to those associated with the simple
one-dimensional structures previously considered (see e.g. \Fig{fig:njlcdw}). 
In particular,
the onset of the inhomogeneous phase where the wave vector $Q$ jumps
to a finite value via a first order phase transition 
lies at values of the chemical potential which is very close to the one
found for the one-dimensional solutions. Once again, as expected from
GL arguments, the
second-order chiral-restoration transition is exactly at the same 
chemical potential value as for the one-dimensional modulations.\footnote{
The exact values of the critical chemical potentials depend of course on the
model parameters, which are, however, the same in Figs.~\ref{fig:njlcdw}, \ref{fig:solDQ},
and \ref{fig:2dMQ}.}

\begin{figure}[htc]
\begin{center}
\includegraphics[width=.42\textwidth]{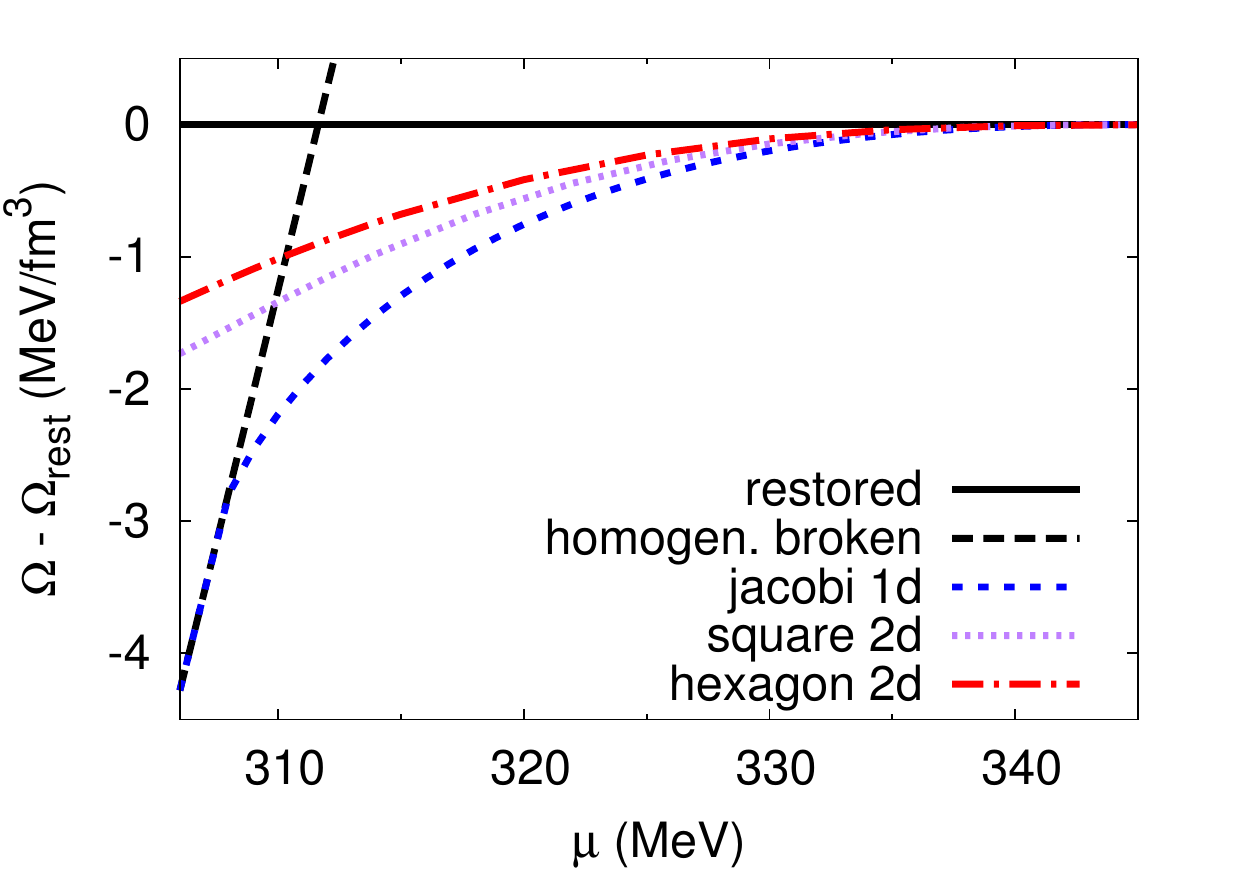}
\end{center}
\caption{
Thermodynamic potential relative to the restored phase
for the two-dimensional modulations of the chiral condensate 
and the solitonic one-dimensional solutions 
at $T=0$. 
The lowest free energy is again 
found for the one-dimensional Jacobi elliptic 
function. 
Among the two-dimensional structures, the egg-carton 
is favored over the honeycomb one.
    From \cite{Carignano:2012sx}.
\label{fig:omegas2d}
}
\end{figure}

Having found out that two-dimensional modulations 
are favored over homogeneous phases in approximately the same chemical-potential window 
as the one-dimensional ones, 
it is once again a matter of comparing the free energies associated 
with the different phases to determine the favored shape for the chiral condensate.
The results are summarized in \Fig{fig:omegas2d}, where one can see that
both two-dimensional structures considered are disfavored 
when compared with the one-dimensional domain-wall solitons. 
In fact, although not shown in the figure, 
they are favored over the CDW solutions, while
they are also disfavored against the one-dimensional cosine, 
which, as we have seen, is very close in free energy to the solitonic solution.
It is therefore unlikely that the result could be changed qualitatively by taking into account
higher harmonics in the ansatz for the two-dimensional modulations.

\begin{figure}[htc]
\begin{center}
\includegraphics[width=.6\textwidth]{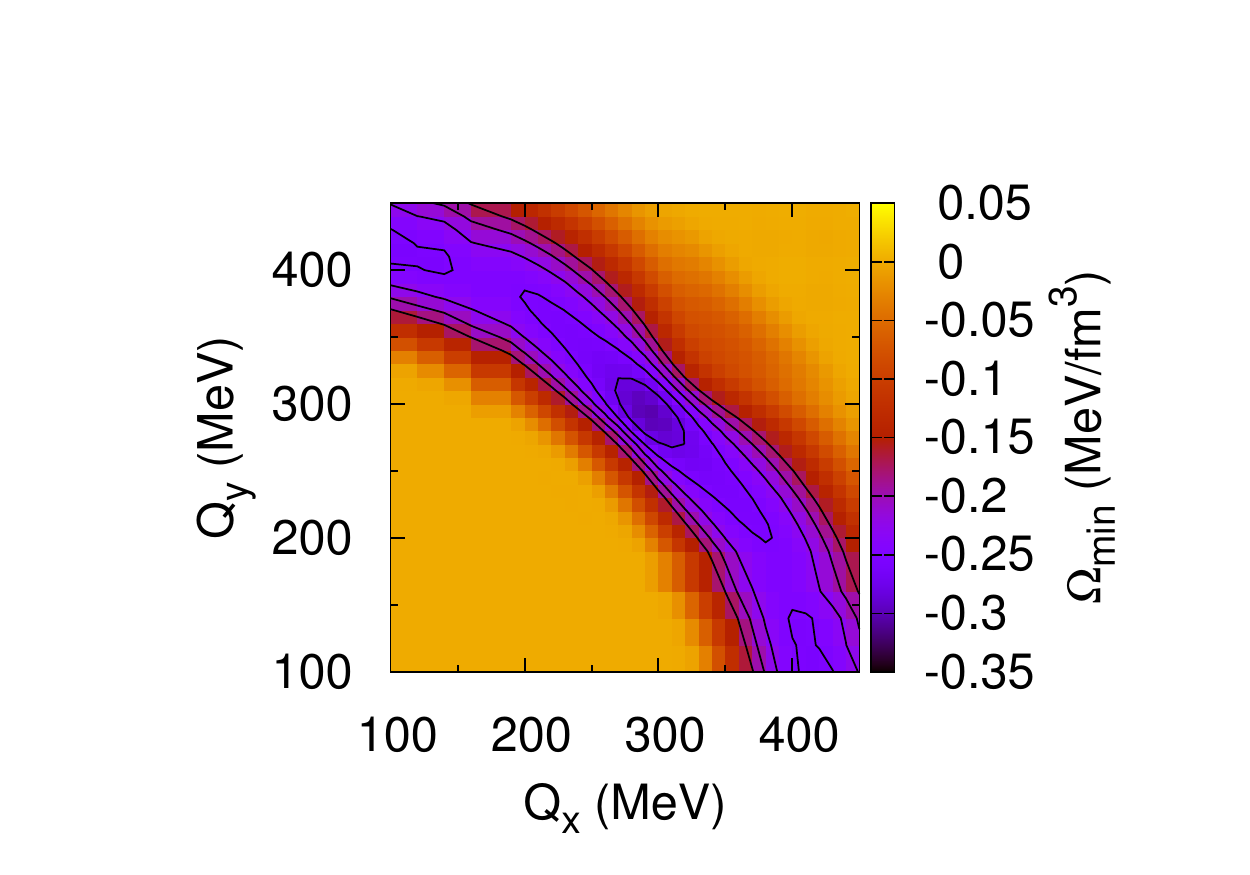}
\end{center}
\caption{Value of the thermodynamic potential at $T=0$ and $\mu=325$~MeV for 
a rectangular lattice, \eq{eq:Mcos2drec}, as a function of $Q_x$ and $Q_y$. 
At each point, $\Omega$ was minimized with respect to $\Delta$.
From \cite{Carignano:2012sx}. 
\label{fig:omegaPxPy}
}
\end{figure}

A complementary check of this behaviour has been performed in \cite{Carignano:2012sx} by considering
a rectangular lattice with the ansatz 

\begin{equation}
  \label{eq:Mcos2drec}
 M(x,y) = \Delta \cos(Q_x x)\cos(Q_y y) \,,
\end{equation}
which reduces to a single cosine varying in one spatial dimension when one 
of the two wave numbers goes to zero, and to the egg-carton ansatz if $Q_x = Q_y$. 
Knowing that the latter is disfavored against the former, the question is whether the optimized
egg-carton solution corresponds to a saddle point or a local minimum in $Q_x-Q_y$ space.
To this end, we show in \Fig{fig:omegaPxPy}
the thermodynamic potential (minimized with respect to the amplitude $\Delta$) evaluated at zero temperature and at a $\mu$ in the middle of the inhomogeneous region as a function of $Q_x$ and $Q_y$. 
It turns out that the solution $Q_x = Q_y$ corresponds to a local minimum, 
although of course disfavored against the one-dimensional cosine, as we have seen before.


Real one-dimensional modulations therefore seem to constitute
the favored shape for the chiral condensate,
both, at $T=0$~\cite{Carignano:2012sx} and near the Lifshitz point~\cite{Abuki:2011}.
An important consideration to be done at this point is that 
all the results presented here have been obtained in the mean-field 
approximation and could be altered if fluctuations are included.
Indeed, as known for a long time~\cite{Landau:1969} and also seen in
studies of pion condensates in nuclear matter~\cite{Baym:1982}
and crystalline superconductors~\cite{Shimahara:1999},
for one-dimensional modulations of the order parameter at finite temperature, 
long-range correlations are destroyed by fluctuations.
Hence, although a quasi-long-range order with power-law correlations may survive (as 
in smectic liquid crystals), these phases are strictly speaking unstable.
Higher-dimensional modulations, on the other hand, are less affected by these effects,
so that they might become thermodynamically favored when fluctuations are included.

\subsubsection{\it Favored phase at high densities} 
\label{sec:cont}

The fact that in \Fig{fig:omegaPxPy} the ``egg-carton'' solution corresponds to a local minimum,
even though not the global one, raises the question whether it could become favored in other regions 
of the phase diagram. 
Indeed, as we will discuss in Sec.~\ref{sec:largenc}, there are arguments suggesting that higher-dimensional
modulations should become relevant at higher densities, at least in 
large-$N_c$ QCD~\cite{Kojo:2010,Kojo:2011}.
In Ref.~\cite{Carignano:2012sx} the analysis of two-dimensional modulations in the NJL model
was therefore extended to higher densities. 

This was possible because it had been found earlier that,
although in all phase diagrams we have shown so far 
the inhomogeneous ``island'' ends at some upper chemical potential, 
in the NJL model a  second inhomogeneous phase appears at higher chemical potentials.
This second inhomogeneous region seems to persist up to arbitrarily high chemical potentials
and has therefore been termed ``inhomogeneous continent''~\cite{CB:2011}.
Of course, since the continent appears in a region where the 
chemical potential is of the order of the cutoff parameter of the model, 
its physical significance is questionable.
More thorough investigations in Refs.~\cite{CB:2011,Carignano:2014jla} have 
revealed a possible connection to known vacuum instabilities~\cite{Broniowski:1990gb},
without clarifying the issue completely.
Here we do not want to enter this rather technical discussion, since anyway 
low-energy effective models are not appropriate to analyze high densities,
which eventually should be done within QCD.
Hoping that some generic features do not depend on details of the model,
we may take, however, the inhomogeneous continent as a schematic ``laboratory'' 
where the competition of one- and two-dimensional chiral crystals can be studied.

\begin{figure}[htc] 
\centering
\includegraphics[angle=270,width=.45\textwidth]{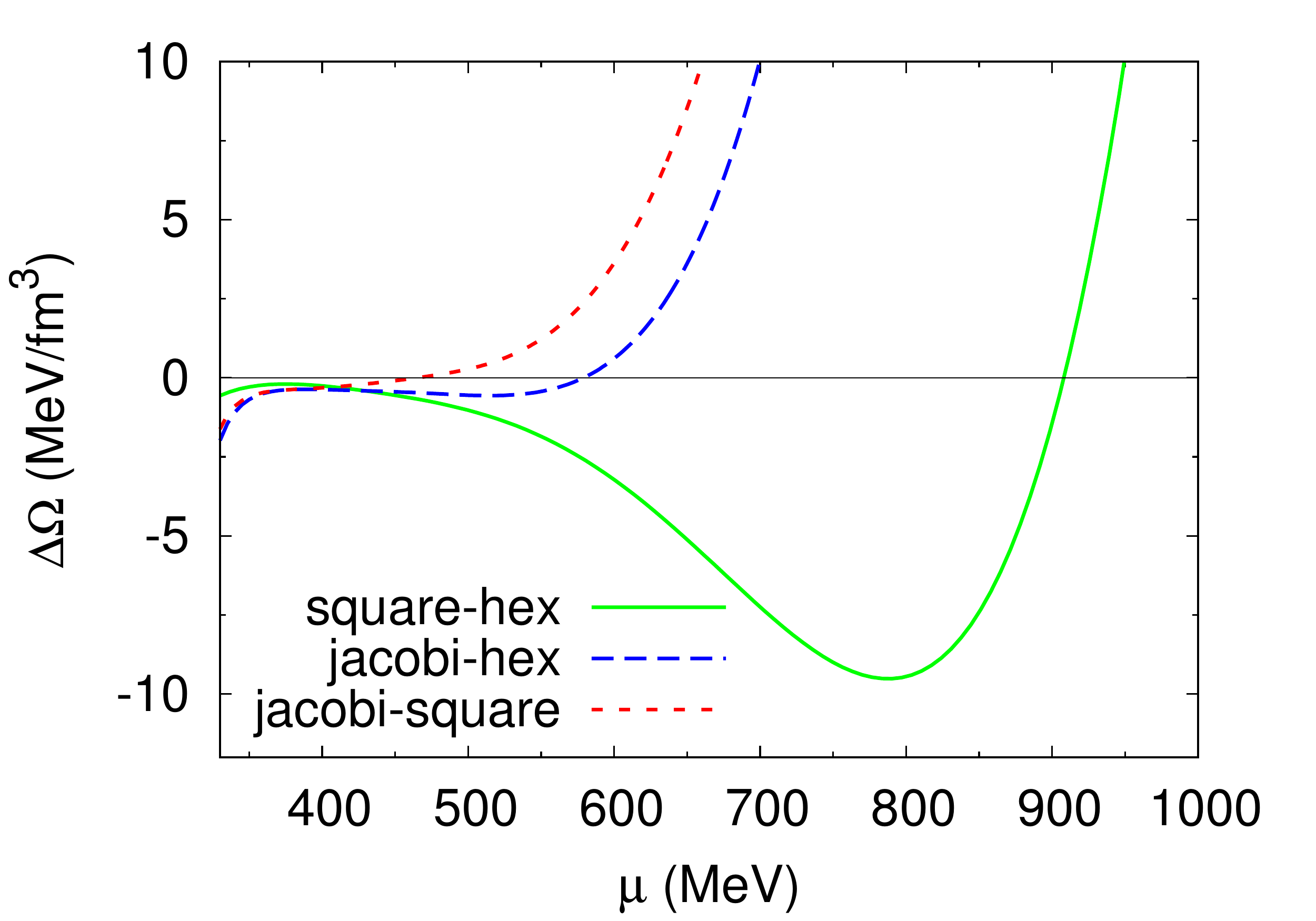}
\caption{Difference in free energy between the two-dimensional structures
considered (square and hexagonal lattices) and the solitonic solutions (``jacobi'') 
at $T=0$ against chemical potential.
From~\cite{Carignano:2012sx}.} 
\label{fig:domhimu}
\end{figure}

Within this spirit, 
the calculations of the ``egg carton'' and ``honeycomb'' modulations have been extended to
higher chemical potentials and compared with the one-dimensional solitonic solution~\cite{Carignano:2012sx}. 
The mutual differences of the free energies are displayed in \Fig{fig:domhimu}.
As we have seen before, at low chemical potentials the two-dimensional 
crystals are disfavored against the one-dimensional Jacobi elliptic 
function. 
Above $\mu \approx 450$~MeV, however, the two-dimensional square lattice 
leads to a lower free energy. 
At $\mu \approx 600$~MeV, also the hexagon surpasses the 
one-dimensional modulation and finally becomes the most favored shape at 
$\mu \approx 900$~MeV. 
Thus, while recalling that the model should not be trusted blindly in this
density region, it is nevertheless remarkable that the same sequence of crystalline 
phases has been predicted in Ref.~\cite{Kojo:2011} for quarkyonic matter, 
cf.~Sec.~\ref{sec:interweaving},
and in Ref.~\cite{Matsuda:2007} for superconductors with cylindrical Fermi surface in an 
external magnetic field.\footnote{The sequence of  Ref.~\cite{Matsuda:2007} contains in addition
a triangular phase, which was not investigated in Ref.~\cite{Carignano:2012sx}.}

\section{Model extensions}
\label{sec:extensions}

So far we have considered the most basic versions of the the NJL model, \eq{eq:LNJL},
and the QM model, \eq{eq:LQM}, consisting of only scalar and pseudoscalar interactions.
In this section, we will discuss how the inhomogeneous phase is affected by 
various model extensions. 

\subsection{\it Vector interactions}
\label{sec:vector}

In view of finite-density investigations, an important extension of the 
original NJL model is given by the inclusion of an isoscalar vector-channel 
interaction term.
This term naturally arises in the NJL model when motivating
the interaction from a one-gluon exchange in QCD \cite{Vogl:1991qt} and was already shown to be of particular
importance in the Walecka model at finite densities \cite{Walecka:1974qa}.
More recently, its influence on the location
and emergence of critical points in the phase diagram has attracted new interest
\cite{Fukushima:2008,Fukushima:2008b,Zhang:2009mk,Bratovic:2012}.
 In particular, it is known that for homogeneous order parameters 
the inclusion of repulsive vector interactions shifts the phase transition line
towards higher chemical potentials and lowers the position of the critical point, leading eventually to its disappearance. 
Since, as seen in the previous sections,
the size of the inhomogeneous phase seems to be intimately related to the
strength of the first-order phase transition for homogeneous matter and the position of the CP, 
it is natural to ask whether the inclusion of vector interactions will strongly influence 
inhomogeneous phases as well. 

The vector-interaction term to be added to the NJL-model Lagrangian is given by
\beq
\label{eq:LV}
{\mathcal L}_{V} = - G_V  \left(\bar{\psi}\gamma^\mu\psi\right)^2
\,,
\eeq
where $G_V$ is a new coupling constant which is expected to be positive
(leading to a repulsive interaction) and of the same order as the scalar coupling $G_S$.
A Fierz transformation of an interaction with the quantum numbers of a one-gluon exchange
gives $G_V = G_S/2$, whereas fits to the omega-meson mass typically yield higher 
values~\cite{Vogl:1991qt,Ebert:1985kz}. An analysis of lattice results, on the other hand,
seems to disfavor large vector couplings~\cite{Steinheimer:2010sp,Steinheimer:2014kka}.
The standard procedure is therefore to keep $G_V$ 
as a free model parameter.
Performing the mean-field approximation along the same lines as discussed in Sec.~\ref{sec:mfa},
the vector interaction is linearized around the expectation values
\beq
        \phi_V^\mu = \langle\bar\psi\gamma^\mu\psi\rangle\,.
\eeq
In homogeneous matter the spatial components of this vector condensate are usually 
argued to be equal to zero, 
since they would break the rotational invariance of the system. 
In inhomogeneous matter, on the other hand, rotational symmetry is already broken by the 
background of the scalar and pseudoscalar condensates, so that 
an induced vector mean-field component along the preferred direction seems to be unavoidable.
It turns out, however, that the assumption of vanishing spatial vector components is still consistent
as long as the background condensates do not have a preferred {\it orientation},
i.e., if they are even functions of parity transformations with respect to an appropriately chosen
point~\cite{Marco:MSc}.
Since this includes the most favored solutions we have found so far, i.e., the solitons, 
we will restrict ourselves in the following to a nonvanishing temporal component,
i.e.,
\beq
       \phi_V^\mu(\x) = n(\x)\, g^{\mu 0}\,, \quad
\eeq       
where $n= \langle\bar\psi\gamma^0\psi\rangle$ corresponds to the quark-number density.

Linearizing \Eq{eq:LV} under this assumption and adding a chemical-potential term yields
\beq
\label{eq:LVMF}
{\mathcal L}_\text{MF}^V + \mu\,\psi^\dagger\psi 
= - 2G_V n \,\psi^\dagger \psi + G_V\,n^2\ + \mu\,\psi^\dagger\psi 
\equiv \tilde\mu\,\psi^\dagger\psi - \mathcal{V}_V\,,
\eeq
where we have defined a ``shifted'' chemical potential
\beq
\label{eq:tildemu}
\tilde{\mu} = \mu-2G_V\, n \,.
\eeq
and a new condensate term
\beq
       \mathcal{V}_V = -G_V n^2
       = -\frac{(\mu-\tilde\mu)^2}{4G_V}\,. 
\eeq
Adding \eq{eq:LVMF} to the mean-field Lagrangian without vector interaction, \eq{eq:mfL},
 the thermodynamic potential for homogeneous phases is then simply obtained 
from the one without vector interaction by replacing $\mu$ with $\tilde\mu$ and
adding $\mathcal{V}_V$,
\beq
\label{eq:Omegashift}
       \Omega(\mu,T) =  \Omega^{G_V=0}(\tilde\mu,T) -\frac{(\mu-\tilde\mu)^2}{4G_V}\,. 
\eeq
From this, $\tilde\mu$ has to be determined self-consistently by solving the 
equation $\partial\Omega/\partial\tilde\mu = 0$.

Unfortunately, while this is a standard procedure for homogeneous phases, 
for general inhomogeneous phases the vector mean-field cannot be absorbed in a redefinition of the
chemical potential because the density $n$ and, hence, $\tilde\mu$ are spatially modulated,
having a nontrivial impact on the spectrum of the Hamiltonian. 
With the exception of the CDW, where $n$ is constant in space (cf.~Sec.~\ref{sec:CDWev}),
the problem must then be tackled by numerical diagonalization of the Hamiltonian
for given trial functions of the scalar and vector mean fields~\cite{Marco:MSc}.
In order to avoid this complication and 
to be able to reuse the known analytical results
for the density of states of the solitonic mass functions,
in Ref.~\cite{CNB:2010} the complete self-consistency was sacrificed by 
approximating the density by its spatial average,
\beq
\label{eq:napproxi}
n(\x) \;\rightarrow\; \nave \; \equiv  \frac{1}{V}\int_V d^3x\, n(\x) \,.
\eeq
As a consequence $\tilde\mu$ becomes constant as well, and the problem
reduces again to the known case without vector interactions at a shifted value
of the chemical potential, \eq{eq:Omegashift}.

\begin{figure}[h]
\begin{center}
\includegraphics[width=.45\textwidth]{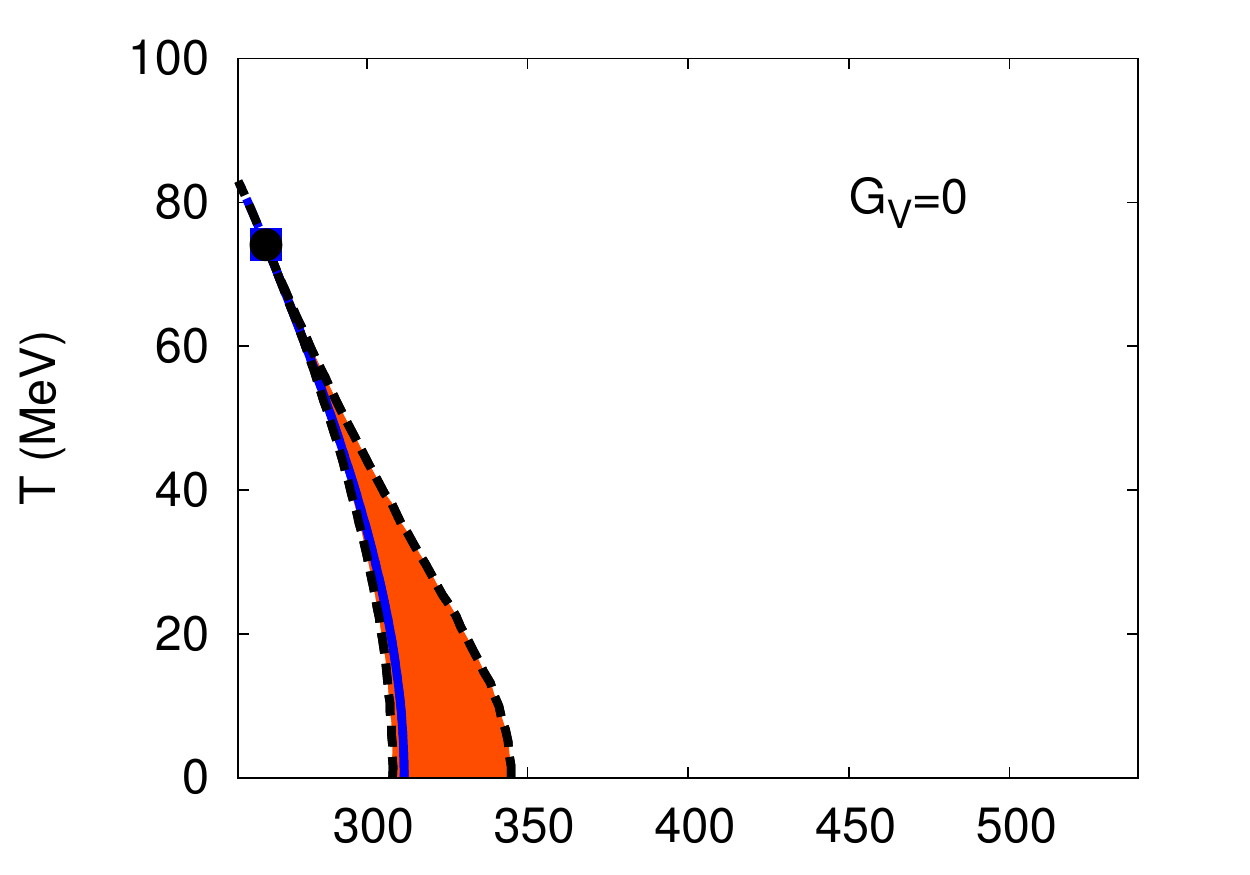}
\includegraphics[width=.45\textwidth]{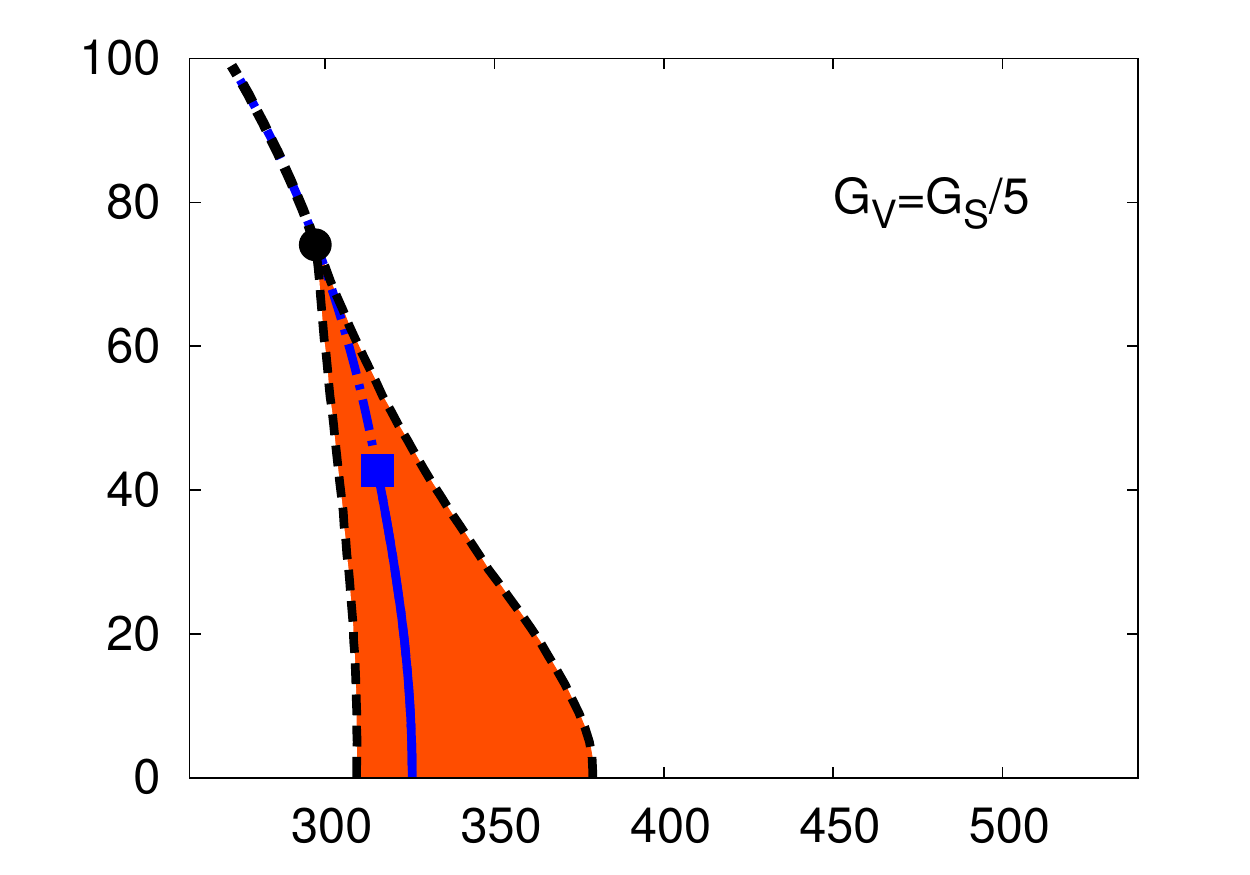}
\includegraphics[width=.45\textwidth]{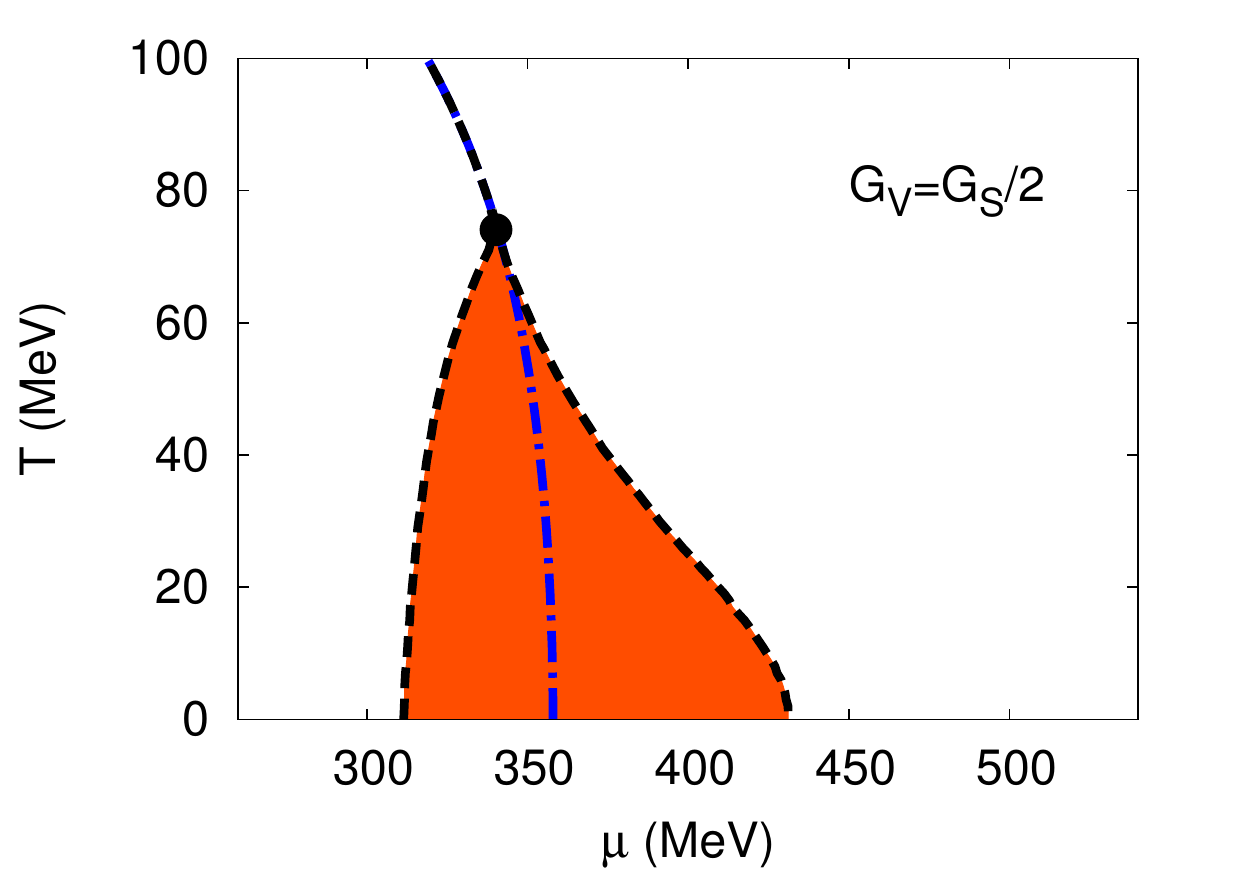}
\includegraphics[width=.45\textwidth]{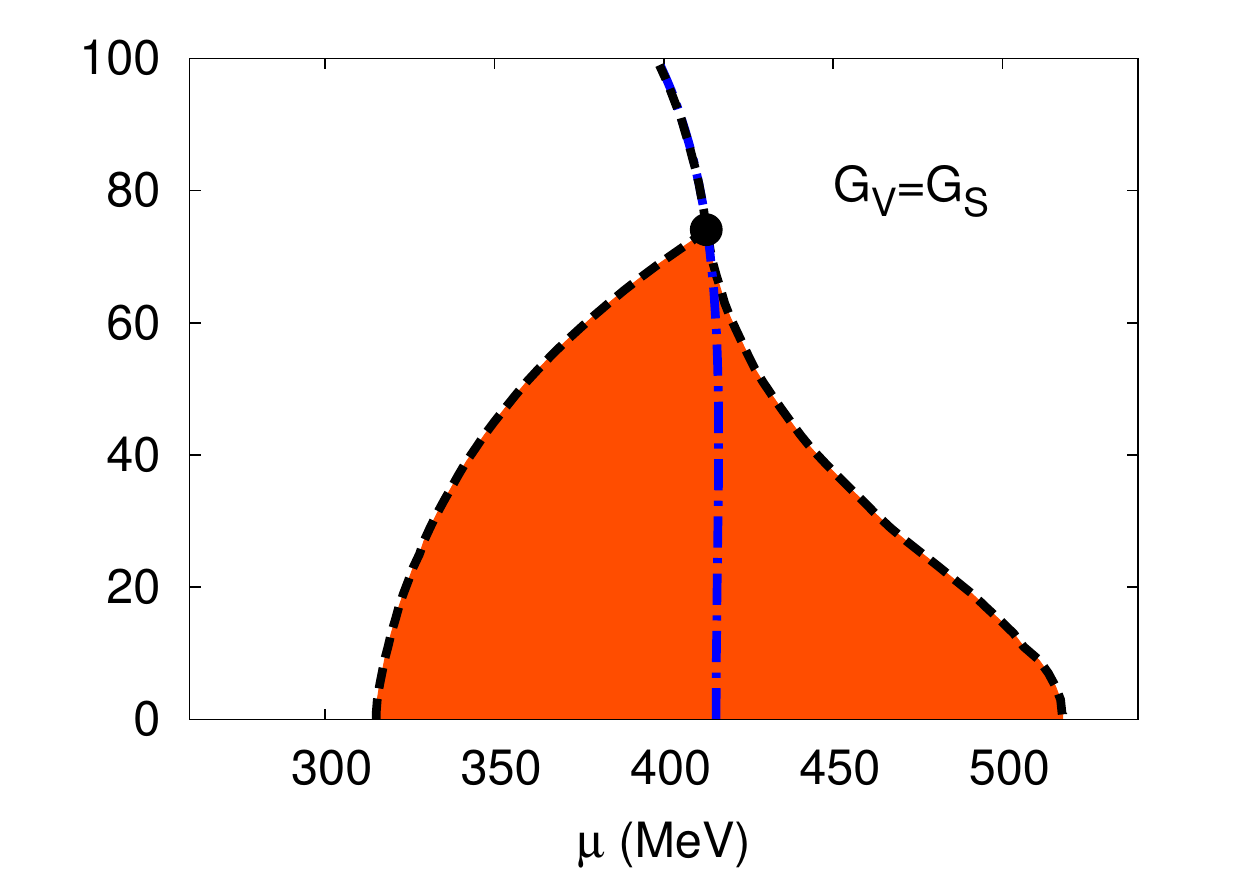}
\caption{The phase diagram in the chiral limit for different values of the 
vector coupling when allowing for the domain-wall soliton lattice. 
The black dashed lines represent the second-order transition lines joining 
at the Lifshitz point  (dot), the shaded region represents the 
inhomogeneous phase. 
The blue solid lines represent the first-order phase 
transition obtained when limiting to homogeneous order parameters, 
which turns to second order (blue dot-dashed lines) at the 
critical point (square). From \cite{CNB:2010}.
\label{fig:pdvector}
}
\end{center}
\end{figure}

The phase diagrams calculated in Ref.~\cite{CNB:2010} in this approximation 
are presented in \Fig{fig:pdvector} for different values of the vector coupling $G_V$.
As previously discussed, 
for $G_V=0$  the Lifshitz point precisely agrees with the 
critical point of the purely
homogeneous analysis.
It turns out, however, that this is no longer true for $G_V>0$:
While with increasing vector coupling the critical point moves downwards in 
temperature and eventually disappears from the phase diagram, we 
observe that the Lifshitz point is only shifted in the
$\mu$-direction, while remaining at the same temperature.
Consequently, unlike the first-order boundary in the purely homogeneous
case, the existence of the inhomogeneous phase is not inhibited by the
vector interaction, but actually turns out to be enhanced, when plotted in the
$\mu-T$ plane. 
In the $\tilde\mu-T$ plane or, equivalently, in the $\nave -T$ plane, on the other hand, 
the boundaries of the inhomogeneous phase can actually be shown to be independent of 
$G_V$, as a consequence of \eq{eq:Omegashift} and the fact that the phase transitions
are second order~ \cite{CNB:2010}.

Recalling the density profiles for the solitonic solutions shown in \Fig{fig:densvsmass},
the replacement \eq{eq:napproxi}
seems to be a reasonable approximation in the intermediate-density region, 
and becomes exact at the boundary to the restored phase, in particular at the Lifshitz point.
On the other hand it is surely questionable close to the phase boundary to the
homogeneous broken phase.
In addition to the solitonic ansatz, the authors of Ref.~\cite{CNB:2010} have therefore also 
studied the effect of vector interactions for a CDW-like modulation, where
\eq{eq:napproxi} is not an approximation but an identity.
It was found that the qualitative behavior was very similar as in \Fig{fig:pdvector}.
Recently the effect of vector interactions has also been investigated by numerical
diagonalization of the Hamiltonian for a sinusoidal mass function together with a constant plus
sinusoidal ansatz for the $n(\x)$~\cite{Marco:MSc}.
There it turned out that for higher values of $G_V$ the CDW becomes favored over the 
sinusoidal mass function, at least in the lower-density part of the inhomogeneous phase,
suggesting that it might also be favored over the solitons if the density modulations are
properly taken into account.

The different behaviour of LP and CP
can also be understood within a Ginzburg-Landau analysis, 
which does not rely on the assumption of a constant density profile~\cite{CNB:2010}.
One finds that while
the GL coefficient multiplying $\vert \nabla M \vert^2$ (and therefore
associated with the location of the LP) remains unmodified 
by the inclusion of vector interactions, 
the $\vert M \vert^4$  coefficient
acquires a contribution which depends explicitly on $G_V$. 
This leads to the displacement of the CP towards lower temperatures
and to the splitting of CP and LP.

\subsubsection{\it Quark number susceptibilities}

The divergence of susceptibilities near a critical point has led to suggestions for how to locate the CP in the 
QCD phase diagram experimentally \cite{Stephanov:1998dy} and therefore attracted significant interest also in model studies~\cite{Sasaki:2006ww,Schaefer:2006ds,Fukushima:2008}.
Having seen that the presence of an inhomogeneous phase could cover the first-order phase boundary
and also the CP, it is a natural and important question how the susceptibilities are altered by the presence 
of an inhomogeneous phase. 

As an example, the behaviour of the quark number susceptibility
\beq
\chi_{nn}
\quad=\quad
-\frac{
\partial^2 \Omega
}{
\partial\mu^2
}
\quad=\quad
\frac{\partial \nave}{\partial \mu}
\eeq
has been investigated in Ref.~\cite{CNB:2010} within the NJL model discussed above.
Looking at the phase diagrams we may already anticipate that there will be a qualitative
difference between the cases with and without vector interaction. 
For $G_V=0$,  the CP is still present in the phase diagram, coinciding with the LP.
Moreover, moving from this point in positive or negative $\mu$ direction, we get into the chirally 
restored or homogeneous broken phase, respectively, i.e., the $\mu$-derivatives and hence the divergence of 
$\chi_{nn}$ at the CP are completely unaffected by the presence of the inhomogeneous phase.
For $G_V > 0$, on the other hand, the CP is covered by the inhomogeneous phase and thus 
disappears from the phase diagram, so that
the homogeneous analysis should be irrelevant for the actual behavior in this region.

\begin{figure}[h]
\begin{center}
\includegraphics[width=.48\textwidth]{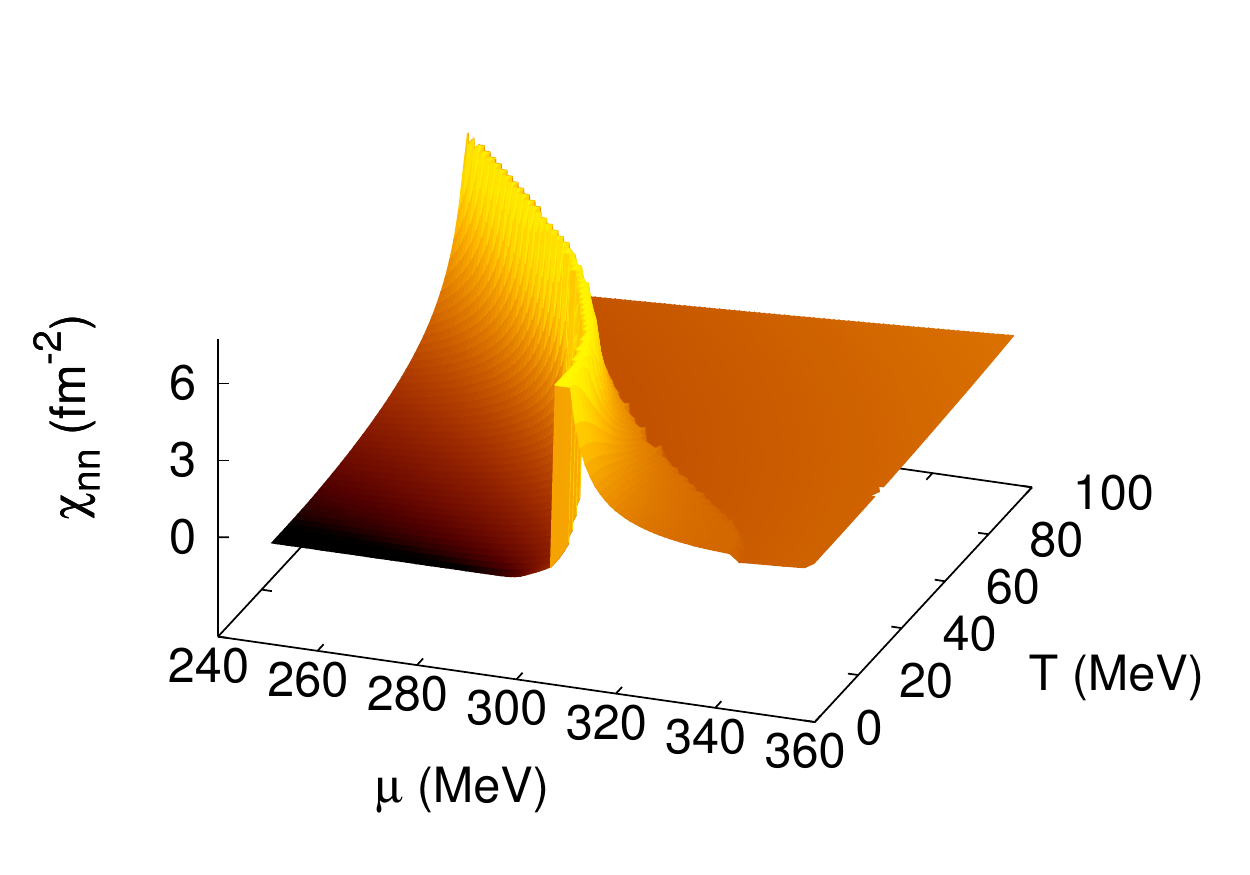}
\includegraphics[width=.48\textwidth]{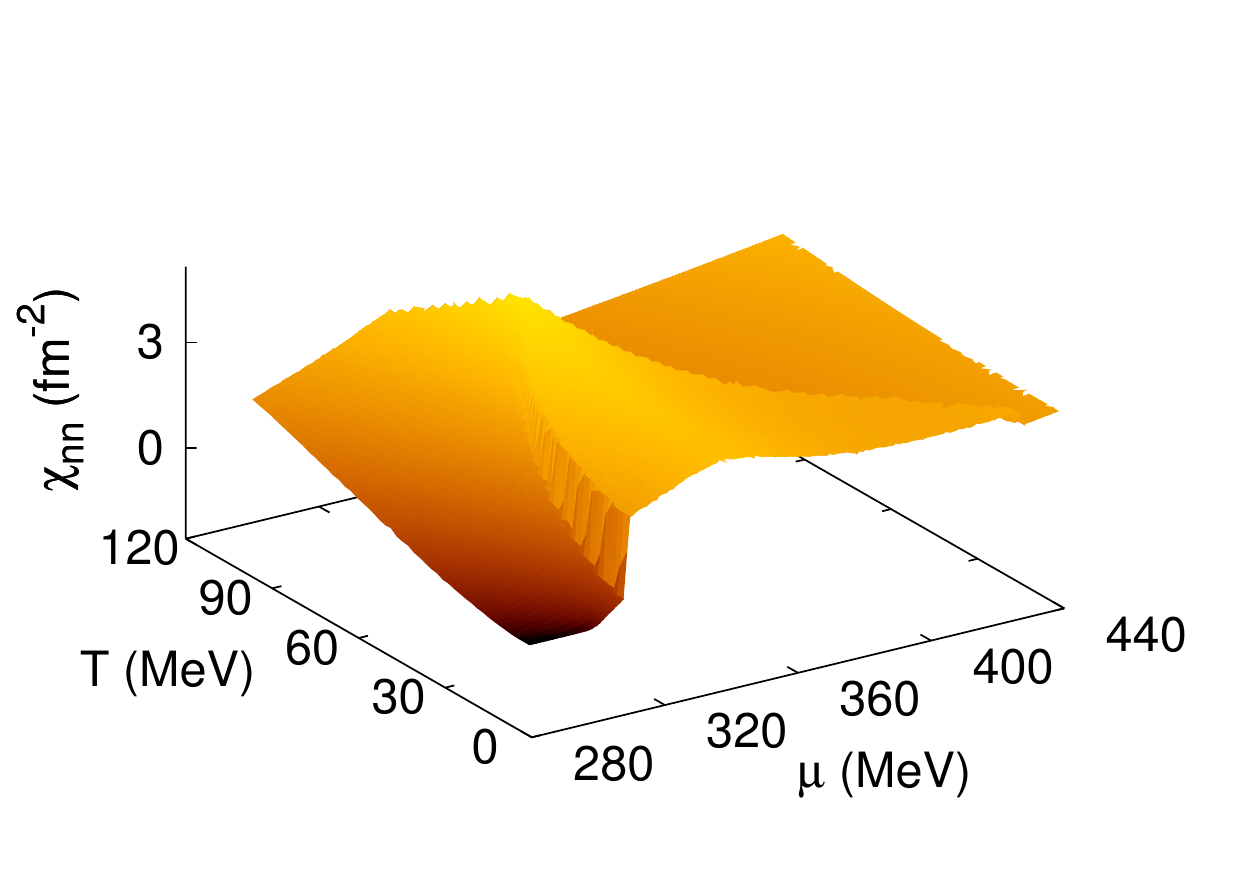}
\caption{
Number susceptibility in the $\mu-T$ plane at $G_V=0$ (left) and $G_V=G_S/2$ (right). For $G_V=0$ the number susceptibility diverges at the transition from the broken to the inhomogeneous phase,
for $G_V>0$ there is only a finite jump. 
From \cite{CNB:2010}.
\label{fig:susc3d}
}
\end{center}
\end{figure}

These expectations have been confirmed by the numerical calculations shown in~\Fig{fig:susc3d}.
It turned out, however, that for $G_V=0$ (left panel) the susceptibility does not only diverge at the CP, 
but along the entire second-order phase boundary to the homogeneous broken phase.
More precisely, the numerical data slightly above the critical chemical potential $\mu_c$ could be 
well described by 
$\chi_{nn} \propto [(\mu-\mu_c)\log^2(\mu-\mu_c)]^{-1}$~\cite{CNB:2010},
which is the same behavior as in the GN model~\cite{Schnetz:2004,Schnetz:2005ih}.
For the NJL model this was confirmed in Ref.~\cite{Abuki:2011} within a GL analysis.

The inclusion of vector interactions, on the other hand (right panel), immediately renders the susceptibility
finite, even at the LP.
This can again be explained by the fact that the model at nonzero $G_V$ and chemical potential $\mu$
essentially corresponds the model with $G_V=0$ at the shifted chemical potential $\tilde\mu$,
cf.~Eqs.~(\ref{eq:Omegashift}) and  (\ref{eq:tildemu}).
Hence
\bea
\chi_{nn}
\quad=\quad
\frac{\partial\nave}{\partial \tilde{\mu}}
\left(\frac{\partial \mu}{\partial \tilde{\mu}}\right)^{-1}
\quad=\quad
\left.
\frac{
1
}{
1+2G_V\frac{\partial\nave}{\partial \tilde{\mu}}
}
\,
\frac{\partial\nave}{\partial \tilde{\mu}}
\right\vert_{\tilde{\mu}(\mu)}
\,,
\eea
and therefore a divergence in $\partial\nave/\partial \mu$ at $G_V=0$ does not result 
in a divergent number susceptibility for $G_V>0$, but merely leads to a finite step
of order $1/2G_V$.

These results should however be taken with care.  
We remind that the calculations have been performed in the mean-field approximation,
and it is possible that the second-order phase transition from the homogeneous broken to the
inhomogeneous phase becomes first order if mesonic fluctuations are included. 
In this case the susceptibility would stay finite along the phase boundary, even for $G_V=0$.
Moreover, the approximation \Eq{eq:napproxi} is extremely questionable near this phase
boundary and treating the vector condensates properly for $G_V>0$ could
also change the order of the phase transition. 
While these arguments point towards finite susceptibilities in both cases,
it remains an important task to investigate these questions explicitly.

\subsection{\it Effects of Polyakov loop dynamics}
\label{sec:Polyakov}

The NJL model describes a system of self-interacting quarks 
and does not include gluonic degrees of freedom.
As such, it does not confine, 
and is known to feature unphysical effects like the
possibility for the decay of a meson
into free quarks. 

In order to mimic some features of confinement, in particular to suppress the 
contribution of free constituent quarks in the confined phase,
and in order to include gluonic contributions to the pressure, 
the NJL model can be coupled to an effective description of the Polyakov 
loop~\cite{Ratti:2005jh,Fukushima:2003fw,Meisinger:1995ih,Megias:2004hj}. 
The resulting model is known
as the Polyakov-loop extended Nambu--Jona-Lasinio (PNJL) model.

\noindent
The Polyakov loop is defined by
\beq
    L(\x) = {\cal P} \exp\left[i\int_0^{1/T} d\tau\, A_4(\tau,\x) \right]\,,
\label{eq:Poly}
\eeq
where $A_4(\tau,\x) = iA_0(t=-i\tau,\x)$ is the temporal part of a gauge 
field $A_\mu = g A_\mu^a \frac{\lambda^a}{2}$ at imaginary time.
In pure Yang-Mills theory, the traced expectation values of $L$ 
and its hermitean conjugate, 
\beq
     \ell=\frac{1}{N_c}\langle{\rm Tr}_c L\rangle 
\,,
     \qquad
     \bar\ell=\frac{1}{N_c}\langle{\rm Tr}_c L^\dagger\rangle
\,,
\label{eq:llbar}
\eeq
can be related to the free energies of a static 
quark or antiquark,
$\ell \sim e^{-F_q/T}$, 
$\bar\ell \sim e^{-F_{\bar q}/T}$~\cite{McLerran:1981pb,Karsch:1985cb},
and are therefore order parameters for the confinement-deconfinement 
transition.

To include the Polyakov-loop dynamics in the NJL model,
the quarks are minimally coupled to a background gauge field
by introducing a covariant derivative $D_\mu$,
\beq
\partial_\mu \rightarrow D_\mu = \partial_\mu + iA_0 \delta_{\mu 0} \,.
\eeq
Furthermore, a local potential $\mathcal{U}(\ell,\bar{\ell})$ is added to 
the thermodynamic potential,
which is essentially constructed to reproduce ab-initio results of pure 
Yang-Mills theory at finite temperature~\cite{Ratti:2005jh,Roessner:2006xn}.

When dealing with inhomogeneous phases, $\ell$ and $\bar{\ell}$ are
naturally expected to be spatially dependent, presumably following the
density profile in some way.\footnote{
Very recently, this expectation was supported by a lattice study
where the Polyakov-loop expectation value was calculated
within an effective model of gluons
on the background of an inhomogeneous chiral condensate field~\cite{Hayata:2014eha}.
}
However, this would not only complicate the calculations, 
it would also create new ambiguities, because the standard parametrizations 
of the Polyakov-loop potential~\cite{Ratti:2005jh,Roessner:2006xn,Fukushima:2008},
do not contain kinetic contributions (i.e., gradient terms) which in principle should
exist but are unknown.
In the analysis of Ref.~\cite{CNB:2010},  $\ell$ and $\bar{\ell}$ have therefore been
treated as space-time independent mean fields.

The effect of the covariant derivative in the kinetic part
amounts to the modification of \Eq{eq:OmegakinRhoSymm} into 
\bea
\label{eq:OmegakinPNJL}
       \Omega_\mathit{kin}^\text{PNJL}
       = 
         -\int_0^\infty dE \, \tilde\rho(E) 
       \Big[ E 
      &+& \frac{1}{N_c}
T\log\left(1 
  +3\,\ell\,\mathrm{e}^{-(E-\mu)/T}
   +3\,{\bar\ell}\,\mathrm{e}^{-2(E-\mu)/T}
     + \mathrm{e}^{-3(E-\mu)/T}
   \right)
   \nonumber\\
&+&\frac{1}{N_c}
T\log\left(1  
   +3\,{\bar\ell}\,\mathrm{e}^{-(E+\mu)/T}
   +3\,\ell\,\mathrm{e}^{-2(E+\mu)/T}
+ \mathrm{e}^{-3(E+\mu)/T} 
   \right) \Big]
\,,
\eea
which reduces to \Eq{eq:OmegakinRhoSymm}
in the deconfined phase where $\ell = \bar\ell = 1$,
while in the confined phase, $\ell = \bar\ell = 0$, the thermal excitation of quarks is
strongly suppressed. 
The total thermodynamic potential of the PNJL model is then given by
\beq
       \Omega_\text{PNJL}
       =
       \Omega_\mathit{kin}^\text{PNJL}
       +
       \Omega_\mathit{cond}      
       +
       \mathcal{U}(\ell,\bar\ell)\,.
\eeq
From this the values of $\ell$ and $\bar\ell$ at given temperature and chemical potential are
determined by the additional gap equations 
${\partial\Omega_\text{PNJL}}/{\partial\ell} ={\partial\Omega_\text{PNJL}}/{\partial\bar\ell} =0$.

\begin{figure}
\begin{center}
\includegraphics[width=.4\textwidth]{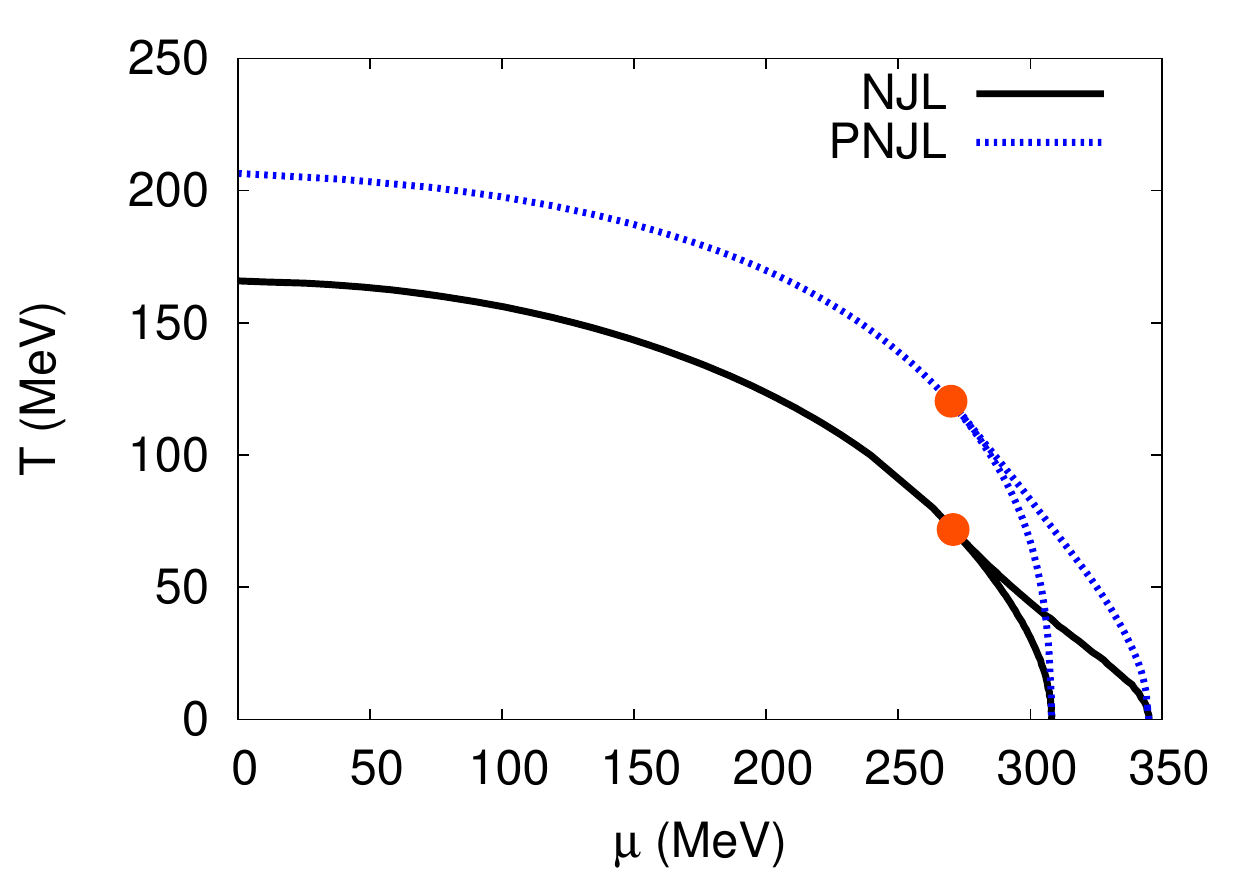}
\caption{Phase diagram of the NJL (solid line) and PNJL (dashed line) 
model allowing for one-dimensional solitonic modulations of the order parameter.
From Ref.~\cite{CNB:2010}.}
\label{fig:pdpnjl}
\end{center}
\end{figure}

In \Fig{fig:pdpnjl} we show the comparison of
NJL and PNJL phase diagrams from \cite{CNB:2010}, allowing for one-dimensional solitonic 
modulations in both cases. 
Similar to what happens for homogeneous phases,
the most notable effect of the coupling with the Polyakov loop
is a stretch of the phase diagram in the $T$-direction,
which is a consequence of the suppression of thermal quarks and antiquarks 
in the regime where $\ell$ and $\bar\ell$ are small.

Although not noted in Ref.~\cite{CNB:2010},  
a Ginzburg-Landau analysis reveals that the coupling with the Polyakov loop 
again leads to a splitting between the critical point and the Lifshitz point.
The shift is opposite to the case of vector interactions, so that a small
piece of the first-order phase boundary between the homogeneous phases
is left which is not covered by the inhomogeneous phase.
The effect is, however, quite small, making its numerical verification is very
difficult. 

 \subsubsection{\it PNJL and large $N_c$}
\label{sec:PNJLNc}

In Ref.~\cite{CB:2011} the PNJL-model analysis discussed above was extended to investigate
the influence of the number of colors on the phase diagram. 
This was motivated by the prediction of an inhomogeneous ``quarkyonic'' phase at large 
$N_c$~\cite{Kojo:2009}, which will be discussed in more detail in Sec.~\ref{sec:largenc}.
A similar analysis for homogeneous phases had been done before in Ref.~\cite{McLerran:2008}, where
the quark and the gluon sector of the model have been generalized to arbitrary 
$N_c$ by assuming that the coupling constant and the Polyakov loop are rescaled as
\beq
       G_S \rightarrow \frac{3}{N_c} G_S \,, \qquad
        \mathcal{U}(\ell,\bar\ell)  \rightarrow \frac{N_c^2-1}{8} \,\mathcal{U}(\ell,\bar\ell)\,.
\eeq
The second replacement means that the Polyakov-loop potential simply scales with the number
of gluons, otherwise leaving the functional dependence unchanged. We note that this is a strong
assumption. In fact, some parametrizations of the potential, 
like those given in Refs.~\cite{Fukushima:2008,Fukushima:2003fw,Roessner:2006xn}, 
have contributions from the Haar measure, which are not simply proportional to the number of gluons.

Without Polyakov loop, the first replacement  leaves the NJL-model gap equation, and
hence the phase diagram $N_c$-independent in mean-field approximation.
The coupling of both sectors is therefore crucial to find a nontrivial effect. 
Here one encounters the conceptual difficulty that the kinetic part of the thermodynamic
potential in principle depends on $N_c-1$ moments of the Polyakov loop 
so that the generalization of \eq{eq:OmegakinPNJL} in terms of only two variables, $\ell$ and $\bar\ell$, 
is not unique. The authors of Ref.~\cite{McLerran:2008} have therefore performed a
leading-order expansion for small $\ell$ ,
\bea
\label{eq:PNJLapprox}
       &&
       \frac{1}{N_c}
       T\log\left(1 +N_c\,\ell\,\mathrm{e}^{-(E-\mu)/T} + \dots  + \mathrm{e}^{-N_c(E-\mu)/T} \right)
       +
       \frac{1}{N_c}
       T\log\left(1 +N_c\,\ell\,\mathrm{e}^{-(E+\mu)/T} + \dots  \right)
\nonumber \\
       &\approx&
       \theta(\mu-E)\,\left[ (\mu-E) + \ell\, T e^{-(\mu-E)/T} \right]      
       +
       \theta(E-\mu)\,\ell \,T \, e^{-(E-\mu)/T} 
       +
       \ell\, T \, e^{-(E+\mu)/T}\,, 
\eea
which is expected to be accurate at large $N_c$ in the confined phase.

\begin{figure}
\begin{center}
\includegraphics[width=.32\textwidth]{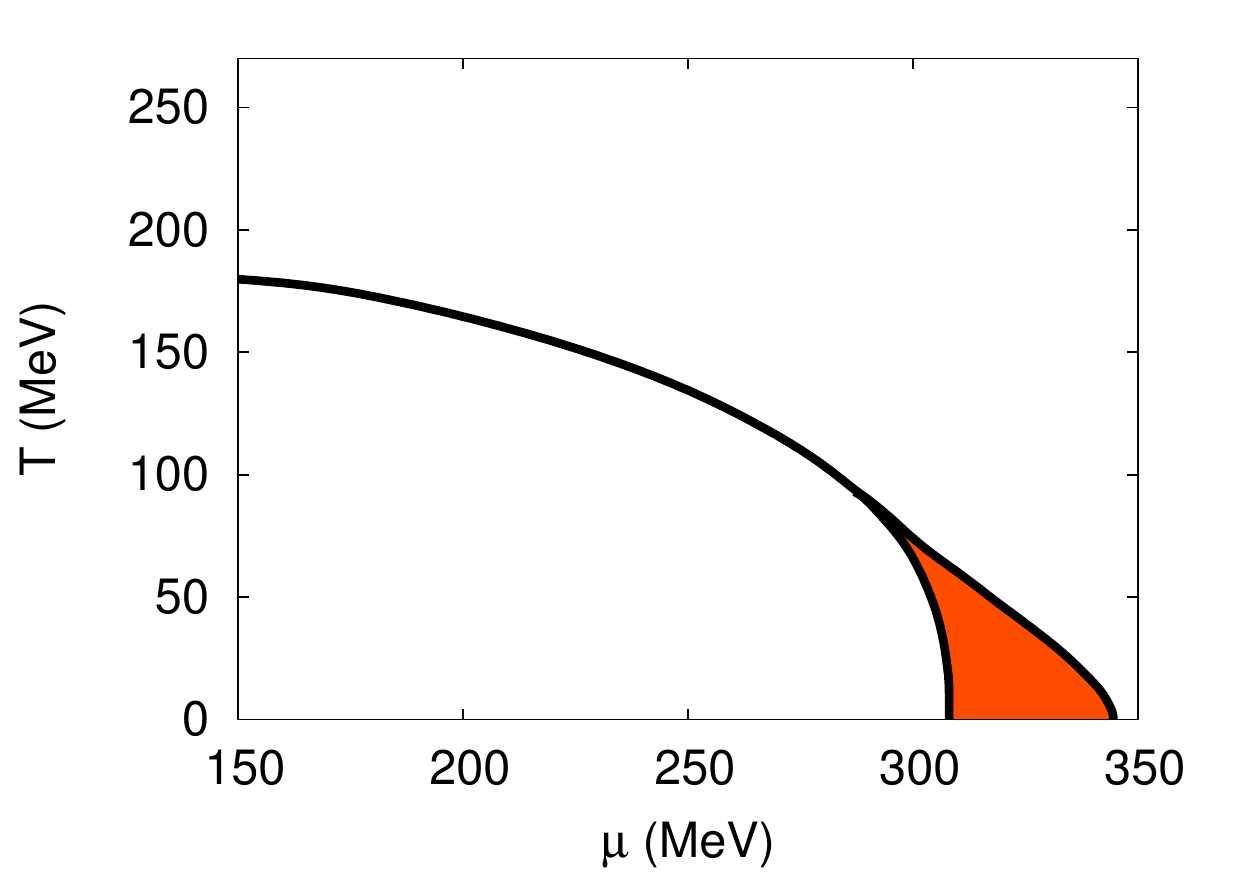}
\includegraphics[width=.32\textwidth]{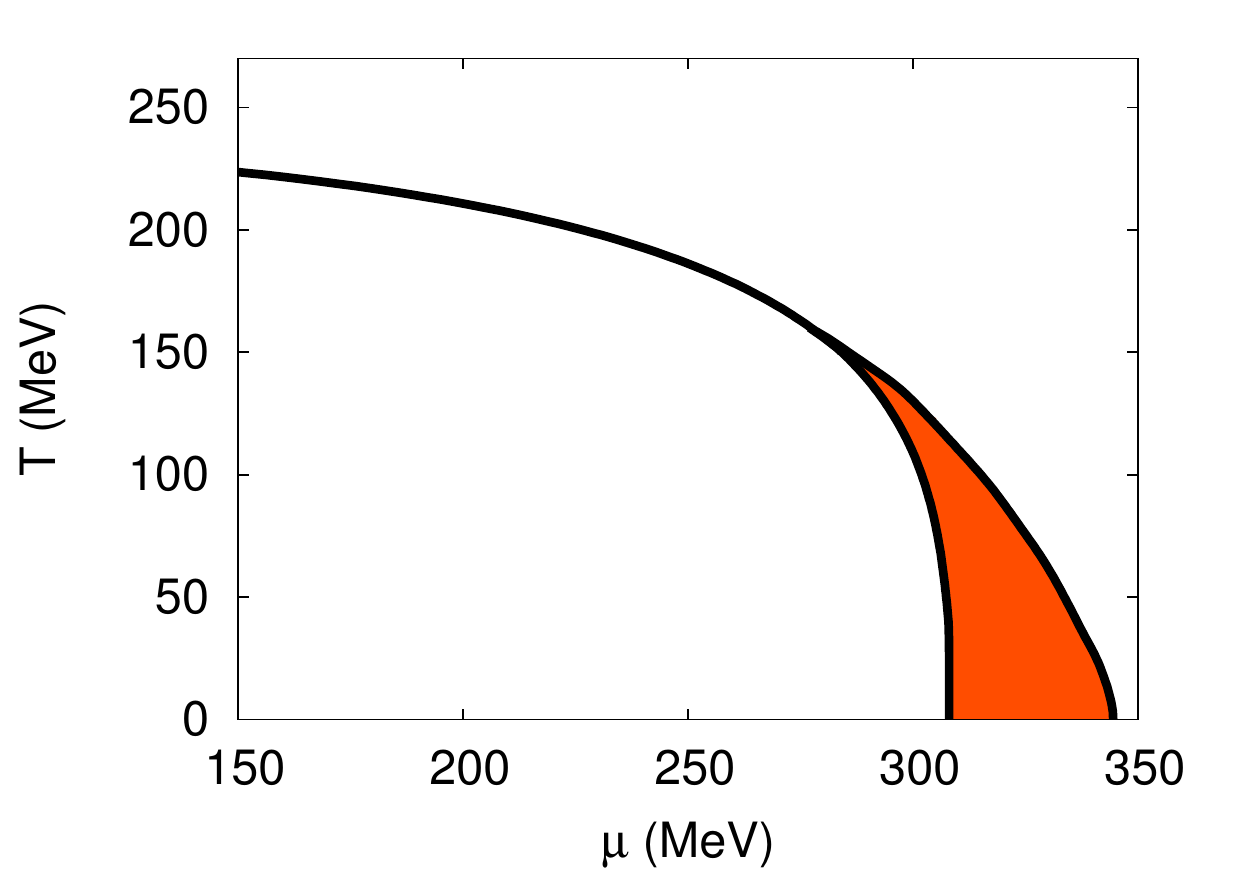}
\includegraphics[width=.32\textwidth]{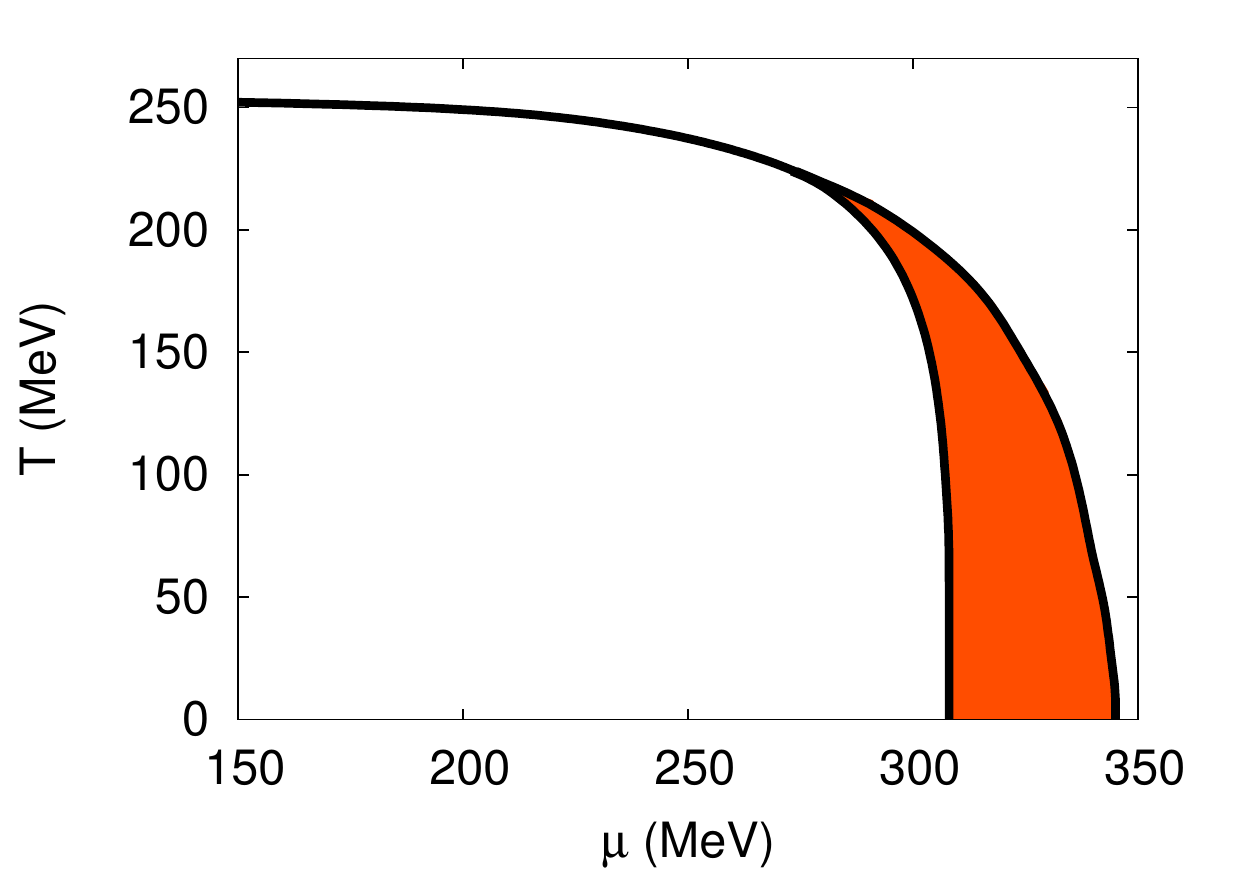}
\end{center}
\caption{PNJL phase diagram varying the number of colors: 
$N_c=3$ (left), $N_c=10$ (center), $N_c=50$ (right).
The shaded region indicates the inhomogeneous phase.
From Ref.~\cite{CB:2011}.}
\label{fig:PNJL}
\end{figure}

This prescription was also adopted in  Ref.~\cite{CB:2011} for studying the inhomogeneous
phase at varying $N_c$.
The results for $N_c =$ 3, 10 and 50 with the solitonic ansatz for the mass function
are shown in \Fig{fig:PNJL}.
By increasing the number of colors, the inhomogeneous phase is enlarged and 
stretches towards higher temperatures, approaching the upper limit given by 
the pure-glue transition temperature
$T_0=270$~MeV, which is encoded in the parametrization of the Polyakov-loop potential.
The transition lines between homogeneous and inhomogeneous phases
become more and more vertical, assuming a shape resembling the expected form 
of the phase diagram in the large $N_c$ limit \cite{McLerran:2007qj}. 
The size of the inhomogeneous phase is not dramatically enhanced in the $\mu$ direction,
since at $T=0$ the Polyakov loop decouples from the quark sector,
as seen in \Eq{eq:PNJLapprox}.
Hence, since the latter is $N_c$-independent in mean-field
approximation, the transitions at $T=0$ are unchanged.

\subsection{\it Nonzero isospin chemical potential}
\label{sec:isospin}

So far we have restricted our discussions to systems of (mostly massless) up and down 
quarks at a common chemical potential $\mu$. 
In most surroundings, however,  where the density is high enough 
that inhomogeneous chiral condensates might be observed in nature,
there is a surplus of down quarks over up quarks
and, hence,  a nonvanishing (negative) isospin chemical potential $\mu_I = \mu_u - \mu_d$.
In compact stars, as long as strange quarks are absent, neutrality requires that the density
of down quarks is roughly twice as big a the density of up quarks,\footnote{The density of electrons in
beta equilibrated matter $n_e \propto -\mu_I^3$ is a minor correction.} 
leading to $\mu_I \approx -0.2~\mu$, being of the order of 100~MeV. 
In heavy-ion collisions there is a nonzero isospin density due to the neutron excess in 
heavy nuclei, although the effect is much smaller. 
Thermal fits to the particle ratios at freeze out give $\mu_I = -5$~MeV for Pb-Pb collisions at 
SPS~\cite{BraunMunzinger:1999qy} and the value decreases with increasing collision energies. 

As discussed in the Introduction,
nonzero isospin chemical potentials are known to have large effects in color superconductivity,
where they hamper the BCS pairing of up with down quarks~\cite{Alford:2002kj},
possibly leading to inhomogeneous pairing patterns~\cite{Alford:2000ze,Anglani:2013gfu}.  
Moreover, it has been suggested that 
an unequal chemical potential  for up and down quarks could lead to two separate chiral phase 
transitions~\cite{Klein:2003fy,Toublan:2003tt,Barducci:2003un},
although this effect was found to be suppressed by the chiral anomaly~\cite{Frank:2003ve}.

At vanishing quark (or baryon) number density, but nonvanishing isospin density, 
QCD has no sign problem and can be investigated on the lattice~\cite{Son:2000xc}. 
When $\mu_I$ exceeds the pion mass, homogeneous charged pion condensation sets in,
which has been studied intensively within chiral perturbation theory~\cite{Son:2000xc}
and effective models~\cite{Barducci:2004tt,He:2005nk,Warringa:2005jh,Andersen:2007qv,
Kamikado:2012bt}.
One main focus of these model calculations is the effect of a nonvanishing baryon chemical potential 
on the pion condensate,
which is analogous to the effect of a nonvanishing isospin chemical potential on a color superconductor
with reversed roles of $\mu$ and $\mu_I$:
In the same way as $\mu_I$ exerts stress on the $u$-$d$ pairs in the latter, 
$\mu$ exerts stress on the $u$-$\bar d$ pairs in the former,
eventually rendering the homogeneous pion-condensed phase unstable.
Restricting the calculations to homogeneous condensates then leads to a first-order phase transition
at this point, which could be replaced by an inhomogeneous pion condensate if this possibility
is taken into account~\cite{He:2006tn}.

On the other hand it is clear that the inhomogeneous chiral condensates we have studied so
far will be influenced by a nonzero isospin chemical potential as well.
For all modulations we have discussed, the wave number is a monotonically rising function of
the chemical potential, which so far was considered to be the same for up and down quarks. 
At nonzero chemical potential up and down quarks would thus favor different values of the wave 
number, leading to stress on the pairing, as long as we force the periodicities to be
the same for both flavors.  
One should therefore expect the emergence of new condensate functions where up and down
quarks are modulated with different wave numbers~\cite{dno}.

\subsubsection{\it Ginzburg-Landau analysis}

Except for the early works by Kutschera et al.~\cite{Kutschera:1990xk}, explicit
model studies of inhomogeneous chiral condensates in isospin-asymmetric matter  
are so far restricted to  $1+1$ dimensions~\cite{Ebert:2011rg}.
In a recent series of papers, however, Abuki and collaborators have approached the problem 
within a GL analysis in the region around the critical point~\cite{Iwata:2012bs,Iwata:2012jy,Abuki:2013vwa,Abuki:2013pla}.
Thereby they did not consider a specific model, but concentrated on the modifications of the 
quark-loop contributions, corresponding to the coefficients $\beta_i$ in Sec.~\ref{sec:GL}.
In order to include the possibility of charged pion condensation,
one must now allow for the flavor-nondiagonal isospin components of the pseudoscalar 
condensate $\vpp$, \Eq{eq:phispdef}. It is then convenient to formulate the GL functional
in a QM-model-like language in terms of a four-component condensate field, 
\beq
      \phi 
      \equiv \left(\! \begin{array}{c} \sigma \\ \vec\pi \end{array} \! \right)\,, 
      \quad \text{with} \quad
      \vec\pi = \left(\begin{array}{c} \pi^1 \\ \pi^2 \\ \pi^3 \end{array} \right)
      \equiv \left(\!\begin{array}{c} \vec\pi_\perp \\ \pi^3 \end{array} \!\right)\,.
\eeq      
At $\mu_I =0$, the GL functional is $O(4)$ symmetric (in the chiral limit)
and takes the form analogous to \eq{eq:GLNJLgen},
basically with the mass function $M$ replaced by $\phi$.
If we now turn on a nonzero $\mu_I$, this has two effects: 
First, the GL coefficients of this part become $\mu_I$ dependent, in addition to
their dependence on $T$ and $\mu$ already determined by Nickel~\cite{Nickel:2009ke}.
Second, since the isospin chemical potential breaks the original isospin symmetry 
(being an $O(3)$ subgroup of the $O(4)$), 
there are new terms allowed, which to 4th order are given by 
\beq
       \delta\Omega_I
       = 
       \frac{1}{V} \int d^3x \left\{
       \frac{\xi_2}{2}\vec\pi_\perp^{\,2} + 
       \frac{\xi_{4,a}}{4}\vec\pi_\perp^{\,4} + 
       \frac{\xi_{4,b}}{4}(\phi^2 - \vec\pi_\perp^{\,2})\vec\pi_\perp^{\,2} +
       \frac{\xi_{4,c}}{4}(\nabla \vec\pi_\perp)^{\,2}   \right\}\,,   
\eeq
with new coefficients $\xi_i$.
Performing an expansion in $\mu_I$,  Abuki et al.\@~\cite{Iwata:2012bs,Abuki:2013vwa},
have explicitly worked out both corrections, those to the ``old'' GL coefficients and the new ones,
basically following the lines presented in Sec.~\ref{sec:GLNJL} for the $\beta$-coefficients.
The results could then be expressed entirely in terms of the old coefficients and powers of
$\mu_I$, so that no new parameters appear. 

Since in the chiral limit the vacuum pion mass vanishes, charged pion condensation
occurs already at an infinitesimal value of $\mu_I$. 
In Ref.~\cite{Abuki:2013pla},  Abuki has therefore included an explicitly chiral-symmetry
breaking term $\sim h\sigma$, to be phenomenologically more realistic.

\begin{figure}
\begin{center}
\includegraphics[width=.5\textwidth]{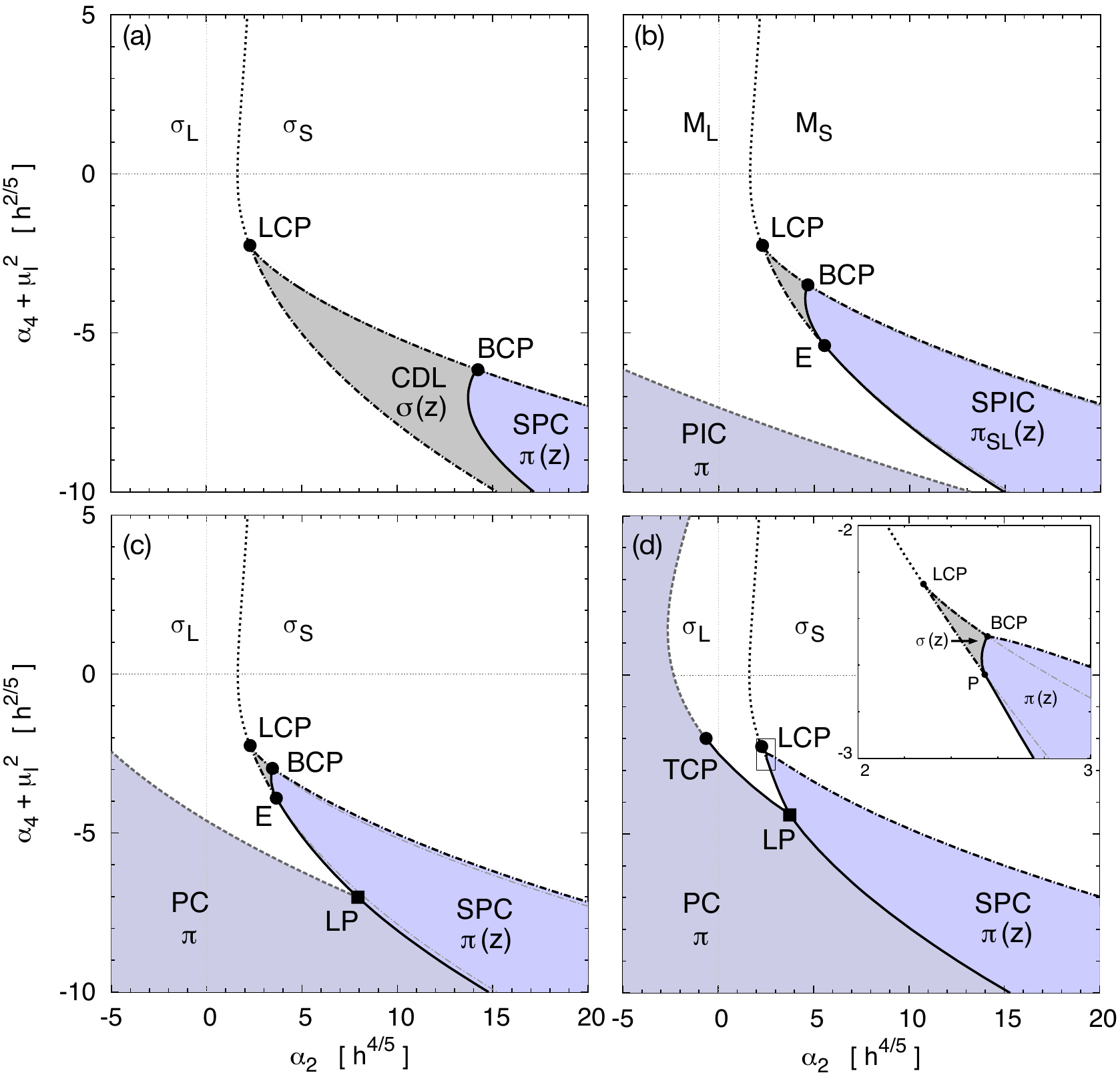}
\end{center}
\caption{GL phase diagrams for 
$\mu_I^2 = 0.01$ (a), $0.1$ (b), $0.2$ (c), and 0.5 (d).
The the energy scale is given by setting the 6th-order coefficient
$\beta_6 = 1$, i.e, $\mu_I^2 \equiv \mu_I^2 \beta_6$. 
In the language of Sec.~\ref{sec:GL}, the axes correspond to
$\alpha_2 \equiv \beta_2\beta_6$ and $\alpha_4 + \mu_I^2 \equiv \beta_4 + \mu_I^2 \beta_6$. 
The figures correspond to $h\beta_6^3 = 1$, and the axes scale with $h$ as indicated. 
The various phases are explained in the text.
Solid, dashed and dotted lines indicate first-order, second-order and crossover transitions,
respectively.
From \cite{Abuki:2013pla}.}
\label{fig:isospin}
\end{figure}

The resulting GL phase diagrams for four different values of $\mu_I$ are displayed in 
Fig.~\ref{fig:isospin}. 
According to a rough estimate, diagram (a) corresponds to 
$\mu_I \sim 40$~MeV~\cite{Abuki:2013pla}, which yields $\mu_I \sim 300$~MeV for diagram (d). 
Abuki obtained these phase diagrams by minimizing the free energies 
considering four possible phases:
(i) a homogeneous phase with $\sigma \neq 0$ and $\vec\pi = 0$, 
(ii) a homogeneous pion-condensed phase (PC) where $\vec\pi_\perp \neq 0$,
(iii) an inhomogeneous phase with a solitonic scalar condensate and $\vec\pi = 0$ (CDL),
and (iv) an inhomogeneous phase with a solitonic charged pion condensate (SPC).
Note that because of the explicit chiral-symmetry breaking, the $\sigma$ field never vanishes,
but is also present as a background in the homogeneous and inhomogeneous pion-condensed phases.
In phase (i) one can distinguish two regions with large ($\sigma_L$) and small ($\sigma_S$) 
chiral-symmetry breaking, which are connected by a crossover.

As seen in the figure, the most interesting effect at moderate values of $\mu_I$ is the emergence of 
an inhomogeneous pion-condensed phase in the ``lower'' part of the inhomogeneous island, i.e.,
away from the Lifshitz point (here: LCP). At high values of $\mu_I$ homogeneous pion condensates
are present as well.

\subsection{\it Three-flavor case}

While all the 
results presented until now have been obtained 
considering only the two lightest quark flavors, we know that 
a realistic description of dense quark matter must take into account 
the presence of strange quarks, whose mass is comparable 
to the chemical potentials where inhomogeneous phases are expected to occur. 

A study of inhomogeneous phases in an NJL model with 
a third 
flavor 
has been performed very recently for the first time by 
Moreira et al.\@ in \cite{Moreira:2013ura}. 
The Lagrangian in this case is given by~\cite{Kunihiro:1989my,Hatsuda:1994pi}
\beq
\mathcal{L} 
=
\overline{\psi}\left(\imath\gamma^\mu\partial_\mu-m\right)\psi\nonumber
+
\frac{G}{2}\sum_{a=0}^8
\left(\left(\overline{\psi}\lambda_a \psi\right)^2+\left(\overline{\psi}\imath\gamma^5\lambda_a \psi\right)^2\right)\nonumber
+ 
\mathcal{L}_{H}\nonumber\,,
\eeq
where 
$m = \mathrm{diag}_f(m_u,m_d,m_s)$ is the current mass matrix,
$\lambda_a$, $a=1,\dots,8$ are the Gell-Mann matrices in flavor space,
$\lambda_0 = \sqrt{\frac{2}{3}} 1\hspace{-1.5mm}1_f$,
and
\beq
\mathcal{L}_{H}=\kappa\left(\mathrm{det}_f\left(\overline{\psi}\frac{1-\gamma^5}{2} \psi\right)+\mathrm{det}_f\left(\overline{\psi}\frac{1+\gamma^5}{2} \psi\right)\right),
\eeq
is a 't Hooft interaction term, which needs to be introduced to break the $U_A(1)$ symmetry. 
Here $\mathrm{det}_f$ denotes a determinant in flavor space.
The coupling strength $\kappa$ can be fitted to the $\eta-\eta'$ mass 
splitting~\cite{Vogl:1991qt,Hatsuda:1994pi}, but is treated as a free model parameter in 
Ref.~\cite{Moreira:2013ura}. 

In order to study inhomogeneous phases in this model,
Moreira et al.\@ make a CDW ansatz in the light (up- and down) quark sector,
\beq
      \ave{\bar\psi_l \psi_l} = \frac{h_l}{2}\cos(\q\cdot\x)\,, \qquad 
      \ave{\bar\psi_l i\gamma^5\tau^3\psi_l} = \frac{h_l}{2}\sin(\q\cdot\x)\,,     
\eeq 
while keeping the strange quark condensate as an homogeneous background,
$\ave{\bar\psi_s \psi_s} = \frac{1}{2}h_s$.
By enforcing this particular ansatz and keeping $m_u = m_d = 0$,
the mean-field Hamiltonian can be diagonalized analytically, with the eigenvalues
in the light-quark sector once again given by \Eq{eq:CDWdispers2}
(with $\Delta$ replaced by the light-quark constituent mass $M_l$)
and the standard dispersion relations for homogeneous matter in the strange quark
sector.

An additional complication is thereby caused by the 't Hooft interaction, which mixes
the various condensate contributions to the dynamical quark masses. 
As in
the homogeneous case, one finds
\beq
      M_i=  m_i - G h_i  - \frac{\kappa}{16}h_j h_k\,,
\eeq
where $(i,j,k)$ is any permutation of $(u,d,s)$. Hence, for $\kappa\neq 0$, the strange-quark
condensate also enters into the up- and down-quark masses and thus into the 
CDW energies. 

\begin{figure}[htc]
\begin{centering}
\includegraphics[width=0.75\textwidth]{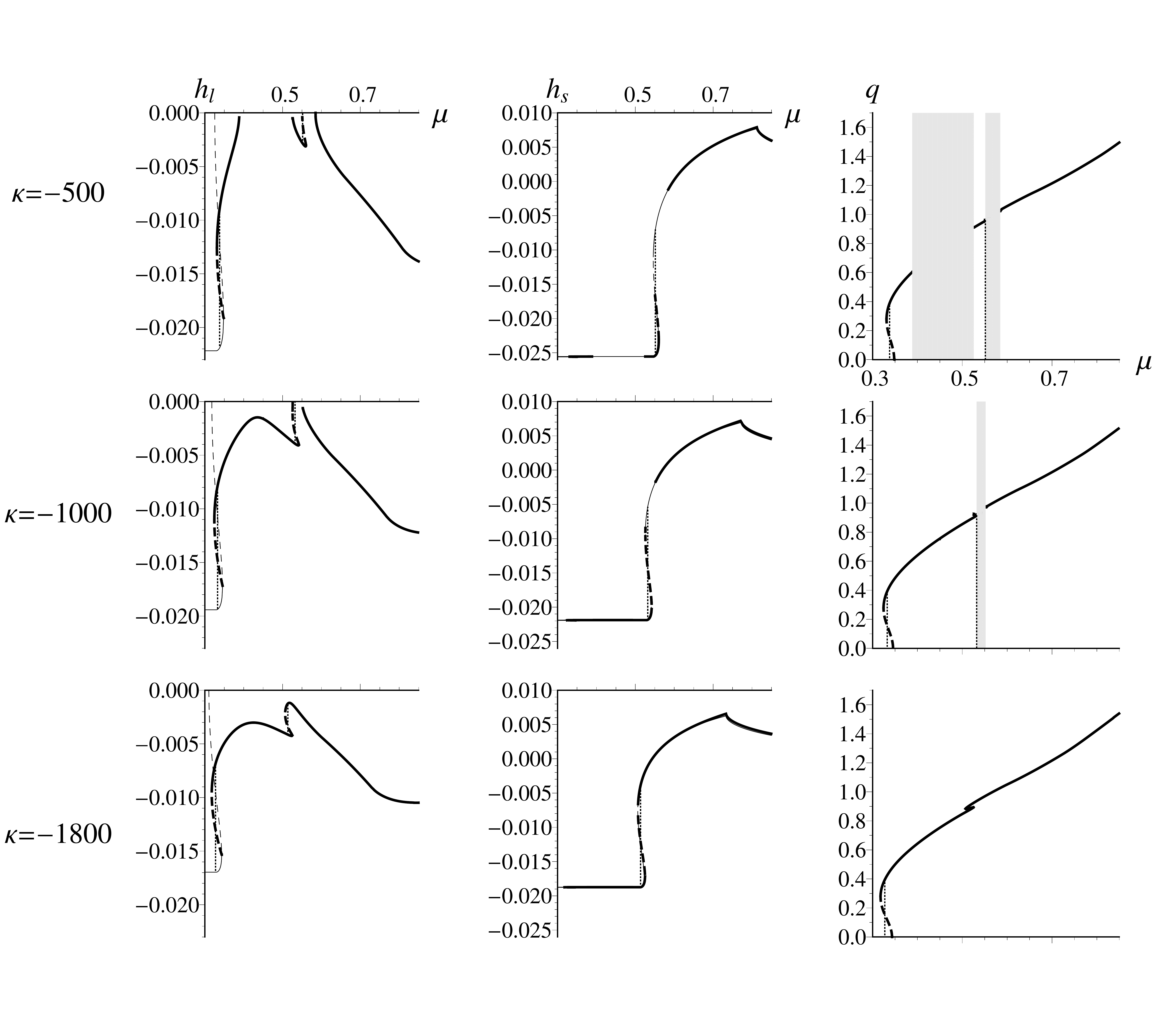}
\caption{Chemical potential dependence of the light and strange quark chiral condensate ($h_l$ 
and $h_s$, respectively) and 
of the wave number
$q$ of the CDW modulation. The three rows correspond 
to different coupling strengths of the
't Hooft determinant ($\left[\kappa\right]=\mathrm{GeV}^5$). Thicker lines
correspond to the finite-$q$ solutions. In the shaded areas 
the light condensate is zero and the value of $q$ is undefined. 
Chemical potential and
$q$ are expressed in GeV, condensates in GeV$^3$. From \cite{Moreira:2013ura}.}
\label{grafpainelNJLHSU3}
\end{centering}
\end{figure}

The results for the light and strange quark condensate as well as for the wave number $q$,
obtained by minimizing the thermodynamic potential, are shown in \Fig{grafpainelNJLHSU3}
for different values of $\kappa$.
The main result is the appearance of a new 
inhomogeneous phase for chemical
potential values below and in the neighborhood of 
the vacuum value of the strange quark mass.
This phase was found both, for physical strange current quark masses, as
well as in the ${\rm SU}(3)$ chiral limit.

For small values of $\kappa$, this
new phase is separated from the usual 
inhomogeneous island, and its onset is characterized
by a smooth growth of the light chiral condensate from zero. The light condensate 
increases with $\mu$ up to a certain value and then drops to zero at a new first-order phase transition, giving rise to a ``shark-fin'' shape (see the upper row of \Fig{grafpainelNJLHSU3}).
 At higher chemical potentials, the inhomogeneous ``continent'' discussed in Sec.~\ref{sec:cont} appears again as the favored ground state for the system.
The interval
between the inhomogeneous island and this new phase 
shrinks with increasing values of $\kappa$, so that above a certain critical value of the 't Hooft coupling the two join, as 
can be seen from the middle row of \Fig{grafpainelNJLHSU3}.
For even bigger couplings, the low-$\mu$ inhomogeneous phases join with the 
``continent'' as well, resulting in a new first-order transition between 
them.

Quantitative considerations aside, it therefore clearly appears that 
the inclusion of strange quarks and flavor mixing acts as a catalyst for the emergence of new spatially modulated solutions, further enlarging the domain
of inhomogeneous phases.

\subsection{\it Magnetic fields}

In most systems where inhomogeneous chiral-symmetry breaking phases might 
be of phenomenological relevance, particularly inside compact stellar
objects and in heavy-ion collision experiments, strong magnetic fields are expected to be present. 
While it is already well known that magnetic fields can influence the 
properties of the 
chiral phase transition
\cite{Suganuma:1990nn,Gusyin:1994,Bali:2011qj}
(for a recent review see e.g.~\cite{MagneticRev:2013}),
their effect on inhomogeneous phases 
might be even more significant. This is related to the effective dimensional
reduction occurring for strong fields at the lowest Landau level (LLL), where momenta transverse to the direction of the magnetic fields 
become restricted
and the problem becomes effectively 1+1-dimensional, leading to 
a significant enhancement of the possibility for inhomogeneous condensation.

A detailed NJL model analysis in the presence of a constant magnetic field $H$ has been performed for the case of a chiral density wave by Frolov, Klimenko and Zhukovsky in \cite{Frolov:2010}. 
In order to include the pairing with the magnetic fields in the NJL Lagrangian, one introduces the covariant derivative
$ D_\mu = \partial_\mu + i Q A_\mu $,  $ A $ being the electromagnetic
field and $ Q = \mathrm{diag} \lbrace \frac{2}{3}\,e,-\frac{1}{3}\,e\rbrace $ the electric charge matrix acting in the flavor space. 
The energy spectrum for the Dirac Hamiltonian in the case of a constant field pointing 
in the same direction as the CDW \footnote{A variational calculation performed in \cite{Frolov:2010} seems to suggest that 
this is the energetically favored configuration, and the authors of \cite{Tatsumi:2014wka} find the same in the limit of weak magnetic fields.} can be calculated analytically and is given by
\begin{equation}
\label{eq:Emag}
    E_{np\zeta\epsilon} =
    \begin{cases}
        \epsilon \sqrt{ \left( \zeta \sqrt{\Delta^{2} + p^{2}} + q/2 \right)^{2} + 2eHn}, & n = 1, 2, \ldots, \\[6pt]
        \epsilon \sqrt{\Delta^{2} + p^{2}} + q/2, & n = 0, \\
    \end{cases}
\end{equation}
where $n$ enumerates the the Landau levels, $ \zeta = \pm 1 $ is the spin quantum number, $ \epsilon = \pm 1
$ is the energy sign (when $ n > 0 $) and $p$
is the quark momentum component parallel to the magnetic
field direction.

The model thermodynamic potential is then obtained as 
\begin{eqnarray} 
    \Omega &=& =
    \frac{\Delta^{2}}{4G_S} +
    N_{c} \, \Omega' \big |_{e \rightarrow \frac{2}{3} e} +
    N_{c} \, \Omega' \big |_{e \rightarrow \frac{1}{3} e},
\nonumber \\
\label{eq:Omega_e}
    \Omega' &=&
    -\frac{1}{2} \frac{eH}{(2 \pi)^{2}}
    \int dp \sum_{n \zeta \epsilon}
    \left[
        | E_{np\zeta\epsilon} - \mu | + {2 T} \log \left( 1 + e^{-| E_{np\zeta\epsilon} - \mu |/T} \right)
    \right].
\end{eqnarray}
Particular care is required when considering the zero temperature contributions to $\Omega$. Indeed,
as noticed by Frolov et al.\@ and successively pointed out by Tatsumi et al.\@ in \cite{Tatsumi:2014wka},
in the presence of an external magnetic field the spectrum of the model Hamiltonian is not symmetric 
because of the $q/2$ offset in the LLL,
as can be seen from the $n=0$ row of \Eq{eq:Emag}.

\begin{figure}[htc]
\includegraphics[width=.33\textwidth]{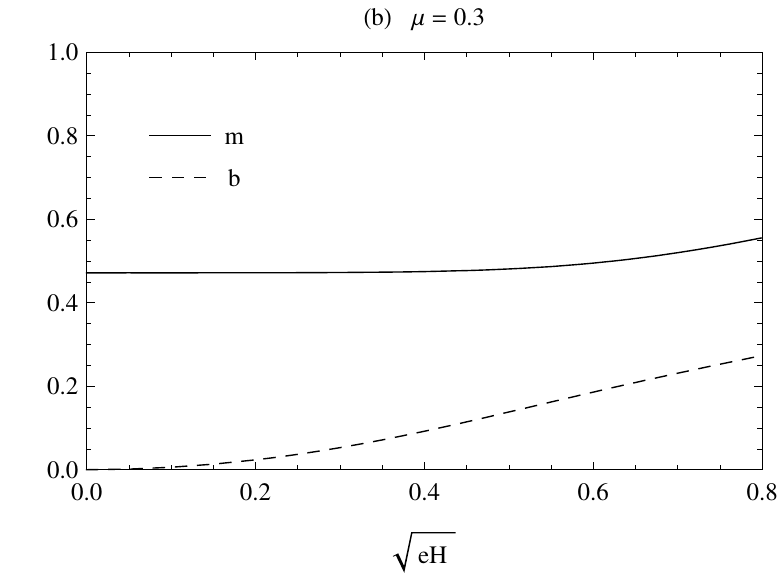}
\includegraphics[width=.33\textwidth]{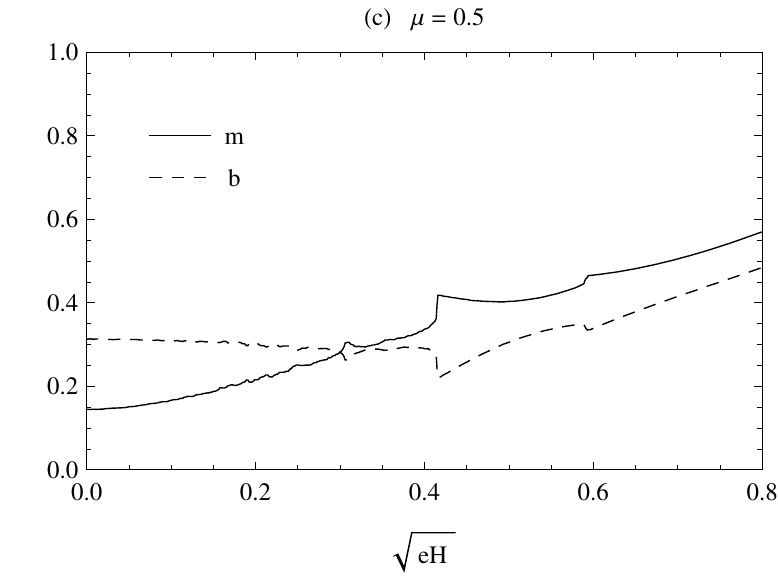}
\includegraphics[width=.33\textwidth]{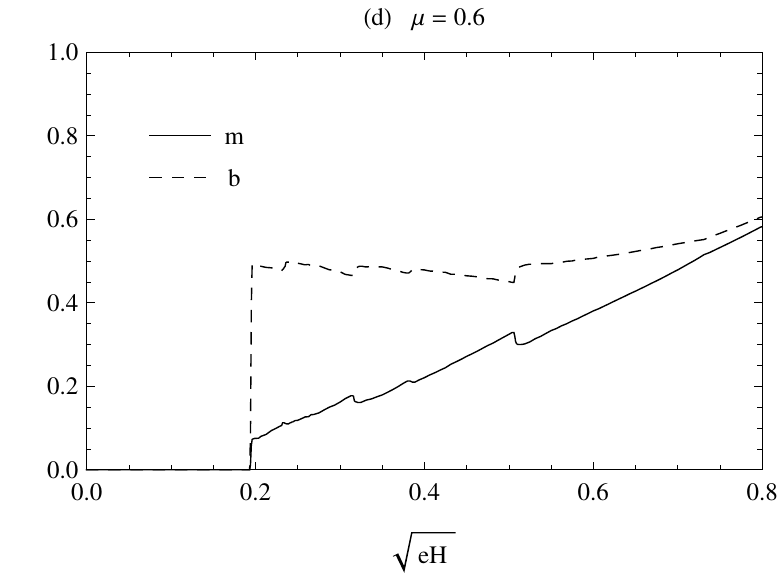}
\caption{The order parameters $m = \Delta/\Lambda$ and $b = q/(2\Lambda)$ as functions of the magnetic field strength
for various values of the chemical potential at zero temperature.
 All quantities are divided by the cutoff and therefore dimensionless. 
 The order parameter oscillation occurring at higher chemical potentials
is a typical  behavior inherent to cold many-body quantum systems in a magnetic field.
From \cite{Frolov:2010}. \label{fig:mb6x}}
\end{figure}

\begin{figure}[htc]
\includegraphics[width=.33\textwidth]{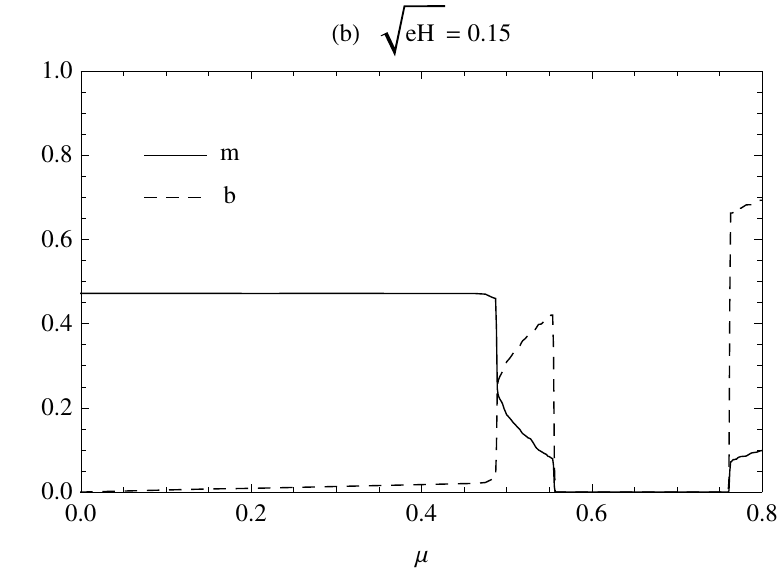}
\includegraphics[width=.33\textwidth]{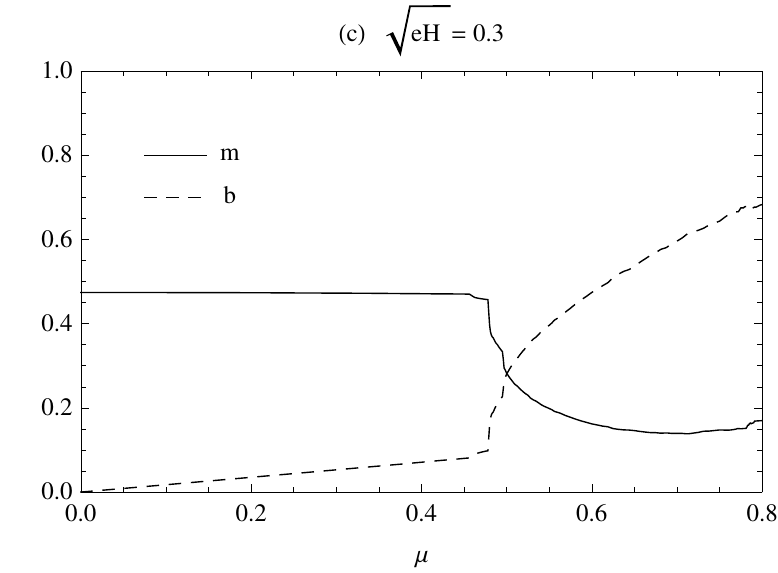}
\includegraphics[width=.33\textwidth]{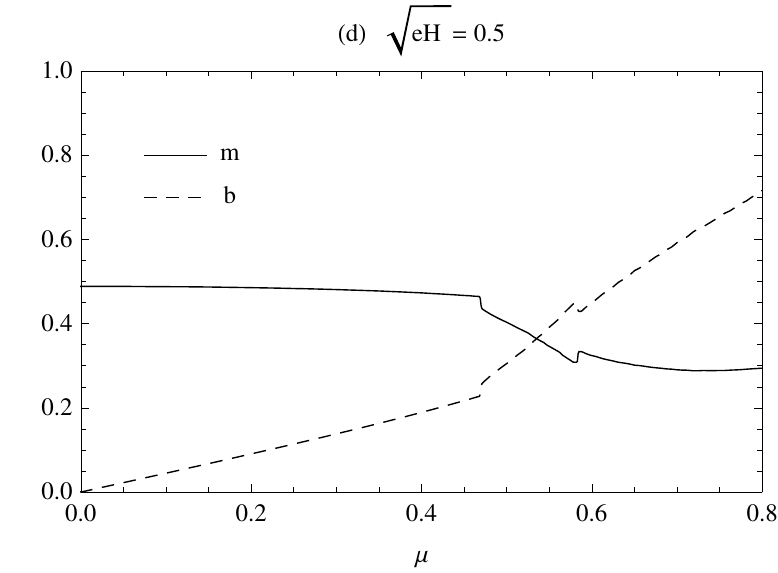}
\caption{
The order parameters as functions of the chemical potential
for various values of the magnetic field strength at zero temperature.
  The wave number grows linearly with $ \mu $ (up to a critical value
where a phase transition occurs), the growth rate being higher in the
stronger field. 
  From \cite{Frolov:2010}. \label{fig:mb6y}}
\end{figure}

Numerical results for the order parameters as a function of $H$ and $\mu$ are reported in \Fig{fig:mb6x} and \Fig{fig:mb6y}.
In the presence of a magnetic field, 
  the system smoothly develops a nonzero $q$ 
  as soon as $\mu > 0$ and 
  the whole region where chiral symmetry is usually broken in a homogeneous way becomes then inhomogeneous.
  For sufficiently high values of $H$, chiral restoration is never reached, even at high $\mu$ (see e.g. the right panels of \Fig{fig:mb6y}). 

\begin{figure}[htc]
\includegraphics{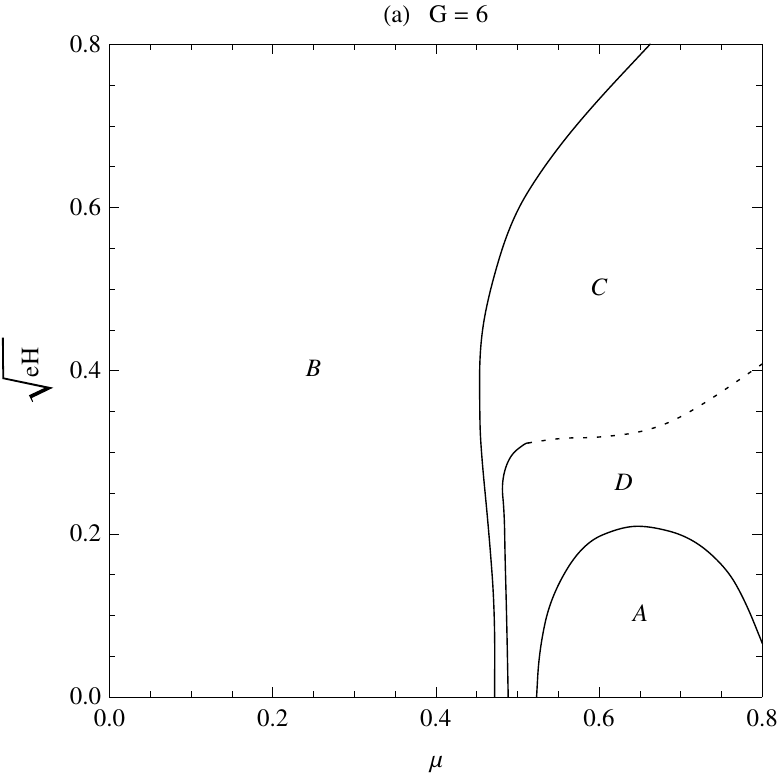}
\qquad 
\includegraphics{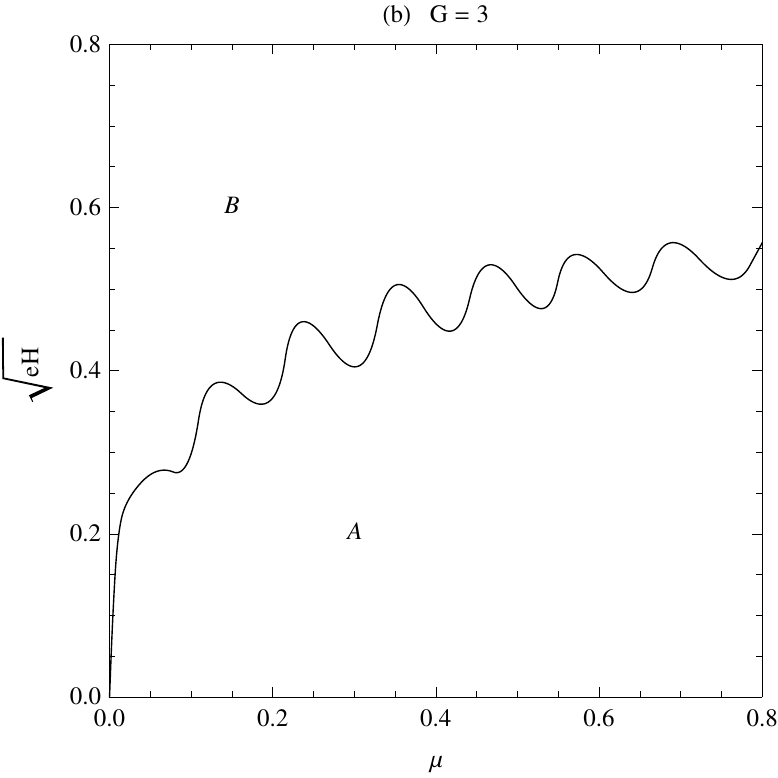}
\caption{Phase diagrams for (a) supercritical and (b) subcritical values of
the coupling constant $ G_S $ at zero temperature. All quantities are
normalized by the cutoff and 
dimensionless.
There is a symmetric massless phase $ A $ with no chiral condensate
and three chirally broken massive phases $ B $, $ C $ and $D$.
Solid lines indicate first-order transitions, while the dashed line is a crossover. From \cite{Frolov:2010}.
\label{fig:phases}}
\end{figure}

In \Fig{fig:phases} the ($\mu-H$) phase diagram for $T=0$ is shown for two different values of the
coupling constant. 
The left panel corresponds to the standard case, where the coupling is strong enough to enable
spontaneous chiral symmetry breaking in vacuum.
Here one can distinguish four different regions: a chirally restored phase $A$ and three phases
with broken chiral symmetry, $ B $, $ C $ and $D$.
At $H=0$ phase $D$ corresponds to the inhomogeneous ``island'', whereas $B$ and $C$ are
homogeneous phases with vanishing and nonvanishing density, respectively.
Turning on the magnetic field all three phases with broken chiral symmetry 
become inhomogeneous, 
so that there is no qualitative difference in terms of an order parameter.
However, there is a first-order phase transition between $B$ and $C$, and, at not too 
large values of $H$, also between $C$ and $D$, 
where the order parameters exhibit a significant jump. 
At higher magnetic fields the discontinuities 
between $C$ and $D$
get smeared out, going over into a region 
where the order parameters oscillate (see, e.g.,  middle and right panel in Fig. \ref{fig:mb6x}).
This is interpreted by Frolov et al.\@ as a crossover transition, see Ref.~\cite{Frolov:2010} for a 
detailed discussion.
The phase transition between $B$ and $C$ is also first order, except for the point at $H=0$, 
which is second order.

In the presence of sufficiently strong magnetic fields, the system
can develop an inhomogeneous chirally broken phase even for subcritical couplings, that is, even 
if $G_S$ is too small to lead to 
spontaneous chiral symmetry breaking at $H=0$. The phase transition
exhibits an oscillating shape in this case as well (see \Fig{fig:phases} right).

The most striking effect of the inclusion of an external magnetic field 
is thus the extension of the domain of inhomogeneous phases to all $\mu > 0$. 
As discussed in \cite{Tatsumi:2014wka}, this can be seen as 
an effect of the spectral asymmetry occurring at the LLL, which
 induces an additional
term proportional to $ \mu q$ in the thermodynamic potential. 
Such kind of term, related to the chiral anomaly, is present in the NJL$_2$ model \cite{Basar:2009fg}, but not in its 3+1-dimensional counterpart. 
The effective dimensional reduction occurring at the LLL can then be seen as responsible for 
bringing this effect into the 3+1-dimensional system.
At nonzero chemical potentials, this additional term, independent of the value of the vacuum 
constituent quark mass, leads to an energy gain as soon as an inhomogeneous condensate ($q \neq 0$)
is formed.

\begin{figure}[htc]
\begin{centering}
\includegraphics[width=.33\textwidth]{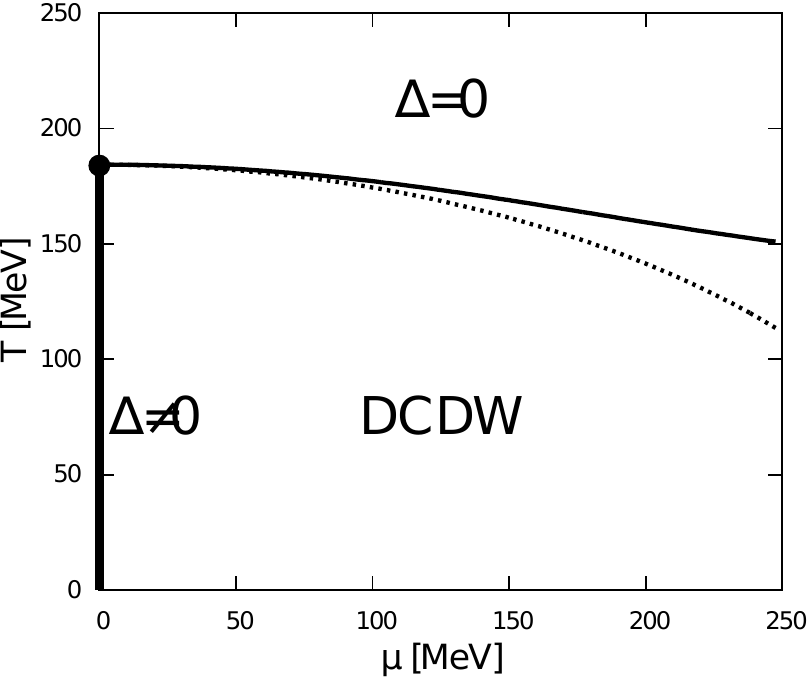}
\caption{Phase diagram obtained within the GL approximation 
for a CDW modulation in the presence of nonzero magnetic fields. 
The solid line denotes the boundary between the CDW and the
symmetric phases, while the dotted line denotes the usual 
chiral restoration transition for homogeneous matter. From \cite{Tatsumi:2014wka}.
\label{fig:pdGLmag}}
\end{centering}
\end{figure}

A recent GL analysis by Tatsumi et al.\@
\cite{Tatsumi:2014wka} has shown that the spectral asymmetry due to the presence of an
external magnetic field  
leads to the appearance of a $\gamma_3$ term in the GL expansion (cf. \Eq{eq:GLNJL2gen})
given by 
\beq
\gamma_3(\mu,T,H)=-\sum_f\frac{N_c|e_fH|}{16\pi^3T}{\rm Im}\psi^{(1)}\left(\frac{1}{2}+i\frac{\mu}{2\pi T}\right),
\eeq
 where $f$ is the flavor index and $\psi^{(1)}$ is the trigamma function. The position of the 
Lifshitz point is now given by the equation $\gamma_2 = \gamma_3 = 0$,
and since $\gamma_3 = 0$ only for $\mu = 0$, this implies that the LP 
now lies on the temperature axis at zero chemical potentials.
The phase boundary obtained in the GL approximation (with the 
further approximation of neglecting the $H$-dependence 
of the $\gamma_2$ and $\gamma_6$ coefficients) in the $T-\mu$ plane 
is shown in \Fig{fig:pdGLmag}. 

\subsection{\it Competition with color superconductivity}

In all the model calculations 
 discussed so far, the possibility of color superconductivity 
has been ignored completely. 
On the other hand it is known that at very high densities the ground state of QCD
is a color superconductor, whereas particle-hole pairing is suppressed in this regime~\cite{Shuster}. 
Moreover, NJL-model calculations for homogeneous matter typically find that the chiral phase
transition at low temperature goes directly from the chirally broken phase into a color
superconducting phase, at least for isospin symmetric 
matter~\cite{Berges:1998rc,Schwarz:1999dj,Buballa:2005}.
It is therefore natural to ask how the chiral phase transition looks like if both possibilities,
particle-hole and particle-particle pairing, are taken into account. 

This question has been investigated by Sadzikowski within an NJL model at zero
\cite{Sadzikowski:2002iy} and nonzero \cite{Sadzikowski:2006jq} temperature.
To that end, the standard NJL Lagrangian,  \eq{eq:LNJL}, was extended by a
quark-quark interaction term
\beq
      {\mL}_D = G_D \left( \bar\psi_c i\gamma^5 \tau^2 \lambda^A\psi\right)
                                \left( \bar\psi i\gamma^5 \tau^2 \lambda^A\psi_c\right)\,,
\eeq
which corresponds to the spin-0 color-antitriplet channel.
This is the most attractive channel for both, one-gluon exchange
and instanton-induced interactions, and thus the channel where diquark condensation is
most likely to occur~\cite{Bailin:1983,Alford:1997,Rapp:1997}.
$\psi_c = C{\bar\psi}^T$ with $C = i\gamma^2\gamma^0$ is the charge conjugated quark field, 
and $\lambda^A$, $A=2,5,7$, denotes the antisymmetric Gell-Mann matrices in color space. 
Like the vector coupling in Sec.~\ref{sec:vector}, the coupling constant $G_D$ is a free parameter.
A typical value is given by $G_D = 3G_S/4$, which is the result of a Fierz transformation, 
starting from an interaction with the quantum numbers of a gluon exchange.

In mean-field approximation we now assume the presence of the diquark condensate
\beq
        \delta = \ave{\bar\psi_c i\gamma^5 \tau^2 \lambda^2\psi}\,, \quad
\eeq
in addition to the scalar and pseudoscalar quark-antiquark condensates $\ps$ and $\pp$,
defined in Eqs.~(\ref{eq:phispdef})  and (\ref{eq:ppa3}).
This choice corresponds to the so-called 2SC phase. Due to $\lambda^2$, only the first two
quark colors (``red'' and ``green'') participate in the condensate, while quarks of the third color
(``blue'') remain unpaired.

After introducing Nambu-Gorkov (NG) bispinors,
\beq
       \Psi(x) = \frac{1}{\sqrt{2}} \left(  \!\! \begin{array}{c} \psi(x) \\ \psi_c(x) \end{array}  \!\! \right)\,,
\eeq
the mean-field Lagrangian can be written as
\beq
       \mL_\text{NJL} + \mL_D + \mu\psi^\dagger\psi 
       \;\longrightarrowover{MF}\;
       \bar\Psi S^{-1} \Psi - \mathcal{V}\,,
\eeq
with the condensate contribution
\beq
       \mathcal{V} = G_S (\ps^2 + \pp^2) + G_D |\delta|^2\
\eeq
and the inverse propagator
\beq
       S^{-1}(x) = 
       \left(  \! \begin{array}{cc}
       i\gamma^\nu \partial_\nu - m + 2G_S(\ps + i\gamma^5\tau^3\pp) + \mu\gamma^0 &
       2G_D i \gamma^5\tau^2\lambda^2\,\delta
       \\
       2G_D  i\gamma^5\tau^2\lambda^2 \delta^*      &
       i\gamma^\nu \partial_\nu - m + 2G_S(\ps + i\gamma^5\tau^3\pp) - \mu\gamma^0       
      \end{array}  \! \right)\,,
\eeq
which is now a $2\times 2$ matrix in NG space.
In addition, the color and flavor structure has become more complicated due to the diquark 
condensates in the off-diagonal NG components.
Conceptually, however, we can proceed as before. 

In Refs.~\cite{Sadzikowski:2002iy,Sadzikowski:2006jq} 
Sadzikowski considered the chiral limit ($m=0$) and made a
CDW ansatz for the quark-antiquark condensates, while assuming that the diquark condensate is
constant in space:
\beq
       2G_S \ps = -M \cos(\q\cdot\x)\,, \quad  2G_S \pp = -M \sin(\q\cdot\x)\,, \quad
       2G_D\delta = \Delta~\,.
\eeq
After removing the explicit space dependence from the inverse propagator in the usual way
by a chiral rotation,  he was then able to solve the eigenvalue problem analytically. 
The resulting thermodynamic potential can be written as~\cite{Sadzikowski:2006jq}\footnote{
We used the identity 
$\frac{\omega}{2} + T\log(1+e^{\omega/T}) = \frac{|\omega|}{2} + T\log(1+e^{|\omega|/T})$
to simplify the expression given in Ref.~\cite{Sadzikowski:2006jq}.
} 
\beq
       \Omega
       =
       -2\sum_{i,j=\pm} \int\frac{d^3p}{(2\pi)^3}\, 
       \left\{
       \varepsilon_{ij}   +\frac{\varepsilon^{(0)}_{ij}}{2} 
       + 2T\log\left(1 + e^{-\varepsilon_{ij}/T}\right)
       + T\log\left(1 + e^{-\varepsilon^{(0)}_{ij}/T}\right)              
       \right\} 
       \,+\, \frac{M^2}{4G_S} + \frac{|\Delta|^2}{4G_D}\,,
\eeq
with the dispersion relations
\beq
        \varepsilon_{-,\pm} = \sqrt{(E_\pm - \mu)^2 + |\Delta|^2}\,, \quad 
        \varepsilon_{+,\pm} = \sqrt{(E_\pm + \mu)^2 + |\Delta|^2}\,, \quad 
\eeq
for the gapped (red and green) quarks and antiquarks, and 
$\varepsilon^{(0)}_{-,\pm} = |E_\pm - \mu|$ and $\varepsilon^{(0)}_{+,\pm} = |E_\pm + \mu|$
for the ungapped (blue) ones. 
Here $E_\pm$ are usual CDW dispersion relations, given in \eq{eq:CDWdispers2}.\footnote{ 
Note that in \eq{eq:CDWdispers2} $\Delta$ denotes the mass gap and should be replaced by $M$ in the
present context.}

\begin{figure}[hbt]
\label{fig:CSC}
\centering
\includegraphics[angle=0,width=.45\textwidth]{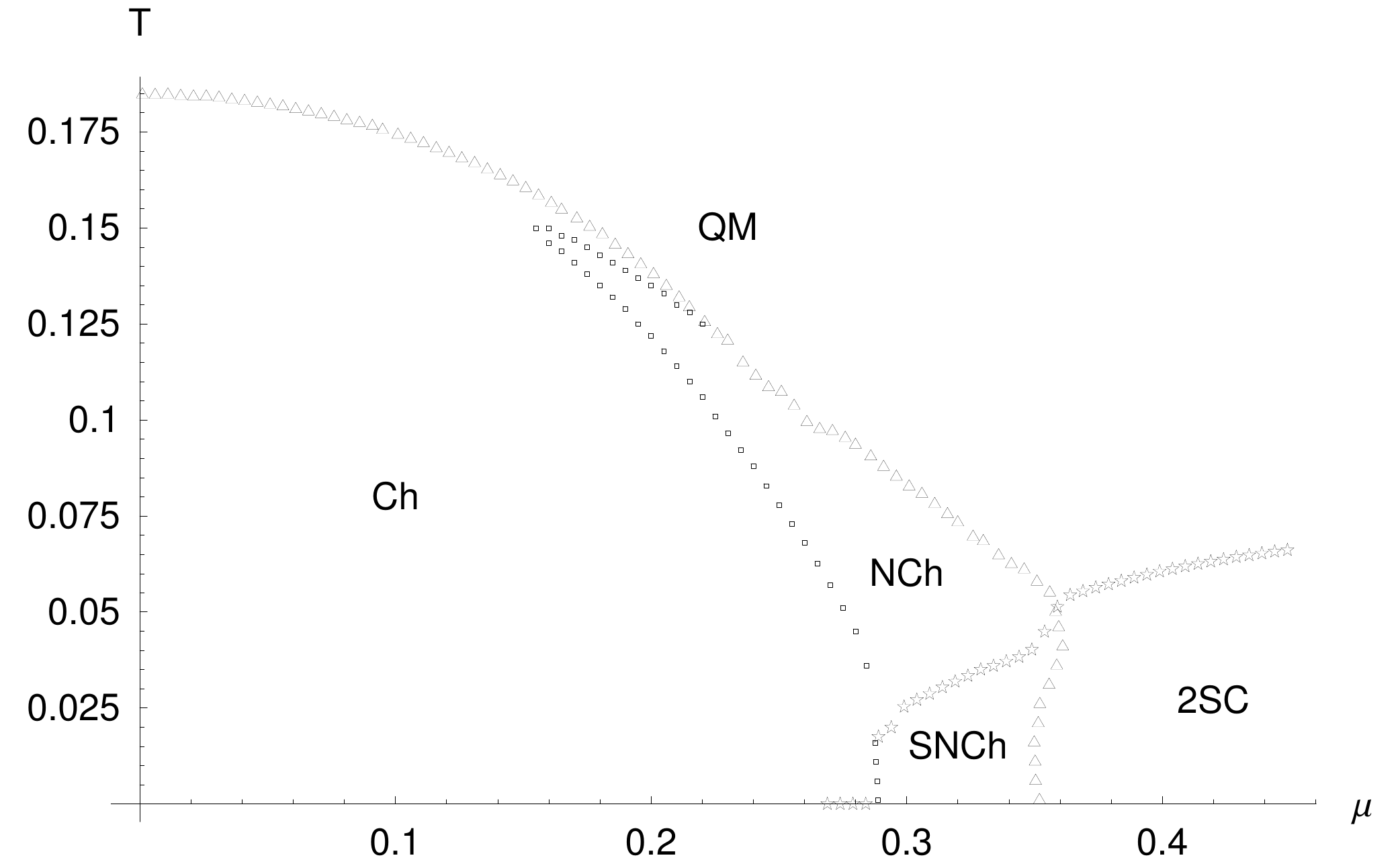}
\qquad
\includegraphics[angle=0,width=.45\textwidth]{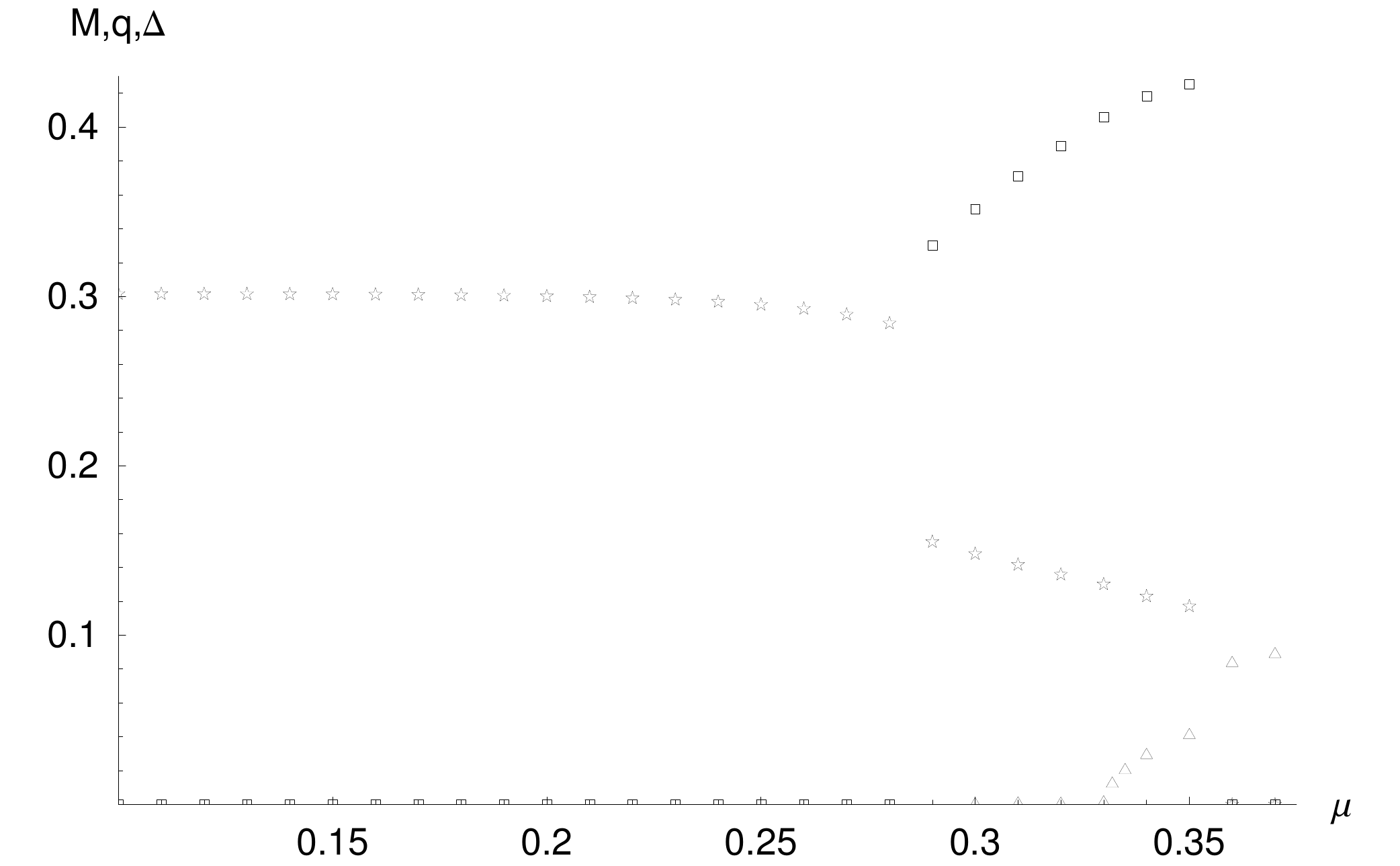}
\caption{
Left: Phase diagram. 
The various phases are defined as follows.
QM: $M = \Delta = q = 0$,
Ch: $M\neq 0$,   $\Delta = q = 0$,
NCh: $M, q\neq 0$, $\Delta = 0$,
2SC: $\Delta\neq 0$, $M = q = 0$,
SNCh: $M, \Delta, q\neq 0$.
Right: Mass amplitude $M$ (stars), wave number $q$ (squares), and diquark gap $\Delta$
(triangles) as functions of the chemical potential $\mu$ at $T= 0.035$~GeV. 
The calculations have been performed with $G_D = 3G_S/4$. All units are in GeV.
From Ref.~\cite{Sadzikowski:2006jq}.
}
\end{figure}

Minimizing $\Omega$ with respect to $M$, $\Delta$, and $q$ for $G_D = 3G_S/4$, 
Sadzikowski obtained the phase diagram shown in the left panel of \Fig{fig:CSC}.
In the right panel the corresponding values of  $M$, $\Delta$, and $q$ are displayed
as functions of $\mu$ at a fixed temperature $T=35$~MeV.
As expected, there is a region at moderate densities and relatively low temperatures where
color superconductivity and the inhomogeneous phase compete with each other. 
Perhaps the most interesting result is that this leads to the presence of a phase (``SNCh'') where 
both pairing modes coexist, i.e., where all three order parameters are nonzero.
At higher chemical potentials one finds a pure 2SC phase, whereas the inhomogeneous phase
without diquark condensation (``NCh'') extends to higher temperatures. 
Extrapolating the boundaries of these phases to the coexistence region, one sees that the 
competition slightly reduces the critical temperature of color superconductivity as well as 
the upper chemical potential of the inhomogeneous regime.

It should be noted that the inhomogeneous phase extends to much higher temperatures
than in most other examples shown in this review. Also, the Lifshitz point, i.e., the point where 
the three non-superconducting phases (``Ch'', ``QM'' and ``NCh'') meet, does not correspond
to the point where the inhomogeneous phase has its highest temperature.
Both features might be related to the particular regularization scheme used in the calculations
(cutoff regularization combined with a Taylor expansion in $q$).
If the highest temperature of the inhomogeneous phase is lower, color superconductivity
obviously affects a larger fraction of the phase. On the other hand, the size of the superconducting 
gap  and, related to this, the critical temperature of the 2SC phase depend strongly on the choice of 
the coupling constant $G_D$.

We should also keep in mind that the calculations have been performed for isospin symmetric
matter. Since the 2SC phase involves the pairing of up with down quarks, it is rather sensitive
to isospin asymmetries, and could therefore be suppressed in compact 
stars~\cite{Alford:2002kj,Steiner:2002gx,Abuki:2004zk,Ruester:2005jc}. 
In this case, inhomogeneous color superconducting phases might become 
favored~\cite{Anglani:2013gfu} and it would be interesting to study this possibility simultaneously 
with inhomogeneous chiral  condensates.

\section{Inhomogeneous chiral condensates in large-$N_c$ QCD}
\label{sec:largenc}

So far, we have discussed various aspects of inhomogeneous chiral-symmetry 
breaking within QCD-inspired models.
In this section we want to make a first step towards real QCD by studying 
inhomogeneous chiral condensates in QCD with a large number of colors,
$N_c \gg 1$.
Strictly speaking, the NJL- and QM-model studies discussed 
before have been performed in the large-$N_c$ limit as well, which simply corresponds
to the mean-field approximation.\footnote{Only in the PNJL model discussed
in Sec.~\ref{sec:PNJLNc}, $N_c$ has the additional effect of controlling the relative strength between the
gluon and the quark sector.}
In QCD, the large-$N_c$ limit has been introduced by 't~Hooft~\cite{tHooft:1973}
and corresponds to the limit $N_c \rightarrow \infty$, while keeping the product
$g^2 N_c$ constant.
Here $g$ is the QCD coupling constant, which thus decreases with increasing $N_c$.
As a consequence, if the number of flavors is kept constant, the screening of the gluon
propagator by quarks can be neglected.

A second assumption which is made in the context of the large-$N_c$ studies we are
going to discuss is that of a large chemical potential,
$\mu \gg \Lambda_\text{QCD}$.
As we will see 
in Sec.~\ref{sec:CS}, together with the infrared singularity of the gluon propagator,
this allows for an effective reduction of the
$3+1$ dimensional problem to a $1+1$ dimensional one. 
It is therefore not surprising that the corresponding solutions bear large similarities to the
$1+1$ dimensional chiral spirals discussed in the Introduction. 

The first analysis of this type has been done in 1992 by Deryagin, Grigoriev, and Rubakov
(DGR)~\cite{DGR},
assuming that the chemical potential is large enough that a perturbative gluon propagator 
can be used. 
In real QCD this weak-coupling regime is the realm of color superconductivity, 
which is, however, suppressed in the large-$N_c$ limit.
The inhomogeneous chiral condensate, on the other hand, survives in this limit
because it is a color singlet.
Since each pair consists of a particle and a hole of the same color,
summing over all colors yields that the condensate is of the order $N_c$,
just like the homogeneous vacuum condensate. 
As a consequence, DGR found that  QCD in the weak-coupling regime 
and $N_c \rightarrow \infty$ has a CDW-like ground state with a wave number
$q=2\mu$.

\begin{figure}[hbt] 
\centering
\includegraphics[angle=0,width=.5\textwidth]{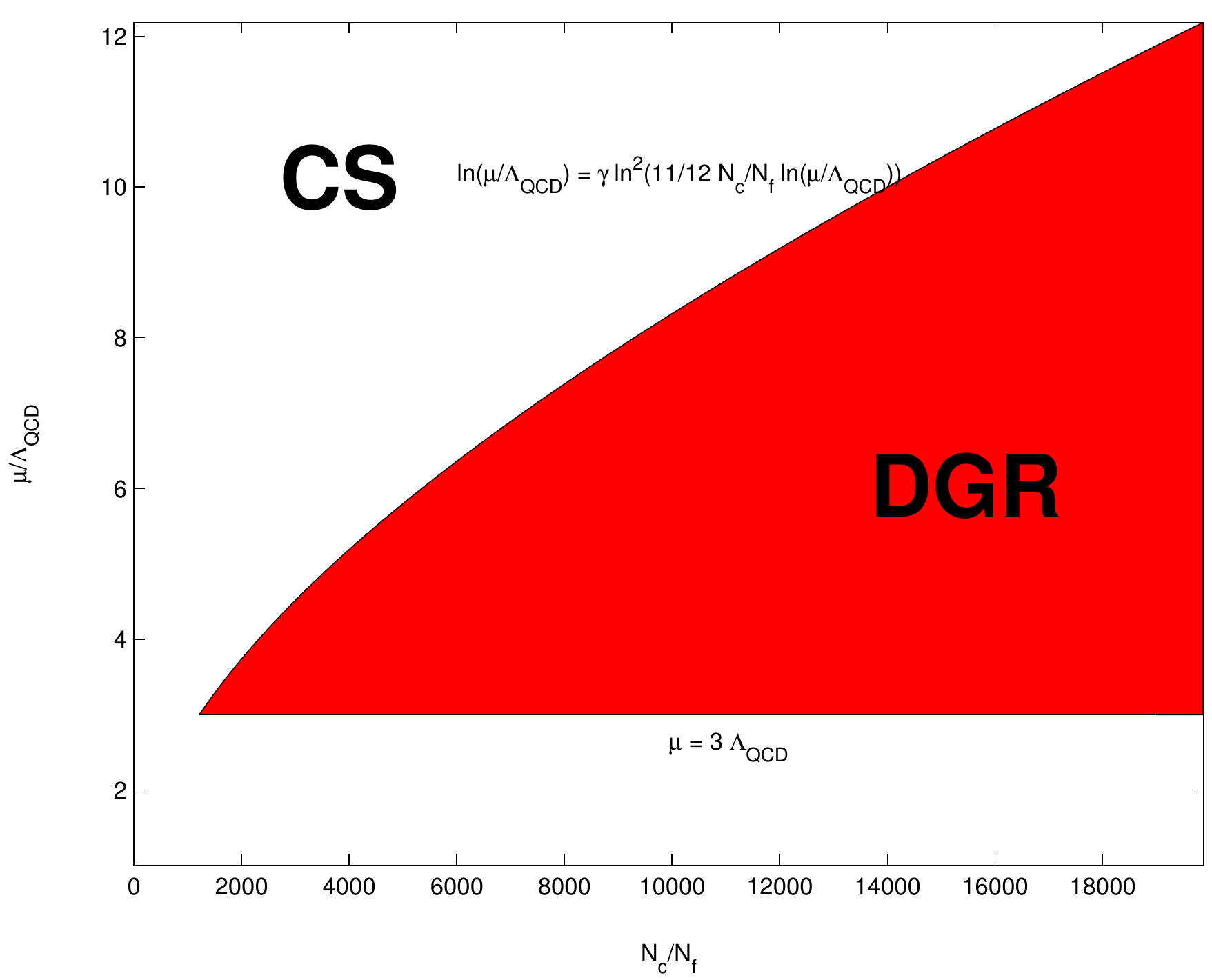}
\caption{Region of the DGR instability in the $N_c-\mu$ plane.
Above the upper limit the ground state is expected to be a color superconductor (CS).
The line $\mu = 3\Lambda_\text{QCD}$ marks the estimated lower limit of the range 
of validity of the approach.
From Ref.~\cite{Shuster}.
\label{fig:shuster_son}
} 
\end{figure}

At any finite value of $N_c$, on the other hand, color superconductivity 
eventually ``wins'' over the density wave when $\mu$ gets sufficiently large.
This has been shown by Shuster and Son within a renormalization-group approach,
where the DGR instability towards the formation of a density wave shows up as 
a Landau pole in the particle-hole scattering channel~\cite{Shuster}.
The result is summarized in \Fig{fig:shuster_son}.
At a fixed value of $N_c/N_f$, the DGR instability only occurs in a finite window of 
chemical potentials, indicated by the red shaded area.  
Above this region, the instability is suppressed by Debye screening and Landau damping
of the gluon propagator, whereas color superconductivity
survives for arbitrarily weak couplings and, hence,
up to arbitrarily high chemical potentials.\footnote{The 
superconducting gap even grows with growing $\mu$, despite the coupling getting 
weaker~\cite{Son:1998uk}.}

Looking at the actual numbers in  \Fig{fig:shuster_son}, we see that the DGR window only 
appears for $N_c/N_f > 1000$, which seems to suggest that the density wave is totally 
irrelevant in the physical regime.
It should be noted, however, that the lower boundary of the window only indicates the 
breakdown of the weak-coupling approach, 
which Shuster and Son estimated to be valid above $\mu \sim  3\Lambda_\text{QCD}$. 
Hence, the analysis does not exclude that for $N_c=3$ a density wave
(or more involved particle-hole condensates) exists in the nonperturbative 
regime~\cite{Park:1999bz,Rapp:2000zd}.
Indeed, using an instanton-induced four-fermion interaction, Rapp, Shuryak, and Zahed
found that at $\mu \simeq 400$~MeV particle-hole and particle-particle pairing are
competitive for $N_c = 3$~\cite{Rapp:2000zd}.

At large $N_c$, on the other hand, $\mu \gg \Lambda_\text{QCD}$ does not necessarily
imply that we are in the perturbative regime since quark effects are suppressed relative
to gluons.
In particular, in the limit $N_c \rightarrow \infty$ the deconfinement temperature should 
become independent of the chemical potential.
Based on this observation, McLerran and Pisarski developed 
the idea of the so-called ``quarkyonic phase''~\cite{McLerran:2007qj}, 
 a phase of cold dense matter where the dominant contribution to the free energy
is due to quarks deep inside the Fermi sphere, whereas the excitations near the Fermi surface are
confined. 
While in the beginning it was speculated that in this phase chiral symmetry might be restored 
(giving rise to the popular but wrong definition of the quarkyonic phase as being 
a chirally restored but confined phase),
it was realized later by Kojo, Hidaka, McLerran and Pisarski (KHMP)
that the same mechanisms which lead to the DGR instability in the
perturbative regime should also give rise to the emergence of CDW-like solutions in the 
quarkyonic phase (``quarkyonic chiral spirals'', QCSs)~\cite{Kojo:2009}.
In the following this will be discussed in more detail.

\subsection{\it Dimensional reduction: chiral spirals in the perturbative and quarkyonic regimes}
\label{sec:CS}

As pointed out above, 
the crucial step, in both, weak-coupling and the quarkyonic-matter calculations,
is the reduction of the $3+1$ dimensional problem to a $1+1$ dimensional one,
which can be done at large chemical potentials
as a consequence of the collinear singularity of the gluon propagator~\cite{Shuster,Kojo:2009}.

In order to demonstrate the idea, consider excitations 
in a small patch around the point $\mu\e_z$ on the Fermi surface, 
\beq
\mk = (\mu + \delta k_z) \e_z + \mk_\perp\,, \quad
\delta k_z, |\mk_\perp| \ll \mu\,.
\eeq
The Euclidean propagator of bare massless quarks then has the form
\beq
        S_0(k) 
        \propto \frac{1}{ik_4+\mu - |\mk|} 
        = \frac{1}{ik_4 - \delta k_z + \frac{\mk_\perp^{\,2}+ \delta k_z^2}{2\mu} + \dots}
        \,,
\eeq
where we neglected the antiparticle contributions and did not write the Dirac structure explicitly.
We see that transverse momenta are suppressed by the large chemical potential and can thus 
be neglected in leading order.
Assuming that this remains true for the dressed propagator $S(k)$ 
(which can be justified a posteriori), we can use this to evaluate the quark selfenergy~\cite{Kojo:2009}
\beq
\label{eq:QCSSigma}
        \Sigma(p) 
       =
        - \int \frac{d^4k}{(2\pi)^4}\, \gamma_\mu^a\,  S(k) \gamma_\nu^b \, D_{\mu\nu}^{ab}(p-k)\,, 
\eeq
which enters the quark Dyson-Schwinger equation $S^{-1} = S_0^{-1} +\Sigma$ in rainbow 
approximation.\footnote{
Using standard conventions for Euclidean $\gamma$-matrices
leads to a plus sign in front of the selfenergy, see also \Eq{eq:qDSE},
in contrast to the minus sign, which appears in Minkowski space,  \Eq{eq:SigmaGL}.
}
Here $\gamma_\mu^a\equiv\gamma_\mu t_a$ corresponds to the (bare) quark-gluon vertex, with $t_a$ being the generators of
$SU(N_c)$, and $D_{\mu\nu}^{ab}$ denotes the gluon propagator.
The important point is that the latter is strongly peaked at small momenta,
which is true for both, the perturbative gluon propagator and the confining one 
which will be introduced below.
This means that the integral is dominated by $\mk \approx \p$, and we can thus neglect 
the $\mk_\perp$ dependence of the quark propagator if $\p$ is close to the Fermi sphere. 
Hence,
\beq
        \Sigma(p) 
        \approx
         - \int \frac{dk_4 dk_z}{(2\pi)^2}\, \gamma_\mu^a S(k_4,k_z, \vec 0_\perp) \gamma_\nu^b 
         \int\frac{d^2k_\perp}{(2\pi)^2}D_{\mu\nu}^{ab}(p-k)\,,        
\eeq
i.e., the $k_\perp$-integral is only over the gluon propagator, but not the quark propagator. 
Therefore, if we define $\tilde S(k_4,k_z) \equiv S(k_4,k_z,\vec 0_\perp)$,
 $\tilde\Sigma(k_4,k_z) \equiv \Sigma(k_4,k_z,\vec 0_\perp)$, and
\beq
\label{eq:Dreduced}
        \tilde D_{\mu\nu}^{ab}(k_4,k_z) \equiv  
        \int\frac{d^2k_\perp}{(2\pi)^2}D_{\mu\nu}^{ab}(k_4,k_z,\mk_\perp)\,, 
\eeq
we get a dimensionally reduced Dyson-Schwinger equation, $\tilde S^{-1} = \tilde S_0^{-1} + \tilde\Sigma$, 
with 
\beq
\label{eq:Sigmareduced}
        \tilde \Sigma(p_4,p_z) 
        \approx
         - \int \frac{dk_4 dk_z}{(2\pi)^2}\, \gamma_\mu^a\tilde S(k_4,k_z) \gamma_\nu^b \,
         \tilde D_{\mu\nu}^{ab}(p_4-k_4,p_z-k_z)\,.        
\eeq

Until this point, there is essentially no difference whether the analysis is done in the weak-coupling 
regime or for quarkyonic matter. 
In the former case, the gluon propagator is just the perturbative one, while in the quarkyonic 
phase it should be confining. To that end KHMP employ the model propagator~\cite{Kojo:2009}
\beq
\label{eq:DGZ}
       D_{44}^{ab	}(k) = -\frac{16\pi N_c}{N_c^2-1} \frac{\sigma}{(\mk\,^2)^2} \delta^{ab}\,, \quad
       D_{4i}^{ab	} =  D_{ij}^{ab}=0 \; \text{for} \; i,j \in \{1,2,3\}\,, 
\eeq
which is valid in Coulomb gauge and corresponds to a linearly rising potential in coordinate space 
with string tension $\sigma$.
This form was originally proposed by Gribov and Zwanziger~\cite{Gribov:1977wm,Zwanziger:2002sh}.
Performing the integration over transverse momenta, \eq{eq:Dreduced}, and inserting the result
into \eq{eq:Sigmareduced}, one finally gets
\beq
        \tilde \Sigma(p_4,p_z) 
        \approx
         2\sigma \int \frac{dk_4 dk_z}{(2\pi)^2}\, \gamma_4\tilde S(k_4,k_z) \gamma_4 \,
         \frac{1}{(p_z-k_z)^2}\,,       
\eeq
which corresponds to QCD in $1+1$ dimensions in axial gauge.

The corresponding equation in the weak-coupling case contains a logarithmic coupling
$\propto g^2 \log(\mu^2/(p_z-k_z)$~\cite{Shuster,Kojo:2009} 
as a result of the integration over the perturbative gluon propagator in \Eq{eq:Dreduced}.
Although milder than the divergence of the confining propagator, 
the logarithmic divergence of this coupling in the infrared is sufficient to guarantee 
the emergence of a nonzero gap, 
quite similar to the BCS instability in (color) superconductivity.\footnote{ 
In Ref.~\cite{Shuster} this instability was derived by studying the Bethe-Salpeter equation (BSE)
for particle-hole scattering, which can be dimensionally reduced in the same way
as we discussed for the Dyson-Schwinger equation.}

In order to see the physical nature of the instability, it is useful to
discuss the dimensional reduction in terms of an effective 
Lagrangian~\cite{Shuster,Kojo:2009}.
Starting from a free Lagrangian in $3+1$ dimensions, 
we neglect again the transverse momenta, 
for describing
quarks near the point $\mu\e_z$ at the Fermi surface, 
\beq
\label{eq:Lkineff}
       \mathcal{L}_\mathit{eff}^\mathit{kin}
       =
       \bar\psi \left(i\gamma^0 \partial_0 + i\gamma^3\partial_3 + \mu\gamma^0\right) \psi\,.
\eeq
In the $3+1$ dimensional theory, the spinors have of course four Dirac
components, whereas in $1+1$ dimensions, since there is no spin,
we need only two, cf.~\eq{eq:NJL2}. 
As a consequence, the original spin degree of freedom decouples from the dynamics in the
dimensionally reduced theory and effectively becomes a new flavor degree of freedom.
This can be made explicit by performing a unitary transformation, such that the components
which correspond to the same spin orientation (relative to the $z$-direction) are grouped
together.
Starting from the chiral representation used earlier, this means that we 
transform~\cite{Kojo:2009}
\beq
        \psi 
        =
        \left(\! \begin{array}{c} \psi_R \\ \psi_L \end{array} \! \right)
        =
        \left(\! \begin{array}{c} 
        \psi_{R\uparrow} \\ \psi_{R\downarrow} \\ \psi_{L\uparrow}  \\ \psi_{L\downarrow} 
        \end{array} \! \right)
        \quad \rightarrow \quad
        \Phi = U\psi 
        =
        \left(\! \begin{array}{c} \varphi_\uparrow \\ \varphi_\downarrow \end{array} \! \right)
        =
        \left(\! \begin{array}{c} 
        \psi_{R\uparrow} \\ \psi_{L\uparrow} \\ \psi_{L\downarrow}  \\ \psi_{R\downarrow} 
        \end{array} \! \right)\,,
\eeq
where $\psi_{R\atop L} = \frac{1}{2}(1\pm\gamma^5)\psi$ are the right- and left-handed
fields, and $\varphi_\uparrow =  \frac{1}{2}(1 + \gamma^5\gamma^0\gamma^3)\psi$ 
and $\varphi_\downarrow =  \frac{1}{2}(1 - \gamma^5\gamma^0\gamma^3)\psi$ 
correspond to the spin projections parallel and antiparallel to the $z$ direction,
respectively.\footnote{
In this section we adopt the conventions of Ref.~\cite{Kojo:2009},
which differ from the conventions used so far, see e.g.~\Eq{eq:Hchrep},
by a minus sign for $\gamma^i$ and $\gamma^5$.
}
\eq{eq:Lkineff}  can then be written as
\beq
\label{eq:Lkinefft}
       \mathcal{L}_\mathit{eff}^\mathit{kin}
       =
       \bar\Phi \left(i\Gamma^0 \partial_0 + i\Gamma^3\partial_3 + \mu\Gamma^0\right) \Phi\,,       
\eeq
with $\bar\Phi = \Phi^\dagger \Gamma^0$ and the transformed gamma matrices
$\Gamma^\mu = U\gamma^\mu U^\dagger$. 
One finds
\beq
       \Gamma^0 = 
        \left(\! \begin{array}{cc} \sigma^1  & 0 \\ 0 & \sigma^1 \end{array} \! \right)\,, \quad       
       \Gamma^3 = 
        \left(\! \begin{array}{cc} -i\sigma^2  & 0 \\ 0 & -i\sigma^2 \end{array} \! \right)\,,      
\eeq        
which is just a ``flavor'' doubling of the gamma matrices 
$\tilde\gamma^0$ and $\tilde\gamma^1$, 
we have defined in the Introduction in the context of $1+1$ dimensional models. 
Hence, \eq{eq:Lkinefft} indeed describes the free propagation of quarks with two 
flavor degrees of freedom in $1+1$ dimensions at chemical potential $\mu$. 
In the same way, we can map a gauge theory in $3+1$ dimensions onto a $1+1$ dimensional 
one if we add gauge fields to the system.

The final step is then to eliminate the chemical potential from the effective Lagrangian.
In complete analogy to the procedure described in the Introduction, cf.~\eq{eq:chirot1p1},
this can be achieved via a local chiral transformation 
\beq
\label{eq:chirot1p1eff}
       \Phi(x) = \exp(-i\mu z \Gamma^5) \Phi'(x)\,,
\eeq
where
\beq   
\label{eq:Gamma5def}     
       \Gamma^5 \equiv \Gamma^0 \Gamma^3 
        =
        \left(\! \begin{array}{cc} \sigma^3  & 0 \\ 0 & \sigma^3 \end{array} \! \right)     
\eeq
is the two-flavor extension of the matrix $ \tilde\gamma^5$.
The existence of a nonvanishing chiral condensate $\ave{\bar\psi \psi} = \ave{\bar\Phi\Phi} \neq 0$ 
in vacuum then implies a 
chiral spiral
of the form
\beq
\label{eq:QSCPhimu}
       \ave{\bar\Phi\Phi}_{\mu} =  \cos(2\mu z) \ave{\bar\Phi\Phi}_{\mu=0}\,, 
       \quad 
       \ave{\bar\Phi i\Gamma^5\Phi}_{\mu} = \sin(2\mu z)\ave{\bar\Phi\Phi}_{\mu=0} 
\eeq
 at chemical potential $\mu$.
Note, however,  that $\Gamma^5$ as defined by \eq{eq:Gamma5def} does not correspond
to $\gamma^5$ of the $3+1$ dimensional theory, but to the combination $\gamma^0\gamma^3$.
Hence, in terms of the original fields, the 
chiral spiral
is given by the oscillation pattern
\beq
       \ave{\bar\psi\psi}_{\mu} =  \cos(2\mu z) \ave{\bar\psi\psi}_{\mu=0}\,, 
       \quad 
       \ave{\bar\psi i\gamma^0\gamma^3\psi}_{\mu} = \sin(2\mu z)\ave{\bar\psi\psi}_{\mu=0} 
\eeq
and is thus different from the CDW discussed in Sec.~\ref{sec:CDW}.

\subsection{\it Chiral spirals in strong magnetic fields}

In Ref.~\cite{Ferrer:2012zq}, Ferrer, de la Incera, and Sanchez have extended the analysis
of KHMP to include the effects of a strong external magnetic field $\vec H$, which was assumed 
to be homogeneous and to point to the $z$-direction. 
As a consequence the quark transverse momenta are quantized, $|\mk_\perp| = \sqrt{2|e_f H| n}$,
where $e_f$ is the electric charge related to the quark flavor $f$ and $n$ enumerates the Landau levels,
so that the radial part of the $k_\perp$-integral in \eq{eq:QCSSigma} is replaced by a sum over $n$. 
For $n>0$ this does not lead to a major difference because the dependence of the quark propagator
on $k_\perp$ is again suppressed by $1/\mu$, as before.

A notable exception is however the lowest Landau level, $n=0$. 
This is 
because there is no spin degeneracy in this level, but all quark spins are oriented parallel to $e_f\vec H$.
At $\mu = 0$ there is therefore a nonvanishing expectation value  
\beq
       \ave{\bar\Phi\tau^3\Phi}_{\mu=0} 
       \equiv
      \ave{\bar\varphi_\uparrow \varphi_\uparrow} - \ave{\bar\varphi_\downarrow \varphi_\downarrow}\,,
\eeq
where $\tau^3$ is the third Pauli matrix acting in the effective ``flavor'' space, 
which corresponds to the spin projection in the $3+1$ dimensional theory.
Thus, in analogy to \eq{eq:QSCPhimu}, we have at nonzero chemical potential
\beq
       \ave{\bar\Phi\tau^3\Phi}_{\mu} =  \cos(2\mu z) \ave{\bar\Phi\tau^3\Phi}_{\mu=0}\,, 
       \qquad 
       \ave{\bar\Phi i\Gamma^5\tau^3\Phi}_{\mu} = \sin(2\mu z)\ave{\bar\Phi\tau^3\Phi}_{\mu=0}\,, 
\eeq
defining a second chiral spiral in addition to \eq{eq:QSCPhimu}.
Translated back to the $3+1$ dimensional spinors this yields~\cite{Ferrer:2012zq}
\beq
       \ave{\bar\psi\, i\gamma^1\gamma^2\psi}_{\mu} =  
       \cos(2\mu z) \ave{\bar\psi\, i\gamma^1\gamma^2\psi}_{\mu=0}\,, 
       \qquad 
       \ave{\bar\psi\, i\gamma^5\psi}_{\mu} = \sin(2\mu z)\ave{\bar\psi\, i\gamma^1\gamma^2\psi}_{\mu=0}\,. 
\eeq

Yet another effect of strong magnetic fields are the ``chiral magnetic spirals'', 
discussed by Basar, Dunne, and Kharzeev in Ref.~\cite{Basar:2010zd}.
While in the previous example the dimensional reduction was taken to be a 
density effect, and the influence of the magnetic field was studied on top of that,
Basar et al.\@ consider the magnetic field as the origin of the dimensional reduction,
as it effectively restricts the motion of the quarks in the lowest Landau level to 
the longitudinal direction. 
The chemical potential then mainly acts as a source for the 
total momentum of the particle-hole pairs.
In addition to $\mu$, the authors also study the influence of a chiral chemical potential $\mu_5$,
which has the opposite effect on left- and right-handed quarks
and is introduced to mimic local chiral-anomaly effects. 

Again using the dimensionally reduced spinors and gamma matrices, they evaluate 
vector and axial currents, $J^\mu = \ave{\bar\psi\gamma^\mu\psi}$ and 
$J^\mu = \ave{\bar\psi\gamma^\mu\gamma^5\psi}$, respectively. 
The key observation is that,  while the longitudinal components $J^0_{(5)}$ and  $J^3_{(5)}$
are given in terms of currents or densities in the $1+1$ dimensional language,  
the transverse components $J^1_{(5)}$ and  $J^2_{(5)}$ involve scalar or pseudoscalar densities.
For instance, one finds
\beq
       J^1 = -\bar\varphi_\uparrow \tilde\gamma^5 \varphi_\downarrow 
                +\bar\varphi_\downarrow \tilde\gamma^5 \varphi_\uparrow\,,
       \qquad  
       J^1_5 =  \bar\varphi_\uparrow \varphi_\downarrow 
                    +\bar\varphi_\downarrow\varphi_\uparrow\,,               
\eeq
where $\tilde\gamma^5\equiv \sigma^3$ is the $1+1$-dimensional version of $\gamma^5$,
as defined in the Introduction.
Hence, performing again local chiral transformations to eliminate $\mu$  (or $\mu_5$) from 
the Lagrangian, 
it follows that the transverse current components also oscillate when going along the $z$
direction.
Basar et al.\@ suggest that this could induce out- and in-plane fluctuating charge asymmetries
in heavy-ion collisions~\cite{Basar:2010zd}.

\subsection{\it Interweaving chiral spirals}
\label{sec:interweaving}

The above analyses were restricted to a single patch on the Fermi sphere, located
around the point $\p = \mu\e_z$.
We have seen that in the vicinity of this point, the $3+1$ dimensional system can be well
approximated by a $1+1$ dimensional one, leading to a QCS solution with a wave vector 
$\q = 2\mu\e_z$.
However, while for a true $1+1$ dimensional system the Fermi ``sphere'' consists of only 
two points, so that a chiral spiral which connects these 
points is maximally efficient, this is obviously not the case in $3+1$ dimensions, where large 
parts of the Fermi sphere are left out. 
It is thus plausible that there could be better configurations where the mass function is a 
superposition of multiple plane waves. 

In the present section we want to approach the problem within the framework introduced
above. The idea is to study not only one single patch but to cover the Fermi sphere with several 
patches. 
This was already suggested by Shuster and Son for the perturbative regime~\cite{Shuster}.
In this context it was studied further by Park et al.~\cite{Park:1999bz},
and applied by Rapp et al.~\cite{Rapp:2000zd} to the nonperturbative region.
For quarkyonic matter it was discussed by Kojo and collaborators in Refs.~\cite{Kojo:2010}
and \cite{Kojo:2011}.
To first approximation, the condensate is then a superposition of several QCSs
with wave vectors $\q_i$, which have all the same length $\sim 2\mu$
but different directions. This excludes solitonic solutions, where higher harmonics are needed,
but includes a one-dimensional cosine as well as
two- and three-dimensional crystalline structures. 

At first glance, one might think that the number of patches should be as high as possible in order 
to have an optimal coverage of the Fermi surface. 
A closer inspection shows however that the difference $|\q_i - \q_j|$
between the wave vectors of two different QCSs must not be too small
in order to avoid destructive interference effects. 
As a consequence there is an optimal number of patches which grows 
with the size of the Fermi surface.

A detailed analysis of this optimization problem has been performed in Ref.~\cite{Kojo:2011} 
within an NJL-type model.
In this model the infrared dominance of the confining gluon propagator, \eq{eq:DGZ},
is mimicked by attaching a form factor to the four-point vertices, cutting off high momentum 
transfers.
A quark of a given momentum $\p$ therefore only interacts with other quarks or condensates
if their momenta $\mk$ lie within a finite region $|\mk -\p| < \Lambda_f$ 
with a cutoff parameter $\Lambda_f \sim \Lambda_\text{QCD}$.
This opens the possibility to have several QCSs present in the the system, 
while most of the quarks or holes near the Fermi surface interact only with one of them.
As a technical simplification, the model was studied in $2+1$ dimensions, so that the noninteracting 
system is characterized by a Fermi circle, instead of a Fermi sphere. 

\begin{figure}[htb]
\centering
\includegraphics[angle=0,width=.35\textwidth]{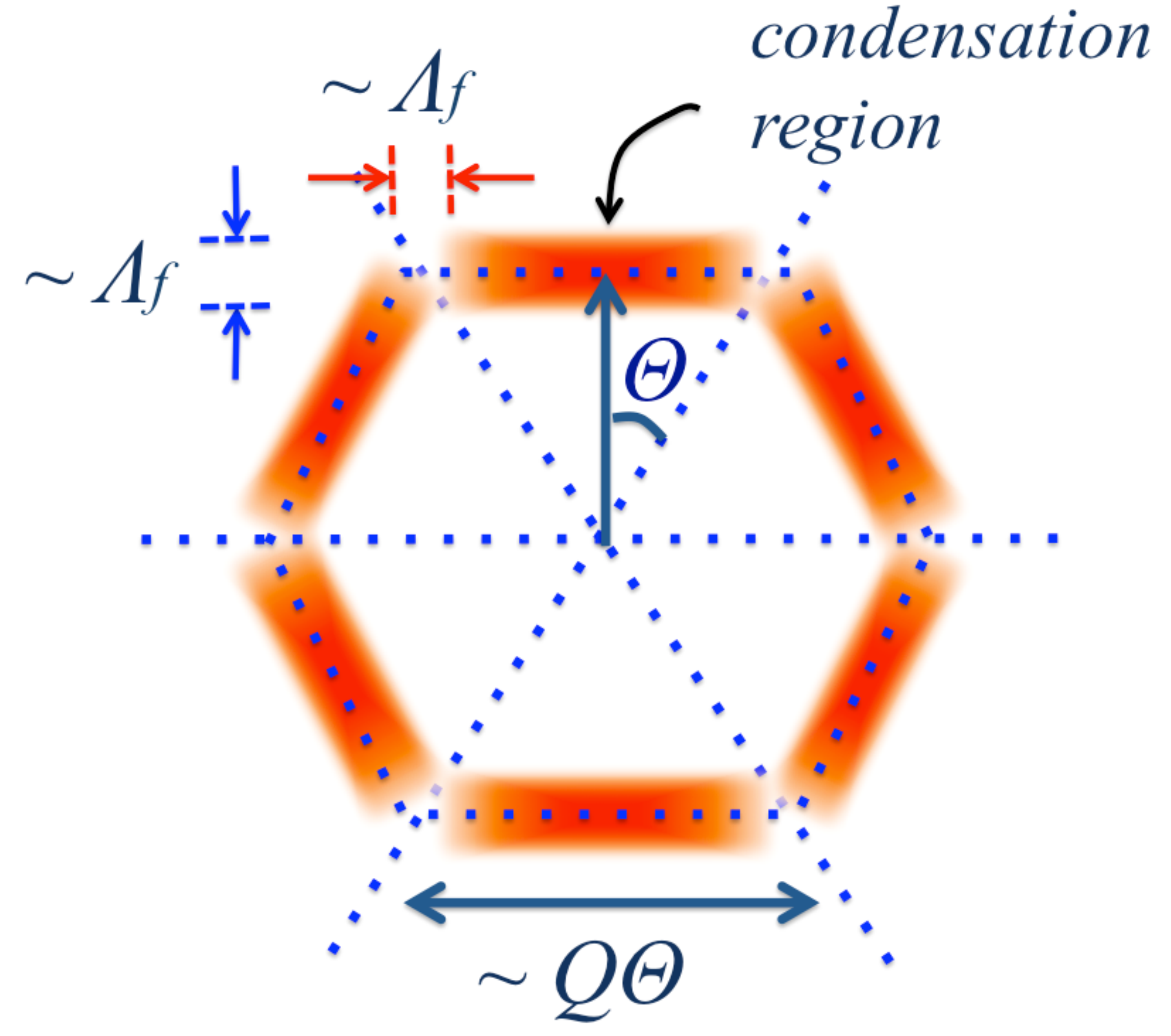}
\hspace{9mm}
\includegraphics[angle=0,width=.5\textwidth]{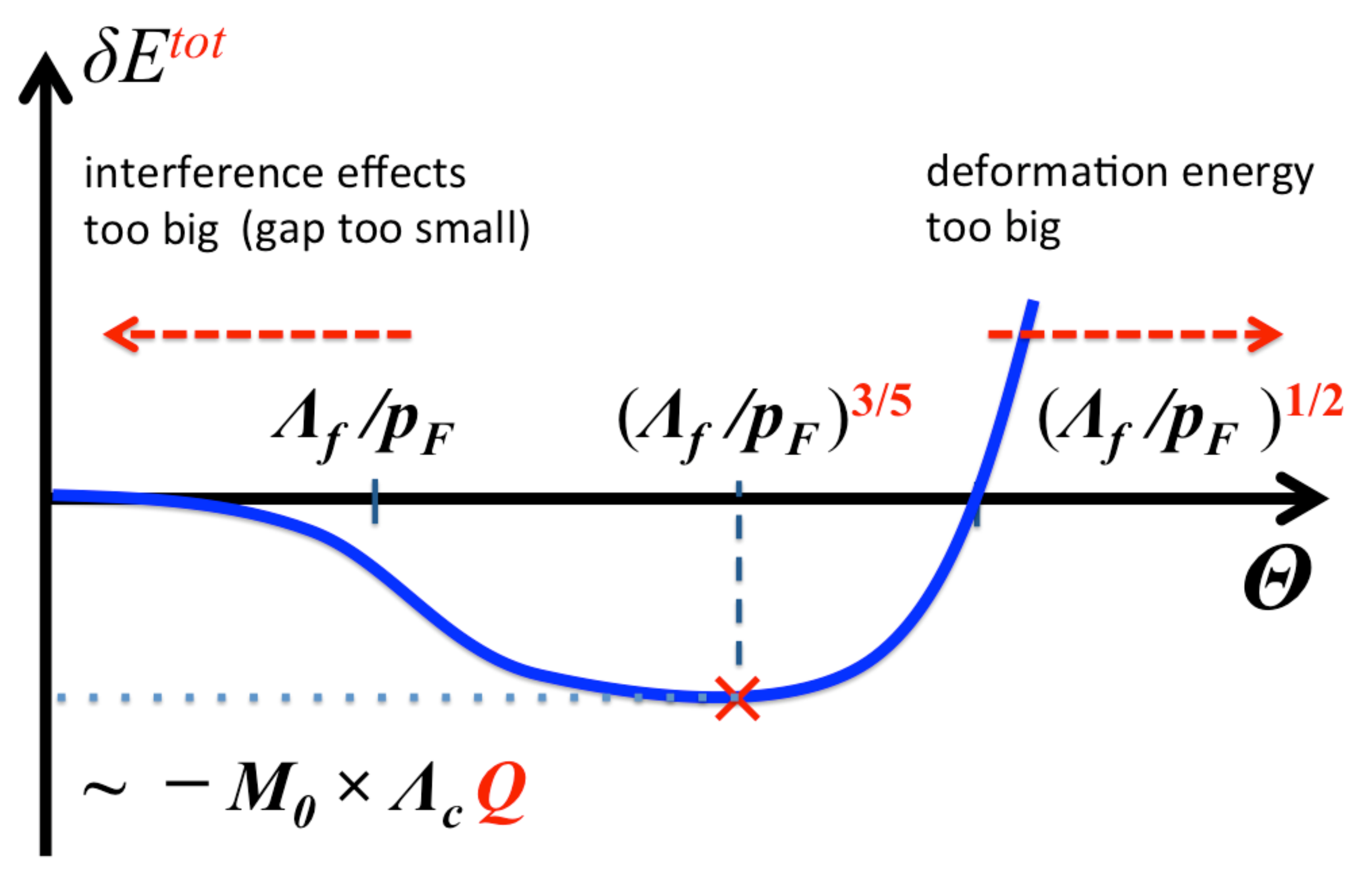}
\caption{Left: Condensation region in the $2+1$ dimensional model of 
Ref.~\cite{Kojo:2011}. Because of the form factor only particles or holes within a region 
of thickness $\sim \Lambda_f$ around the deformed Fermi surface contribute to the
condensate. 
Quarks or holes with a distance $\lesssim\Lambda_f$ to the patch boundary
interact also with the neighboring chiral spiral,
leading to destructive interference. 
Right: Corresponding energy landscape (schematic) as a function of the angle
$\theta \propto 1/N_p$.
Both illustrations have been taken from~\cite{Kojo:2011}.
}
\label{fig:kojo_condreg}
\end{figure}

Now consider the presence of $N_p$ chiral spirals, characterized by the wave vectors $\q_i$, 
$i = 1, \dots, N_p$, with length $|\q_i| = 2Q$.
Since each of these QCSs consists of particle-hole pairs with total momentum $\q_i$, 
an effective pairing corresponds to a deformation of the Fermi circle to a polygon of degree $N_p$,
so that for each point on the new Fermi surface there exists a point on the opposite side, 
which differs just by $\q_i$.
In this way, basically all particles and holes near the Fermi surface can participate in a condensate
and thus contribute to the condensation energy.  
On the other hand, the deformation of the Fermi surface away from a circle costs of course 
kinetic energy.
From this point of view it would be preferable to a have a large number of small patches,
$N_p \gg 1$, so that the polygon is very close to the original Fermi circle. 
A large number of small patches is, however, disfavored by destructive interference effects,
which take place in the boundary regions of neighboring patches.
This is illustrated on the left of \Fig{fig:kojo_condreg}. 
Recalling that the form factor restricts the quark interactions to regions of size
$\Lambda_f$ around its momentum, we see that both, the radial thickness of the condensation region
 around the Fermi surface as well as the transverse size of the boundary regions where the quarks are 
 affected by two QCSs, are of the order $\sim\Lambda_f$.

From this picture the optimal number of patches can be estimated in a semiquantitative way. 
Introducing the angle $\theta \propto 1/N_p$ as shown in
 \Fig{fig:kojo_condreg} and inspecting the $\theta$ dependence
of the various contributions to the energy, one finds that the difference of the energy density
relative to the uncondensed system is of the form~\cite{Kojo:2011}
\beq
\label{eq:deltaeps}
        \delta\varepsilon(M,\theta)
        \sim
        N_c \left(
        \frac{\Lambda_f^2 f(M_B)}{\theta} - c_0 M_0\Lambda_f p_F - c_2 M_0\Lambda_f p_F \theta^2
        + c_4 p_F^3 \theta^4 + \dots \right)\,,
\eeq
which is sketched on the right-hand side of the figure.
Here $p_F$ is the Fermi momentum, and $M$ is the mass gap, which takes the value $M_0$ 
in the condensation regions far away from the patch boundaries and $M_B$ in the boundary regions. 
The $c_i$ are positive dimensionless coefficients.
The $c_0$ and $c_2$ terms are thus negative and correspond to the condensation energy, which 
is proportional to the gap $M_0$, the thickness $\Lambda_f$ of the condensation region and the 
circumference of the polygon. The latter depends only weakly on $\theta$ if we take into
account that the particle number and, hence, the 
phase-space volume must be kept constant when the Fermi circle is deformed to the polygon. 
The first term in parentheses describes the destructive interference effects in the boundary regions
of two patches and is thus positive.  Obviously this term is proportional to the number of patches,
$N_p\propto\theta^{-1}$, times the energy loss related to an individual intersection domain, which in turn is 
proportional to its size $\Lambda_f^2$ times some function of the gap $f(M_B)$.
The last term in \Eq{eq:deltaeps} describes the positive deformation energy, which turns out to be of 
the order $\theta^4$ if one simply evaluates the kinetic energy density for a noninteracting system. 

\begin{figure}[htb]
\centering
\includegraphics[angle=0,width=.4\textwidth]{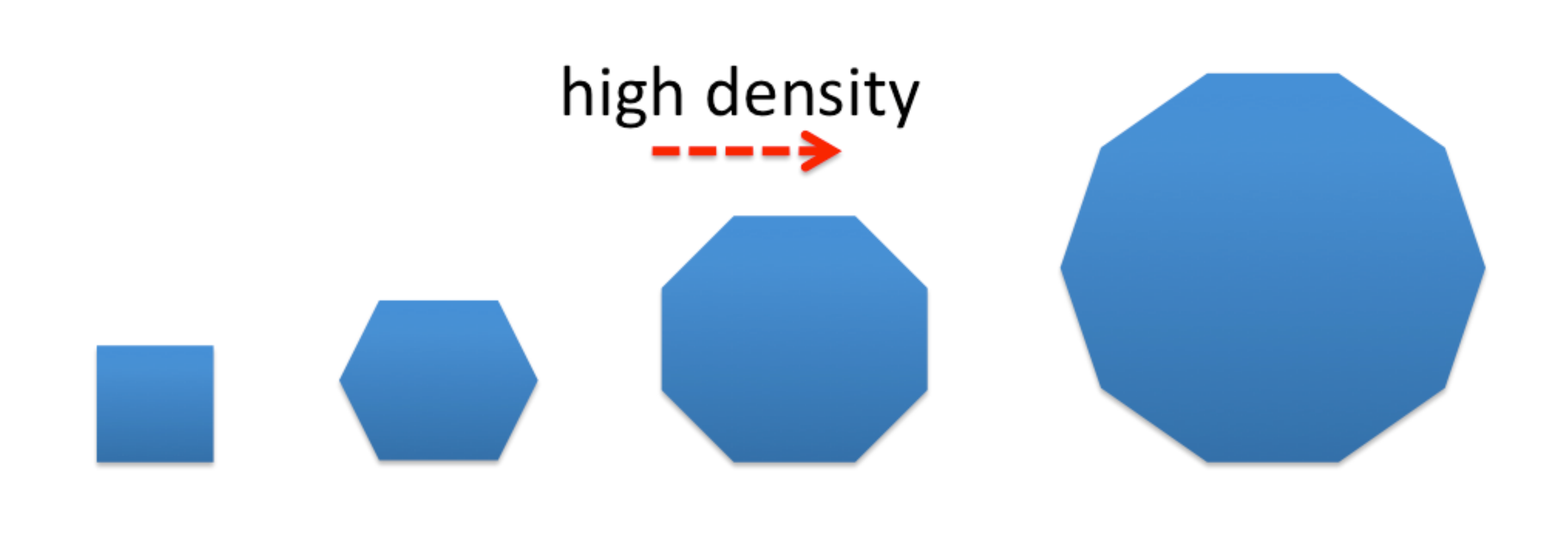}
\hspace{20mm}
\includegraphics[angle=0,width=.35\textwidth]{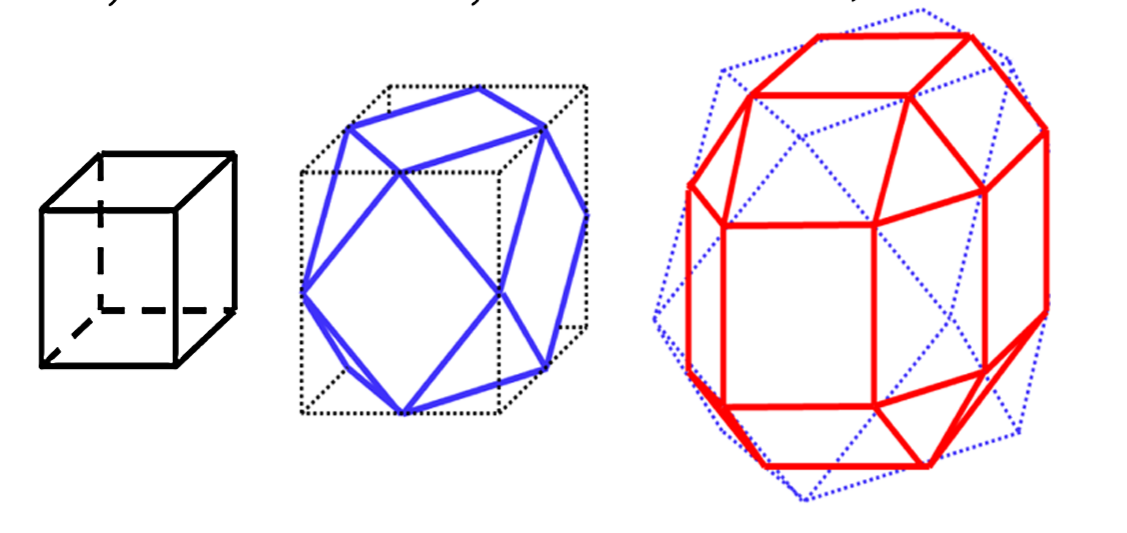}
\caption{Suggested shapes of the Fermi surface, corresponding to multi-patch solutions 
of interweaving quarkyonic chiral spirals in $2+1$ (left) and $3+1$ (right) dimensional systems. 
In both cases the density increases from left to right.
The figures have been taken from Refs.~\cite{Kojo:2011} (left) and
\cite{Kojo:2010} (right).
}
\label{fig:kojo_polygons}
\end{figure}

One thus finds that the deformation energy prevents $\theta$ from getting too big, while the patch-patch
interaction prevents it from getting too small. Neglecting the other terms and assuming that 
$f(M_B)$ is of the order of $\Lambda_f$, one finds that the energy is minimized by an angle
$\theta \propto (\Lambda_f/p_F)^{3/5}$~\cite{Kojo:2011}.
Accordingly, the number of patches grows with the Fermi momentum,
$N_p \propto p_F^{3/5} \simeq \mu^{3/5}$, predicting a series of phase transitions between 
phases which have different discrete symmetries under rotation, see \Fig{fig:kojo_polygons} (left). 
At high densities, this series eventually terminates, when particle-particle pairing (color superconductivity)
becomes more favored.

At the low-density end of the series, a full coverage of the Fermi surface with patches 
might not be favorable at all, so that a one-dimensional modulation could be realized.
With increasing chemical potential, this should then be followed by a square and a 
hexagonal lattice.
Interestingly, this is in qualitative agreement with the NJL-model results discussed in 
Sec.~\ref{sec:cont}.
Notice, however, that all solutions with more than six patches do not correspond to a 
periodic structure in coordinate space, although they still have discrete rotational 
symmetry. 
Therefore they cannot be ordinary crystals, and it was suggested in Ref.~\cite{Kojo:2011} 
that they might belong to the class of ``quasi-crystals''~\cite{Shechtman1984,Levine1984}
which, besides crystals and amorphous structures,
is known from condensed matter physics as a third category of solids.

Studying multi-patch solutions in $3+1$ dimensional systems is obviously technically more involved,
but the qualitative arguments should remain similar as for $2+1$ dimensional systems. 
In particular, we expect that the number of patches increases again with density, leading to a series
of polyhedrons, see right-hand side of \Fig{fig:kojo_polygons}~\cite{Kojo:2010}.

\section{Dyson-Schwinger approach to chiral density waves}
\label{sec:DSE}

In this section we want to discuss inhomogeneous chiral condensates in QCD,
without referring to the large-$N_c$ limit. 
As we have seen in the previous section, the DGR instability is then suppressed in the weak-coupling 
regime at high densities, where the QCD ground state is a color superconductor~\cite{Shuster}.
In Ref.~\cite{Muller:2013tya} M\"uller et al.\@
have therefore studied the possible emergence of an inhomogeneous phase
in the non-perturbative regime, employing Dyson-Schwinger equations (DSEs).

In general, DSEs are an infinite set of integro-differential equations, 
which are as such the exact equations for the $n$-point functions of a given
quantum-field theoretical system.
In practice, one has to make truncations in order to obtain a closed set of 
equations and to solve it. 
The truncated system is then no longer exact, but it is systematically amendable, and one 
may hope to get convergence or at least qualitative insights already at a relatively low level of the 
truncation.
This strategy has been applied to QCD already for a long time, 
initially mostly concentrating on aspects of vacuum properties and confinement 
(for reviews see, e.g., Refs.~\cite{Alkofer:2000wg, Fischer:2006ub}).
More recently DSEs have also been employed to study the QCD phase diagram
at zero~\cite{Fischer:2009gk,Fischer:2010fx} and nonzero~\cite{Fischer:2011mz,Fischer:2012vc} 
chemical potential, including color superconductivity~\cite{Nickel:2006,Nickel:2006kc,Muller:2013pya}.

In coordinate space with Euclidean metric, 
the DSE for the dressed quark propagator $S$
is given by
\beq
\label{eq:qDSE}
       S^{-1}(x,x') = Z_2\left( S_0^{-1}(x,x') + \Sigma(x,x')\right)\,,
\eeq       
depending on two space-time variables, $x$ and $x'$.
$S_0$ denotes the bare propagator, $\Sigma$ the selfenergy, 
and $Z_2$ the wave-function renormalization constant of the quark field.

In homogeneous, i.e., translationally invariant matter, the propagator depends only on the 
relative coordinate $x-x'$. In momentum space, this translates into a dependence on a single
4-momentum, and the inverse propagator at temperature $T$ and chemical potential $\mu$
can be parametrized as
\beq
S_\mathit{hom}^{-1}(p) = -i\omega_n\gamma_4 C(p)-i{\vec\gamma}\cdot {\vec p}A(p) + B(p)\,,
\label{eq:Shom}
\eeq
with $p:=(\vec p, p_4=\omega_n+i\mu)$, the Matsubara frequencies $\omega_n=(2n+1)\pi T$,
and three dressing functions $A$, $B$, and $C$.
In vacuum, due to Lorentz covariance, these functions depend on $p^2$ only and $A(p) = C(p)$. 
The wave-function renormalization constant $Z_2$ is then fixed by the condition that 
$A(p)|_{p^2=\nu^2} =1$ at an arbitrary renormalization point $\nu$.
This prescription remains valid for the analysis of inhomogeneous phases
since $Z_2$ is always fixed in vacuum, which is homogeneous.

In inhomogeneous matter, the quark selfenergy (and hence the propagator)
depend separately on both coordinates $x$ and $x'$, or, equivalently, on the relative coordinate
$x-x'$ and the center of momentum coordinate $(x+x')/2$.
In momentum space they thus depend on two momenta, and the DSE reads
\beq
\label{eq:qdse}
S^{-1}(p,p') = Z_2\left( S_0^{-1}(p,p') + \Sigma(p,p')\right)\,,
\eeq
where $p$ and $p'$ correspond to the out- and ingoing momenta of the quark.
As we have seen already for the NJL model, the quark momenta do not need to be 
conserved since the quarks can scatter off the nonuniform condensates.
However, while in the NJL model the Fourier modes of the mass functions could be
labelled by a single momentum, the selfenergy $\Sigma$ in the expression above 
depends on two momentum variables as well.
This difference is due to the fact that in the NJL model the selfenergy in mean-field 
approximation is local, whereas in QCD, where 
interactions are mediated by gluon exchange, it is not. 
As a consequence, the selfenergy in QCD is already momentum dependent in 
homogeneous matter, cf.~\Eq{eq:Shom}, whereas it is constant in the NJL model.
A comparison of these cases (local vs.\@ nonlocal, homogeneous vs.\@
inhomogeneous) is presented in Table~\ref{tab:selfenergy}.

\begin{table}[bt]
  \centering
  \begin{tabular}{l l l}
    \toprule
     & coordinate space  & momentum space   \\
    \midrule
    &&\\[-4mm]
    nonlocal  & $\Sigma(x,x')$ & 
    $ \Sigma(p,p') = \int\limits_{V_4}d^4x\int\limits_{V_4}d^4x'\, e^{i(p\cdot x - p'\cdot x')}\,\Sigma(x,x')$  
    \\[5mm]
    nonlocal homogeneous & $\Sigma(x,x') = \Sigma_{hom}(x-x')$ &
    $\Sigma(p,p') = \Sigma_{hom}(p)\,V_4 \delta_{p,p'}$,
    \\[5mm]
    &&
    $\Sigma_{hom}(p) \equiv  \int\limits_{V_4}d^4x\, e^{ip\cdot x}\, \Sigma_{hom}(x)$
    \\[5mm]
    local  & $\Sigma(x,x') = M(x) \delta(x-x')$ &
    $ \Sigma(p,p') = \int\limits_{V_4}d^4x \,e^{i(p-p')\cdot x}\, M(x) \equiv V_4 M_{p-p'}$
    \\[5mm]
    local homogeneous & $\Sigma(x,x') = M \delta(x-x')$ &
    $ \Sigma(p,p') = M\,V_4 \delta_{p,p'}$ \quad $\Rightarrow$ \quad $\Sigma_{hom} = M = \mathit{const.}$
    \\
    \bottomrule
  \end{tabular}
  \caption{Comparison of nonlocal (QCD) and local (NJL) selfenergies in inhomogeneous
  and homogeneous matter. 
  The quantities $M_{p-p'}$ in the local inhomogeneous case correspond to the Fourier 
  coefficients of the mass function defined earlier in \Eq{eq:Mxq}. 
  }
  \label{tab:selfenergy}
\end{table}

\begin{figure}[htb]
	\centering
		\includegraphics[scale=0.9]{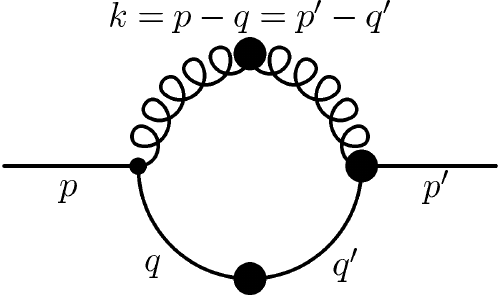}	
\caption{Quark-selfenergy in momentum space. 
The plain lines represents the quark propagator, the curly line the gluon propagator. 
Thick dots represent dressed quantities.
In inhomogeneous matter the momentum of the dressed quark is not conserved,
whereas momentum conservation is assumed for the gluon propagator and at the 
vertices. From~\cite{DanielM:PhD}.
}
\label{feyn_qdse}
\end{figure}

Explicitly, the quark selfenergy is given by (see \Fig{feyn_qdse})
\beq
\label{eq:qsig}
Z_2\Sigma(p,p') = 
g^2 \frac{1}{V_4}\sum\limits_{q}\Gamma_\mu^{a,0}\ S(q,q+p'-p) D_{\mu\nu}(p-q)
\Gamma^a_\nu(q+p'-p,p'),
\eeq
where a finite quantization volume was introduced, so that all momenta are discrete.\footnote{
Similar as in Sec.~\ref{sec:general}, this was done in Ref.~\cite{Muller:2013tya} for conceptional 
reasons, e.g., when inverting matrices in momentum space.
Here we keep this notation mainly for notational convenience.
In the end the infinite-volume limit will be taken.}
Besides the dressed quark propagator, which makes the simultaneous solution of Eqs.~(\ref{eq:qdse})
and (\ref{eq:qsig}) a selfconsistency problem, 
the selfenergy depends on the dressed gluon propagator $D_{\mu\nu}$ and
the bare and dressed quark-gluon vertices, $g\Gamma_\mu^{a,0}$ and 
$g\Gamma^a_\nu$, respectively, where $g$ is the QCD coupling constant.
At this point M\"uller et al.\@ 
truncated the system by using a semi-phenomenological model input,
instead of calculating $D_{\mu\nu}$ and $\Gamma^a_\nu$ from higher-order DSEs.
The gluon propagator is based on a parametrization of lattice data for the Yang-Mills 
system~\cite{Fischer:2010fx}
which was then corrected for quark effects by perturbatively adding a polarization loop in 
hard-thermal-loop--hard-dense-loop approximation~\cite{LeBellac}.
In this way screening effects at high densities or temperatures have been included, 
but not in a selfconsistent way. 
In particular, possible modifications with respect to the homogeneous case have been 
neglected for both, the gluon propagator and the vertex.

The most difficult problem is now to generalize the structure 
of the quark propagator in a homogeneous ground state, \eq{eq:Shom},
in such a way that it describes a quark in an inhomogeneous background 
and still allows to find a selfconsistent solution of the DSE.
To this end, M\"uller et al.\@ considered a CDW-like modulation,
i.e., scalar and pseudoscalar condensates of the form
\beq
\label{eq:cond}
\langle \bar \psi \psi \rangle \propto \cos (Qz)\,,\quad
\langle \bar \psi i\gamma^5 \tau^3 \psi \rangle \propto \sin (Qz)\,.
\eeq
Starting from the definition $S(x,x') = Z_2^{-1}\langle{\cal T} (\psi(x) \bar \psi(x'))\rangle$
of the Euclidean propagator, where ${\cal T}$ denotes the imaginary time ordering operator,
the condensates are related to the propagator as\footnote{
The generalization to nonlocal condensates, e.g., 
$\langle \bar \psi(x) \psi(x') \rangle$, which was discussed in the weak-coupling regime by
Deryagin, Grigoriev and Rubakov~\cite{DGR}, is straightforward but has not been done in Ref.~\cite{Muller:2013tya}.
}
\beq
\langle \bar \psi(x) \,{\cal O}\, \psi(x) \rangle = 
-Z_2 \frac{1}{V_4^2}\sum\limits_{p,p'}e^{i (p-p')\cdot x} ~ \mathrm{tr}\left[ {\cal O} S(p, p')\right ],
\eeq
with the trace in Dirac, flavor and color space.
Comparing this with \eq{eq:cond}, one finds that the desired spatial behavior is obtained if
\beq
\mathrm{tr}\left[ (\one\pm\gamma^5\tau^3) S(p, p')\right ]  \propto \delta_{p,p'\mp Q}\,,
 \eeq
 with  the wave vector $Q \equiv Q e_3$ being a 4-vector of length $Q$, pointing to the 
 3-direction.
This suggests to generalize the dressing function $B$, which is the only 
chiral-symmetry breaking term in \eq{eq:Shom}, in a similar way.
M\"uller et al.\@ therefore made the ansatz that the inverse propagator $S^{-1}(p,p')$ contains a 
term~\cite{Muller:2013tya} 
\beq
\label{eq:binh}
B(p,p') = \frac{B(p) + B(p')}{2} \sum\limits_{s=\pm}
                \frac{\one+s\gamma^5\tau^3}{2} \,V_4\delta_{p,p'-sQ}\,,
\eeq
where $B(p)$ is closely related to the $B$ function of the homogeneous case.
In fact,  when we take $Q=0$, the matrix $B(p,p')$ becomes purely scalar and
diagonal in momentum space, with the diagonal matrix elements essentially given by 
$B(p)$.
On the other hand, one can recover the structure of the CDW mass function in the 
NJL model, if the dressing functions $B(p)$ are replaced by a constant $\Delta$.

In order to determine $B(p)$ through the DSE,
the inverse propagator with the dressing function \eq{eq:binh} must be 
inverted and inserted into \eq{eq:qsig}. It turns out that this induces further structures,
and additional dressing functions are needed to achieve a self-consistent solution. 
For instance, since the wave vector defines a preferred direction, 
the $A$ function in \eq{eq:Shom} must be replaced by two independent functions, 
corresponding to the momentum component in $Q$ direction and to the perpendicular
part.
 As in the quarkyonic chiral spirals, there are also terms proportional to $\gamma^0\gamma^3$, 
 but their amplitudes turned out to be rather small. 
 
Altogether, the complete selfconsistent ansatz contains 10 dressing functions, 
which are listed in Ref.~\cite{Muller:2013tya}. 
However, despite these tremendous complications in the Dirac and flavor structure,
the most important feature is that the inverse propagator in momentum space
still has a block structure which is simple enough that it can be inverted analytically.
As a consequence, selfconsistent solutions of the DSE can be found iteratively, starting from
some trial ansatz for the dressing functions.

The last missing step is then to determine the most favored solution at given $T$ and $\mu$.
In particular the value of $Q$ is not fixed by the DSE.
We therefore have to maximize the pressure, 
which corresponds to the effective action $\Gamma$ in CJT formalism~\cite{Cornwall:1974vz},
and is
a functional of the dressed propagator.
Although the numerical evaluation of $\Gamma$ is always demanding, the stationarity condition
$\frac{d\Gamma}{dQ} = 0$ yields an additional gap equation from which $Q$ can be 
determined~\cite{Muller:2013tya}.
\begin{figure}[H]
\centering
\includegraphics[scale=1.0]{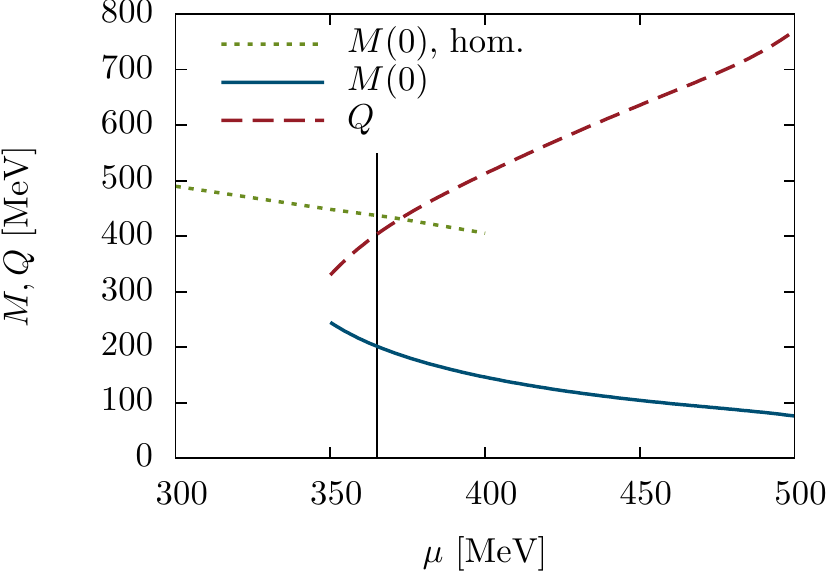}
\quad
\includegraphics[scale=1.0]{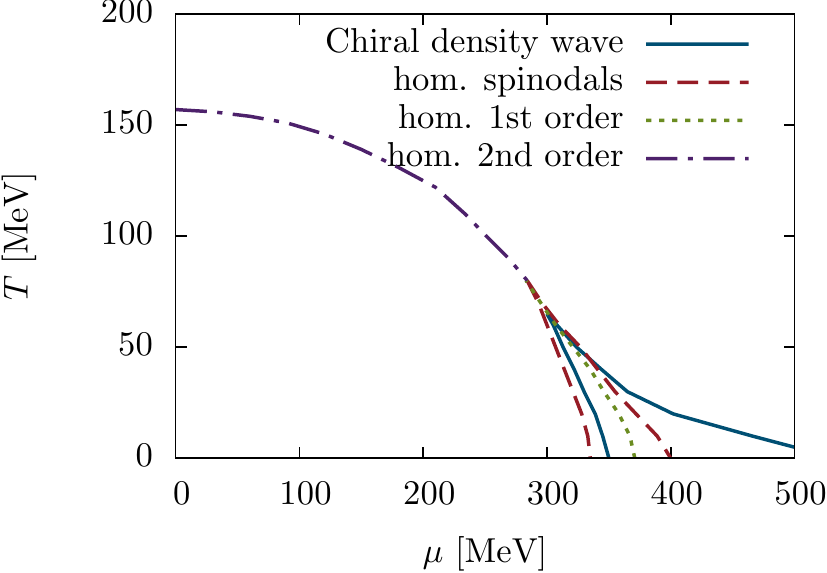}
\caption{Left: 
Mass amplitude and wave number as functions of $\mu$ 
at $T=0$. The position of the first-order phase transition is indicated
by the thin vertical line.
Right: Phase diagram in the $\mu-T$ plane. 
The blue solid line indicates the boundaries of the region where the inhomogeneous solution exists.
From Ref.~\cite{Muller:2013tya}.}
\label{fig:DSEresults}
\end{figure}

Numerical results from Ref.~\cite{Muller:2013tya} are shown in \Fig{fig:DSEresults}.
In the left panel the wave number $Q$ and
the combination $M(0) \equiv (B(p)/C(p))|_{\p=\vec 0,n=0}$, 
which can be interpreted as a mass amplitude, 
are plotted as functions of $\mu$ at $T=0$.
At low chemical potential, there is only a homogeneous solution, 
whereas above $\mu \approx 350$~MeV,  also an inhomogeneous 
solution exists, which becomes favored above $\mu \approx 365$~MeV
where a first-order phase transition takes place.
While the general behavior is quite similar to the NJL model, cf.~\Fig{fig:njlcdw},
an apparent difference is that in the inhomogeneous phase the amplitude 
decreases with decreasing slope, suggesting that this phase survives 
up to arbitrarily high chemical potentials. 
In fact, within numerical accuracy, the chirally restored phase was always
found to be disfavored against the inhomogeneous phase at $T=0$. 
On the other hand, color superconducting phases, which should eventually become favored, 
have been neglected in the analysis.

The phase diagram in the $\mu$-$T$ plane is displayed in the right panel of 
\Fig{fig:DSEresults}. The blue solid lines indicate the limits of the region where 
inhomogeneous solutions are found. 
The right one also corresponds to the second-order phase boundary to the 
restored phase, while the phase transition to the homogeneous chirally broken
phase is first order and takes place somewhere between the left solid 
and the dotted line. 
The latter indicates the first-order phase boundary between the homogeneous 
phases for the case when the inhomogeneous solution is ignored. 
Hence, within numerical accuracy, this phase boundary is again completely 
covered by the inhomogeneous phase.
Moreover, as in the NJL model, the Lifshitz point seems to coincide with the critical 
point. 
Of course, the numerical determination of these points is quite difficult,
and it would be interesting to investigate this question within a GL approach to QCD.

\section{Conclusions and Outlook}
\label{sec:conclusions}

The study of finite-density QCD remains one of the biggest theoretical
and experimental challenges in contemporary nuclear physics. 
In this article we have discussed the possibility of inhomogeneous chiral-symmetry
breaking,
reviewing results obtained
in Ginzburg-Landau theory,
within effective models, 
and directly from QCD,
employing large-$N_c$ expansions or Dyson-Schwinger techniques.
All these approaches find strong indications that nonuniform chiral condensates
might indeed exist, 
possibly covering the first-order chiral phase boundary in the standard picture
of the QCD phase diagram, and in particular the critical endpoint.
Although not discussed in this review,
complementary support also comes from holographic models,
where the favored state of cold nuclear matter in the strong-coupling limit 
is inhomogeneous~\cite{Rozali:2007}.
Moreover, the inhomogeneous phase turned out to be quite robust to 
variations and extensions of the employed models,
most notably vector interactions, magnetic fields, the inclusion
of strange quarks and the coupling to the Polyakov loop,
which have all been found to enhance the size of the inhomogeneous window.

Although the question about the most favored shape of the condensate
modulations is still open, a Ginzburg-Landau analysis as well as
explicit calculations with various trial functions seem to favor
one-dimensional real modulations, in particular in the vicinity of the 
Lifshitz point.
In this context we should recall that this is the result of a mean-field 
analysis, and might thus change if fluctuations are taken into account. 
Indeed, it is known that one-dimensional crystalline structures are washed out
at finite temperature by thermal fluctuations and are therefore,
strictly speaking, unstable \cite{Landau:1969}. 
On the other hand, this statement refers
to the fact that the modulation is not rigid, but fluctuates locally,
as in a liquid crystal. 
Quasi-one dimensional correlations could therefore still survive over 
distances much bigger than the length scale of interest, e.g., the size of the
fireball in heavy-ion collisions (which is small but the temperature is high)
or the radius of a compact star (which is large but the temperature is low)~\cite{Baym:1982}.

It would nevertheless be of extreme interest
to study inhomogeneous phases beyond the mean-field approximation.
Besides the dimensionality of the favored modulation, fluctuations can also change the
order of the phase transition. 
For instance, for a superconductor in a magnetic field it was found 
that fluctuations of the chemical potential remove the second-order phase transition 
between the inhomogeneous superconducting and the normal conducting phase~\cite{Ohashi:2006},
thus turning it into first order if the inhomogeneous phase is stable at all.
Similar effects may also occur at the phase boundary between the inhomogeneous 
and the restored phase.

Unfortunately,
the extension of the formalism beyond mean field is a highly non-trivial task.
One possibility is the expansion of the thermodynamic potential around the mean-field
solutions to include the fluctuating fields up to quadratic order. 
This is known as the ``Nozi\`eres--Schmitt-Rink approach'' in the context of
homogeneous
superconductors~\cite{Nozieres:1985zz} and was extended to inhomogeneous 
(Fulde-Ferrell-type) superconductors in Ref.~\cite{Ohashi:2006}.
Although in principle straightforward, the method becomes very involved in our case, since it requires the 
evaluation of the quasiparticle-quasihole polarization function, which depends on the dressed
quark propagator in the inhomogeneous phase.
An alternative and very efficient approach to include fluctuations in quantum field theory 
is the functional renormalization group (FRG)~\cite{Pawlowski:2005xe,Gies:2006wv}. 
For homogeneous phases, this method has successfully been applied to the QM model~\cite{Schaefer:2006sr},
but its extension to inhomogeneous phases has yet to be developed.

NJL-model studies find higher-dimensional modulations favored at higher densities, 
which is consistent with large-$N_c$ arguments.
We should keep in mind, however, that this is the realm of color-superconductivity 
which was not taken into account in both cases. 
Focusing on dense matter under compact-star conditions requires additionally 
the consideration of a non-vanishing isospin chemical potential,
as well as strange quarks. 
While single aspects of this list have already been studied in the past, a complete
calculation, taking all these aspects into account simultaneously is still lacking,
even in simplified models. 

At the lower-density end, on the other hand, confinement aspects should become
relevant, which are inherently contained in the quarkyonic-matter picture but not,
e.g., in the NJL and QM models. 
Here it would be extremely interesting to see whether nucleons and 
nuclear matter can emerge as three-dimensional inhomogeneities when going 
from quark matter to lower densities~\cite{Buballa:2012vm}.
In this context it would certainly be rewarding to reconsider the chiral soliton models 
of the nucleon~\cite{Alkofer:1994ph,Christov:1995vm,Ripka}
and work out their relation to the domain-wall solitons in the inhomogeneous phases. 

Of course lattice QCD could in
principle have the last word on all these questions. Unfortunately,
being a density-driven effect, 
the phenomenon of inhomogeneous chiral symmetry breaking is extremely challenging 
to access in this way.
The first study of this kind, performed by Bringoltz in Ref. \cite{Bringoltz:2006pz},
relied on a simulation in the strong-coupling limit, employing an action expressed 
in terms of hadronic degrees of freedom.
 Due to artifacts related to the implementation of the fermions 
on the lattice, leading to very large baryon masses however, inhomogeneous chiral condensation
was found never to be favored in 3+1 dimensions. 
A more recent analysis by Fukuda et al.\@ \cite{Fukuda:2013ada}
showed that while in 1+1 dimensions the sign problem should not arise,
particular care in the implementation of the quarks is required 
in order to avoid the appearance of spurious lattice artifacts
when studying inhomogeneous chiral condensation. 
The authors of Ref. \cite{Fukuda:2013ada} nevertheless developed a promising procedure
for implementing inhomogeneous solutions in the fermion determinant
which could in principle be employed within higher-dimensional calculations as well.

Another important task is to work out possible empirical
signatures related to 
inhomogeneous chiral symmetry breaking. 
As discussed above, the critical properties 
of the chiral phase transition are significantly altered by the appearance of
inhomogeneous phases, and quantities like susceptibilities 
are consequently strongly modified as well. Such effects
could thus lead to novel experimental signatures and must therefore be
taken into account when trying to access the phase structure of QCD 
at finite densities in heavy-ion collision experiments.
On the other hand, the finite size and the finite lifetime of the fireball may wash out
many effects one finds in infinite static systems.
In this context it would be interesting to extend the present studies to finite volumes,
and keep the time dependence of the condensates as well.

Crystalline condensates could also be relevant for the phenomenology of 
compact stars.
The typical energy gains found in model calculations are, however,
too small to have a significant impact on the equation of state and,
hence, the mass-radius relation. 
It is therefore certainly more promising to study the possible effects of 
inhomogeneous chiral condensates on cooling and transport properties 
in compact stars. 
As a first step in this direction, the Goldstone modes of the various inhomogeneous
phases should be worked out.
Indeed,
since the condensates break chiral symmetry and the translational
invariance simultaneously, this is an interesting problem by itself.

\section*{Acknowledgments}

We are very grateful to   
W.~Broniowski,
K.~Fukushima,
T.~Kojo,
L.~McLerran,
D.~Nickel,
M.~Thies, 
and
J. Wambach
for numerous fruitful discussions and  their encouragement.
We also thank 
H.~Abuki,
T.~Brauner,
E.~Ferrer,
B.~Friman,
H.~Gies,
A.~Heinz,
Y.~Hidaka,
E.-M.~Ilgenfritz,
V.~de la Incera,
D.~M\"uller,
D.~Nowakowski,
J.~Pawlowski,
D.~Rischke,
M.~Schramm,
B.-J.~Schaefer,
L.~von Smekal,
T.~Tatsumi,
as well as our referees for helpful comments and suggestions.

\end{document}